\documentclass[12pt,twoside,longbibliography,nofootinbib,preprintnumbers]{puthesis}

\title{\LARGE{\lettrine[lines=1]{E}{xtragalactic} Searches} \\ \Large{for} \\ \LARGE{Dark Matter Annihilation}}

\submitted{September 2018}  
\copyrightyear{2018}  
\author{{Siddharth Mishra-Sharma}}
\adviser{Mariangela Lisanti}  
\department{Physics}

\usepackage{verbatim}
\usepackage{longtable}

\ifdefined\printmode

\usepackage{url}

\else

\usepackage{hyperref}
\usepackage[dvipsnames]{xcolor}
\usepackage{cleveref}

\newcommand\myshade{85}
\colorlet{mylinkcolor}{violet}
\colorlet{mycitecolor}{YellowOrange}
\colorlet{myurlcolor}{Aquamarine}

\hypersetup{colorlinks,bookmarksnumbered,
  linkcolor  = mylinkcolor!\myshade!black,
  citecolor  = mycitecolor!\myshade!black,
  urlcolor   = myurlcolor!\myshade!black,
}

\fi

\usepackage{Zallman,lettrine}

\PassOptionsToPackage{svgnames}{xcolor}
\usepackage[object=vectorian]{pgfornament}
\usepackage{lipsum,tikz}

\newcommand{\sectionline}{
  \noindent
  \begin{center}
  {
    \resizebox{0.5\linewidth}{1ex}
    {{%
    {\begin{tikzpicture}
    \node  (C) at (0,0) {};
    \node (D) at (9,0) {};
    \path (C) to [ornament=88] (D);
    \end{tikzpicture}}}}}%
    \end{center}
  }

\usepackage{ae,aecompl}
\usepackage{graphicx}   
\usepackage{amsmath}    
\usepackage{graphicx}
\usepackage{dcolumn}
\usepackage{bm}
\usepackage{graphics}
\usepackage{afterpage}
\usepackage{float}
\usepackage{subfigure}
\usepackage{rotating}
\usepackage{multirow}
\usepackage{fancyheadings}
\usepackage{theorem}
\usepackage{moreverb}
\usepackage{euscript}
\usepackage{psfrag}
\usepackage{slashed}
\usepackage{mathtools}

\usepackage{listings}
\usepackage{color,xcolor}

\definecolor{deepblue}{rgb}{0,0,0.9}
\definecolor{deepred}{rgb}{0.85,0,0}
\definecolor{deepgreen}{rgb}{0,0.95,0}
\definecolor{mygray}{gray}{0.97}

\lstdefinestyle{python}{
  belowcaptionskip=1\baselineskip,
  breaklines=true,
  frame=L,
  xleftmargin=\parindent,
  language=Python,
  showstringspaces=false,
  basicstyle=\small\ttfamily,
  morekeywords={models, lambda, forms,True,False,None},
  keywordstyle=\bfseries\color{deepgreen!40!black},
  commentstyle=\itshape\color{gray},
  identifierstyle=\color{black},
  stringstyle=\color{deepred},
  rulecolor=\color{gray},
  backgroundcolor=\color{mygray},
}

\newcommand{\es}[2] {\begin{equation} \label{#1} \begin{split} #2 \end{split} \end{equation}}

\usepackage{amssymb,amsmath,latexsym,graphics, graphicx,epsfig,epstopdf,multirow,comment,hyperref,appendix,slashed,xcolor,afterpage, makecell} 
\usepackage{booktabs}
\usepackage{tabularx}

\newcommand{\abs}[1]{\left\lvert #1 \right\rvert}
\newcommand{\overbar}[1]{\mkern 2mu\overline{\mkern-2mu#1\mkern-2mu}\mkern 2mu}

\newcommand {\be} {\begin {equation}}
\newcommand {\ee} {\end {equation}} 

\newcommand {\bes} {\begin {equation*}}
\newcommand {\ees} {\end {equation*}}

\usepackage{csvsimple}

\newcommand\Tstrut{\rule{0pt}{2.6ex}}         
\newcommand\Bstrut{\rule[-0.9ex]{0pt}{0pt}}   

\newcommand{\beq}{\begin{equation}}
\newcommand{\eeq}{\end{equation}}

\usepackage[perpage]{footmisc}

\usepackage{booktabs}
\newcolumntype{C}[1]{>{\centering\let\newline\\\arraybackslash\hspace{0pt}}m{#1}}

\usepackage{inconsolata}
\usepackage{aas_macros}

\makeatletter
\hypersetup{pdftitle=Extragalactic Searches for Dark Matter Annihilation,
        pdfauthor=\@author}
\makeatother

\ifodd 0

\renewcommand{\maketitlepage}{}
\renewcommand*{\makecopyrightpage}{}
\renewcommand*{\makeabstract}{}

\else

\abstract{\lettrine[lines=3]{W}{e} are at the dawn of a data-driven era in astrophysics and cosmology. A large number of ongoing and forthcoming experiments combined with an increasingly open approach to data availability offer great potential in unlocking some of the deepest mysteries of the Universe. Among these is understanding the nature of dark matter (DM)---one of the major unsolved problems in particle physics. Characterizing DM through its astrophysical signatures will require a robust understanding of its distribution in the sky and the use of novel statistical methods. 

The first part of this thesis describes the implementation of a novel statistical technique which leverages the ``clumpiness'' of photons originating from point sources (PSs) to derive the properties of PS populations hidden in astrophysical datasets. This is applied to data from the \emph{Fermi} satellite at high latitudes ($|b|\geq 30^\circ$) to characterize the contribution of PSs of extragalactic origin. We find that the majority of extragalactic gamma-ray emission can be ascribed to unresolved PSs having properties consistent with known sources such as active galactic nuclei. This leaves considerably less room for significant dark matter contribution.

The second part of this thesis poses the question: ``what is the best way to look for annihilating dark matter in extragalactic sources?'' and attempts to answer it by constructing a pipeline to robustly map out the distribution of dark matter outside the Milky Way using galaxy group catalogs. This framework is then applied to \emph{Fermi} data and existing group catalogs to search for annihilating dark matter in extragalactic galaxies and clusters.}

\fi

\begin{document}

\makefrontmatter

\chapter{Introduction}
\label{ch:intro}

\lettrine[lines=3]{T}{he} nature of dark matter (DM) remains one of the major unsolved problems in physics. Originally inferred through its gravitational influence on galaxies and clusters, a rich body of evidence has accumulated over the last four decades firmly establishing its existence. All of the evidence, however, comes from inferring dark matter's presence solely through its gravitational effects. Many open questions remain: Does dark matter consist  of a fundamental particle? If so, what is its mass? Could there be an entire dark sector, akin to the Standard Model (SM)? How does dark matter interact with the SM? The quest to answer these questions drives a huge collective effort that draws from a rich body of theoretical and experimental work, as well as major input from computational and numerical studies. We are currently at the dawn of a data-driven era in astrophysics and cosmology---a large number of ongoing and forthcoming experiments, both in the lab and in the sky, combined with an increasingly open approach to data availability, offer great potential in elucidating the nature of dark matter. 

Dark matter plays a central role in many subfields of particle physics, astrophysics and cosmology. Understanding its nature and interactions would have far reaching consequences in those fields by providing major insights into fundamental physics beyond the Standard Model as well as elucidating the evolution of our Universe and the formation of structures within it. 


This introduction is organized as follows. In Sec.~\ref{sec:evidence}, I will summarize the large body of evidence pointing to the existence of dark matter, occasionally touching upon relevant historical developments. In Sec.~\ref{sec:particledm}, I will describe possible explanations for the particle nature of dark matter and various detection schemes, focusing on DM thermally produced in the early Universe and specifically Weakly Interacting Massive Particles (WIMPs). Section~\ref{sec:astrodm} will focus on the effort to detect and characterize WIMPs through their astrophysical signatures, in particular using gamma-ray data. I will briefly summarize the theoretical and experimental tools available to us in these searches. Finally, in Sec.~\ref{sec:summary}, I will describe the organization of the rest of this thesis. This chapter partially draws from a number of excellent review articles on the topic which the reader is referred to for further details. Refs.~\cite{Lisanti:2016jxe,Plehn:2017fdg} provide recent, comprehensive reviews of dark matter physics. Ref.~\cite{Slatyer:2017sev} reviews indirect detection, which will be the main focus of this thesis. Finally, Ref.~\cite{Bertone:2016nfn} provides a thorough overview of the history of the field.


\section{Evidence for Dark Matter}
\label{sec:evidence}

Although the study of dark matter had its inception and development in the 20th century, the interplay between theory and observation in making the unknown knowable goes back much earlier. For example, the Aristotelian view of an immutable Universe with the Earth at its center offered a clean framework that did not call for additional celestial objects, and was the orthodox viewpoint until Renaissance astronomers conclusively refuted it with observations. Galileo was able to leverage new technological developments and make observations that arguably played the largest role in this. After pioneering the development of the telescope, he was able to understand the make-up of the Milky Way as consisting of individual stars rather than diffuse clouds, observe Saturn's rings and discover Jupiter's four largest moons. These observations are very much in the spirit of modern dark matter searches---demonstrating that the Universe can contain invisible forms of matter, and that scientific inquiry and technological developments can play a big role in revealing them to us.


Evidence for some yet-unknown form of matter started piling up in the early 19th century. 
In 1922, Dutch astronomer Jacobus Kapteyn wrote down for the first time a predictive model for the distribution of matter in the Milky Way, describing the stars as particles in a virialized system~\cite{1922ApJ....55..302K} and using this model to obtain the local matter density in terms of the observed stellar mass. Kapteyn's student Jan Oort~\cite{1932BAN.....6..249O} and others~\cite{1922MNRAS..82..122J} were able to derive estimates for the local matter density, in some cases seeing excesses above the observed luminous mass. Astronomers during this time reckoned with the existence of missing matter in the Universe, in some cases explicitly using the term \emph{dark matter}~\cite{1922ApJ....55..302K} and positing that it could potentially be accounted for by the extrapolation of the stellar luminosity function down to very faint stars~\cite{1932BAN.....6..249O}.

In 1933, Swiss-American astronomer Fritz Zwicky studied redshift data for galaxy clusters collected by Hubble and Humason~\cite{1931ApJ....74...43H}, using estimates of the velocity dispersions in eight galaxies within the Coma cluster to estimate its mass through the virial theorem~\cite{1933AcHPh...6..110Z}. Zwicky obtained a theoretical prediction for the dispersion by using the number of observed galaxies, average mass of a galaxy and its extent, finding a value of $\sim$80 km\,s$^{-1}$. This was in stark conflict with the observed line-of-sight velocity dispersion of $\sim$1000 km\,s$^{-1}$. Although Zwicky's work made use of an estimate of the Hubble constant that was a factor of $\sim$8 too big compared to the current accepted value, the large discrepancy between the observed and expected values pointed to the existence of unaccounted-for matter in the Coma system. Zwicky himself concluded that ``If this would be confirmed, we would get the surprising result that dark matter is present in much greater amount than luminous matter.'' An analysis of the Virgo cluster by Sinclair Smith in 1936 again pointed to a very high mass-to-light ratio in that system. In either case, the astronomers put forward potential explanations in terms of diffuse clouds of internebular material~\cite{1937ApJ....86..217Z}.

\begin{figure}[htbp] 
\hspace{-0.9 cm} 
\includegraphics[width=0.5185\textwidth]{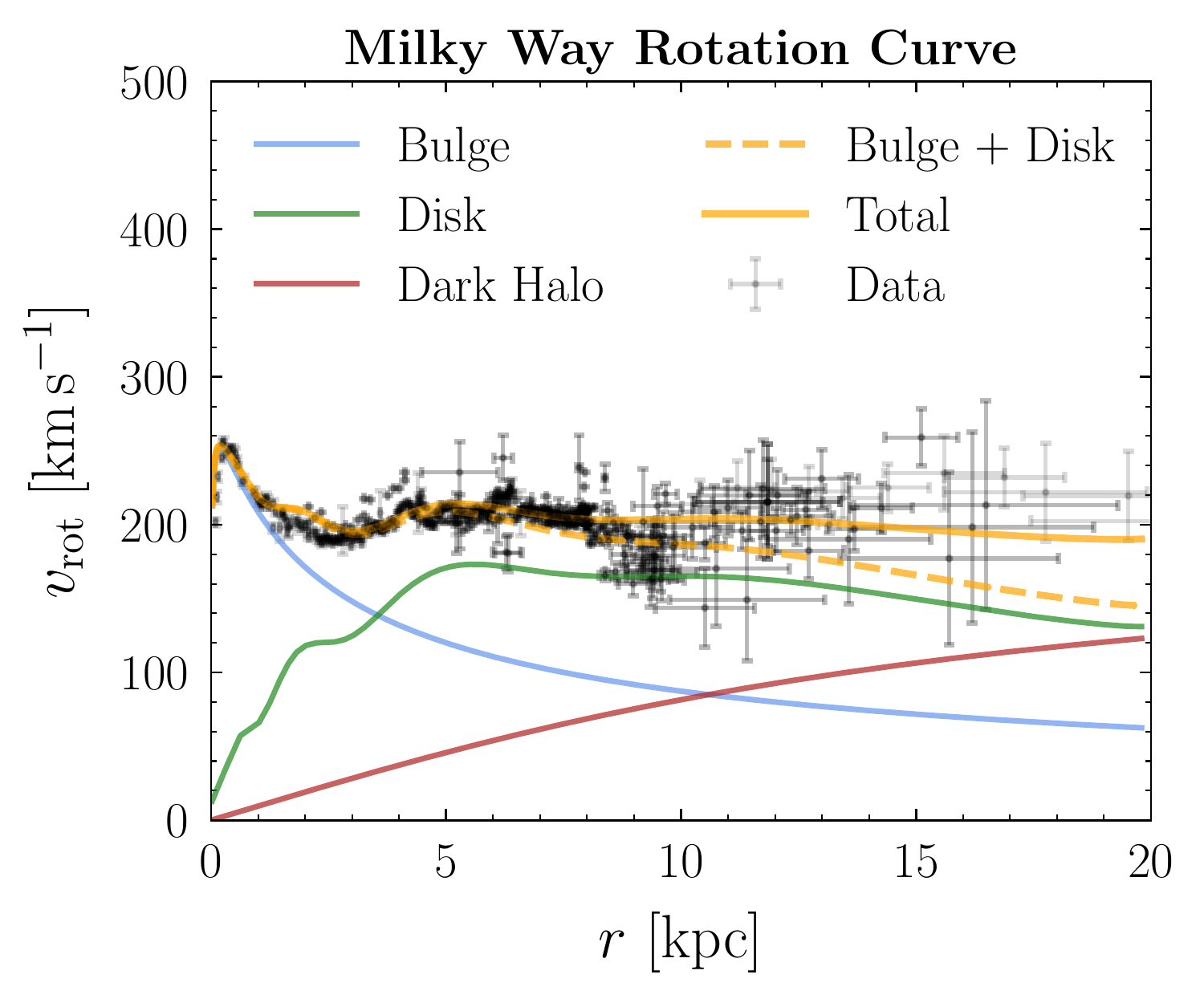}
 \includegraphics[width=0.528\textwidth]{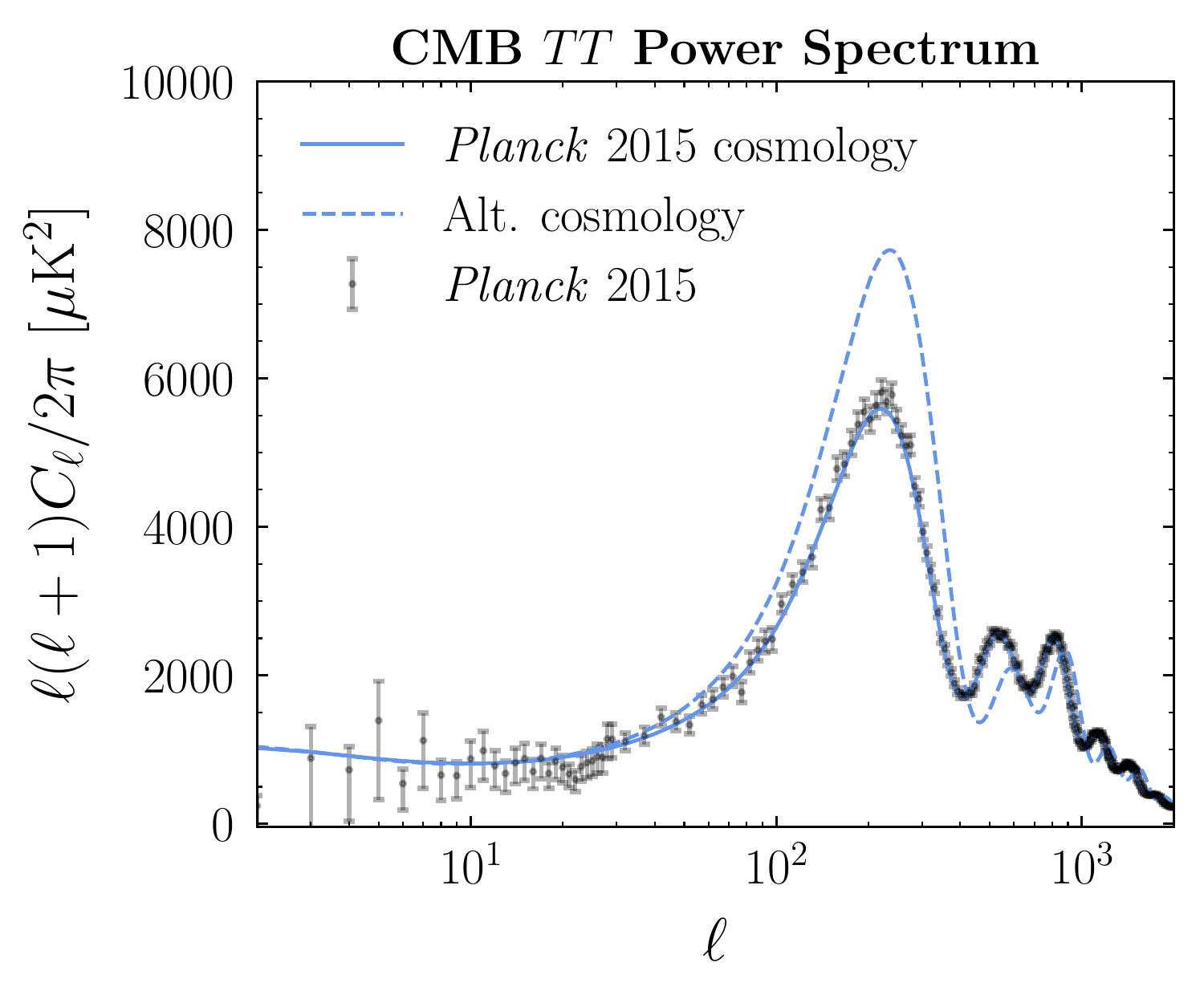}  
\caption{\textbf{(Left)} The measured rotation curves for the Milky Way compiled in Ref.~\cite{2009PASJ...61..227S}, and theoretical expectations from bulge- and disk-like components (blue and green lines, respectively) inferred from baryonic matter~\cite{2009PASJ...61..227S}, as well as an additional dark matter component from a spherical, isothermal halo (red line). The rotation curve for the baryonic-only component (disk + bulge) is shown as the dashed yellow line, and the total rotation curve including the dark halo is shown as the solid yellow line. The dark halo component is required to match the observed data at larger radii $r \gtrsim 15$ kpc. \textbf{(Right)} The \emph{Planck} $TT$ spectrum~\cite{Ade:2015xua} along with the best-fit theoretical predictions (solid blue line, computed with \texttt{CAMB}~\cite{Lewis:1999bs}), as well as predictions for a slightly altered cosmology with $\sim$10\% less non-baryonic (dark) matter (dashed blue line) where  striking differences from the observed spectrum can be seen.}  
\label{fig:evidence}
\end{figure}

Although this presented a conundrum, there was widespread consensus within the astronomical community that more information would be needed to understand what was going on. Historically, velocity rotation curves---the circular velocity profiles of stars in a galaxy as a function of the distance from the galactic center---did the most to convince the scientific community of the existence of large amounts of non-luminous matter in galaxies. The basic idea here is as follows. Standard Newtonian theory dictates that the circular velocity of stars is given by $v_c(r) = \sqrt{GM(r)/r}$, where $r$ is the radial distance, $M(r)$ the mass enclosed within radius $r$ and $G$ the universal gravitational constant. In the region beyond the galactic disk (which defines the observed extent of a given galaxy), we expect the enclosed mass to be constant, and consequently the circular velocity to fall as $v_c \propto r^{-1/2}$. Measurements started in the late 1930s with Babcock's observations of the rotation curve of M31 (Andromeda) out to about 20 kpc from its center~\cite{1939LicOB..19...41B}. Technological advancements over the next few decades enabled more accurate measurements. 
In the 1970s, Kent Ford, Vera Rubin and others observed in galaxies such as M31 and M33 as well as the Milky Way the approximate flattening of rotation curves at distances extending well beyond the baryonic disk~\cite{1970ApJ...159..379R,1973A&A....26..483R}. The implications of these observations for the missing mass problem were realized soon after~\cite{1974Natur.250..309E,1974ApJ...193L...1O}. Flat rotation curves indicated that the mass contained in a galaxy continues to increase as $M \propto r$ beyond the extent of the visible matter, in the form of unobserved ``dark'' matter whose density can be inferred to roughly scale as $\rho(r) \propto 1/r^2$.
The left panel of Fig.~\ref{fig:evidence} shows the measured rotation curves for the Milky Way compiled in Ref.~\cite{2009PASJ...61..227S} compared with theoretical expectations from bulge- and disk-like components (blue and green lines, respectively) inferred from baryonic matter, as well as an additional dark matter component from a spherical, isothermal dark matter halo (red line). The rotation curve for the baryonic-only component (disk + bulge) is shown as the dashed yellow line, and the total rotation curve including the dark halo is shown as the solid yellow line. It can clearly be seen that the additional dark halo component is required to match the observed data at larger radii $r \gtrsim 15$ kpc. The descriptions of the individual components shown are provided in Ref.~\cite{2009PASJ...61..227S}.

While astrophysical observations played a significant role historically in motivating the study of dark matter, modern cosmological data provides substantial evidence supporting its existence in our Universe. $\Lambda$CDM, a phenomenological framework often referred to as the standard model of cosmology, contains dark energy ($\Lambda$) and cold dark matter (CDM) as essential ingredients. It is able to account for a plethora of cosmological observations, including the existence and structure of the cosmic microwave background (CMB) radiation, large-scale distribution of matter, accelerating expansion of the Universe and relic elemental abundances~\cite{Dodelson:1282338,Kolb:1990vq}. In particular, the CMB, which is the imprint of photons that decoupled from the baryon-photon fluid in the Universe about 370,000 years ago and have been free-streaming ever since, provides irrefutable evidence for (non-baryonic) dark matter. The primary relevant observable is the angular scale of inhomogeneities in the temperature distribution (the $TT$ angular power spectrum) of the CMB. The power spectrum largely consists of a set of peaks, each indicating an angular scale with a particularly large contribution to the temperature fluctuations. The leading physical effect behind these are acoustic oscillations in the baryon-photon fluid during photon decoupling. Early on, photons and baryons were electromagnetically coupled, and non-baryonic dark matter was responsible for generating gravitational potential wells that could pull in the baryon-photon fluid. The photon pressure acting against these wells gave rise to a tower of acoustic modes, imprinted in the CMB as characteristic peaks. While the detailed physics is somewhat nuanced\footnote{See Wayne Hu's CMB tutorials for an excellent introduction: \url{http://background.uchicago.edu/index.html}.}, the relative heights of these peaks can provide information about the energy content of our Universe, including the relative composition of baryonic and non-baryonic (dark) matter. Very heuristically, the position of the first peak provides information about the curvature of the universe (and hence how much total ``stuff'' there is in it), while the second peak tells us how much of the matter is baryonic (ordinary matter). The third peak and its relative height can shed insights into the abundance of non-baryonic dark matter. Historically, the WMAP satellite, while not able to fully resolve the third peak, was already able to conclusively say that dark matter makes up the majority of the matter budget in the Universe, finding the baryon density $\Omega_b h^2=0.02264\pm0.00050$ and cold dark matter density $\Omega_c h^2=0.1138\pm0.0045$~\cite{2013ApJS..208...19H}. Since then, \emph{Planck} has been able to precisely measure eight peaks of the $TT$ spectrum, finding $\Omega_b h^2= 0.02225\pm0.00016$ and $\Omega_c h^2=0.1198\pm0.0015$ when additionally including the CMB $E$-mode polarization auto- and cross-spectra ($EE$ and $TE$). The right panel of Fig.~\ref{fig:evidence} shows the \emph{Planck} $TT$ spectrum~\cite{Ade:2015xua} along with the best-fit theoretical predictions (solid blue line), as well as predictions for a slightly altered cosmology $\Omega_b h^2= 0.042$ and $\Omega_c h^2=0.10$ with a reduced dark matter density (dashed blue line), where striking differences from the measured spectrum can be seen.

The above classes of observational evidence or the existence of DM are by no means exhaustive---many other observations over a large range of scales support the existence of dark matter, including observations of the distribution of galaxies on large scales~\cite{Alam:2016hwk}, weak~\cite{Abbott:2017wau} and strong lensing~\cite{2010MNRAS.408.1969V,2012Natur.481..341V} of background galaxies by foreground structure, and observations of merging clusters~\cite{Clowe:2006eq}.

\section{(Particle) Nature of Dark Matter}
\label{sec:particledm}

Although there exists a great deal of evidence for the existence of dark matter, its nature largely remains a mystery. These days, it is often implicitly assumed that when people are talking about detecting dark matter, say at a Xenon direct detection experiment or in gamma-ray data, they are referring to a dark matter \emph{particle}. As touched upon above, this has by no means always been the case---early usage and references to dark matter usually referred to the existence of generic dark objects that would be too faint to be observed, such as dim stars or internebular material~\cite{1937ApJ....86..217Z}. The transition in usage was a result of sociological changes within the particle physics and astrophysics communities, bringing the two closer after the missing mass problem had been firmly accepted in the 1970s. All evidence amassed since then is consistent with dark matter being a fundamental particle, or even the existence of an entire dark sector consisting of many particles with a rich set of properties and interactions. It should be noted however that there exist alternatives to particle dark matter that seek to explain the dynamical observations suggesting the existence of missing mass in the Universe. In particular, MOdified Newtonian Dynamics (MOND)~\cite{1983ApJ...270..365M,1983ApJ...270..371M,1983ApJ...270..384M} posits an alteration of Newtonian gravitation on larger scales and is successful in explaining the observed rotation curves as well as the empirical Tully-Fisher relation between the intrinsic luminosities and angular velocities of spiral galaxies~\cite{1977A&A....54..661T}. While having some observational success, MOND and related theories~\cite{Bekenstein:2004ne} are (arguably) less successful at explaining observations on cluster and cosmological scales. See the reviews in Refs.~\cite{2009CQGra..26n3001S,Famaey:2011kh} for further details.

Within the Standard Model, neutrinos---by virtue of being stable (or very long-lived), electrically neutral particles that do not interacting strongly---contain some of the essential attributes for a particle dark matter candidate, and were considered a promising DM candidate from early on. Cosmological effects of neutrinos were explored throughout the 1960s and 1970s, pioneered by the work of Zeldovich and others~\cite{Gershtein:1966gg,PhysRevLett.29.669}, and implications of massive neutrinos for the missing mass observed on (super-)galactic scales were discussed in the the late 1970s~\cite{PhysRevLett.39.165,1978ApJ...223.1015G}. Early simulations during the 1980s eventually showed that hot (relativistic) and cold (non-relativistic) particle dark matter would lead to very different outcomes for structure formation: in the former case leading to formation and collapse of larger structures (known as ``top-down'' structure formation), where in the latter case overdensities would seed larger structures, leading to hierarchical (known as ``bottom-up'') structure formation. Neutrinos, by virtue of being very light thermal relics, would be extremely relativistic during structure formation and, combined with these simulations, early surveys of the local Universe were able to quickly discount them as dark matter candidates~\cite{1983ApJ...274L...1W}. Nevertheless, neutrinos served as a gateway to understanding how potential new particles could affect observations on galactic, cluster and cosmological scales.

With no reason to be confined to the Standard Model, people turned to theories beyond the Standard Model that could explain DM. Supersymmetry (SUSY) posits that nature may contain a spacetime symmetry relating bosons and fermions, requiring that for every boson there must exist a fermion with the same quantum numbers (and vice versa)~\cite{Wess:1974tw,PhysRevLett.48.223}. This leads to the prediction of several new electrically neutral particles that are uncharged under the strong force. If some of these were stable, they could have played an important role in the history of our Universe and could conceivably make up (some portion of) the dark matter~\cite{Jungman:1995df}. Supersymmetry took its modern form in a paper by Dimopolous and Georgi, who introduced the Minimal Supersymmetric Standard Model (MSSM)~\cite{Dimopoulos:1981zb}. Here, superpartners of the $Z$ boson, photon and two Higgses mix to form four particles, known today as neutralinos. Neutralinos have arguably been the most-discussed (particle) dark matter candidate~\cite{Bertone:2004pz}, in part because supersymmetry---able to achieve gauge coupling unification and to solve the electroweak hierarchy problem---is motivated in its own right independent of the dark matter problem, and the existence of a viable DM candidate within SUSY is often seen as a desirable bonus.

Outside of SUSY, there is no shortage of viable particle DM candidates, including but not limited to axions~\cite{PhysRevLett.40.223,Peccei:1977hh}, sterile neutrinos~\cite{Abazajian:2001nj,Seljak:2006qw}, light (sub-GeV) dark matter~\cite{Feng:2008ya,Lin:2011gj} and fuzzy dark matter~\cite{Hui:2016ltb}. Such a wealth of possibilities exists in part because the most general observational constraints on the properties of particle DM are relatively mild. For example, the mass of the dominant DM component has only been constrained with $\sim70$ orders of magnitude. 
In particular, observations constrain $m_\text{boson} \gtrsim 10^{-22}$\,eV for bosonic dark matter~\cite{Zhang:2017chj} and $m_\text{fermion} \gtrsim 0.7$\,keV for fermionic dark matter~\cite{Horiuchi:2013noa}. This is obtained from observations of DM halos around dwarf galaxies, imposing the requirement for particles to occupy a minimum phase-space volume according to the uncertainty principle for bosons and the Pauli exclusion principle for fermions. An upper limit of $\sim$10$^{48}$\,GeV comes from searches for microlensing signatures of MACHOS (Massive Astrophysical Compact Halo Objects) in our Galaxy~\cite{Griest:2013aaa}.

\subsection{Thermal Dark Matter and WIMPs}

Assumptions about dark matter's role in the cosmological history of the Universe can further impose constraints on its particle properties. A specific scenario is that of thermal dark matter, where it is assumed that dark matter particles were in equilibrium with the thermal bath of matter and radiation in the early Universe. The cooling and expansion of the Universe reduced its density and consequently suppressed its interaction rates. DM fell out of chemical equilibrium (a process known as freeze-out) when the forward process in $\chi\,\chi\leftrightarrow\mathrm{SM}\,\mathrm{SM}$ (where $\chi$ is a DM particle) could no longer be maintained, establishing the DM relic density. The turning off of the elastic process $\chi\,\mathrm{SM}\rightarrow\chi\,\mathrm{SM}$, known as kinetic decoupling, set a scale after which the DM could free-stream (see~\cite{Bringmann:2006mu} for further details).

There are several general arguments that apply to dark matter particles in thermal equilibrium with the Standard Model in the early Universe. As already mentioned in the context of Standard Model neutrinos, thermal relics that are sufficiently relativistic at decoupling (corresponding to light particle masses) would strongly suppress structure formation at small scales~\cite{Bringmann:2006mu}, and the DM mass is accordingly constrained to be $\gtrsim 3.3$\,keV from measurements of the power spectrum in the non-linear regime~\cite{Viel:2013apy}. Unitarity arguments place an upper bound of $\lesssim340$\,TeV on the mass of a stable particle that was once in thermal equilibrium with the SM~\cite{Griest:1989wd}, although this is model-dependent and assumes that there are no states heavier than the DM. Additionally, a weak-scale self-annihilation cross section of $\langle\sigma v\rangle\sim 3\times10^{-26}$\,cm$^3$\,s$^{-1}$ and GeV--TeV particle masses can reproduce the observed DM density through thermal freeze-out in the early Universe (see Refs.~\cite{1991NuPhB.360..145G,Jungman:1995df,Lisanti:2016jxe} for further details). This fact holds for a large variety of electroweak-scale DM candidates, including those naturally arising from SUSY~\cite{Bertone:2004pz,Jungman:1995df}, and combined with the theoretical arguments for the existence of new physics at electroweak scales these particles---known as Weakly Interacting Massive Particles (WIMPs)---have been the dominant particle dark matter paradigm over the last three decades and have motivated an extensive search program. 

Searches for WIMPs are generally organized into three categories depending on the experimental detection paradigm. Direct detection experiments look for the energy deposited when dark matter particles recoil against nuclei through the process $\text{SM}\,\chi\rightarrow\text{SM}\,\chi$, where $\chi$ is a DM particle. While the flux of WIMPs through a terrestrial detector can be large, the expected deposited energies and interaction rates would be very small, requiring large amounts of target material and exquisite control over backgrounds~\cite{Lisanti:2016jxe}. Direct detection experiments have been able to set very strong limits on WIMP scenarios~\cite{Aprile:2018dbl,Agnes:2018ves} and have been able to exclude several attractive baseline models~\cite{Escudero:2016gzx}. The second class of searches involves production of WIMPs at particle colliders like the Large Hadron Collider (LHC) through the process $\mathrm{SM}\,\mathrm{SM}\rightarrow\chi\,\chi$, usually in association with additional visible particles emitted by initial or intermediate SM particles that can used to detect the event along with the missing energy characterizing the WIMP. Dedicated collider searches can also target specific scenarios, such as neutralino production~\cite{Patrignani:2016xqp}. See Ref.~\cite{Kahlhoefer:2017dnp} for a recent review of collider searches for dark matter.

The final strategy and the focus of this thesis is indirect detection, which looks for the annihilation of DM particles into SM particles through the process $\chi\,\chi\rightarrow\mathrm{SM}\,\mathrm{SM}$ by looking for its signature in astrophysical data. The nature of the SM particles depends on the specific DM model and interaction properties considered. The basic idea behind indirect detection is that annihilation processes will be taking place at higher rates in regions of the Universe that have more dark matter, leading to an excess in production of SM particles from those regions. These would then cascade onto photons, electrons, positrons, (anti)protons and neutrinos, some of which could eventually reach us and be detected with appropriate telescopes. 

It is worth noting that the WIMP scenario, while well-motivated, relies on several assumptions that can easily be relaxed~\cite{PhysRevD.43.3191}. The possibility of the DM relic density set by annihilations into heavier states (``Forbidden'' DM)~\cite{DAgnolo:2015ujb,PhysRevD.43.3191} or $3\rightarrow2$ annihilations of Strongly Interacting Massive Particles (SIMPs)~\cite{Hochberg:2014dra,Hochberg:2014kqa} are representative examples where relatively small modifications to the WIMP paradigm can lead to very different ranges of allowed masses and cross sections. See Refs.~\cite{Berlin:2016gtr,Berlin:2016vnh,Berlin:2017ftj,Bernal:2017mqb,Cline:2017tka,DAgnolo:2015nbz,DAgnolo:2017dbv,DEramo:2010keq,Dror:2016rxc,Farina:2016llk,Kopp:2016yji,Kuflik:2015isi,Pappadopulo:2016pkp,Pospelov:2007mp} for further examples of such scenarios.

\section{Indirect Detection of Annihilating Dark Matter}
\label{sec:astrodm}

As noted above, for thermal WIMP scenarios where the DM can self-annihilate, the late-time DM abundance is set by the coupling of the DM particle to the Standard Model. In this case the DM would have an electroweak-scale cross section around $\langle\sigma v\rangle\sim 3\times10^{-26}$\,cm$^3$\,s$^{-1}$ and a particle mass of $m_\chi\sim\mathcal O($GeV--TeV). When DM particles in this mass range annihilate to SM particles, the resulting photons fall dominantly in the gamma-ray energy range. This regime is well-probed by gamma-ray telescopes, including the \emph{Fermi} Large Area Telescope (\emph{Fermi}-LAT)~\cite{Atwood:2009ez}, data from which will be used in the analyses presented in this thesis. Terrestrial gamma-ray observatories such as HAWC~\cite{Abeysekara:2014ffg}, H.E.S.S.~\cite{Abdallah:2018qtu}, MAGIC~\cite{Ahnen:2017pqx}, VERITAS~\cite{Archambault:2017wyh} and the upcoming CTA~\cite{Doro:2012xx} can typically achieve better sensitivity at higher photon energies (and correspondingly higher DM masses $m_\chi\gtrsim 100$\,GeV) due to their much larger effective area. In certain cases (\emph{e.g.} leptonic final states), experiments like AMS-02 can be sensitive probes of DM annihilation via observations of charged cosmic ray spectra. See Ref.~\cite{Slatyer:2017sev} for a comprehensive recent review of indirect dark matter searches.   

\subsection{Tools for Indirect Detection}
\label{subsec:tools}

A major challenge for indirect detection searches is to calculate the expected dark matter annihilation flux from a given astrophysical target or source population. The basic prescription for doing so is as follows. If we denote the DM (particle) number density at coordinate $r(l,\psi)$ (parameterized by the angle away from the Galactic plane $\psi$ and line-of-sight distance from us $l$) by $n[r(l,\psi)]$ and the velocity-averaged self-annihilation cross section by $\langle\sigma v\rangle$, then the annihilation rate per particle is given by
\begin{equation}
n[r(l,\psi)]\langle\sigma v\rangle = \frac{\rho[r(l,\psi)]}{m_\chi}\langle\sigma v\rangle,
\end{equation}
where $\rho[r(l,\psi)]$ is the DM density and $m_\chi$ its particle mass. The annihilation rate in a volume element $dV = l^2\,dl\,d\Omega$ is given by multiplying this quantity by the number of particles in the volume:
\begin{equation}
\frac{\rho[r(l,\psi)]}{m_\chi}\langle\sigma v\rangle \frac{\rho[r(l,\psi)]}{2m_\chi}dV.
\end{equation}
The factor of $2$ in the denominator is to avoid double counting since two particles are involved in the annihilation process. The observed annihilation flux (in units of photons\,cm$^{-2}$\,s$^{-1}$) is obtained by inserting the area factor $(4\pi l^2)^{-1}$ and integrating over the desired volume:
\begin{equation}
\frac{d\Phi}{dE}(E,\psi) = \frac{1}{4\pi}\int\,d\Omega\,dl\,\rho[r(l,\psi)]^2\frac{\langle\sigma v\rangle}{2m_\chi^2}\frac{dN}{dE}
\end{equation}
where the photon energy spectrum $dN/dE$ gives the number of photons produced per annihilation for a given 2-body final state, and can be obtained with parton shower tools like \texttt{Pythia8}~\cite{Sjostrand:2007gs} or from tabulated values for certain specific cases~\cite{Cirelli:2010xx}. While there are many possibilities for the annihilation final states, the resulting spectra can be broadly classed into a few categories: \emph{(i)} Annihilation directly to photons, which would show up as a spectral line and allow for bump hunts. However, since DM is not expected to be electrically charged, such interactions would generically be loop-suppressed. \emph{(ii)} Annihilation to gauge bosons or quarks and their subsequent hadronization, which would produce pions that would dominantly decay to photons. This would result in a broad continuum photon spectrum. \emph{(iii)} Annihilation to electrons and muons, which would produce photons through final-state radiation and/or radiative decays. This would result in a narrower spectrum and suppressed rate compared to \emph{(ii)}. Annihilation to taus, which have both hadronic and leptonic decays, would result in a spectrum intermediate to \emph{(ii)} and \emph{(iii)}. As a benchmark and for comparison purposes, limits in the literature are often presented for annihilation into $b$-quarks ($\chi\chi\rightarrow b \overline b$). 

The annihilation cross section can be taken out of the integral, and the annihilation flux factorizes as 
\begin{equation}
\frac{d\Phi}{dE}(E,\psi) = \underbrace{\frac{\langle\sigma v\rangle}{8\pi m_\chi^2}\frac{dN}{dE}}_{d\Phi_\mathrm{PP}/dE}\underbrace{\int\,d\Omega\,dl\,\rho[r(l,\psi)]^2}_J
\end{equation}
where $d\Phi_\mathrm{PP}/dE$ encapsulates the particle physics assumptions, and $J\equiv\int\,d\Omega\,dl\,\rho[r(l,\psi)]^2$ is the so-called $J$-factor, which captures the astrophysical dependence of the flux. Objects with higher $J$-factors over some localized region typically make for more interesting indirect detection targets. However, a high $J$-factor by itself does not guarantee a good annihilation target, since the figure of merit is the signal-to-noise ratio. This must additionally be balanced with how well the systematic uncertainties on  the potential signal, astrophysical backgrounds and Galactic foregrounds can be accounted for and controlled.

\subsection{Sources of Gamma Rays from Annihilating Dark Matter}
\label{subsec:dmsources}

\begin{figure}[htbp] 
\centering
 \includegraphics[width=1.0\textwidth]{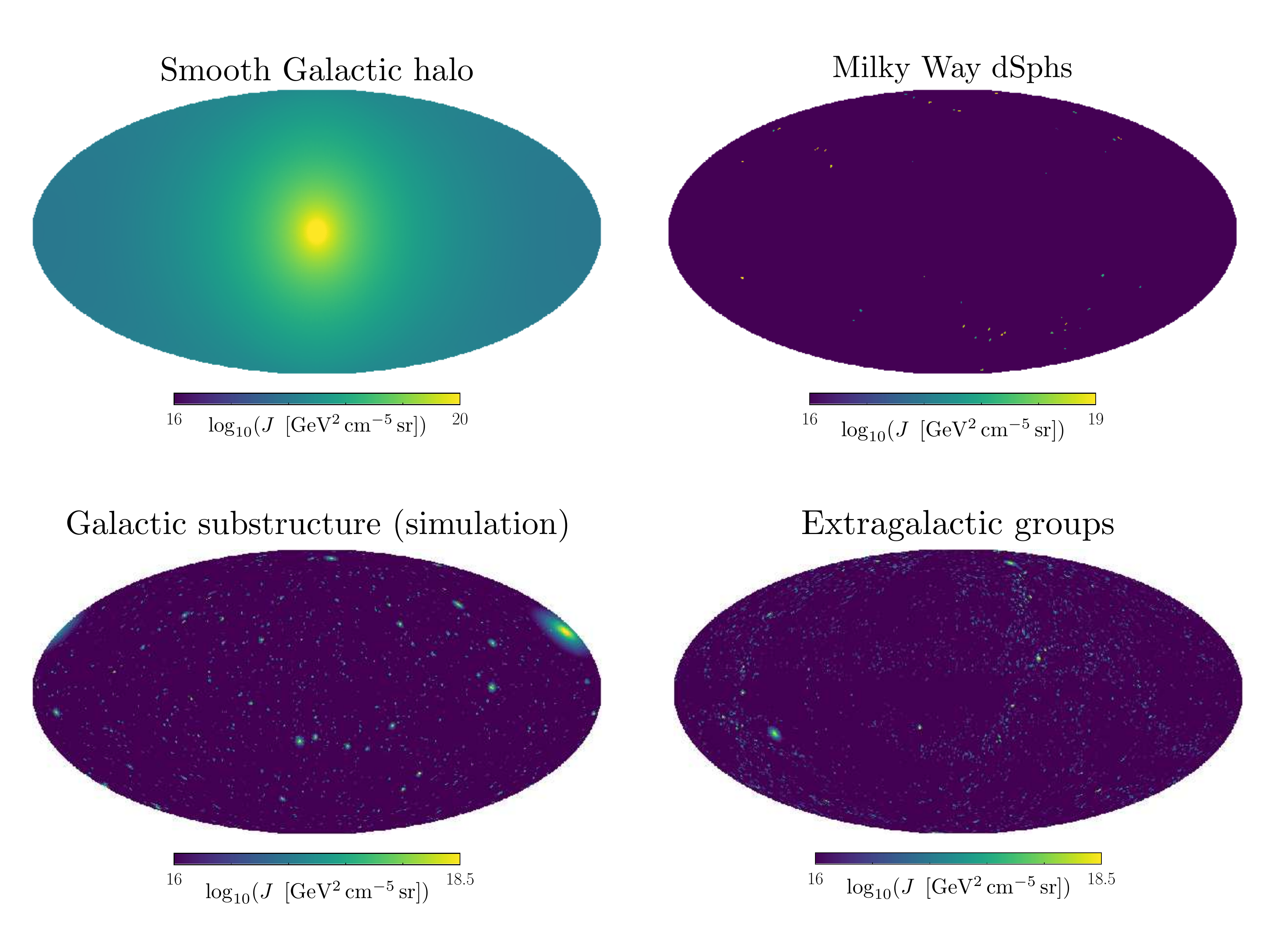}
\caption{Maps of annihilation $J$-factors for some commonly considered gamma-ray search targets. \textbf{(Top left)} The smooth Galactic halo, assuming a canonical NFW dark matter profile $\rho_\text{NFW}(r)=\frac{\rho_{s}}{r/r_{s}\,(1+r/r_{s})^{2}}$, where $r_s=17$ kpc is the Milky Way scale radius and $\rho_s$ is the normalization chosen to reproduce the local DM density $\rho_\text{NFW}(r_\odot) = 0.4$ GeV$\,$cm$^{-3}$~\cite{2015ApJ...814...13M,Sivertsson:2017rkp} at the Solar radius $r_\odot = 8$~kpc~\cite{Read:2014qva}. \textbf{(Top right)} Milky Way dwarf spheroidal galaxies (dSphs) as considered in Ref.~\cite{Fermi-LAT:2016uux}. Following that study, the dSphs are assumed to be point-like sources since the properties of the corresponding DM halos are not currently well constrained. \textbf{(Bottom left)} A simulated realization of $J$-factors for Galactic substructure (subhalos) following the prescription in~\cite{Hutten:2016jko}. Subhalos are spatially distributed according to the results of the Aquarius simulation~\cite{Springel:2008cc} and a halo mass distribution of $dN/dm\propto m^{-1.9}$ is assumed. The concentration-mass parameterization from Ref.~\cite{Sanchez-Conde:2013yxa} is used and DM in the subhalos is assumed to be NFW-distributed. The number of subhalos is calibrated to give 300 objects between $10^8$--$10^{10}$\,M$_\odot$. The bright source in the top right corner of the map would likely show up as a resolved unassociated source in \emph{Fermi} point source catalogs such as 3FGL~\cite{Bertoni:2015mla}. \textbf{(Bottom right)} $J$-factors of extragalactic groups derived using properties compiled in the group catalogs of Refs.~\cite{Tully:2015opa} and~\cite{2017ApJ...843...16K} and the prescription presented in Chs.~\ref{ch:groups_sim} and \ref{ch:groups_data}.}  
\label{fig:sources}
\end{figure}

An important ingredient in indirect detection is the accurately characterization of the DM signal and its associated uncertainties. This often involves input from astrophysics, observations at other wavelengths and $N$-body simulations. Given the typically sizable systematic uncertainties in both signal and background modeling, it is crucial to have the ability to probe the same DM parameter space using multiple complementary targets and search strategies. The following sources have been and continue to be used as gamma-ray targets in annihilation searches:  
\begin{itemize}
\item \emph{Milky Way dwarf galaxies}: Dwarf spheroidal satellite galaxies (dSphs) of the Milky Way are expected to be dark matter dominated and thus to have relatively low expected astrophysical backgrounds. As such, dSphs have traditionally been considered excellent targets for DM annihilation searches. There have been about 45 dSphs candidates discovered recently by surveys like optical SDSS and DES (see Ref.~\cite{Fermi-LAT:2016uux} and references therein), and searches for gamma-ray emission from these have been able to place strong constraints on annihilation scenarios, excluding thermal WIMPs at masses below $\lesssim 70$~GeV at 95\% confidence level for the case of annihilation into  the $b\bar b$ final state~\cite{Fermi-LAT:2016uux,Ackermann:2015zua}. However, the relevant $J$-factors are far from well-characterized---assumptions about \emph{e.g.}, the dSph halo shape~\cite{Geringer-Sameth:2014qqa,Sanders:2016eie} and stellar membership criteria used to infer the halo properties~\cite{2016MNRAS.462..223B,Geringer-Sameth:2014yza} can lead to significant uncertainties on the predicted annihilation signal and the corresponding annihilation limit. Figure~\ref{fig:sources} (top right) shows a map of the inferred $J$-factors of dSphs considered in Ref.~\cite{Fermi-LAT:2016uux}. As in that study, the dSphs are assumed to be point-like since the shape of the corresponding DM halos is not very well constrained.
\item \emph{The Milky Way halo}: Because of its proximity to us, the DM halo surrounding our own Galaxy is the brightest source of DM emission in the sky. Figure~\ref{fig:sources} (top left) shows the expected annihilation $J$-factor for the smooth component of the Milky Way halo (see caption for further details). 

Searches in the inner Galaxy ($|b|\lesssim20^\circ$), where the signal is expected to be the brightest, have yielded an excess emission whose spatial and spectral properties can be consistent with those of a DM annihilation signal (\emph{e.g.}, a $\sim$40\,GeV WIMP annihilating to $b\bar b$ with an approximately thermal cross section), often called the Galactic Center Excess~\cite{Daylan:2014rsa,Goodenough:2009gk,Hooper:2010mq,Calore:2014xka,Gordon:2013vta,TheFermi-LAT:2015kwa,Karwin:2016tsw}. This region of the sky is however plagued by the presence of substantial and difficult-to-characterize Galactic foregrounds, which complicates the interpretation of any signal and/or constraint from it. In addition, recent results based on analyzing the statistics of photons in the region~\cite{Lee:2015fea,Bartels:2015aea} (see also Ch.~\ref{ch:nptfit}) indicate that the excess is more consistent with emission from an unresolved population of point sources rather than a dark matter signal, which is expected to be more diffuse in nature. There is also some evidence that the morphology of the excess emission preferentially traces the stellar overdensity in the Galactic bulge~\cite{Bartels:2017vsx,Bartels:2018eyb,Macias:2016nev}, suggesting association with an underlying stellar population.

Another class of searches focus on looking for DM emission from the Milky Way halo over larger regions of the sky at higher Galactic latitudes ($|b|\gtrsim 20^\circ$), where the signal is still appreciable but Galactic foregrounds are much lower. These studies necessitate being able to accurately characterize the Galactic foreground emission over larger regions of the sky, and a careful consideration of potential foreground mismodeling effects yields stringent limits, excluding thermal WIMPs at masses below $\lesssim 70$~GeV at 95\% confidence level for the case of annihilation into $b\bar b$~\cite{Chang:2018bpt}.

\item \emph{Galactic substructure}: By definition, hierarchical bottom-up structure formation implies the existence of substructure (``subhalos'') within galactic DM halos, and these have the potential to be attractive DM annihilation targets. Unlike the dwarf galaxies mentioned above, low-mass subhalos with virial mass $M_\mathrm{vir}\lesssim 10^8$\,M$_\odot$ would be mostly dark and have highly suppressed stellar activity~\cite{2017MNRAS.467.2019R,Fitts:2016usl}. This makes it difficult to localize them and look for their gamma-ray emission. Figure~\ref{fig:sources} (bottom left) shows a simulated realization of $J$-factors for Galactic substructure (subhalos) following the prescription in~\cite{Hutten:2016jko} (see caption for further details).

Traditional searches rely on assuming that the emission from unassociated gamma-ray sources detected by \emph{Fermi} is coming from DM annihilation in individual subhalos, and comparing this to expectations from $N$-body simulations~\cite{Bertoni:2015mla,Calore:2016ogv,Hooper:2016cld,Schoonenberg:2016aml}. The bright source in the top right corner of the substructure map in Fig.~\ref{fig:sources}, for example, would likely show up as a resolved unassociated source in \emph{Fermi} point source catalogs such as 3FGL~\cite{Acero:2015hja}.

An orthogonal approach is to study the statistics of photons coming from DM annihilation within dim subhalos. While these subhalos may not be detectable individually, their collective emission could be detected statistically as a heightened level of ``clumpiness'' in the photon map. Statistical methods described in Chs.~\ref{ch:nptfit} and~\ref{ch:igrb} of this thesis can be applied to search for such signals structure in gamma-ray data, and this approach is currently a topic of ongoing study.

\item \emph{Extragalactic galaxies and clusters}: Searches for DM annihilation in extragalactic targets have traditionally been complicated by the difficulty in characterizing the DM properties of extragalactic halos and the presence of potentially significant astrophysical emission. Searches for emission from individual, nearby clusters~\cite{Ackermann:2015fdi}; the integrated, isotropic emission from background halos~\cite{Ackermann:2015tah,Ajello:2015mfa,Cholis:2013ena,DiMauro:2015tfa}; and cross-correlation between gamma ray emission and catalogs of galaxies or large-scale structure~\cite{Ando:2013xwa,Ando:2014aoa,Ando:2016ang,Cuoco:2015rfa,Regis:2015zka,Xia:2015wka,Shirasaki:2014noa,Shirasaki:2015nqp,Shirasaki:2016kol,Troster:2016sgf} have yielded constraints on DM annihilation properties. These searches typically do not attain sensitivity to thermal WIMPs for realistic astrophysical assumptions. Chapter~\ref{ch:groups_sim} of this thesis focuses on developing methods to systematically characterize the dark matter emission and associated uncertainties from a large number of nearby extragalactic galaxies and clusters~\cite{Lisanti:2017qoz}. Figure~\ref{fig:sources} (bottom right) shows the extragalactic $J$-factor map derived using this prescription and the group catalogs from Refs.~\cite{Tully:2015opa} and~\cite{2017ApJ...843...16K}. Chapter~\ref{ch:groups_data} presents a search for gamma-ray emission using this map, which results in stringent limits on annihilating DM and excludes thermal WIMPs at masses below $\lesssim 40$~GeV at 95\% confidence level for the case of annihilation into $b\bar b$~\cite{Lisanti:2017qlb}.
\end{itemize}

\subsection{Template Methods for Gamma-Ray Searches}
\label{subsec:statmethods}

\begin{figure}[htbp] 
\centering
 \includegraphics[width=0.8\textwidth]{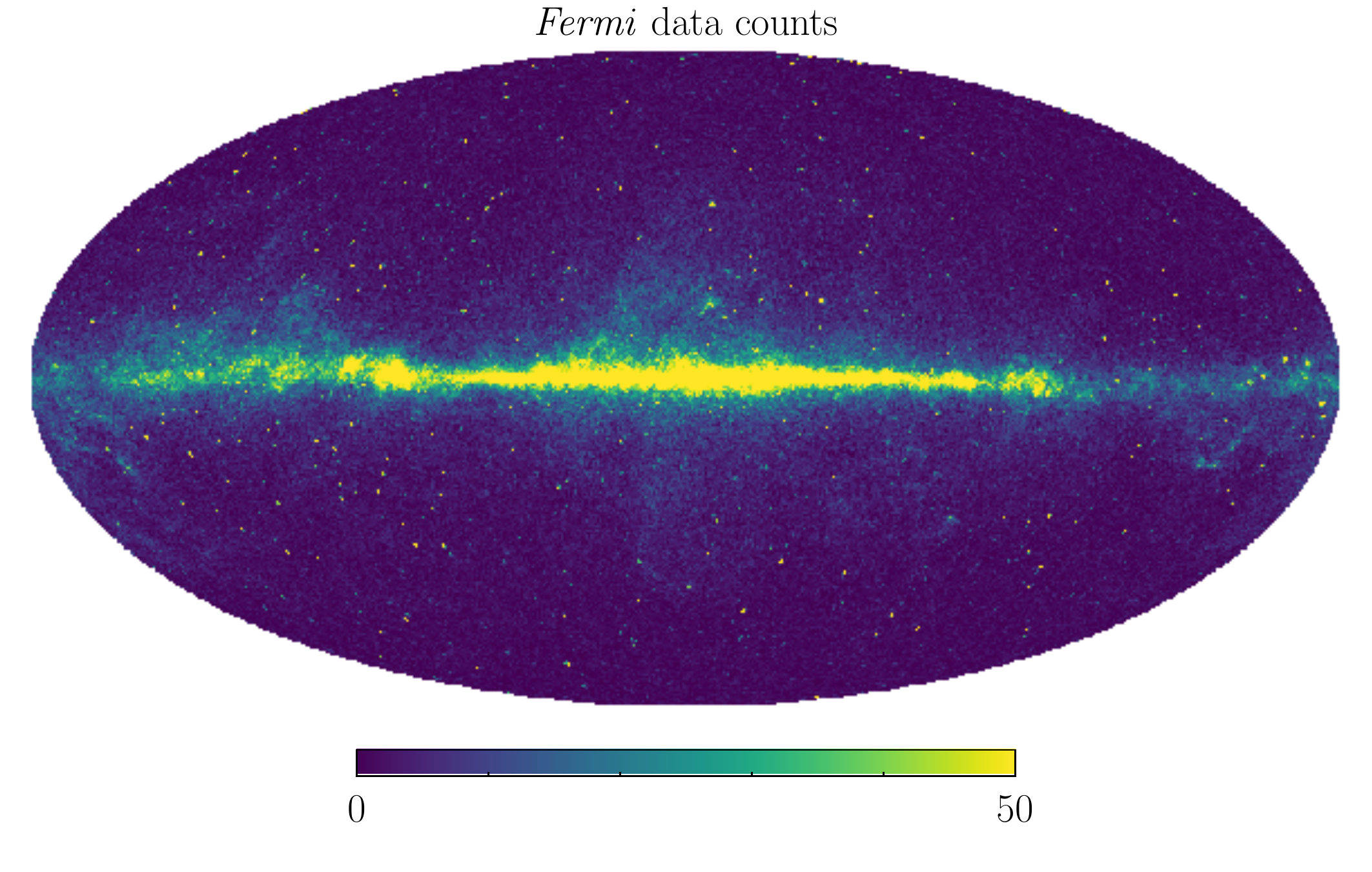}
\caption{A subset of the photons collected by \emph{Fermi}-LAT between August 4, 2008 and July 7, 2016, in the energy range 2--20 GeV. The visualization is of the top quartile of the UltracleanVeto event class (PSF3) as ranked by angular resolution, with the recommended quality cuts applied (see Ch.~\ref{ch:groups_sim} for further details).}  
\label{fig:data}
\end{figure}

Data from gamma-ray detectors such as \emph{Fermi}-LAT is typically a series of sky maps, representing the number of photons binned spatially as well as in energy. Figure~\ref{fig:data} shows a subset of a typical \emph{Fermi}-LAT dataset. In analyzing such data within the context of dark matter indirect detection, the challenge lies in have contributions from large-scale structures such as the smooth Galactic halo as well as point/extended sources like dwarf galaxies, from various astrophysical backgrounds. The most common technique for characterizing the various potential sources that contribute to gamma-ray data is Poissonian template fitting, which is briefly described here; a detailed description will be given in Ch.~\ref{ch:groups_sim}. 

\begin{figure}[htbp] 
\centering
 \includegraphics[width=1.0\textwidth]{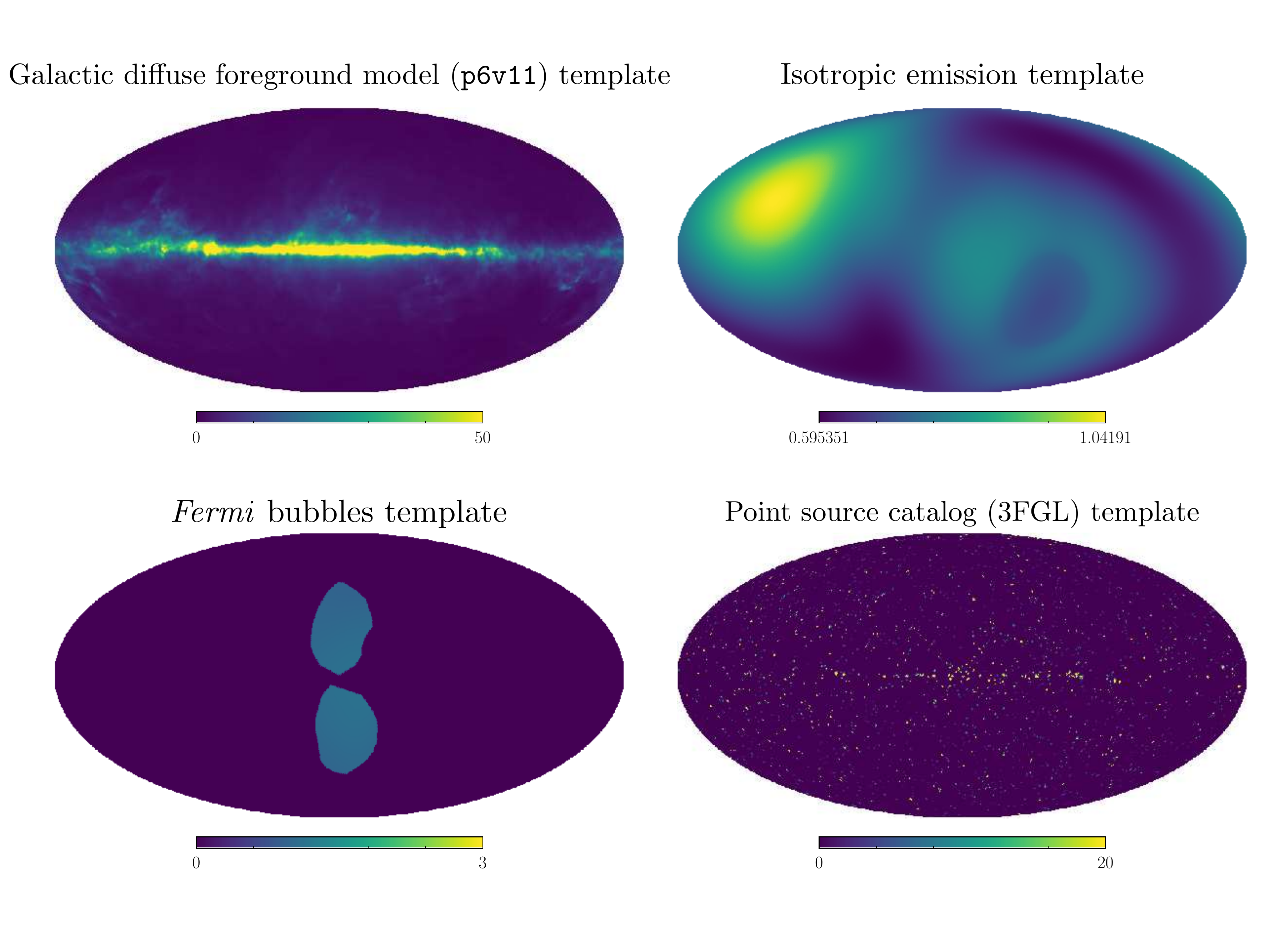}
\caption{Representative templates commonly considered in \emph{Fermi}-LAT gamma-ray analyses. The normalizations of the templates correspond to the best-fit values to the data shown in Fig.~\ref{fig:data}. \textbf{(Top left)} Template for the Galactic diffuse foreground emission, as modeled by the  {\it Fermi} \texttt{p6v11} model. \textbf{(Top right)} Isotropic template, intended to account for emission from unresolved extragalactic point sources. This template is not perfectly uniform due to the non-uniform exposure of the LAT instrument. \textbf{(Bottom left)} Template for the \emph{Fermi} bubbles, two lobe-like structures likely of astrophysical origin~\cite{Su:2010qj,Fermi-LAT:2014sfa}. \textbf{(Bottom right)} Template for resolved point sources as compiled in the \emph{Fermi} 3FGL catalog~\cite{Acero:2015hja}.}  
\label{fig:templates}
\end{figure}

A template is a spatial map which traces the modeled contribution of a particular source or class of sources to the data, \emph{e.g.} the expected emission from the diffuse Galactic foreground or resolved astrophysical point sources. Figure~\ref{fig:templates} shows some templates commonly used in \emph{Fermi} gamma-ray analyses (see caption for descriptions). Templates for DM emission can be constructed as described in Secs.~\ref{subsec:tools} and \ref{subsec:dmsources}. 

Within a single energy bin, if we denote the value of a given template $i$ in pixel $p$ by $T_i^p$, then the total expected counts in pixel $p$ is given by 
\begin{equation}
\mu^p(\boldsymbol \theta) = \sum_i A_i\,T_i^p,
\end{equation}
where $\boldsymbol \theta$ represents the signal and background model parameters ${A_i}$, which in this case are the normalizations of the corresponding templates. The observed data in pixel $p$ should therefore be a Poisson realization of the sum of modeled components. It follows that the likelihood function for the parameters $\boldsymbol \theta$ given the data $d$ is a product over all pixels in the region-of-interest of the Poisson probabilities associated with observing $n^{p}$ counts in each pixel $p$:
\begin{equation}
\mathcal{L}(d | {\boldsymbol \theta}) = \prod_p \frac{\mu^{p}({\boldsymbol \theta})^{n^{p}} e^{-\mu^{p}({\boldsymbol \theta})}}{n^{p}!}\,.
\label{eq:pi}
\end{equation}
With the likelihood in hand, we can quantify the contribution of various components using conventional inference methods, \emph{e.g.} obtaining posterior distributions within a Bayesian framework or building up a likelihood surfaces using frequentist profile likelihood techniques. The latter is more commonly used in DM searches---typically, we are more interested in the parameters associated with the DM model (\emph{e.g.} its particle mass $m_\chi$ and annihilation cross section $\langle\sigma v\rangle$, which are in 1-to-1 correspondence with the normalization of the DM template) than those corresponding to the astrophysical backgrounds. A likelihood surface $\mathcal{L}(d|\mathcal M, \{m_\chi, \langle\sigma v\rangle\})$ for the signal parameters corresponding to a given DM model $\mathcal M$ can be obtained by maximizing the likelihood with respect to the background parameters at each signal parameter point. This can be generalized to the cases of analyzing several energy bins and/or stacking multiple sources (\emph{e.g.} several extragalactic halos), where the total likelihood would be given by the product of the individual likelihoods.

For inferring dark matter properties, a log-likelihood difference test statistic (TS) can be defined for a given mass $m_\chi$ as
\begin{equation}\begin{aligned}
{\rm TS}(\mathcal{M}, \{ \langle\sigma v\rangle, m_\chi\}) \equiv 2 &\left[ \log \mathcal{L}(d | \mathcal{M}, \{\langle\sigma v\rangle, m_\chi \}) \right.\\
&\left.- \log \mathcal{L}(d | \mathcal{M}, \{{\langle\sigma v\rangle=0}, m_\chi \}) \right]\,,
\label{eq:TSdef_darksky}
\end{aligned}\end{equation}
where $\langle\sigma v\rangle=0$ corresponds to the null signal hypothesis. Wilks' theorem guarantees that in the asymptotic limit of a large sample size, the TS is $\chi^2$-distributed, allowing us to discover (if we're lucky) or exclude a DM signal in the data to a desired statistical significance in accordance with $\chi^2$ statistics. A TS value of $-2.71$, for example, corresponds to exclusion at a confidence level of 95\%. Modified versions of this statistical procedure will be used in Chs.~\ref{ch:groups_sim} and \ref{ch:groups_data} to look for DM annihilation in extragalactic galaxies and clusters.

A fundamental limitation of Poissonian template fitting is that while resolved point sources can be either modeled with templates or masked, this is not possible for dim, sub-threshold point sources that cannot be detected individually. Depending on their spatial distribution, emission from these unresolved point sources is typically absorbed by other extended templates, \emph{e.g.} isotropic (in the case of extragalactic sources) or Galactic dark matter (in the case of an approximately spherically symmetric population of unresolved sources in the Galactic center). Chapter~\ref{ch:nptfit} will be dedicated to extending traditional Poissonian template fitting methods to statistically account for the presence of unresolved point sources in the data.

\section{Thesis Organization}
\label{sec:summary}

The rest of this thesis is organized as follows. Chapter~\ref{ch:nptfit} describes the implementation of a novel statistical method, first introduced in Ref.~\cite{Lee:2015fea}, which leverages the ``clumpiness'' of photons associated with populations of unresolved point sources (PSs) in astronomical datasets to derive their contribution and properties. In Ch.~\ref{ch:igrb}, this method is applied to the gamma-ray sky at higher latitudes as seen by \emph{Fermi} to characterize the contribution of PSs to the extragalactic gamma-ray sky over three order of magnitude in energy, from 2 to 2000 GeV. Chapter~\ref{ch:groups_sim} poses the question ``what is the best way to look for annihilating dark matter in extragalactic sources?'' and attempts to answer it by constructing a pipeline to robustly map out the distribution of dark matter outside the Milky Way using galaxy group catalogs. Uncertainties involved in inferring various dark matter parameters are discussed in detail. In Ch.~\ref{ch:groups_data}, this framework is then applied to \emph{Fermi} data and existing group catalogs to search for annihilating dark matter in extragalactic galaxies and clusters.

\sectionline



\chapter{Non-Poissonian Template Fitting: Fundamentals and Code}
\label{ch:nptfit}

This chapter is based on an edited version of \emph{NPTFit: A code package for Non-Poissonian Template Fitting},  \href{http://iopscience.iop.org/article/10.3847/1538-3881/aa6d5f/meta}{Astron.J. \textbf{153} (2017) no.6, 253}  \href{https://arxiv.org/abs/1612.03173}{ [arXiv:1612.03173]} with Nicholas Rodd and Benjamin Safdi~\cite{Mishra-Sharma:2016gis}.  

\section{Introduction}

\lettrine[lines=3]{A}{strophysical} point sources (PSs), which are defined as sources with angular extent smaller than the resolution of the detector, play an important role in virtually every analysis utilizing images of the cosmos.  It is useful to distinguish between resolved and unresolved PSs; the former may be detected individually at high significance, while members of the latter population are by definition too dim to be detected individually.  However, unresolved PSs---due to their potentially large number density---can be a leading and sometimes pesky source of flux across wavelengths.
  Recently, a novel analysis technique called the non-Poissonian template fit (NPTF) has been developed for characterizing populations of unresolved PSs at fluxes below the detection threshold for finding individually-significant sources~\cite{Lee:2014mza,Lee:2015fea}.  The technique expands upon the traditional fluctuation analysis technique (see, for example,~\cite{Miyaji:2001dp,Malyshev:2011zi}), which analyzes the aggregate photon-count statistics of a data set to characterize the contribution from unresolved PSs, by additionally incorporating spatial information both for the distribution of unresolved PSs and for the potential sources of non-PS emission.  In this work, we present a code package called \texttt{NPTFit} for numerically implementing the NPTF in \texttt{python} and \texttt{cython}.
  
The most up-to-date version of the open-source package \texttt{NPTFit} may be found at
\begin{center}
\url{https://github.com/bsafdi/NPTFit}
\end{center}
and the latest documentation at
\begin{center}
\url{http://nptfit.readthedocs.io}.
\end{center}

The NPTF generalizes traditional astrophysical template fits.  Template fitting is useful for pixelated data sets consisting of some number of photon counts $n_p$ in each pixel $p$, and it typically proceeds as follows.   Given a set of model parameters $\bm{\theta}$, the mean number of predicted photon counts $\mu_p({\bm{\theta}})$ in the pixel $p$ may be computed.  More specifically, $\mu_p({\bm{\theta}}) = \sum_{\ell} T^{(S)}_{p,\ell}({\bm{\theta}})$, where $\ell$ is an index of the set of templates $T^{(S)}_{p, \ell}$, whose normalizations and spatial morphologies may depend on the parameters $\bm{\theta}$. These templates may, for example, trace the gas-distribution or other extended structures that are expected to produce photon counts.  Then, the probability to detect $n_p$ photons in the pixel $p$ is simply given by the Poisson distribution with mean $\mu_p({\bm{\theta}})$.  By taking a product of the probabilities over all pixels, it is straightforward to write down a likelihood function as a function of ${\bm{\theta}}$.  

The NPTF modifies this procedure by allowing for non-Poissonian photon-count statistics in the individual pixels.  That is, unresolved PS populations are allowed to be distributed according to spatial templates, but in the presence of unresolved PSs the photon-count statistics in individual pixels, as parameterized by ${\bm{\theta}}$, no longer follow Poisson distributions.  This is heuristically because we now have to ask two questions in each pixel: first, what is the probability, given the model parameters ${\bm{\theta}}$ that now also characterize the intrinsic source-count distribution of the PS population, that there are PSs within the pixel $p$, then second, given that PS population, what is the probability to observe $n_p$ photons?

It is important to distinguish between resolved and unresolved PSs.  Once a PS is resolved---that is once its location and flux is known---that PS may be accounted for by its own Poissonian template.  Unresolved PSs are different because their locations and fluxes are not known.  When we characterize unresolved PSs with the NPTF, we characterize the entire population of unresolved sources, following a given spatial distribution, based on how that population modifies the photon-count statistics.
  
The NPTF has played an important role recently in addressing various problems in gamma-ray astroparticle physics with data collected by the \emph{Fermi}-LAT gamma-ray telescope.\footnote{\url{http://fermi.gsfc.nasa.gov/}} The NPTF was developed to address the excess of gamma rays observed by \emph{Fermi} at $\sim$GeV energies originating from the inner regions of the Milky Way~\cite{Goodenough:2009gk,Hooper:2010mq,Boyarsky:2010dr,Hooper:2011ti,Abazajian:2012pn,Hooper:2013rwa,Gordon:2013vta,Abazajian:2014fta,Daylan:2014rsa,Calore:2014xka,Abazajian:2014hsa,TheFermi-LAT:2015kwa,Macias:2016nev,Clark:2016mbb}.  The GeV excess, as it is commonly referred to, has received a significant amount of attention due to the possibility that the excess emission arises from dark matter (DM) annihilation.  However, it is well known that unresolved PSs may complicate searches for annihilating DM in the Inner Galaxy region due to, for example, the expected population of dim pulsars~\cite{Abazajian:2014fta,Abazajian:2010zy,Hooper:2013nhl,Calore:2014oga,Cholis:2014lta,Petrovic:2014xra,Yuan:2014yda,OLeary:2015gfa,Brandt:2015ula}.  In~\cite{Lee:2015fea} (see also~\cite{Linden:2016rcf}) it was shown, using the NPTF, that indeed the photon-count statistics of the data prefer a PS over a smooth DM interpretation of the GeV excess.  The same conclusion was also reached by~\cite{Bartels:2015aea} using an unrelated method that analyzes the statistics of peaks in the wavelet transformation of the \emph{Fermi} data.  

In the case of the GeV excess, there are multiple PS populations that may contribute to the observed gamma-ray flux and complicate the search for DM annihilation.  These include isotropically distributed PSs of extragalactic origin, PSs distributed along the disk of the Milky Way such as supernova remnants and pulsars, and a potential spherical population of PSs such as millisecond pulsars.  Additionally, there are various identified PSs that contribute significantly to the flux as well as a variety of smooth emission mechanisms such as gas-correlated emission from pion decay and bremsstrahlung.  The power of the NPTF is that these different source classes may be given separate degrees of freedom and constrained by incorporating the spatial morphology of their various contributions along with the difference in photon-count statistics between smooth emission and emission from unresolved PSs.  Although the origin of the GeV excess is still not completely settled, as even if the excess arises from PSs as the NPTF suggests the source class of the PSs remains a mystery at present, the NPTF has emerged as a powerful tool for analyzing populations of dim PSs in complicated data sets with characteristic spatial morphology.

The NPTF and related techniques utilizing photon-count statistics have also been used recently to study the contribution of various source classes to the extragalactic gamma-ray background (EGB)~\cite{Malyshev:2011zi,Zechlin:2015wdz,TheFermi-LAT:2015ykq,Zechlin:2016pme,Lisanti:2016jub}.\footnote{The complementary analysis strategy of probabilistic catalogues has also been applied to this problem \cite{Daylan:2016tia}.}  In these works it was shown that unresolved blazars would predominantly show up as PS populations under the NPTF, while other source classes such as star-forming galaxies would show up predominantly as smooth emission.  For example, in~\cite{Lisanti:2016jub} (described in Ch.~\ref{ch:igrb}) it was shown using the NPTF that blazars likely account for the majority of the EGB from $\sim$2 GeV to $\sim$2 TeV.  These results set strong constraints on the flux from more diffuse sources, such as star-forming galaxies, which has significant implications for, among other problems, the interpretation of the high-energy astrophysical neutrinos observed by IceCube~\cite{Aartsen:2013bka,Aartsen:2013jdh,Aartsen:2015knd,Aartsen:2015rwa} (see, for example,~\cite{Bechtol:2015uqb,Murase:2016gly}).  This is because certain sources that contribute gamma-ray flux at {\it Fermi} energies, such as star forming galaxies and various types of active galactic nuclei, may also contribute neutrino flux observable by IceCube.   

Another promising application of the NPTF is to searches of annihilating dark matter from a population of subhalos in our Galaxy. Annihilation emission from Milky Way subhalos would be characterized by three distinctive features: their spatial distribution, energy spectrum, and non-Poissonian photon-count distribution.  These three features taken together can be used to effectively distinguish subhalos from more standard extragalactic sources.  This approach is quite different from traditional subhalo searches that look for resolved subhalo candidates in the \emph{Fermi} point-source catalog~\cite{Calore:2016ogv,Hooper:2016cld,Schoonenberg:2016aml}.  When the spectrum of an isolated source resembles DM, it is difficult to confirm the exotic nature of the emission~\cite{Bertoni:2016hoh,Bertoni:2015mla}. The NPTF-based proposal relies on looking for a population of subhalos, rather than isolated objects, and is therefore less sensitive to the variations between individual sources.

The NPTF originates from the older fluctuation analysis technique, which is sometimes referred to as the $P(D)$ analysis.  This technique has been used extensively to study the flux of unresolved X-ray sources~\cite{hasinger1993,1993MNRAS.262..619G,Gendreau:1997di,Perri:2000fv,Miyaji:2001dp}.  In these early works, the photon-count probability distribution function (PDF) was computed numerically for different PS source-count distributions using Monte Carlo (MC) techniques.  The fluctuation analysis was first applied to gamma-ray data in~\cite{Malyshev:2011zi},\footnote{The fluctuation analysis has more recently been applied to both gamma-ray \cite{Feyereisen:2015cea} and neutrino \cite{Feyereisen:2016fzb} datasets.} and in that work the authors developed a semi-analytic technique utilizing probability generating functions for calculating the photon-count PDF.  The code package $\texttt{NPTFit}$ presented in this work uses this formalism for efficiently calculating the photon-count PDF.  The specific form of the likelihood function for the NPTF, while reviewed in this work, was first presented in~\cite{Lee:2015fea}.  The works~\cite{Lee:2015fea,Linden:2016rcf,Lisanti:2016jub} utilized an early version of $\texttt{NPTFit}$ to perform their numerical analyses. 

The $\texttt{NPTFit}$ code package has a \texttt{python} interface, though the likelihood evaluation is efficiently implemented in \texttt{cython}~\cite{behnel2010cython}.  The user-friendly interface allows for an arbitrary number of PS and smooth templates.  The PS templates are characterized by pixel-dependent source-count distributions $dN_p/dF = T_p^{({\rm PS})} dN/dF$, where $T_p^{({\rm PS})}$ is the spatial template tracking the distribution of point sources on the sky and $dN/dF$ is the pixel-independent source-count distribution.  The distribution $dN_p/dF$ quantifies the number of sources $dN_p$ that contributes flux between $F$ and $F + dF$ in the pixel $p$.  The $dN/dF$ are parameterized as multiply broken power-laws, with an arbitrary number of breaks.  The code is able to account for both an arbitrary exposure map (accounting for the pointing strategy of an instrument) as well as an arbitrary point spread function (PSF, accounting for the instrument's finite angular resolution) in translating between flux $F$ (in units of photons\,cm$^{-2}$\,s$^{-1}$) and photon counts $S$. 

 \texttt{NPTFit} has a built-in interface with \texttt{MultiNest}~\cite{Feroz:2008xx,Buchner:2014nha}, which efficiently implements nested sampling of the posterior distribution and Bayesian evidence for the user-specified model, given the specified data and instrument response function, in the Bayesian framework~\cite{Feroz:2013hea,Feroz:2007kg,skilling2006}.  The interface handles the Message Passing Interface (MPI), so that inference may be performed efficiently using parallel computing.  A basic analysis package is provided in order to facilitate easy extraction of the most relevant data from the posterior distribution and quick plotting of the \texttt{MultiNest} output.  The preferred format of the data for \texttt{NPTFit} is \texttt{HEALPix}~\cite{Gorski:2004by} (a nested equal-area pixelation scheme of the sky), although the the code is also able to handle non-\texttt{HEALPix} data arrays. Note that the code package may also be used to simply extract the NPTF likelihood function so that \texttt{NPTFit} may be interfaced with any numerical package for Bayesian or frequentist inference.

A large set of example \texttt{Jupyter}~\cite{PER-GRA:2007} notebooks and \texttt{python} files are provided to illustrate the code.  The examples utilize 413 weeks of processed \emph{Fermi} Pass 8 data in the UltracleanVeto event class collected between August 4, 2008 and July 7, 2016 in the energy range from 2 to 20 GeV. We restrict this dataset to the top quartile as graded by PSF reconstruction in order to reduce cosmic-ray contamination and further apply the standard quality cuts \texttt{DATA\_QUAL==1 \&\& LAT\_CONFIG==1}, as well as restricting the zenith angle to be less than $90^\circ$. This data is made available in the code release.  Moreover, the example notebooks illustrate many of the main results in~\cite{Lee:2015fea,Linden:2016rcf,Lisanti:2016jub}.

In addition to the above, the base \texttt{NPTFit} code makes use of the \texttt{python} packages \texttt{corner}~\cite{dan_foreman_mackey_2016_53155}, \texttt{matplotlib}~\cite{Hunter:2007}, \texttt{mpmath}~\cite{mpmath}, \texttt{GSL}~\cite{galassi2015gnu} and  \texttt{numpy}~\cite{oliphant2006guide}.

The rest of this chapter is organized as follows.  Section~\ref{NPTF} outlines in more detail the framework of the NPTF. Sections~\ref{details} and ~\ref{algorithms} describe further details behind the mathematical framework of the NPTF. Section~\ref{NPTFit-orientation} highlights the key classes and features in the \texttt{NPTFit} code package and usage instructions.  In Sec.~\ref{NPTFit-example} we present an example of how to perform an NPTF scan using \texttt{NPTFit}, looking at the Galactic Center with \emph{Fermi} data to reproduce aspects of the main results of~\cite{Lee:2015fea}. We conclude in Sec.~\ref{Conclusion}. 

\section{The Non-Poissonian Template Fit}
\label{NPTF}

In this section we review the NPTF, which was first presented in~\cite{Lee:2015fea} and described in more detail in~\cite{Linden:2016rcf,Lisanti:2016jub} (see also~\cite{Malyshev:2011zi,Lee:2014mza,Zechlin:2015wdz,Zechlin:2016pme} and Ch.~\ref{ch:igrb}).  The NPTF is used to fit a model $\mathcal{M}$ with parameters $\bm{\theta}$ to a data set $d$ consisting of counts (\emph{i.e.}, number of photon) $n_p$ in each pixel $p$.  The likelihood function for the NPTF is then simply
\es{eq:likelihood}{
p(d |{\bm \theta}, \mathcal{M}) = \prod_p p_{n_p}^{(p)}({\bm \theta}) \,,
}
where $p_{n_p}^{(p)}( {\bm \theta})$ gives the probability of drawing $n_p$ counts in the given pixel $p$, as a function of the parameters $\bm{\theta}$.  The main computational challenge, of course, is in computing these probabilities.

It is useful to divide the model parameters into two different categories: the first category describes smooth templates, while the second category describes PS templates.  We describe each category in turn, starting with the smooth templates.  

For most applications, the data has the interpretation of being a two-dimensional pixelated map consisting of an integer number of counts in each pixel.  The smooth templates may be used to predict the mean number of counts $\mu_p({\bm \theta})$ in each pixel $p$:   
\es{mean_pixel}{
\mu_p( {\bm \theta}) = \sum_\ell \mu_{p, \ell} ({\bm \theta}) \,.
}
Above, $\ell$ is an index over templates and $\mu_{p, \ell} ({\bm \theta})$ denotes the mean contribution of the $\ell^\text{th}$ template to pixel $p$ for parameters ${\bm \theta}$.  In principle, ${\bm \theta}$ may describe both the spatial morphology as well as the normalization of the templates.  However, in the current implementation of the code, the Poissonian model parameters simply characterize the overall normalization of the templates: $\mu_{p, \ell} ({\bm \theta}) = A_{\ell}( {\bm \theta}) T^{(S)}_{p,\ell}$.  Here, $A_{\ell}$ is the normalization parameter and $T^{(S)}_{p,\ell}$ is the $\ell^\text{th}$ template, which takes values over all pixels $p$ and is independent of the model parameters.  The superscript $(S)$ implies that the template is a counts templates, which is to be contrasted with a flux template, for which we use the symbol $(F)$. The two are related by the exposure map of the instrument $E_{p}$: $T^{(S)}_p = E_p T^{(F)}_p$. In the case where we only have smooth, Poissonian templates, the probabilities are then given by the Poisson distribution:
\es{poisson}{
p_{n_p}^{(p)}({\bm \theta}) = \frac{{\mu_p^{n_p}( {\bm \theta})}}{n_p !} e^{- \mu_p( {\bm \theta}) } \,.
}
In the presence of unresolved PS templates, the probabilities $p_{n_p}^{(p)}({\bm \theta})$ are no longer Poissonian functions of the model parameters ${\bm \theta}$.  Each PS template is characterized by a pixel-dependent source-count distribution $dN_p/dF$, which describes the differential number of sources per pixel per unit flux interval.  In this work, we model the source-count distribution by a multiply broken power-law:  
\es{mbpl}{
\frac{dN_p}{dF} (F; {\bm \theta}) = A ( {\bm \theta}) T^{({\rm PS})}_p \left\{ \begin{array}{lc} \left( \frac{F}{F_{b,1}} \right)^{-n_1}, & F \geq F_{b,1} \\ \left(\frac{F}{F_{b,1}}\right)^{-n_2}, & F_{b,1} > F \geq F_{b,2} \\ \left( \frac{F_{b,2}}{F_{b,1}} \right)^{-n_2} \left(\frac{F}{F_{b,2}}\right)^{-n_3}, & F_{b,2} > F \geq F_{b,3} \\ \left( \frac{F_{b,2}}{F_{b,1}} \right)^{-n_2} \left( \frac{F_{b,3}}{F_{b,2}} \right)^{-n_3} \left(\frac{F}{F_{b,3}}\right)^{-n_4}, & F_{b,3} > F \geq F_{b,4} \\ \\
\ldots & \ldots \\ \\
\left[ \prod_{i=1}^{k-1} \left( \frac{F_{b,i+1}}{F_{b,i}} \right)^{-n_{i+1}} \right] \left( \frac{F}{F_{b,k}} \right)^{-n_{k+1}}, & F_{b,k} > F \end{array} \right. .
}
Above, we have parameterized the source-count distribution with an arbitrary number of breaks $k$, denoted by $F_{b,i}$ with \mbox{$i \in [1,2, \ldots, k]$}, and $k+1$ indices $n_i$ with \mbox{$i \in [1,2, \ldots , k+1]$}.  The spatial dependence of the source-count distribution is accounted for by the overall factor $A ( {\bm \theta}) T_p^{({\rm PS})}$, where $A ( {\bm \theta})$ is the pixel-independent normalization, which is a function of the model parameters, and $T_p^{({\rm PS})}$ is a template describing the spatial distribution of the PSs.  More precisely, the number of sources $N^\text{PS}_p = \int dF dN_p / dF$ (and the total PS flux $F^\text{PS}_p = \int dF F dN_p / dF$) in pixel $p$, for a fixed set of model parameters ${\bm \theta}$, follows the template $T_p^{({\rm PS})}$.  On the other hand, the locations of the flux breaks and the indices are taken to be fixed between pixels.\footnote{In principle, the breaks and indices could also vary between pixels.  However, in the current version of \texttt{NPTFit}, only the number of sources (and, accordingly, the total flux) is allowed to vary between pixels.} 

To summarize, a PS template described by a broken power-law with $k$ breaks has $2 (k+1)$ model parameters describing the locations of the breaks, the power-law indices, and the overall normalization.  For example, if we take a single break then the PS model parameters may be denoted as $\{ A, F_{b,1}, n_1, n_2 \}$.  Additionally, a spatial template $T^{({\rm PS})}$ must be specified, which describes the distribution of the number of sources (and total flux) with pixel $p$.     

Notice that when we discussed the Poissonian templates we used the counts templates $T^{(S)}$ and talked directly in terms of counts $S$, while so far in our discussion of the unresolved PS templates we have used the point source distribution template $T^{({\rm PS})}$ and written the source-count distribution $dN/dF$ in terms of flux $F$. Of course as the total flux from a distribution of point sources is also proportional to the template $T^{({\rm PS})}$, it can be thought of as a flux template, however conceptually it is being used to track the distribution of the sources rather than the flux they produce. For this reason we have chosen to distinguish the two. Moreover, in the presence of a non-trivial PSF, $T^{(S)}$ should also be smoothed by the PSF to account for the instrument response function.  That is, $T^{(S)}$ is a template for the observed counts taking into account the details of the instrument, while $T^{({\rm PS})}$ ($T^{(F)}$) is a map of the physical point sources (flux), which is independent of the instrument.  In photon-counting applications, the exposure map $E_p$ often has units of $\text{cm}^2 \text{s}$ and flux has units of $\text{counts}\, \text{cm}^{-2}\text{s}^{-1}$.

For the unresolved PS templates, we also need to convert the source-count distribution from flux to counts.  This is done by a simple change of variables:
\es{dNdS}{
{\frac{dN_p}{dS}} (S; {\bm \theta}) = \frac{1}{E_p} {\frac{dN_p}{dF}} (F = S / E_p; {\bm \theta}) \,,
}
which implies that for a non-Poissonian template the spatial dependence of $dN_p/dS$ is given by $T^{({\rm PS})}_p/E_p$. This inverse exposure scaling may seem surprising, but it is straightforward to confirm that the mean number of counts in a given pixel, $ \int dS S dN_p / dS$, is given by $E_p T^{({\rm PS})}_p$, as expected, up to pixel independent factors.

As an important aside, the template $T^{(S)}$ used by the Poissonian models needs to be smoothed by the PSF.  Incorporating the PSF into the unresolved PS models, on the other hand, is more complicated and is not accomplished simply by smoothing the spatial template.  Indeed, $T^{({\rm PS})}_p$ should remain un-smoothed by the PSF when used for non-Poissonian scans. Accounting for PSF effects in the non-Poissonian likelihood will be described in detail in Sec.~\ref{subsec:psf}.

In the remainder of this section we briefly overview the mathematic framework behind the computation of the $p_{n_p}^{(p)}({\bm \theta})$ with \texttt{NPTFit}; however, details of the algorithms used to calculate these probabilities in practice, along with more in-depth explanations, are given in Secs.~\ref{details} and ~\ref{algorithms}.  We use the probability generating function formalism, following~\cite{Malyshev:2011zi}, to calculate the probabilities.   
For a discrete probability distribution $p_k$, with $k=0,1,2,\ldots$, the generating function is defined as:
\begin{equation}
P(t) \equiv \sum_{k=0}^{\infty} p_k t^k \,,
\label{prob-gen}
\end{equation}
from which we can recover the probabilities:
\es{deriv}{
p_k = \frac{1}{k!} \left. \frac{d^k P(t)}{dt^k} \right|_{t=0} \,.
}
The key feature of generating functions exploited here is that the generating function of a sum of two independent random variables is simply the product of the individual generating functions. 

The probability generating function for the smooth templates, as a function of ${\bm \theta}$, is simply given by
\es{P-PGF}{
P_{\rm P}(t; {\bm \theta}) = \prod_p \text{exp}\left[ \mu_p( {\bm \theta}) (t - 1) \right]  \,.
}
The probability generating function for an unresolved PS template, on the other hand, takes a more complicated form (derived in Sec.~\ref{details}):
\es{NP-PGF}{
P_{\rm NP}(t; {\bm \theta}) = \prod_p \exp \left[ \sum_{m=1}^{\infty} x_{p,m}( {\bm \theta}) ( t^m - 1) \right] \,,
}
where
\es{xm-def}{
x_{p,m}( {\bm \theta}) =\int_0^{\infty} dS \frac{dN_p}{dS}(S;{\bm \theta}) \int_0^1 df \rho(f) \frac{(fS)^m}{m!} e^{-fS} \,.
}
Above, $\rho(f)$ is a function that takes into account the PSF, which we describe in more detail in Sec.~\ref{details}.  In the presence of a non-trivial PSF, the flux from a single source is smeared among pixels.  The distribution of flux fractions among pixels is described by the function $\rho(f)$, where $f$ is the flux fraction.  By definition $\rho(f) df$ equals the number of pixels which, on average, contain between $f$ and $f+ df$ of the flux from a PS; the distribution is normalized such that $ \int_0^1 df f \rho(f) = 1$.  If the PSF is a $\delta$-function, then $\rho(f) = \delta(f-1)$.

Putting aside the PSF correction for the moment, the $x_{p,m}$ have the interpretation of being the average number of $m$-count PSs within the pixel $p$, given the distribution $ dN_p(S;{\bm \theta})/dS $.  The generating function for $x_m$ $m$-count sources is simply $e^{x_m(t^m - 1)}$ (see~\cite{Malyshev:2011zi} or Sec.~\ref{details}), which then leads directly to~\eqref{NP-PGF}.  The PSF correction, through the distribution $\rho(f)$, incorporates the fact that PSs only contribute some fraction of their flux within a given pixel.

\section{Mathematical Foundations of \texttt{NPTFit}}
\label{details}

In this section we present the mathematical foundation of the NPTF and the evaluation of the non-Poissonian likelihood in more detail that what was shown in Sec.~\ref{NPTF}. Note that many of the details presented in this section have appeared in the earlier works of \cite{Malyshev:2011zi,Lee:2014mza,Lee:2015fea}, however we have reproduced these here in order to have a single clear picture of the method.

The remainder of this section is divided as follows. Firstly we outline how to determine the generating functions for the Poissonian and non-Poissonian case. We then describe how we account for finite PSF corrections.

\subsection{The (non-)Poissonian Generating Function}

There are two reasons why the evaluation of the Poissonian likelihood for traditional template fitting can be evaluated rapidly. The first of these is that the functional form of the Poissonian likelihood is simple. Secondly, and more importantly, is the fact that if we have two discrete random variables $X$ and $Y$ that follow Poisson distributions with means $\mu_1$ and $\mu_2$, then the random variable $Z = X + Y$ again follows a Poisson distribution with mean $\mu_1 + \mu_2$. This generalizes to combining an arbitrary number of random Poisson distributed variables and is why we were able to write $\mu_{p,\ell}(\bm{\theta}) = A_{\ell}(\bm{\theta})T_{p,\ell}^{(S)}$ in Sec.~\ref{NPTF}. This fact is not true when combining arbitrary random variables, and in particular if we add in a template following non-Poissonian statistics. 

An elegant solution to this problem was introduced in~\cite{Malyshev:2011zi}, using the method of generating functions. As we are always dealing with pixelized maps containing discrete counts (of photons or otherwise), for any model of interest there will always be a discrete probability distribution $p_k$, the probability of observing $k=0, 1, 2, \ldots$ counts. In terms of these, we then define the probability generating function as in~\eqref{prob-gen}. The property of probability generating functions that make them so useful in the present context is as follows. Consider two random processes $X$ and $Y$, with generating functions $P_X(t)$ and $P_Y(t)$, that follow arbitrary and potentially different statistical distributions. Then the generating function of $Z = X + Y$ is simply given by the product $P_X(t) \cdot P_Y(t)$. In this subsection we will derive the appropriate form of $P(t)$ for Poissonian and non-Poissonian statistics.

To begin with, consider the purely Poissonian case. Here and throughout this section we consider only the likelihood in a single pixel; the likelihood over a full map is obtained from the product of the pixel-based likelihoods. Then for a Poisson distribution with an expected number of counts $\mu_p$ in a pixel $p$:
\begin{equation}
p_k = \frac{\mu_p^k e^{-\mu_p}}{k!}\,.
\end{equation}
Note that the variation of the $\mu_p$ across the full map will be a function of the model parameters, such that $\mu_p = \mu_p(\bm{\theta})$. In order to simplify the notation in this section however, we leave the $\bm{\theta}$ dependence implicit. Given the $p_k$ values, we then have:
\begin{equation}\begin{aligned}
P_{\rm P}(t) &= \sum_{k=0}^{\infty} \frac{\mu_p^k e^{-\mu_p}}{k!} t^k \\
&= e^{-\mu_p} \sum_{k=0}^{\infty} \frac{\left( \mu_p t \right)^k}{k!} \\
&= \exp \left[ \mu_p(t-1) \right]\,.
\label{eq:Pgen}
\end{aligned}\end{equation}
From this form, it is clear that if we have two Poisson distributions with means $\mu_p^{(1)}$ and $\mu_p^{(2)}$, the product of their generating functions will again describe a Poisson distribution, but with mean $\mu_p^{(1)} + \mu_p^{(2)}$.

Next we work towards the generating function in the non-Poissonian case. At the outset, we let $x_{p,m}$ denote the average number of sources in a pixel $p$ that emit exactly $m$ counts. In terms of this, the probability of finding $n_m$ $m$-count sources in this pixel is just a draw from a Poisson distribution with mean $x_{p,m}$, i.e.
\begin{equation}
p_{n_m} = \frac{x_{p,m}^{n_m} e^{-x_{p,m}}}{n_m!}\,.
\end{equation}
Given this, the probability to find $k$ counts from a population of $m$-count sources is
\begin{equation}
p_k^{(m)} = \left\{ \begin{array}{lc} p_{n_m}, & {\rm if}~k=m \cdot n_m~{\rm for~some~}n_m, \\ 0, & {\rm otherwise} \end{array} \right.\,.
\end{equation}
We can then use this to derive the non-Poissonian $m$-count generating function as follows:
\begin{equation}\begin{aligned}
P_{\rm NP}^{(m)}(t) &= \sum_k p_k t^k \\
&= \sum_{n_m} t^{m \cdot n_m} \frac{x_{p,m}^{n_m}e^{-x_{p,m}}}{n_m!} \\
&= \exp \left[ x_{p,m} (t^m - 1) \right]\,.
\end{aligned}\end{equation}
However this is just the generating function for $m$-count sources, to get the full non-Poissonian generating function we need to multiply this over all values of $m$. Doing so we arrive at
\begin{equation}\begin{aligned}
P_{\rm NP}(t) &= \prod_{m=1}^{\infty} \exp \left[ x_{p,m} (t^m - 1) \right] \\
&= \exp \left[ \sum_{m=1}^{\infty} x_{p,m} (t^m - 1) \right]\,,
\end{aligned}\end{equation}
justifying the form given in Sec.~\ref{NPTF}. Again recall for the full likelihood we can just multiply the pixel based likelihoods and that  $x_{p,m} = x_{p,m}(\bm{\theta})$.

So far we have said nothing of how to determine $x_{p,m}$, the average number of $m$-count source in pixel $p$. This value depends on the source-count distribution $dN_p/dS$, which specifies the distribution of sources as a function of their expected number of counts, $S$. Of course the physical object is $dN/dF$, where $F$ is the flux. This distinction was discussed in Sec.~\ref{NPTF}, and can be implemented in \texttt{NPTFit} to arbitrary precision. Nevertheless $dN_p/dS$ does not fully determine $x_{p,m}$---we need to account for the fact that a source that is expected to give $S$ photons could Poisson fluctuate to give $m$. As such any source can in principle contribute to $x_{p,m}$, and so integrating over the full distribution we arrive at:
\begin{equation}
x_{p,m} = \int_0^{\infty} dS \frac{dN_p}{dS}(S) \frac{S^m e^{-S}}{m!}\,.
\label{eq:xpm}
\end{equation}

An important part of implementing the NPTF in a rapid manner, which is a central feature of \texttt{NPTFit}, is the analytic evaluation of the integral in this equation. In order to do this, we need to have a specific form of the source-count distribution. For this purpose, we allow the source count distribution to be a multiply broken power-law and evaluate the integral for any number of breaks. 

Putting the evaluation of the integral aside for the moment then, we have arrived at the full non-Poissonian generating function:
\begin{equation}\begin{aligned}
P_{\rm NP}(t) &= \exp \left[ \sum_{m=1}^{\infty} x_{p,m} (t^m - 1) \right]\,, \\
x_{p,m} &= \int_0^{\infty} dS \frac{dN_p}{dS}(S) \frac{S^m e^{-S}}{m!}\,.
\label{eq:NPgen}
\end{aligned}\end{equation}
Contrasting this with Eq.~\eqref{eq:Pgen}, we see that whilst the Poissonian likelihood is specified by a single number $\mu_p$, the non-Poissonian likelihood is instead specified by a distribution $dN_p/dS$.

In the case of multiple PS templates, we should multiply the independent probability generating functions.  However, this is equivalent to summing the $x_{p,m}$ parameters.  This is how multiple PS templates are incorporated into the \texttt{NPTFit} code:
\begin{equation}
x_{p,m} \to x_{p,m}^{\rm total} = \sum_{\ell=1}^{N_{\rm NPT}} x_{p,m}^{\ell}\,,
\end{equation}
where the sum over $\ell$ is over the contributions from individual PS templates. 

\subsection{Correcting For a Finite Point Spread Function}
\label{subsec:psf}

The next factor to account for is the fact that in any realistic dataset there will be a non-zero PSF.  Here, we closely follow the discussion in~\cite{Malyshev:2011zi}. The PSF arises due to the inability of an instrument to perfectly reconstruct the original direction of the photon, neutrino, or quantity making up the counts. In practice, a finite PSF means that a source in one pixel can contribute counts to nearby pixels as well. To implement this correction, we modify the calculation of $x_{p,m}$ given in Eq.~\eqref{eq:NPgen}, which accounts for the distribution of sources as a function of $S$ and the fact that each one could Poisson fluctuate to give us $m$ counts. The finite PSF means that in addition to this, we also need to draw from the distribution $\rho(f)$, that determines the probability that a given source contributes a fraction of its flux $f$ in a given pixel. Once we know $\rho(f)$, this modifies our calculation of $x_{p,m}$ in Eq.~\eqref{eq:NPgen}---now a source that is expected to contribute $S$ counts, will instead contribute $f S$, where $f$ is drawn from $\rho(f)$. As such we arrive at the result in~\eqref{xm-def}.

In \texttt{NPTFit} we determine $\rho(f)$ using Monte Carlo. To do this we place a number of PSs appropriately smeared by the PSF at random positions on a pixelized sphere. Then integrating over all pixels we can determine the fraction of the flux in each pixel $f_p$, $p=1,\ldots, N_{\rm pix}$, defined such that $f_1+f_2+\ldots = 1$. Note in practice one can truncate this sum at some minimal value of $f$ without impacting the argument below. From the set $\left\{f_p \right\}$, we then denote by $\Delta n(f)$ the number of fractions for $n$ point sources that fall within some range $\Delta f$. From these quantities, we may determine $\rho(f)$ as
\begin{equation}
\rho(f) = \lim_{\substack{\Delta f \to 0 \\ n \to \infty}} \frac{\Delta n(f)}{n \Delta f}\,,
\end{equation}
which is normalized such that $\int df~f \rho(f) = 1$. From this definition we see that the case of a vanishing PSF is just $\rho(f) = \delta(f-1)$ - i.e. the flux is always completely in the pixel with the PS.

\section{\texttt{NPTFit}: Algorithms}
\label{algorithms}

The generating-function formalism for calculating the probabilities $p_{n_p}^{(p)}({\bm \theta})$ is described at the end of Sec.~\ref{NPTF} and in more detail in Sec.~\ref{details}.  In particular---given the generating function $P(t)$---we are instructed to calculate the probabilities by taking $n_p$ derivatives as in~\eqref{deriv}.  However, taking derivatives is numerically costly, and so instead we have developed recursive algorithms for computing these probabilities.  In the same spirit, we analytically evaluate the $x_{p,m}$ parameters defined in~\eqref{xm-def} for the multiply-broken source-count distribution in order to facilitate a fast evaluation of the NPTF likelihood function.  In this section, we overview these methods that are essential to making \texttt{NPTFit} a practical software package.   

In general we may write the full single pixel generating function for a model containing an arbitrary number of Poissonian and non-Poissonian templates as:
\begin{equation}
P(t) = e^{f(t)}\,,
\end{equation}
where we have defined
\begin{equation}
f(t) \equiv \mu_p(t-1) + \sum_{m=1}^{\infty} x_{p,m} (t^m - 1)\,.
\end{equation}
Above, $x_{p,m} $ represents the average number of $m$-count source in pixel $p$. The remaining task is to efficiently calculate the probabilities $p_k$, which are formally defined in terms of derivatives through~\eqref{deriv}. Nevertheless, derivatives are slow to implement numerically, so we instead use a recursion relation to determine $p_k$ in terms of $p_{< k}$.

To begin with, note that
\begin{equation}
f^{(k)} \equiv \left. \frac{d^k}{dt^k} f(t) \right|_{t=0} = \left\{ \begin{array}{lc} -(\mu_p + \sum_{m=1}^{\infty} x_{p,m}), & k=0\,, \\ \mu_p + x_{p,1}, & k=1\,, \\ k! x_{p,k}, & k > 1\,. \end{array} \right.
\label{fk}
\end{equation}
For the rest of this discussion, we suppress the pixel index $p$, though one should keep in mind that this process must be performed independently in every pixel.
From~\eqref{fk}, we can immediately write down
\begin{equation}\begin{aligned}
p_0 &= e^{f^{(0)}}\,, \\
p_1 &= f^{(1)} e^{f^{(0)}}\,.
\end{aligned}\end{equation}
Given $p_0$ and $p_1$, we may write our recursion relation for $k > 1$ as
\es{recursion}{
p_k = \sum_{n=0}^{k-1} \frac{1}{k(k-n-1)!} f^{(k-n)} p_n\,,
}
which as mentioned requires the knowledge of all $p_{< k}$. 
To derive~\eqref{recursion}, we first define
\begin{equation}
F^{(k)}(t) \equiv \frac{d^k}{dt^k} e^{f(t)}\,.
\end{equation}
Then, for example,
\begin{equation}
F^{(1)}(t) = f^{(1)}(t) e^{f^{(0)}(t)}\,.
\end{equation}
From here to determine $F^{(k)}(t)$ we simply need $k-1$ more derivatives. Using the generalized Leibniz rule, we have
\begin{equation}\begin{aligned}
F^{(k)}(t) &= \frac{d^{k-1}}{dt^{k-1}} \left( f^{(1)}(t) e^{f^{(0)}(t)} \right) \\
&= \sum_{n=0}^{k-1} \begin{pmatrix} k-1 \\ n \end{pmatrix} \frac{d^{k-1-n}}{dt^{k-1-n}} f^{(1)}(t) \frac{d^n}{dt^n} e^{f^{(0)}(t)} \\
&= \sum_{n=0}^{k-1} \begin{pmatrix} k-1 \\ n \end{pmatrix} f^{(k-n)}(t) F^{(n)}(t)\,.
\end{aligned}\end{equation}
Then setting $t=0$ and recalling the definition of $p_k$, this yields
\begin{equation}\begin{aligned}
p_k &= \sum_{n=0}^{k-1} \frac{n!}{k!} \begin{pmatrix} k-1 \\ n \end{pmatrix} f^{(k-n)} p_n \\
&= \sum_{n=0}^{k-1} \frac{1}{k(k-n-1)!} f^{(k-n)} p_n\,,
\end{aligned}\end{equation}
as claimed.

To calculate the $f^{(k)}$ in a pixel $p$, we need to calculate the $x_{p,k}$ and the sum $\sum_{m=1}^\infty x_{p,m}$.  We may calculate these expressions analytically using the general source-count distribution in~\eqref{mbpl}.  To calculate the sums, we make use of the relation 
\begin{equation}\begin{aligned}
\sum_{m=1}^{\infty} x_{p,m} = &\int_0^{\infty} dS \frac{dN_p}{dS} e^{-S} \sum_{m=1}^{\infty} \frac{S^m}{m!} \\
= &\int_0^{\infty} dS \frac{dN_p}{dS} - \int_0^{\infty} dS \frac{dN_p}{dS} e^{-S} \\
= &\int_0^{\infty} dS \frac{dN_p}{dS} - x_{p,0}\,.
\end{aligned}
\label{eq:xmsum}
\end{equation}
Finiteness of the total flux, and also the probabilities, requires $n_1 > 2$ and $n_{k+1} < 2$.  However,  
both the integral and $x_{p,0}$, appearing in the last line above, may be divergent individually if $1 < n_{k+1} < 2$.  In this case, we analytically continue in $n_{k+1}$, evaluate the contributions individually, and then sum the two expressions to get a result that is finite across the whole range of allowable parameter space. 

\section{\texttt{NPTFit}: Orientation}
\label{NPTFit-orientation}

\texttt{NPTFit} implements the NPTF, as described above, in \texttt{python}.  In this section we give a brief orientation to the code package and its main classes.  A more thorough description of the code and its uses is available in the \href{http://nptfit.readthedocs.io}{online documentation}.

\subsection*{ \lstinline{class NPTFit.nptfit.NPTF} }

This is the main class used to set up and perform non-Poissonian and Poissonian template scans.  It is initialized by   
\begin{lstlisting}
nptf = NPTF(tag='Untagged',work_dir=None)
\end{lstlisting}
with keywords 
\begin{center}
\begin{tabular}{ cccc}
\toprule
Argument &  Default & Purpose & \lstinline!type! \\ 
\midrule
\lstinline!tag! & \lstinline!'Untagged'! & Label of scan & \lstinline!str! \\  
\lstinline!work_dir! & \lstinline!None! & Output directory & \lstinline!str!  \\  
\bottomrule
\end{tabular} \,.
\end{center}

If no \lstinline{work_dir} is specified, the code will default to the current directory.  This is the directory where all output is stored.  Specifying a \lstinline{tag} will create an additional folder, with that name, within the \lstinline{work_dir} for the output.

The data, exposure map, and templates are loaded into the \lstinline{nptfit.NPTF} instance after initialization (see the example in Sec.~\ref{NPTFit-example}).  The data and exposure map are loaded by 
\begin{lstlisting}
nptf.load_data(data, exposure)
\end{lstlisting}
Here, \lstinline{data} and \lstinline{exposure} are 1-D \texttt{numpy} arrays.  The recommended format for these arrays is the \texttt{HEALPix} format, so that all pixels are equal area, although the code is able to handle arbitrary data and exposure arrays so long as they are of the same length.  The templates are added by 
\begin{lstlisting}
nptf.add_template(template, key, 
    units='counts')
\end{lstlisting}  
Here, \lstinline{template} is a 1-D  \texttt{numpy} array of the same length as the data and exposure map, \lstinline{key} is a string that will be used to refer to the template later on, and \lstinline{units} specifies whether the template is a counts template (keyword \lstinline{'counts'}) or a flux template (keyword \lstinline{'flux'}) in units  $\text{counts}\,text{cm}^{-2}\,\text{s}^{-1}$.  The default, if unspecified, is \lstinline{units = 'counts'}.  The template should be pre-smoothed by the PSF if it is going to be used for a Poissonian model.  If the template is going to be used for a non-Poissonian model, either choice for \lstinline{units} is acceptable, though in the case of \lstinline{'counts'} the template should simply be the product of the exposure map times the flux template and not smoothed by the PSF. 

The user also has the option of loading in a mask that reduces the region of interest (ROI) to a subset of the pixels in the data, exposure, and template arrays.  This is done through the command
\begin{lstlisting}
nptf.load_mask(mask)
\end{lstlisting}
where \lstinline{mask} is a boolean \texttt{numpy} array of the same length as the data and exposure arrays.  Pixels in \lstinline{mask} should be either \texttt{True} or \texttt{False}; by convention, pixels that are \texttt{True} will be masked, while those that are \texttt{False} will not be masked.  Note if performing an analysis with non-Poissonian templates, regions where the exposure map is identically zero should be explicitly masked. 

Afterwards, Poissonian and non-Poissonian models may be added to the instance using the available templates. An arbitrary number of Poissonian and non-Poissonian models may be added to the scan. Moreover, each non-Poissonian model may be specified in terms of a multiply broken power law with a user-specified number of breaks, as in~\eqref{mbpl}.  

Poissonian models are added sequentially using the syntax
\begin{lstlisting}
nptf.add_poiss_model(template_name, model_tag, prior_range=[], log_prior=False, fixed=False, fixed_norm=1.0)
\end{lstlisting}
where the keywords are
\vspace{+0.02in}
\begin{center}
\begin{tabular}{ cccc}
\toprule
Argument &  Default & Purpose & \lstinline!type! \\ 
\midrule
\lstinline!template_name! & - & \lstinline!key! of template & \lstinline!str! \\ 
\lstinline!model_tag! & - & \LaTeX-ready label & \lstinline!str!  \\ 
\lstinline!prior_range! & \lstinline![]! & Prior [min, max ] & [\lstinline!float!, \lstinline!float!]  \\ 
\lstinline!log_prior! & \lstinline!False! & Log/linear-flat prior & \lstinline!bool!  \\ 
\lstinline!fixed! & \lstinline!False! & Is template fixed & \lstinline!bool!  \\ 
\lstinline!fixed_norm! & \lstinline!1.0! & Norm if  \lstinline!fixed! & \lstinline!float!  \\ 
\bottomrule
\end{tabular} 
\end{center}
Any of the model parameters may be fixed to a user specified value instead of floated in the scan.  For those parameters that are floated in the scan, a prior range needs to be specified along with whether or not the prior is flat or log-flat.
Note that if \lstinline{log_prior = True}, then the prior range is set with respect to $\log_{10}$ of the linear prior range.\footnote{More complicated priors will be incorporated in future releases of \texttt{NPTFit}.}  For example, if we want to scan the normalization of a template over the range from $[0.1,10]$ with a log-flat prior, then we would set \lstinline{log_prior = True} and \lstinline{prior_range = [-1,1]}.  In this case, it might make sense to label the model with {\footnotesize\ttfamily \textcolor{black}{model\_tag =}} {\footnotesize\ttfamily \textcolor{deepred}{'\$\textbackslash log\_\{10\}A\$'}}  to emphasize that the actual model parameter is the log of the normalization; 
this label will appear in various plots made using the provided analysis class for visualizing the posterior.

The non-Poissonian models are added with a similar syntax:
\begin{lstlisting}
nptf.add_non_poiss_model(template_name, model_tag, prior_range=[], log_prior=False, dnds_model='specify_breaks', fixed_params=None, units='counts')
\end{lstlisting} 
The \lstinline{template_name} keyword is the same as for the Poissonian models.  The rest of the keywords are

\begin{center}
\renewcommand{\arraystretch}{1.4}
\begin{tabular}{ccC{4cm}C{4cm}}
\toprule
Argument &  Default & Purpose &\lstinline!type! \\ 
\midrule
\lstinline!model_tag! & - & \LaTeX-ready label & \lstinline![str, str, ...]! \\  
\lstinline!prior_range! & \lstinline![]! & Prior [[min, max], ...] & \lstinline![[float, float], ...]!  \\  
\lstinline!log_prior! & \lstinline![False]! & Log/linear-flat prior & \lstinline![bool,bool, ...]!  \\  
\lstinline!dnds_model! & \lstinline!'specify_breaks'! & How to specify multiple breaks & \lstinline!str!  \\  
\lstinline!fixed_params! & \lstinline!None! & Fix certain parameters & \lstinline![[int,float], ...]!  \\  
\lstinline!units! & \lstinline!'counts'! & \lstinline!'flux'! or \lstinline!'counts'! units for breaks  & \lstinline!str!  \\ 
\bottomrule
\end{tabular} 
\end{center}

The syntax for adding non-Poissonian models is that the model parameters are specified by $[A, n_1, n_2, \ldots, n_{k+1}, S_{b,1}, S_{b,2}, \ldots, S_{b,k}]$ for a broken power-law with $k$ breaks.  As such, the \lstinline{model_tag}, \lstinline{prior_range}, and \lstinline{log_prior} are now arrays where each entry refers to the respective model parameter.  The code automatically determines the number of breaks by the length of the \lstinline{model_tag} array.  The arrays \lstinline{prior_range} and \lstinline{log_prior} should only include entries for model parameters that will be floated in the scan.  Any model parameter may be fixed using the \lstinline{fixed_params} array, with the syntax such that \lstinline{fixed_params = [[i,c_i],[j,c_j]]} would fix the $i^\text{th}$ model parameter to $c_i$ and the $j^\text{th}$ to $c_j$, where the parameter indexing starts from 0.     

The \lstinline{units} keyword determines whether the priors for the breaks in the source-count distribution (and also the fixed parameters, if any are given) will be specified in terms of \lstinline{'flux'} or \lstinline{'counts'}.  The relation between flux and counts varies between pixels if the exposure map is non-trivial.  For this reason, it is more appropriate to think of the breaks in the source-count distribution in terms of flux.  The keyword \lstinline{'counts'} still specifies the breaks in the source-count distribution in terms of flux, with the relation between counts and flux given through the mean of the exposure map $\text{mean}(E)$: $F_{b,i} = S_{b,i} / \text{mean}(E)$.  

The \lstinline{dnds_model} keyword has the options \mbox{\lstinline{'specify_breaks'}} and  \lstinline{'specify_relative_breaks'}.  If \lstinline{'specify_breaks'} is chosen, which is the default, then the breaks are the model parameters.  If instead \mbox{\lstinline{'specify_relative_breaks'}} is chosen, the full set of model parameters is given by $[A, n_1, n_2, \ldots, n_{k+1}, S_{b,1}, \lambda_{2}, \ldots, \lambda_{k}]$.  Here, $S_{b,1}$ is the highest break and the lower breaks are determined by $S_{b,i} = \lambda_i S_{b,i-1}$.  Note that the prior ranges for the $\lambda$'s should be between $0$ and $1$ (for linear flat), since $S_{b,i} < S_{b,i-1}$.  

After setting up a scan, the configuration is finished by executing the command 
\begin{lstlisting}
nptf.configure_for_scan(f_ary=[1.0], df_rho_div_f_ary=[1.0], nexp=1)
\end{lstlisting}
For a purely Poissonian scan, none of the keywords above need to be specified.  For non-Poissonian scans, \lstinline{f_ary} and \lstinline{df_rho_div_f_ary} incorporate the PSF correction.  In particular, \lstinline{f_ary} is a discretized list of $f$ values between $0$ and $1$, while  \lstinline{df_rho_div_f_ary} is a discretized list of $df \rho(f) / f$ at those $f$ values.  A class is provided for computing these lists; it is described later in this section.  If no keywords are given for these two arrays they default to the case of a $\delta$-function PSF.  

The keyword \lstinline{nexp}, which defaults to $1$, is related to the exposure correction in the calculation of the source-count distribution $dN_p/dS$ from $dN_p/dF$.  In many applications, it is computationally too expensive to perform the mapping in~\eqref{dNdS} in each pixel.  The overall pixel-dependent normalization factor $T_p^{({\rm PS})} / E_p$ factorizes from many of the internal computations, and as a result this contribution to the exposure correction is performed in every pixel.  However, it is useful to perform the mapping from flux to counts, which should be performed uniquely in each pixel $F = S / E_p$, using the mean exposure within small sub-regions.  Within a given sub-region, we map flux to counts using $F = S / \text{mean}(E)$, where the mean is taken over all pixels in the sub-region.  The number of sub-regions is given by \lstinline{nexp}, and all sub-regions have approximately the same area.  As \lstinline{nexp} approaches the number of pixels, the approximation becomes exact; however, for many applications the approximation converges for a relatively small number of exposure regions.  We recommend verifying, in any application, that results are stable as \lstinline{nexp} is increased.        

After configuring the \lstinline{NPTF} instance, the log-likelihood may be extracted, as a function of the model parameters, in addition to the prior range.  The log-likelihood and prior range may then be used with any external package for performing Bayesian or frequentist inference.  This is particularly useful if the user would like to combine likelihood functions between different energy bins or otherwise add to the default likelihood function, for example, incorporating nuisance parameters beyond those associated with individual templates.   The package \texttt{MultiNest}, however, is already incorporated into the \lstinline{NPTF} class and may be run immediately after configuring the \lstinline{NPTF} instance.  This is done simply by executing the command
\begin{lstlisting}
nptf.perform_scan(run_tag=None,nlive=500)
\end{lstlisting}
where \lstinline{nlive} is an integer that specifies the number of live points used in the sampling of the posterior distribution.  \texttt{MultiNest} recommends an \lstinline{nlive} $\sim$500-1000, though the parameter defaults to $100$ if unspecified for quick test runs.  Additional \texttt{MultiNest} arguments may be passed as a dictionary through the optional \lstinline{pymultinest_options} keyword (see the \href{http://nptfit.readthedocs.io}{online documentation} for more details).  The optional keyword \lstinline{run_tag} is used to create a sub-folder for the \texttt{MultiNest} output with that name.

After a scan has been run (or if a scan has been run previously and saved), the results may be loaded through the command
\begin{lstlisting}
nptf.load_scan(run_tag=None)
\end{lstlisting}
The \texttt{MultiNest} chains, which give a discretized view of the posterior distribution, may then be accessed through, for example, \lstinline{nptf.samples}.  An instance of the \texttt{PyMultiNest} analyzer class may be accessed through \lstinline{nptf.a}.  A small analysis package, described later in this section, is also provided for performing a few common analyses.  
 
\subsection*{ \lstinline{class NPTFit.psf_correction.PSFCorrection} }

This is the class used to construct the arrays \lstinline{f_ary} and \lstinline{df_rho_div_f_ary} for the PSF correction.  An instance of \lstinline{PSFCorrection} is initialized through 
\begin{lstlisting}
pc_inst = PSFCorrection.PSFCorrection(psf_dir=None, num_f_bins=10, n_psf=50000, n_pts_per_psf=1000, f_trunc=0.01, nside=128, psf_sigma_deg=None, delay_compute=False)
\end{lstlisting}
with keywords

\begin{center}
\begin{tabular}{ ccC{8.5cm}c }
\toprule
Argument &  Default & Purpose & \lstinline!type! \\ 
\midrule
\lstinline!psf_dir! & \lstinline!None!  & Where PSF arrays are stored  & \lstinline!str!  \\  
\lstinline!num_f_bins! & \lstinline!10! & Number of linear-spaced points in\lstinline!f_ary!  & \lstinline!int!  \\  
\lstinline!n_psf! & \lstinline!50000! & Number of MC simulations for determining \lstinline!df_rho_div_f_ary!  & \lstinline!int!  \\  
\lstinline!n_pts_per_psf! & \lstinline!1000! & Number of points drawn for each MC simulation & \lstinline!int! \\  
\lstinline!f_trunc! & \lstinline!0.01! & Minimum $f$ value & \lstinline!float!  \\  
\lstinline!nside! & \lstinline!128! & \lstinline!HEALPix! parameter for size of map  & \lstinline!int!  \\ 
\lstinline!psf_sigma_deg! & \lstinline!None! & Standard deviation $\sigma$ of 2-D Gaussian PSF  & \lstinline!float!  \\ 
\lstinline!delay_compute! & \lstinline!False! & If \lstinline!True!, PSF not Gaussian and will be specified later  & \lstinline!bool!  \\ 
\bottomrule
\end{tabular}
\end{center}

Note that the arrays  \lstinline{f_ary} and \lstinline{df_rho_div_f_ary} depend both on the PSF of the detector as well as the pixelation of the data; at present the \lstinline{PSFCorrection} class requires the pixelation to be in the \texttt{HEALPix} pixelation.

The keyword \lstinline{psf_dir} points to the directory where the \lstinline{f_ary} and \lstinline{df_rho_div_f_ary} will be stored; if unspecified, they will be stored to the current directory.  The \lstinline{f_ary} consists of \lstinline{num_f_bins} entries linear spaced between $0$ and $1$.  The PSF correction involves placing many \mbox{(\lstinline{n_psf})} PSFs at random positions on the \texttt{HEALPix} map, drawing \lstinline{n_pts_per_psf} points from each PSF, and then looking at the distribution of points among pixels.  The larger \lstinline{n_psf} and \lstinline{n_pts_per_psf}, the more accurate the computation of \lstinline{df_rho_div_f_ary} will be.  However, the computation time of the PSF arrays also increases as these parameters are increased.  

By default the \lstinline{PSFCorrection} class assumes that the PSF is a 2-D Gaussian distribution:
\es{2DGaussian}{
\text{PSF}(r) = {\frac{1}{2 \pi \sigma^2}} \exp\left[ - {\frac{r^2}{2 \sigma^2}} \right] \,.
}     
Here, $\text{PSF}(r)$ describes the spread of arriving counts with angular distance $r$ away from the arrival direction.  The parameter \lstinline{psf_sigma_deg} denotes $\sigma$ in degrees.  Upon initializing \lstinline{PSFCorrection} with \lstinline{psf_sigma_deg} specified, the class automatically computes the array \lstinline{df_rho_div_f_ary} and stores it in the \lstinline{psf_dir} with a unique name related to the keywords.  If such a file already exists in the \lstinline{psf_dir}, then the code will simply load this file instead of recomputing it.  After initialization, the relevant arrays may be accessed by \lstinline{pc_inst.f_ary} and  \lstinline{pc_inst.df_rho_div_f_ary}. 

The \lstinline{PSFCorrection} class can also handle arbitrary PSF functions.  In this case, the class should be initialized with \lstinline{delay_compute = True}.  Then, the user should manually set the function \lstinline{pc_inst.psf_r_func} to the desired function $\text{PSF}(r)$.  This function will be discretized with \lstinline{pc_inst.psf_samples} points out to \lstinline{pc_inst.sample_psf_max} degrees from $r=0$.  These two quantities also need to be manually specified.  The user also needs to set \lstinline{pc_inst.psf_tag} to a string that will be used for saving the PSF arrays.  After these four attributes have been set manually by the user, the PSF arrays are computed and stored by executing \mbox{\lstinline{pc_inst.make_or_load_psf_corr()}}.

\subsection*{ \lstinline{def NPTFit.create_mask.make_mask_total} }
 
This function is used to make masks that can then be used to reduce the data and templates to a smaller ROI when performing the scan.  While these masks can always be made by hand, this function provides a simple masking interface for maps in the \texttt{HEALPix} format.  The \lstinline{make_mask_total} function can mask pixels by latitude, longitude, and radius from any point on the sphere.  See the \href{http://nptfit.readthedocs.io}{online documentation} for more specific examples.
 
\subsection*{ \lstinline{class NPTFit.dnds_analysis.Analysis} }
  
The analysis class may be used to extract useful information from the results of an NPTF performed using \texttt{MultiNest}.  The class also has built-in plotting features for making many of the most common types of visualizations for the parameter posterior distribution.  An instance of the analysis class can be instantiated by 
\begin{lstlisting}
an = Analysis(nptf, mask=None, pixarea=0.)
\end{lstlisting}
where \lstinline{nptf} is itself an instance of the \lstinline{NPTF} class that already has the results of a scan loaded.  The keyword arguments \lstinline{mask} and \lstinline{pixarea} are optional. The user should specify a \lstinline{mask} if the desired ROI for the analysis is different that that used in the scan.  The user should specify a \lstinline{pixarea} if the data is not in the \texttt{HEALPix} format.  The code will still assume the pixels are equal area with area \lstinline{pixarea}, which should be specified in sr.   

After initialization, the intensities of Poissonian and non-Poissonian templates, respectively, may be extracted from the analysis class by the commands 
\begin{lstlisting}
an.return_intensity_arrays_poiss(comp) 
\end{lstlisting}
and
\begin{lstlisting}
an.return_intensity_arrays_non_poiss(
    comp)
\end{lstlisting}
Here, \lstinline{comp} refers to the template key used by the Poissonian or non-Poissonian model.  The arrays returned give the mean intensities of that model in the ROI in units of  $\text{counts}\, \text{cm}^{-2}\text{s}^{-1}$, assuming the exposure map was in units of cm$^2$s.  The arrays computed over the full set of entries in the discretized posterior distribution output by \texttt{MultiNest}.  Thus, these intensity arrays may be interpreted as the 1-D posteriors for the intensities.  For additional keywords that may be used to customize the computation of the intensity arrays, see the \href{http://nptfit.readthedocs.io}{online documentation}.

The source-count distributions may also be accessed from the analysis class.  Executing
\begin{lstlisting}
an.return_dndf_arrays(comp, flux)
\end{lstlisting}
will return the discretized 1-D posterior distribution for $\text{mean}_\text{ROI} dN_p(F)/dF$ at flux $F$ for the PS model with template key \lstinline{comp}.  Note that the mean is computed over pixels $p$ in the ROI.  

The 1-D posterior distributions for the individual model parameters may be accessed by 
\begin{lstlisting}
A_poiss_post = an.return_poiss_parameter_posteriors(
    comp)
\end{lstlisting}
for Poissonian models, and 
\begin{lstlisting}
A_non_poiss_post, n_non_poiss_post, Sb_non_poiss_post = an.return_non_poiss_parameter_posteriors(comp)
\end{lstlisting}
for non-Poissonian models.  Here \lstinline{A_poiss_post} is a 1-D array of the discretized posterior distribution for the Poissonian template normalization parameter.  Similarly, \lstinline{A_non_poiss_post} is the posterior array for the non-Poissonian normalization parameter.  The arrays \lstinline{n_non_poiss_post} and \lstinline{Sb_non_poiss_post} are 2-D, where---for example---\mbox{\lstinline{n_non_poiss_post = [n_1_array, n_2_array, ...]}} and \lstinline{n_1_array} is a 1-D array for the posterior for $n_1$.   

Another useful piece of information that may be extracted from the scan is the Bayesian evidence:
\begin{lstlisting}
l_be, l_be_err = an.get_log_evidence()
\end{lstlisting}
returns the log of the Bayesian evidence along with the uncertainty on this estimate based on the resolution of the MCMC.

For information on the plotting capabilities in the analysis class, see the \href{http://nptfit.readthedocs.io}{online documentation} or the example in the following section.

\section{\texttt{NPTFit}: An Example}
\label{NPTFit-example}

In this section we give an example for how to perform an NPTF using \texttt{NPTFit}.  Many more examples are available in the \href{http://nptfit.readthedocs.io}{online documentation}.  This particular example reproduces aspects of the main results of~\cite{Lee:2015fea}, which found evidence for a spherical population of unresolved gamma-ray PSs around the Galactic Center.  The example uses the processed, public {\it Fermi} data made available with the release of the \texttt{NPTFit} package.  The data set consists of 413 weeks of \emph{Fermi} Pass 8 data in the UltracleanVeto event class (top quartile of events as ranked by PSF) from 2 to 20 GeV.  The map is binned in \texttt{HEALPix} with $\mathtt{nside} = 128$.  The data, along with the exposure map and background templates, may be downloaded from 
\begin{center}
\url{http://hdl.handle.net/1721.1/105492}.
\end{center}

In the example we will perform an NPTF on the sub-region where we mask the Galactic plane at latitude $|b| < 2^\circ$ and mask pixels with angular distance greater than $30^\circ$ from the Galactic Center.  We also mask identified PSs in the 3FGL PS catalog~\cite{Acero:2015hja} at 95\% containment using the provided PS mask, which is added to the geometric mask.  We include smooth templates for diffuse gamma-ray emission in the Milky Way (using the {\it Fermi} \texttt{p6v11} diffuse model), isotropic emission (which can also absorb instrumental backgrounds), and emission following the {\it Fermi} bubbles, which are taken to be uniform in flux following the spatial template in~\cite{Su:2010qj}.  We also include a dark matter template, which traces the line of sight integral of the square of a canonical NFW density profile.

We additionally include point source (non-Poissonian) models for the DM template, as well as for a disk template which corresponds to a doubly exponential thin-disk source distribution with scale height 0.3 kpc and radius 5 kpc. The source-count distributions for these are parameterized by singly-broken power laws, each described by four parameters $\{ A, F_{b,1}, n_1, n_2 \}$.

\subsection{Setting Up the Scan}

We begin the example by loading in the relevant modules, described in the previous section, that we will need to setup, perform, and analyze the scan.
\begin{lstlisting}
import numpy as np
# module for performing scan
from NPTFit import nptfit
# module for creating the mask
from NPTFit import create_mask as cm 
# module for determining the PSF correction
from NPTFit import psf_correction as pc 
# module for analyzing the output
from NPTFit import dnds_analysis 
\end{lstlisting}
Next, we create an instance of the \lstinline{NPTF} class, which is used to configure and perform a scan.
\begin{lstlisting}
n = nptfit.NPTF(tag='GCE_Example')
\end{lstlisting}
We assume here that the supplementary {\it Fermi} data has been downloaded to a directory \lstinline{'fermi_data'}.  Then, we may load in the data and exposure maps by 
\begin{lstlisting}
fermi_data = np.load('fermi_data/fermidata_counts.npy').astype(int)
fermi_exposure = np.load('fermi_data/fermidata_exposure.npy')
n.load_data(fermi_data, fermi_exposure)
\end{lstlisting}
Importantly, note that the exposure map has units of cm$^2$s.  Next, we use the \lstinline{create_mask} class to generate our ROI mask, which consists of both the geometric mask and the PS mask loaded in from the \lstinline{'fermi_data'} directory:
\begin{lstlisting}
pscmask=np.array(np.load('fermi_data/fermidata_pscmask.npy'), dtype=bool)
mask = cm.make_mask_total(band_mask = True, band_mask_range = 2, mask_ring = True, inner = 0, outer = 30, custom_mask = pscmask)
n.load_mask(mask)
\end{lstlisting}
The templates may also be loaded in from this directory,
\begin{lstlisting}
dif = np.load('fermi_data/template_dif.npy')
iso = np.load('fermi_data/template_iso.npy')
bub = np.load('fermi_data/template_bub.npy')
gce = np.load('fermi_data/template_gce.npy')
dsk = np.load('fermi_data/template_dsk.npy')
\end{lstlisting}
These templates are counts map (i.e. flux maps times the exposure map) that have been pre-smoothed by the PSF (except for the disk-correlated template labeled \lstinline{dsk}).  We then add them to our \lstinline{NPTF} instance with appropriately chosen keywords:
\begin{lstlisting}
n.add_template(dif, 'dif')
n.add_template(iso, 'iso')
n.add_template(bub, 'bub')
n.add_template(gce, 'gce')
n.add_template(dsk, 'dsk')

# remove the exposure correction for PS templates
rescale = fermi_exposure/np.mean(fermi_exposure)
n.add_template(gce/rescale, 'gce_np', units='PS')
n.add_template(dsk/rescale, 'dsk_np', units='PS')

\end{lstlisting}

Note that templates \lstinline{'gce_np'} and \lstinline{'dsk_np'} intended to be used in non-Poissonian models should trace the underlying PS distribution, without exposure correction, and are added with the keyword \lstinline{units='PS'}.

\subsection{Adding Models}
Now that we have loaded in all of the external data and templates, we can add models to our \lstinline{NPTF} instance.  First, we add in the Poissonian models,
\begin{lstlisting}
n.add_poiss_model('dif', '$A_\mathrm{dif}$', False, fixed=True, fixed_norm=14.88)
n.add_poiss_model('iso', '$A_\mathrm{iso}$', [0,2], False)
n.add_poiss_model('gce', '$A_\mathrm{gce}$', [0,2], False)
n.add_poiss_model('bub', '$A_\mathrm{bub}$', [0,2], False)
\end{lstlisting}
All Poissonian models are taken to have linear priors, with prior ranges for the normalizations between 0 and 2.  However, the normalization of the diffuse background has been fixed to the value $14.67$, which is approximately the correct normalization in these units for this template, in order to provide an example of this syntax.  Next, we add in the two non-Poissonian models:
\begin{lstlisting}
n.add_non_poiss_model('gce_np', ['$A_\mathrm{gce}^\mathrm{ps}$','$n_1^\mathrm{gce}$','$n_2^\mathrm{gce}$','$S_b^{(1), \mathrm{gce}}$'], [[-6,1],[2.05,30],[-2,1.95],[0.05,40]], [True,False,False,False])
n.add_non_poiss_model('dsk_np', ['$A_\mathrm{dsk}^\mathrm{ps}$','$n_1^\mathrm{dsk}$','$n_2^\mathrm{dsk}$','$S_b^{(1), \mathrm{dsk}}$'], [[-6,1],[2.05,30],[-2,1.95],[0.05,40]], [True,False,False,False])
\end{lstlisting}
We have added in the models for disk-correlated and NFW-correlated (line of sight integral of the the NFW distribution squared) unresolved PS templates.  Each of these models takes singly-broken power-law source-count distributions.  In this configuration, the normalization parameters are taken to have a log-flat prior while the indices and breaks are taken to have linear priors (relevant for the Bayesian posterior sampling).  The units of the breaks are specified in terms of counts. 

\subsection{Configure Scan with PSF Correction}
In this energy range and with this data set, the PSF may be modeled by a 2-D Gaussian distribution with $\sigma = 0.1812^\circ$.  From this, we are able to construct the PSF-correction arrays:\footnote{For an example of how to construct these arrays with a more complicated, non-Gaussian PSF function, see the \href{http://nptfit.readthedocs.io}{online documentation}.} 
\begin{lstlisting}
pc_inst = pc.PSFCorrection(psf_sigma_deg=0.1812)
f_ary, df_rho_div_f_ary = pc_inst.f_ary, pc_inst.df_rho_div_f_ary
\end{lstlisting}
These arrays are then passed into the \lstinline{NPTF} instance when we configure the scan:
\begin{lstlisting}
n.configure_for_scan(f_ary, df_rho_div_f_ary, nexp=1)
\end{lstlisting}
Note that since our ROI is relatively small and the exposure map does not change significantly over the region, we have a single exposure region with \lstinline{nexp=1}.

\subsection{Performing the Scan With \texttt{MultiNest}}
We perform the scan using \texttt{MultiNest} with \lstinline{nlive=500} as an example to demonstrate the basic features and conclusions of this analysis while being able to perform the scan in a reasonable amount of time on a single processor, although ideally \lstinline{nlive} should be set to a higher value for more reliable results:
\begin{lstlisting}
n.perform_scan(nlive=500)
\end{lstlisting}

\subsection{Analyzing the Results}
Now, we are ready to analyze the results of the scan.  First we load in relevant modules:
\begin{lstlisting}
import corner
import matplotlib.pyplot as plt
\end{lstlisting}
and then we load in the results of the scan (configured as above),
\begin{lstlisting}
n.load_scan()
\end{lstlisting}
The chains, giving a discretized view of the posterior distribution, may be accessed simply through the attribute \lstinline{n.samples}.  However, we will analyze the results by using the analysis class provided with \texttt{NPTFit}.  We make an instance of this class simply by
\begin{lstlisting}
an = dnds_analysis.Analysis(n)
\end{lstlisting}

\subsubsection{Making Corner Plots}

\begin{figure*}[htb]
\leavevmode
\begin{center}
\includegraphics[width=.98\textwidth]{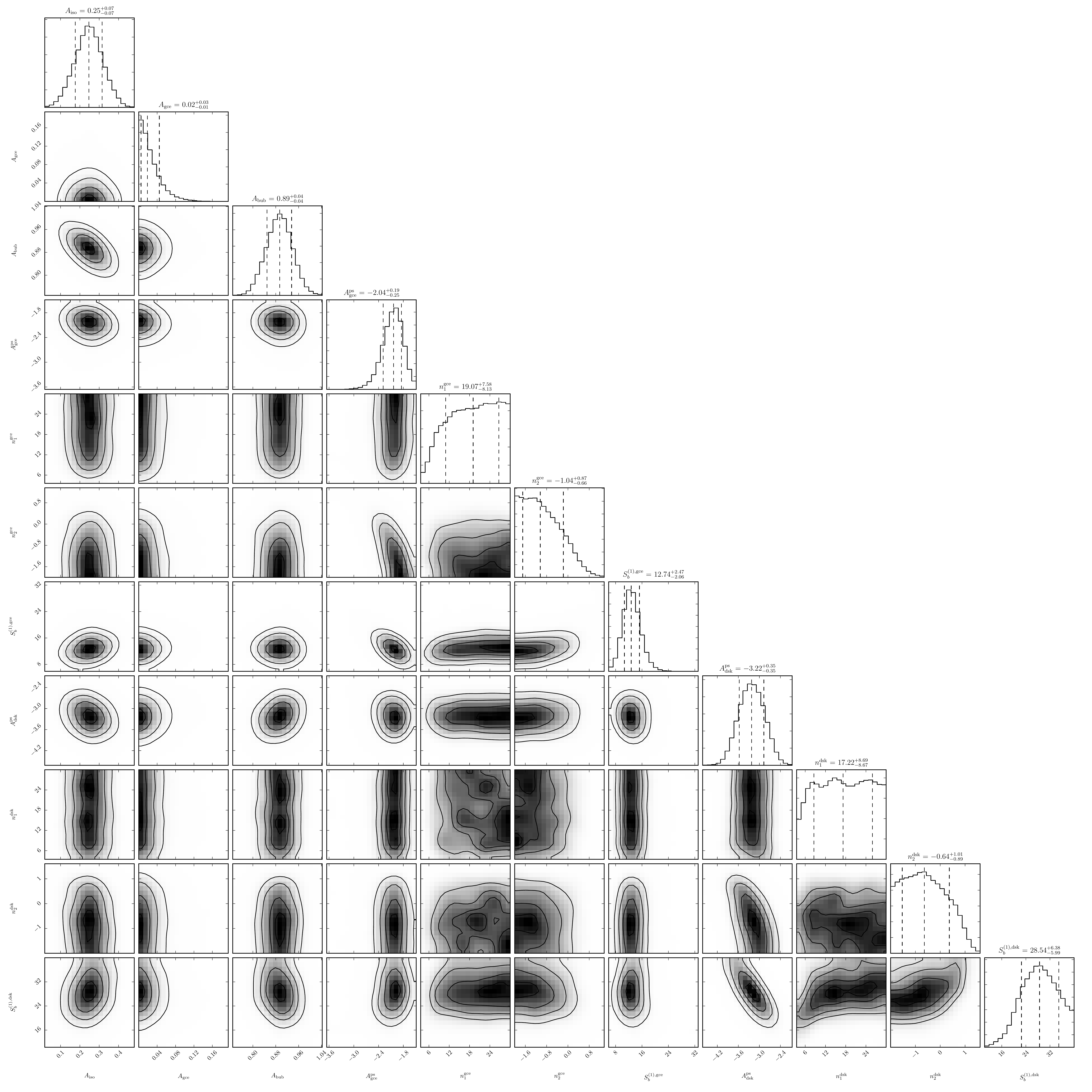}
\end{center}
\vspace{-.50cm}
\caption{The corner plot obtained by analyzing the results of an NPTF in the Galactic Center, showing the one and two dimensional posteriors of the 11 parameters floated in the fit corresponding to three Poissonian and two non-Poissonian templates. For this analysis 3FGL point sources have been masked at 95\% containment. See text for details.}
\label{fig:gc_triangle}
\end{figure*}

Corner (or triangle) plots are a simple and quick way of visualizing correlations in the posterior distribution.  Such plots may be generated through the command
\begin{lstlisting}
an.make_triangle()
\end{lstlisting}
which leads to the plot in Fig.~\ref{fig:gc_triangle}.

\subsubsection{Plotting Source-count Distributions}

\begin{figure}[htb]
\leavevmode
\begin{center}
\includegraphics[width=.8\textwidth]{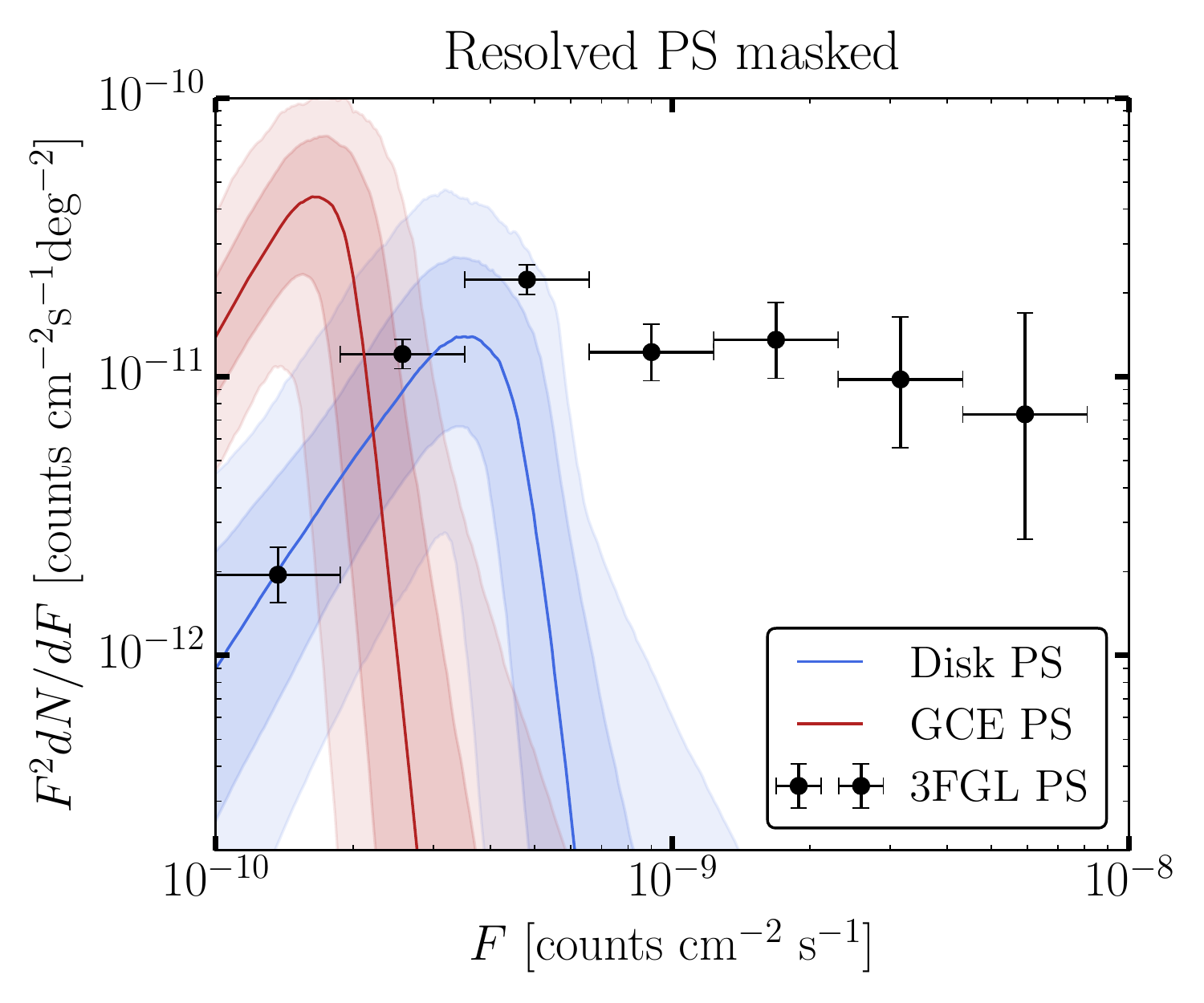}
\end{center}
\vspace{-.50cm}
\caption{The source-count distribution as constructed from the analysis class, for the example NPTF described in the main text.  This scan looks for  disk-correlated PSs along with PSs correlated with the expected DM template (GCE PSs).  Since all resolved PSs are masked in this analysis, the source-count distributions are seen to contribute dominantly below the 3FGL detection threshold.  A histogram of resolved 3FGL sources is also shown.}
\label{fig:gc_dndf}
\end{figure}

The source-count distributions for NFW- and disk-correlated point source models may be plotted with
\begin{lstlisting}
an.plot_source_count_median('dsk',smin=0.01,smax=1000,nsteps=1000,color='cornflowerblue',spow=2,label='Disk')
an.plot_source_count_band('dsk',smin=0.01,smax=1000,nsteps=1000,qs=[0.16,0.5,0.84],color='cornflowerblue',alpha=0.3,spow=2)
an.plot_source_count_median('gce',smin=0.01,smax=1000,nsteps=1000,color='forestgreen',spow=2,label='GCE')
an.plot_source_count_band('gce',smin=0.01,smax=1000,nsteps=1000,qs=[0.16,0.5,0.84],color='forestgreen',alpha=0.3,spow=2)
\end{lstlisting}
along with the following \texttt{matplotlib} plotting options. 
\begin{lstlisting}
plt.yscale('log')
plt.xscale('log')
plt.xlim([5e-11,5e-9])
plt.ylim([2e-13,1e-10])
plt.tick_params(axis='x', length=5, width=2, labelsize=18)
plt.tick_params(axis='y', length=5, width=2, labelsize=18)
plt.ylabel('$F^2 dN/dF$ [counts/cm$^2$/s/deg$^2$]', fontsize=18)
plt.xlabel('$F$  [counts/cm$^2$/s]', fontsize=18)
plt.title('Galactic Center NPTF', y=1.02)
plt.legend(fancybox=True)
plt.tight_layout()
\end{lstlisting}
This is shown in Fig.~\ref{fig:gc_dndf}. Contribution from both NFW- and disk-correlated PSs may be seen, with NFW-correlated sources contributing dominantly at lower flux values.  In that figure, we also show a histogram of the detected 3FGL sources within the relevant energy range and region, with vertical error bars indicating the 68\% confidence interval from Poisson counting uncertainties only.\footnote{The data for plotting these points is available in the \href{http://nptfit.readthedocs.io}{online documentation}.}  Since we have explicitly masked all 3FGL sources, we see that the disk- and NFW-correlated PS templates contribute at fluxes near and below the 3FGL PS detection threshold, which is $\sim$$5 \times 10^{-10}$ counts cm$^{-2}$ s$^{-1}$ in this case.

\subsubsection{Plotting Intensity Fractions}

The intensity fractions for the smooth and PS NFW-correlated models may be plotted with

\begin{lstlisting}
an.plot_intensity_fraction_non_poiss('gce', bins=800, color='cornflowerblue', label='GCE PS')
an.plot_intensity_fraction_poiss('gce', bins=800, color='lightsalmon', label='GCE DM')
plt.xlabel('Flux fraction (%)')
plt.legend(fancybox = True)
plt.xlim(0,6)
\end{lstlisting}

\begin{figure}[htbp]
\centering
\includegraphics[width=0.8\textwidth]{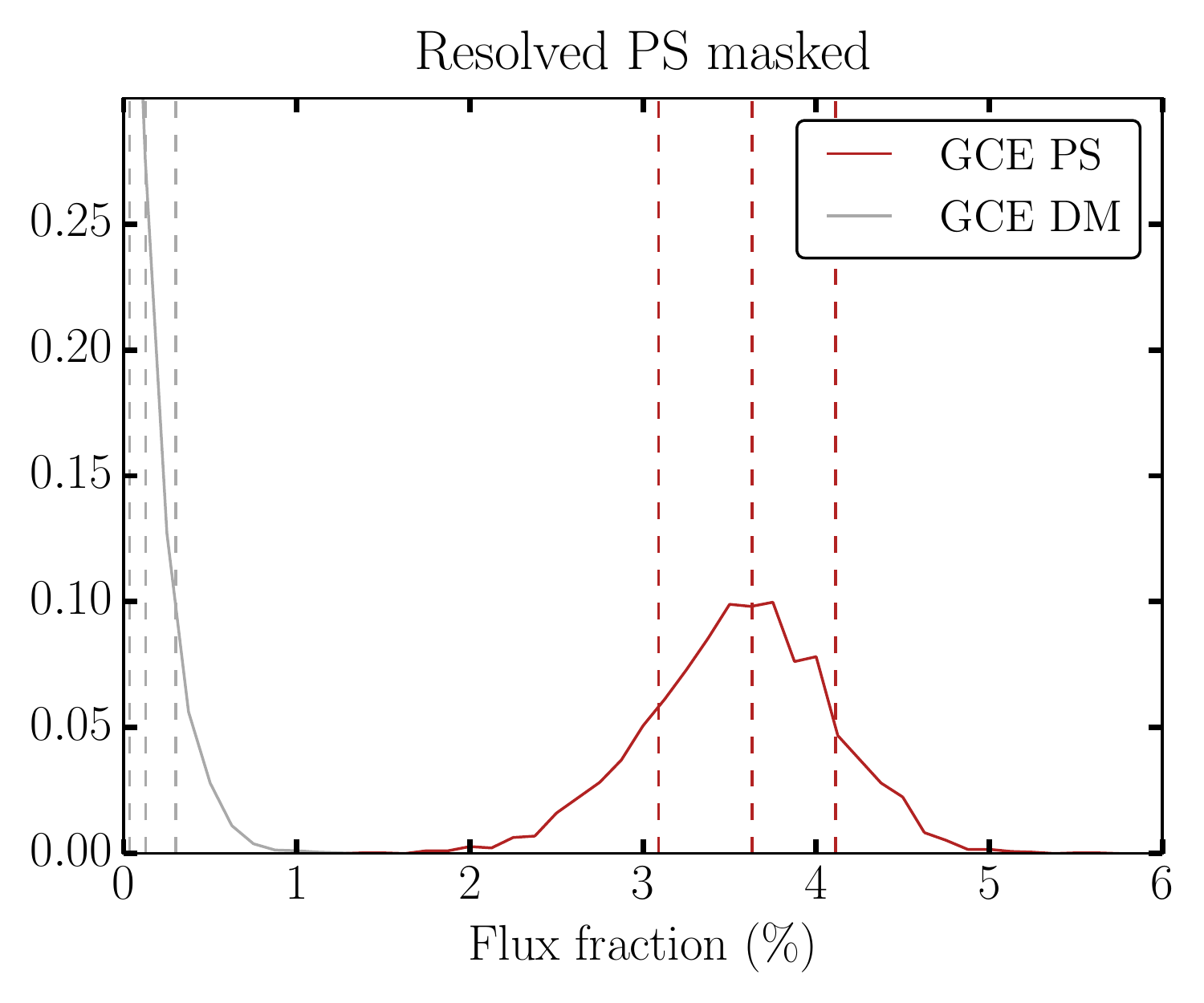} 
\caption{Intensity fractions for the smooth (green) and point source (red) templates correlating with the DM template, obtained by analyzing the results of an NPTF in the Galactic Center with 3FGL point sources masked at 95\% containment.}
\label{fig:gc_intensity}
\end{figure}

This is shown in Fig.~\ref{fig:gc_intensity}. We immediately see a preference for NFW-correlated point sources over the smooth NFW component.

\subsubsection{Further Analyses}

The example above may easily be pushed further in many directions, many of which are outlined in~\cite{Lee:2015fea}.  For example, a natural method for performing model comparison in the Bayesian framework is to compute the Bayes factor between two models.  Here, for example, we may compute the Bayes factor between the model with and without NFW-correlated PSs.  This involves repeating the scan described above but only adding in disk-correlated PSs.  Then, by comparing the global Bayesian evidence between the two scans (see Sec.~\ref{NPTFit-orientation} for the syntax on how to extract the Bayesian evidence), we find a Bayes factor $\sim$$10^3$ in preference for the model with spherical PSs.

\begin{figure}[htb]
\leavevmode
\begin{center}
\includegraphics[width=.8\textwidth]{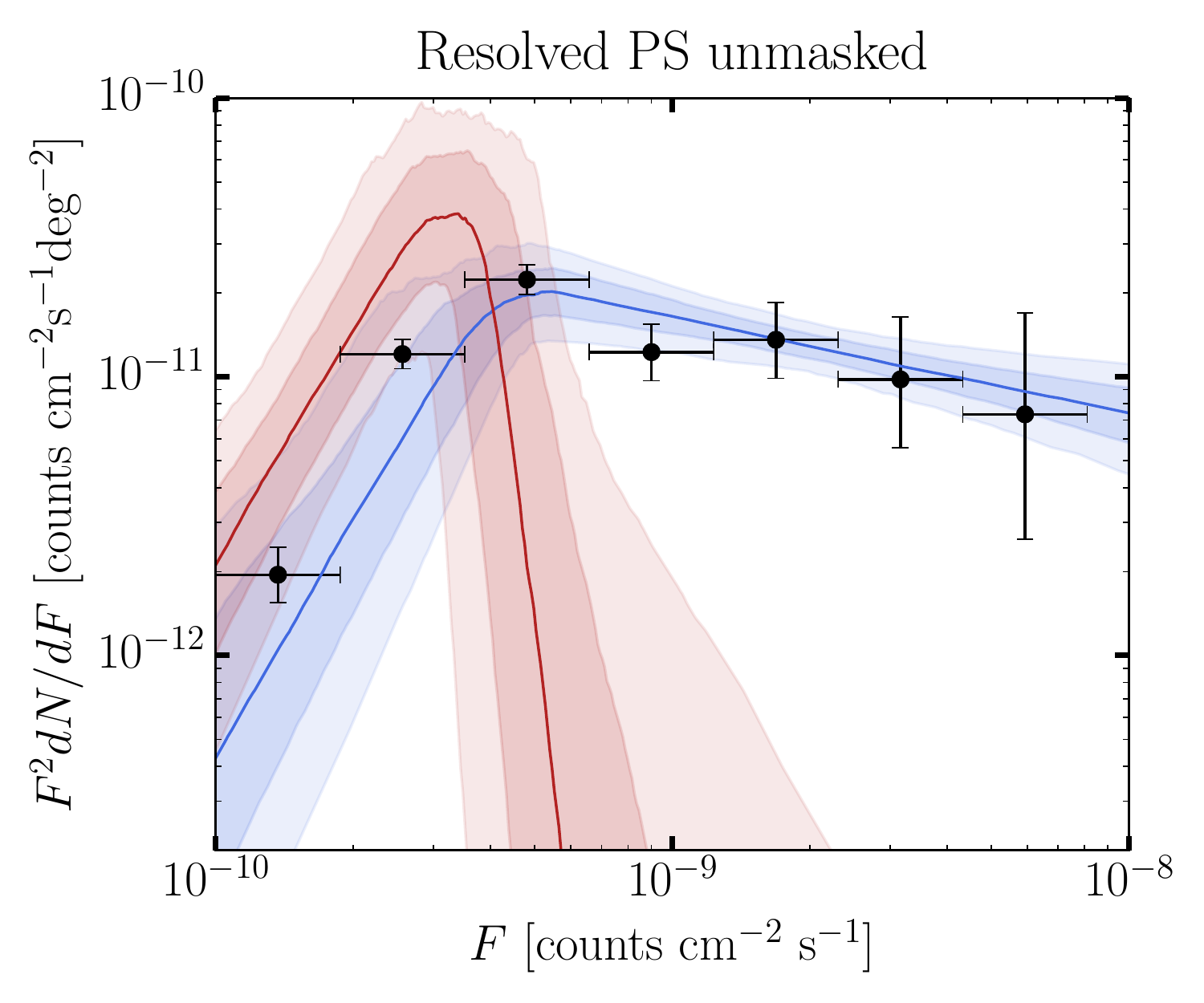}
\end{center}
\vspace{-.50cm}
\caption{As in Fig.~\ref{fig:gc_dndf}, but in this case the resolved 3FGL sources were not masked.  The disk-correlated template accounts for the majority of the resolved PS emission. }
\label{dNdF: unmasked}
\end{figure}   

Another straightforward generalization of the example described above is simply to leave out the PS mask, so that the NFW- and disk-correlated PS templates must account for both the resolved and unresolved PSs.  The likelihood evaluations take longer, in this case, since there are pixels with higher photon counts compared to the 3FGL-masked scan.  The result for the source-count distribution from this analysis is shown in Fig.~\ref{dNdF: unmasked}.  In this case, the disk-correlated PS template accounts for the resolved 3FGL sources, while the NFW-correlated PS template contributes at roughly the same flux range as in the 3FGL masked case.  The Bayes factor in preference for the model with NFW-correlated PSs over that without---as described above---is found to be $\sim$$10^{10}$ in this case.

\section{Conclusion}
\label{Conclusion}

We have presented an open-source code package for performing non-Poissonian template fits.   We strongly recommend referring to the \href{http://nptfit.readthedocs.io}{online documentation}---which will be kept up-to-date---in addition to this chapter accompanying the initial release.  There are many way in which \texttt{NPTFit} can be improved in the future.  For one, the  \texttt{NPTFit} package only handles a single energy bin at a time.  In a later version of the code we plan to incorporate the ability to scan over multiple energy bins simultaneously.  Additionally, there are a few areas---such as the evaluation of the incomplete gamma functions---where the \texttt{cython} code may still be sped up.  Such improvements to the computational cost are relevant for analyses of large data sets with many model parameters.  Of course, we welcome additional suggestions for how we may improve the code and better adapt it to applications beyond the gamma-ray applications it has been used for so far. 

\sectionline


\chapter{Application of Non-Poissonian Template Fitting to the Extragalactic Gamma-Ray Background}
\label{ch:igrb}

\setcounter{footnote}{0}

This chapter is based on an edited version of \emph{Deciphering Contributions to the Extragalactic Gamma-Ray Background from 2 GeV to 2 TeV},  \href{http://iopscience.iop.org/article/10.3847/0004-637X/832/2/117/meta}{Astrophys.J. \textbf{832} (2016) no.2, 117} \href{https://arxiv.org/abs/1606.04101}{[arXiv:1606.04101]} with Mariangela Lisanti, Lina Necib and Benjamin Safdi~\cite{Lisanti:2016jub}. The results of this chapter have been presented at the following conferences and workshops: \emph{Gamma Rays and Dark Matter} in Obergurgl, Austria (December 2015), \emph{TeV Particle Astrophysics (TeVPA) 2016} in Geneva, Switzerland (September 2016) and \emph{APS April Meeting 2017} in Washington, DC (January 2017).

\section{Introduction}

\lettrine[lines=3]{T}{he} Extragalactic Gamma-Ray Background (EGB) is the nearly isotropic all-sky emission that arises from sources outside of the Milky Way.  The OSO-3~\cite{1968ApJ...153L.203C, 1972ApJ...177..341K} and SAS-2 satellites~\cite{1975ApJ...198..163F,1978ApJ...222..833F} were the first to see hints of the EGB and have since been followed by EGRET~\cite{1998ApJ...494..523S,2004ApJ...613..956S} and, most recently, the \emph{Fermi} Large Area Telescope\footnote{\url{http://fermi.gsfc.nasa.gov/}}~\cite{Ackermann:2014usa,TheFermi-LAT:2015ykq}.  The origin of the EGB remains an open question.  The dominant contributions are likely due to blazars~\cite{Stecker:1993ni, Stecker:1996ma, Muecke:1998cs, Narumoto:2006qg,Dermer:2007fg, Pavlidou:2007dv, Ajello:2009ip, Collaboration:2010gqa, Abazajian:2010pc,  Stecker:2010di,Singal:2011yi,  Ajello:2011zi, Ajello:2013lka,DiMauro:2013zfa, Ajello:2015mfa, Ackermann:2015uya}, star-forming galaxies (SFGs)~\cite{Soltan:1998jg,Pavlidou:2002va, Bhattacharya:2009yv, Ando:2009nk, Fields:2010bw, Makiya:2010zt, Ackermann:2012vca, Chakraborty:2012sh, Lacki:2012si, Tamborra:2014xia}, and misaligned active galactic nuclei (mAGN)~\cite{Stawarz:2005tq, Inoue:2011bm, Massaro:2011ww, DiMauro:2013xta, Hooper:2016gjy}.  Understanding the relative contributions of these source components to the EGB has taken on a new sense of importance in light of IceCube's observation of ultra-high-energy extragalactic neutrinos~\cite{Aartsen:2013bka,Aartsen:2013jdh,Aartsen:2015knd,Aartsen:2015rwa}, the origin of which still remains a mystery.  For instance, the same sources that dominate the extragalactic neutrino background at $\sim$PeV energies may also contribute significantly to the EGB from $\sim$GeV--TeV energies~\cite{Murase:2013rfa, Tamborra:2014xia,Hooper:2016gjy}.  In addition, the EGB may harbor the imprints of more exotic physics such as dark matter annihilation or decay~\cite{Bengtsson:1990xf,Bergstrom:2001jj,Ullio:2002pj,Bottino:2004qi,Bertone:2004pz,Bringmann:2012ez,Ajello:2015mfa, DiMauro:2015tfa, Ackermann:2015tah}, as well as contributions from truly diffuse processes such as propagating ultra-high-energy cosmic rays~\cite{Loeb:2000aa,Kalashev:2007sn,Ahlers:2011sd,Murase:2012df,Taylor:2015rla} and structure formation shocks in clusters of galaxies~\cite{Murase:2008yt,Zandanel:2014pva}.
Given the potential wealth of information that can be extracted from the EGB, deciphering its constituents remains a high priority.     

Most recently, \emph{Fermi} presented a measurement of the EGB intensity from 100~MeV to 820~GeV~\cite{Ackermann:2014usa}. The total EGB intensity is the \textit{sum of all resolved point sources (PSs) and smooth isotropic emission}. The smooth emission, referred to as the Isotropic Gamma-Ray Background (IGRB), arises from \textit{PSs that are too faint to be resolved individually as well as other truly diffuse processes.}    It is also important to note that both the EGB and IGRB may be contaminated by cosmic rays that are mis-identified as gamma rays; this emission is expected to be smoothly distributed across the sky.
Of the known gamma-ray emitting PSs at high latitudes, which are captured by Fermi's 3FGL~\cite{Acero:2015hja} catalog from 0.1--300 GeV and the more recent 2FHL~\cite{Ackermann:2015uya} catalog from 50--2000 GeV, the dominant source class is blazars.

In this chapter, we use the analysis method Non-Poissonian Template Fitting (NPTF) introduced in the previous chapter, to study the source populations that contribute to the EGB in a data-driven manner.  The method relies on photon-count statistics to illuminate the aggregate properties of a source population, even when its constituents are not individually resolvable~\cite{Malyshev:2011zi,Lee:2014mza, Lee:2015fea}.  This allows us to constrain the contribution of PSs to the EGB whose flux is too dim to be detected individually.  While at very low fluxes the NPTF also loses the ability to distinguish PSs from smooth emission, the threshold for PS detection is lower for the NPTF than it is for other techniques that rely on finding individually-significant sources.  This is because the NPTF only measures the aggregate properties of a PS population. 

Using the NPTF, we are able to recover, for the first time, the source-count distribution (\emph{e.g.}, flux distribution) for isotropically distributed PSs at high Galactic latitudes, as a function of energy from 1.89 GeV to 2~TeV.  This builds on previous studies that use related methods to obtain the source-count distributions in single energy bins from $\sim$2--12~GeV~\cite{Zechlin:2015wdz,Lee:2015fea} and from 50--2000~GeV~\cite{TheFermi-LAT:2015ykq}. 

The source-count distribution for a given astrophysical population convolves information about its cosmological evolution.  For a flat, non-expanding universe, a uniformly distributed population of galaxies has a differential source-count distribution $dN/dF \propto F^{-5/2}$, where $F$ is the source flux at Earth and $dN$ is the differential number of sources~\cite{Sandage}.  This is the well-known Euclidean limit.  However, the power-law index changes when one takes the standard $\Lambda$CDM cosmology and more realistic assumptions for the redshift evolution of source-dependent observables such as luminosity.  Therefore, the features of the source-count distribution---especially, its power-law indices and/or flux breaks---encode information about the number of source classes contributing to the EGB as well as their cosmological evolution.

These source-count distributions provide the keys for interpreting the GeV--TeV sky.  For example, they enable us to obtain the intensity spectrum for PSs, down to a certain flux threshold, as a function of energy.  We find that while the EGB is dominated by PSs, likely blazars, in the entire energy range from 1.89--2000~GeV, there is also room for other source classes, which contribute flux more diffusely, to produce a sizable fraction of the EGB.    
Our findings may therefore leave open the possibility that IceCube's PeV neutrinos~\cite{Aartsen:2013bka,Aartsen:2013jdh,Aartsen:2015knd,Aartsen:2015rwa} can be explained by $pp$ hadronic interactions in \emph{e.g.}, SFGs~\cite{Murase:2013rfa,Tamborra:2014xia,Ando:2015bva} or mAGN~\cite{Hooper:2016jls}, which---as we show in Sec.~\ref{sec:simulations}---show up as smooth isotropic emission under the NPTF.  Additionally, the high-energy source-count distributions allow us to make predictions for the number of blazars, which dominate the high-energy data, that will be resolved by upcoming TeV observatories such as the Cherenkov Telescope Array (CTA)~\cite{2011arXiv1111.2183C, Dubus:2012hm}.  
While our analysis does not let us conclusively identify the locations of these sources, we provide maps showing the locations on the sky where, statistically, there are most likely to be PSs.

This chapter is organized as follows.  We begin in Sec.~\ref{sec:methodology} by reviewing the analysis methods.  Sec.~\ref{sec:simulations} then applies these methods to simulated sky maps.  We cannot stress the importance of these simulated data studies enough; they are crucial for proving the stability of the analysis methods and laying the foundation for the data results that follow.  Our data study is divided into two separate analyses for low (1.89--94.9~GeV) and high (50--2000~TeV) energies, described in Sec.~\ref{sec:lowenergy} and \ref{sec:highenergy}, respectively. The global fits to the full energy range, as well as their implications, are discussed in Sec.~\ref{sec:conclusions_igrb}.  Further details on the creation of the simulated data maps and supplementary analysis plots are provided in the Appendix.  The main results of this chapter are summarized in a few key figures.  In particular, the source-count distributions for the low and high-energy analyses are shown in Figs. \ref{fig:dndsdata}, \ref{fig:dndsdata_HE}, and \ref{fig:dnds_0_10}, respectively, while Fig. \ref{fig:global} presents a spectral fit to the PS intensity from 2 GeV to 2 TeV.

\section{Methodology} 
\label{sec:methodology}

In this chapter, we make use of both Poissonian and non-Poissonian template-fitting techniques.  
Poissonian template fitting is a standard tool in astrophysics for decomposing a sky map into component ``templates" with different spatial morphologies.  The NPTF builds upon this technique by allowing for the addition of templates whose spatial morphology traces the distribution of a PS population, even if the exact position of the sources that make up that population are not known.  More precisely, in both template-fitting procedures one starts with a data set $d$ that consists of counts $n_p$ in each pixel $p$.\footnote{We will only work with a single energy bin at a time for simplicity, though in principle model parameters may be shared between energy bins.  In this case, the likelihood function over the full energy range may be written as the product of the likelihood functions in the energy sub-bins.}  One then fits a model ${\cal M}$ with parameters $\theta$ to the data by calculating the likelihood function 
\be
p(d |\theta, \mathcal{M}) = \prod_p p_{n_p}^{(p)}(\theta) \,,
\label{eq:likelihood_igrb}
\ee
where $p_{n_p}^{(p)}(\theta)$ denotes the probability of observing $n_p$ photons in pixel $p$ with model parameters $\theta$.

In Poissonian template fits, the probabilities $p_{n_p}^{(p)}(\theta)$ are Poisson distributions, with the model parameters $\theta$ only determining the means of the distributions.  That is, the mean expected number of photon counts at each pixel $p$ may be written as 
\es{}{
\mu_p(\theta) = \sum_\ell \mu_{p, \ell} (\theta) \,,
}
where the sum is over template components and $\mu_{p, \ell} (\theta)$ denotes the mean of the $\ell^\text{th}$ component for model parameters $\theta$.  The $\theta$ may parameterize, for example, the overall normalization of the templates or the shapes of the templates.  Then, the probability $p_{n_p}^{(p)}(\theta)$ is simply given by the Poisson distribution with mean $\mu_p$.  

In the NPTF, the situation is more complicated because we do not know where the PSs are.  As a result, if we want to calculate the probability of observing $n_p$ photons in a given pixel $p$, we must first calculate the probability that a PS (or a collections of PSs) exists in the vicinity of the pixel $p$, with a given flux (or set of fluxes).  Then, for that PS population, we calculate the probability of $n_p$ photons being produced in pixel $p$.  Convolving these two calculations together leads to distinctly non-Poissonian probabilities.  In particular, the probability distributions in the presence of unresolved PSs tend to be broader than Poisson distributions, if both distributions have the same mean expected number of photon counts.  The intuition behind this fact is that relative to a diffuse source, a collection of PSs leads to more ``hot'' pixels with many photons (where there are PSs) and more ``cold'' pixels with very few photons (where there are no PSs).  

\subsection{The Templates}
We include three Poissonian templates for (1) diffuse gamma-ray  emission in the Milky Way, assuming the \emph{Fermi} \texttt{p8r2} (\emph{gll\_iem\_v06.fits}) foreground model, (2) uniform emission from the \emph{Fermi} bubbles~\cite{Su:2010qj}, and (3) smooth  isotropic emission.  Each of these templates is associated with a single model parameter describing its overall normalization.  Variations to the choice of foreground model and bubbles template will be discussed in Sec.~\ref{sec:systematictests}.  

The model parameters specific to the isotropic-PS population enter into the source-count distribution $dN/dF$, which we characterize as a triply-broken power law:

\be
\frac{dN}{dF} = A^\text{PS}_\text{iso} \,\begin{cases} 
 \left(  \frac{F}{F_{b,3}}\right)^{-n_4} &  F < F_{b,3}\\ 
\left( \frac{F}{F_{b,3}}\right)^{-n_3}  & F_{b,3} \leq F< F_{b,2}  \\ 
\left( \frac{F_{b,2}}{F_{b,3}}\right)^{-n_3} \left( \frac{F}{F_{b,2}}\right)^{-n_2}   & F_{b,2} \leq F< F_{b,1}  \\ 
\left( \frac{F_{b,2}}{F_{b,3}}\right)^{-n_3} \left( \frac{F_{b,1}}{F_{b,2}}\right)^{-n_2}  \left(  \frac{F}{F_{b,1}}\right)^{-n_1} & F_{b,1} \leq F \end{cases} \,.
\label{eq:sourcecount3break}
\ee

In particular, there are three breaks, $F_{b,1...3}$, along with four indices, $n_{1..4}$, and the overall normalization, $A^\text{PS}_\text{iso}$.\footnote{Note that the NPTF can also handle PS templates with non-trivial spatial distribution, as was done in the Inner Galaxy analyses in~\cite{Lee:2015fea,Linden:2016rcf}, though in this work we will only consider the isotropic-PS template.}
  The justification for a triply-broken power law is that $F_{b,1}$ designates the high-flux loss of sensitivity, beyond which $dN/dF$ cannot be probed because no sources exist with such high flux.  The break $F_{b,3}$ designates the low-flux sensitivity, below which PS emission cannot be distinguished from smooth emission.  This leaves $F_{b,2}$ to probe any physical break in the source-count distribution in the flux region where the NPTF can constrain it.  We have verified, however, that the results do not change significantly if the source-count distribution is fit by a doubly broken power law. 

It is important to stress that the photon-count probabilities are non-Poissonian in the presence of unresolved PSs because their locations are unknown.  Once we know where a PS is, we can fix its location and describe it through a Poissonian template with a free parameter for the overall normalization of the source.  However, even  resolved sources with known locations may be characterized by the non-Poissonian template if we do not also put down Poissonian templates at their locations.  This is the approach that we take throughout this chapter; that is, we model both the resolved (in the 3FGL and 2FHL catalogs) and unresolved PS populations through a single $dN/dF$ distribution, without individually specifying the locations of any sources.    

The point-spread function (PSF) must be properly accounted for in the template-fitting procedure.  The diffuse models are smoothed according to the PSF using the {\it Fermi} Science Tools routine \texttt{gtsrcmaps}.  The bubbles template is smoothed with a Gaussian approximation to the PSF, with width set to give the correct 68\% containment radius in each energy bin.  We follow the prescription developed in~\cite{Malyshev:2011zi} to account for the PSF in the calculation of the non-Poissonian photon-count probabilities; for this, we use the King function parameterization of the PSF provided with the instrument response function for the given data set.  In Sec.~\ref{sec:systematictests}, however, we show that consistent results are obtained when using a Gaussian approximation to the PSF instead.    
 
\subsection{Bayesian Fitting Procedure} 
 
The formalism developed in~\cite{Malyshev:2011zi,Lee:2014mza,Lee:2015fea} (see also ~\cite{Zechlin:2015wdz} and ~\cite{Linden:2016rcf}) is used to calculate the photon-count probability distributions in each pixel as a function of the Poissonian and non-Poissonian model parameters $\theta$.  Then, Bayesian techniques are used to construct a posterior distribution $p(\theta | d, {\cal M})$ for the parameters $\theta$ and the likelihood function in~\eqref{eq:likelihood_igrb}.  We construct the posterior distribution numerically using the \texttt{MultiNest} package~\cite{Feroz:2008xx,Buchner:2014nha} with 700 live points, importance nested sampling and constant efficiency mode disabled, and sampling efficiency set for model-evidence evaluation.  
 
 All prior distributions are taken to be flat except for $A_\text{iso}^\text{PS}$, which is taken to be log-flat.  The prior ranges for the model parameters are shown in Tab.~\ref{tab:priors}.  
 \begin{table*}[t]
\renewcommand{\arraystretch}{1.4}
\setlength{\tabcolsep}{5.2pt}
\begin{center}
\begin{tabular}{  cc | cc | cc}
\toprule
Parameter	 & Prior Range &Parameter	& Prior Range & Parameter	& Prior Range   \Tstrut\Bstrut	\\   
\midrule 
$A_\text{diff}$  & $[0, 2]$  & $\log_{10}A_\text{iso}^{\text{PS}}$  & [-10, 20] & $n_1 $ & $[2.05, 5]$\\
$A_\text{bub}$ & $[0, 2]$  & $S_{b,3}$ & $[0.1, 1]$ ph  &$n_2$ & $[1.0, 3.5]$ \Tstrut\Bstrut \\ 
$A_\text{iso}$  & $[0, 2]$  &$S_{b,2}$ & $[1, 30]$ ph &$n_3 $ & $[1.0, 3.5]$\Tstrut\Bstrut \\ 
& &$S_{b,1}$ & $[30, 2 \times S_{b,\text{max}}]$ ph &$n_4$ & $[-1.99, 1.99]$ \Tstrut\Bstrut \\
\bottomrule
\end{tabular}
\end{center}
\caption{
Parameters and associated prior ranges for the templates used in the NPTF.   The priors on the breaks $S_{b,1...3}$ are given in terms of counts, defined relative to the mean exposure $\langle \overbar{ \mathcal{E}}^{(p)} \rangle$ in the ROI.  $S_\text{b,max}$ is the maximum number of photons in the 3FGL~\cite{Acero:2015hja} (2FHL~\cite{Ackermann:2015uya}) catalog in the energy bin of interest for the low (high)-energy analysis.  Note that all prior distributions are linear-flat, except for that of $A_\text{iso}^\text{PS}$, which is log-flat.  The baseline normalizations of the $A_\ell$ are described in the text.  }
\label{tab:priors}
\end{table*}  
These prior ranges successfully reconstruct the source-count distributions of simulated data sets, as discussed in Sec.~\ref{sec:simulations}.  Variations to the prior ranges in Tab.~\ref{tab:priors} are considered in Sec.~\ref{sec:systematictests}.

In Tab.~\ref{tab:priors}, the parameter $A_\ell$ denotes the normalization of the $\ell^\text{th}$ template, which is defined in terms of a baseline value.    The baseline value is obtained by first performing a Poissonian template fit over 17 (10) log-spaced energy sub-bins between $1.89$ and $94.9$ GeV (50 and 2000 GeV) for the low (high)-energy analysis.  When this procedure is applied to the low-energy analysis where the known PSs are very bright, we mask the 300 brightest and most variable 3FGL sources, at 95\% containment.  At both high and low energies, we include a PS model constructed from the 3FGL catalog.\footnote{Importantly, we do not include the PS model or mask any PSs in the NPTF analyses.}  The fitting procedure then allows us to recover the normalizations for the diffuse background, bubbles, and isotropic templates in each energy sub-bin.

The actual energy bins used for the NPTF studies presented in this study are larger than the sub-bins described above.  Therefore, the baseline normalizations used to define the NPTF priors in the energy range $[E_\text{min}, E_\text{max}]$ are found by applying the best-fit Poissonian normalizations from the individual sub-bins to the corresponding templates, which are then combined.\footnote{In practice, however, this prescription for combining the templates between energy sub-bins does not significantly affect our results.  
}
  Therefore, $A_\ell = 1$ in the NPTF analysis implies that the normalization of the $\ell^\text{th}$ template is the same as that computed from the Poissonian scans.  The benefit of this approach is that it allows one to keep track of how the individual Poissonian templates react to the addition of non-Poissonian ones.  For example, the normalization of the diffuse-background template should remain consistent between a standard template analysis, where PSs are accounted for by the 3FGL model, and the NPTF analysis, where PSs are accounted for by the non-Poissonian template; indeed, we find that is the case in all of the analyses we perform. 
  
\subsection{Exposure Correction}  
  
While the source-count distribution $dN/dF$ is defined in terms of flux, $F$, with units of $\text{ph}/\text{cm}^2/\text{s}$, the priors for the breaks in Tab.~\ref{tab:priors} are written in terms of counts, $S_{b,1...3}$.  To convert from flux to counts, we multiply by the exposure of the instrument, with units of cm$^2$ s.  However, the relation between flux and counts is complicated by the fact that the exposure of the instrument varies both with energy and position in the sky. Below, we describe how we deal with both complications, starting first with the energy dependence.  
  
The exposure map in the $i^\text{th}$ energy sub-bin is given by $\mathcal{E}^{(p)}_i$.  To construct the exposure map $\overbar{ \mathcal{E}}^{(p)}$ in the larger energy range from $[E_\text{min}, E_\text{max}]$, which contains multiple energy sub-bins, we average over the $\mathcal{E}^{(p)}_i$ of the individual sub-bins, weighted by a power-law spectrum $dN/dE \sim E^{-2.2}$, as this is generally consistent with the isotropic spectrum over most of our energy range.   
This procedure introduces a source of systematic uncertainty in going from counts to flux, as not all source components have an energy spectrum consistent with this spectrum.  However, we have checked that variations to this procedure---such as weighting the exposures in the sub-bins by power laws of the form $E^{-n}$, with $n$ varying between $1$ and $3$---do not significantly change the results.\footnote{We have also checked that weighting the exposures in the sub-bins by the intensities  computed from the Poissonian template scans gives consistent results.}   The weighting procedure is most important at very high energies, on the order of hundreds of GeV, where the exposure map varies strongly across the energy sub-bins.   
 
The breaks $S_{b,1...3}$ in Tab.~\ref{tab:priors}, with units of counts, are defined relative to the mean exposure $\langle \overbar{ \mathcal{E}}^{(p)} \rangle$, averaged over all pixels in the region of interest (ROI).  Because the NPTF is performed at the level of counts and not flux, we must also convert the source-count distribution $dN/dF$ to a distribution $dN^{(p)}/dS$, which is unique to each pixel $p$:
 \es{dNdFdS}{
 {dN^{(p)} \over dS} (S) = {1 \over \overbar{\mathcal{E}}^{(p)}} \left. {dN \over dF} \right\vert_{F = S / \overbar{\mathcal{E}}^{(p)}} \,.
 }   
 Then, the photon-count probability distribution must be computed uniquely at each pixel.  In practice, however, it is numerically expensive to perform this procedure for every pixel in the ROI.  Instead, we follow~\cite{Zechlin:2015wdz} and break the ROI up into $N_\text{exp}$ regions by exposure.  Within each region, we assume that all pixels have the same exposure, which is taken to be the mean over all pixels in the sub-region.  The likelihood function is then computed uniquely in each exposure region, and the total likelihood function for the ROI is the product of the likelihoods across exposure regions.  In practice, we find that our results are convergent for $N_\text{exp} \geq 10$.  We will take $N_\text{exp} = 15$ throughout this study, though we have checked that our main results are consistent with those found using $N_\text{exp} = 25$.
 
 \subsection{Data Samples}
 \label{sec:data}
 
 We run the NPTF analysis, as described above, on \emph{Fermi} data, considering low (1.89--94.9~GeV) and high (50--2000~GeV) energies separately.  The former is discussed in Sec.~\ref{sec:lowenergy}, while the latter is the focus of Sec.~\ref{sec:highenergy}.  The primary difference between the data sets used in these studies is the data-quality cuts; moving to higher energies requires loosening these criteria to avoid being limited by statistics.  The overlap in energy between the two studies allows us to compare the consistency of the results when transitioning between analyses.  

The low-energy study uses the Pass 8 \emph{Fermi} data from $\sim$August 4, 2008 to June 3, 2015.  The primary studies use the top quartile of the \emph{ultracleanveto} event class (PSF3) as ranked by angular resolution, although the top-three quartiles (PSF1--3) are also studied separately.\footnote{The PSF quartiles indicate the quality of the reconstructed photon direction, with `PSF3' being the best and `PSF0' being the worst.}  As a systematic check, we also consider the top-three quartiles of {\it source} data.  The \emph{ultracleanveto} event class is the cleanest event class released with the Pass 8 data and is recommended for studies of the EGB.  However, the {\it source} event class has an enhanced exposure and thus may be advantageous at high energies where statistics become limited.  On the other hand, we expect the {\it source} data to have additional cosmic-ray contamination relative to the \emph{ultracleanveto} data.    

The recommended\footnote{\url{http://fermi.gsfc.nasa.gov/ssc/data/analysis/documentation/Cicerone/Cicerone_Data_Exploration/Data_preparation.html}} event quality cuts are applied, requiring that all photons have a zenith angle less than $90^\circ$ and satisfy ``\texttt{DATA\_QUAL==1 \&\& LAT\_CONFIG==1 \&\& ABS(ROCK\_ANGLE)$ < 52$}.''   A HEALPix~\cite{Gorski:2004by} pixelation is used with \emph{nside}=128, which corresponds to pixels roughly $0.5^\circ$ to a side.   We consider four separate energy bins:  $[1.89, 4.75]$, $[4.75, 11.9]$, $[11.9, 30]$, and $[30, 94.9]$~GeV. 

In the low-energy analysis with {\it ultracleanveto} PSF3 data, the means of the weighted exposure maps in the four increasing energy bins are $[5.78 \times 10^{10}, 5.40 \times 10^{10}, 5.18 \times 10^{10}, 5.38 \times 10^{10}]$ cm$^2$ s over the region of interest with $|b| \geq 30^\circ$.  The 68\% containment radii for the PSF, averaged over the isotropic spectra in the energy sub-bins, are $[0.20, 0.11, 0.06, 0.04]$ degrees.  Going to PSF1--3 data, the exposures increase to $[1.69 \times 10^{11}, 1.66 \times 10^{11}, 1.63 \times 10^{11}, 1.67 \times 10^{11}]$ cm$^2$ s, while the 68\% containment radii of the PSF degrade to $[0.32, 0.16, 0.10, 0.08]$ degrees.  Going to {\it source} data with PSF1--3, the exposures ($[2.10 \times 10^{11}, 2.07 \times 10^{11}, 2.07 \times 10^{11},2.15 \times 10^{11}]$ cm$^2$ s) increase further, while the 68\% containment radii ($[0.32,0.16,0.10,0.08]$ degrees) are essentially the same as in the {\it ultracleanveto} case.

The high-energy analysis uses the Pass 8 \emph{Fermi} data from $\sim$August 4, 2008 to May 2, 2016 and all PSF quartiles of either the \emph{ultracleanveto} or \emph{source} event class.  The ROI is also extended to $|b| >10^\circ$.  We include more data in the high-energy analysis as there are far fewer photons than at lower energies.  We employ the recommended event-quality cuts as in the low-energy analysis and also choose \emph{nside}=128 HEALPix pixelation.  Results are presented for the three energy bins $[50, 151], [151, 457]$, and $[457, 2000]$~GeV.  With {\it ultracleanveto} data, the weighted exposures in the energy bins are $[2.48 \times 10^{11}, 2.31 \times 10^{11}, 1.69 \times 10^{11}]$ cm$^2$ s, while with {\it source} data the exposures become $[3.23 \times 10^{11}, 3.20 \times 10^{11}, 2.87 \times 10^{11}]$ cm$^2$ s.  For both data sets, the 68\% containment radii are approximately $[0.14, 0.12, 0.11]$ degrees.  We will also discuss results of analyses performed over a single wide-energy bin from $[50, 2000]$~GeV.

\section{Simulated Data Studies}
\label{sec:simulations}

To study the behavior of the NPTF, we apply it to simulated data sets of the gamma-ray sky.  These results are crucial both for understanding systematics associated with the NPTF as well as for interpreting the results of the NPTF in terms of evidence for or against the existence of these source populations. 

A simulated data map can be created starting from a particular source population that contributes to the EGB.  Using a theory model for the energy spectrum and luminosity function, the source-count distribution for that population can be derived in a specified energy range---see Sec.~\ref{sims} for further details on this procedure.  The appropriate number of sources is then drawn from this function and randomly distributed across the sky, with counts chosen to follow the intensity spectrum. Sources are then smeared with the appropriate Gaussian PSF to mimic the desired {\it Fermi} data set
bin-wise in energy, and Poisson counts are drawn to obtain the simulated map for the population.  This is then combined with the simulated contribution of the \texttt{p8r2} foreground model and the \emph{Fermi} bubbles, whose normalizations are determined from the Poissonian template fits to the real data, as described in Sec.~\ref{sec:methodology}.  

For most of this section, we simulate data corresponding to the PSF3 event type (best PSF quartile) of the \emph{ultracleanveto} event class and focus on the following four energy bins: [1.89, 4.75], [4.75, 11.9], [11.9, 30], and [30, 94.9]~GeV.
However, we also simulate data corresponding to the PSF1--3 (top 3 PSF quartiles) instrument response function to illustrate potential advantages in going to the more inclusive data set, albeit with a slightly worse PSF.   Once the simulated data maps are created, we run them through the NPTF analysis pipeline.  First, we analyze the case where either blazars or SFGs fully account for the EGB, and then we analyze a perhaps more realistic scenario where both populations contribute significantly to the flux.  The particular blazar and SFG models used here are merely meant for illustration.  They are chosen as examples that span the range of possibilities between smooth and PS isotropic contributions.  As mAGN are fainter and more numerous then blazars, they likely act similarly to SFGs in the context of the NPTF and so we do not consider them separately here.  A detailed analysis of how the NPTF responds to the broader class of theoretical models for these source classes is beyond the scope of this study.

\subsection{Simulating Energy-Binned Source-Count Distributions}
\label{sims}

We generate simulated maps directly from the source-count distribution $dN/dF_{\gamma}$. To obtain this, we need two inputs: the gamma-ray luminosity function, $\Phi(L_{\gamma},z,\Gamma)$, and the source energy spectrum, $dF/dE$~\cite{DiMauro:2014wha}.  Typically, the luminosity function (LF) is given by
\begin{equation}
\Phi(L_{\gamma},z,\Gamma)=\frac{d^3N}{dL_\gamma\,dV\,d\Gamma} \,  ,
\end{equation}
where $V$ is the comoving volume, $\Gamma$ is the photon spectral index, $z$ is the redshift, $N$ is the number of sources, and $L_\gamma$ is the rest-frame luminosity for energies from 0.1--100~GeV in units of GeV\,s$^{-1}$.     

The photon flux in this energy range, $F_\gamma$, 
is defined in terms of the source energy spectrum,
\begin{equation}
F_\gamma(\Gamma) = \int_{E_\text{min}}^{E_\text{max}}\frac{dF}{dE} \, dE \, ,
\label{eq: Fgamma}
\end{equation}
where the units are cm$^{-2}$\,s$^{-1}$, and $E_\text{min(max)} = 0.1(100)$~GeV.

The source-count distribution is then given by
\begin{equation}
\frac{dN}{dF_\gamma} = \frac{1}{4\pi} \int_{\Gamma_\text{min}}^{\Gamma_\text{max}} d\Gamma \int_{z_\text{min}}^{z_\text{max}} dz \, \Phi(L_\gamma,z,\Gamma) \, \frac{dV}{dz} \, \frac{dL_\gamma}{dF_{\gamma}} \, ,
\label{eq: dNdFexact}
\end{equation}
which can be accurately estimated as
\begin{equation}
\frac{dN}{dF_\gamma} \approx \frac{1}{\Delta F_\gamma} \, \frac{1}{4\pi} \int_{\Gamma_\text{min}}^{\Gamma_\text{max}} d\Gamma \int_{z_\text{min}}^{z_\text{max}} dz \, \int_{L_\gamma(F_\gamma, \Gamma,z)}^{L_\gamma(F_\gamma+\Delta F_\gamma, \Gamma,z)} dL_\gamma \, \Phi(L_\gamma,z,\Gamma) \, \frac{dV}{dz} \, ,
\label{eq: dNdF}
\end{equation}
where $4\pi$ is the full-sky solid angle, $dV/dz$ is the comoving volume slice for a given redshift and $\Delta F_\gamma$ is sufficiently small.  To calculate $dN/dF_\gamma$, we need the following expression, which relates the luminosity to the energy flux:
\begin{equation} 
L_\gamma(F_\gamma, \Gamma,z) = \frac{4\pi d_L^2}{(1+z)^{2-\Gamma}}  \, \int_{E_\text{min}}^{E_\text{max}} \, E\,\frac{dF}{dE}\, dE \, ,
\label{eq:lumi}
\end{equation}
where $d_L$ is the luminosity distance. For a given $F_\gamma$ and $\Gamma$, one can use~\eqref{eq: Fgamma} to solve for the normalization of $dF/dE$, which can be substituted into~\eqref{eq:lumi}, along with $z$ and $\Gamma$, to obtain the associated value of the luminosity. The photon flux, $F_\gamma$, is related to the photon count, $S_\gamma$, via the mean exposure $\langle \bar{\mathcal{E}} \rangle$, which is averaged over 0.1--100~GeV and the ROI.  This allows us to finally obtain $dN/dS_\gamma$ from \eqref{eq: dNdF}.

The procedure outlined above allows one to obtain the source-count distributions based on models of luminosity functions and spectral energy distributions provided in the literature. For the AGN and SFG examples we consider in detail in this work, the luminosity functions correspond to photon energies from 0.1--100~GeV.  However, we also need the source-count distributions in subset energy ranges corresponding to our energy bins of interest, with $E'_\text{min, max}\in [0.1, 100]$~GeV.  We rescale the fluxes for these individual energy bins of interest to those in the provided 0.1--100~GeV range using a procedure similar to~\cite{DiMauro:2014wha}. Denoting quantities associated with this energy bin with a prime, we can write the new source-count distribution as
\begin{equation}
\frac{dN}{dF'_\gamma}\approx \frac{1}{\Delta F'_\gamma} \, \frac{1}{4\pi} \int_{\Gamma_\text{min}}^{\Gamma_\text{max}} d\Gamma \int_{z_\text{min}}^{z_\text{max}} dz \int_{L_\gamma(F_\gamma(F'_\gamma,\Gamma),\Gamma,z)}^{L_\gamma(F_\gamma(F'_\gamma+\Delta F'_\gamma,\Gamma),\Gamma,z)} dL_\gamma \, \Phi(L_\gamma,z,\Gamma) \, \frac{dV}{dz} \, ,
\label{eq: dNdFprime}
\end{equation}
where $\Delta F'_\gamma$ is again sufficiently small---we set $\Delta F'_\gamma \equiv 10^{-3} F'_\gamma$, and verify that the answer is robust to this choice. Note that the integral must still be done over $L_\gamma$ (unprimed) because the luminosity function is explicitly defined in terms of it.  So, we must solve for the photon flux over the full energy, $F_\gamma$, in terms of the value in the sub-bin, $F_\gamma'$.  The two are related via a proportionality relation
\begin{equation}
F_\gamma(F'_\gamma,\Gamma) = F'_\gamma \, \frac{\int_{E_\text{min}}^{E_\text{max}}  dE \int_{L_{\gamma,\text{min}}}^{L_{\gamma,\text{max}}} dL_{\gamma}\int_{z_\text{min}}^{z_\text{max}} dz \,  \Phi(L_\gamma,z,\Gamma) \, \frac{dV}{dz} \, \frac{dF}{dE}\,e^{-\tau_\text{\tiny{EBL}}(E,z)}}{\int_{E'_\text{min}}^{E'_\text{max}}  dE \int_{L_{\gamma,\text{min}}}^{L_{\gamma,\text{max}}} dL_{\gamma}\int_{z_\text{min}}^{z_\text{max}} dz \, \Phi(L_\gamma,z,\Gamma) \, \frac{dV}{dz} \, \frac{dF}{dE}\,e^{-\tau_\text{\tiny{EBL}}(E,z)}} \, ,
\end{equation}
where the exponential factor accounts for the attenuation due to extragalactic background light (EBL)~\cite{Gould:1966pza, Fazio:1970pr, 1992ApJ...390L..49S, Franceschini:2008tp, 2012Sci...338.1190A,Abramowski:2012ry,Dominguez:2013lfa}.  It arises from pair annihilation of high-energy gamma-ray photons with other background photons in infrared, optical, and/or ultraviolet, and is described by the optical depth, $\tau_\text{\tiny{EBL}}$.  We use the EBL attenuation model from~\cite{2010ApJ...712..238F}.  

Additionally, the expected gamma-ray spectrum can be calculated from the luminosity function as
\begin{equation}
\frac{dN}{dE} = \frac{1}{4\pi} \int_{\Gamma_\text{min}}^{\Gamma_\text{max}} d\Gamma \int_{z_\text{min}}^{z_\text{max}} dz \, \int_{L_{\gamma,\text{min}}}^{L_{\gamma,\text{max}}} dL_\gamma \, \Phi(L_\gamma,z,\Gamma) \, \frac{dV}{dz}\,\frac{dF}{dE} \, e^{-\tau_\text{\tiny{EBL}}(E,z)} \, .
\label{eq: dIdE}
\end{equation}
We use this equation to appropriately weight the number of photons per energy sub-bin for the individual sources when creating simulated maps. This ensures that the variations in PSF and exposure within the larger energy bins used in the NPTF analyses are properly accounted for in the simulation procedure.

\subsection{Blazars}

Active galactic nuclei (AGN) are the highly luminous central regions of galaxies where emission is dominated by accretion onto a supermassive black hole~\cite{Urry:1995mg}.  If the black hole is spinning, then relativistic jets may also form.  Blazars are a subclass of AGN in which the jet is oriented within $14^\circ$ of the line-of-sight~\cite{Angel:1980}.  The spectral energy distribution of these objects is bimodal with a peak in the ultraviolet due to synchrotron radiation of electrons in the jet, and another peak in the gamma band from inverse Compton scattering of the same electrons \cite{Fossati:1998zn,Ghisellini:1998it,Ghisellini:2009fj,Boettcher:2013wxa}.  There is also the possibility of a hadronic contribution to blazar gamma-ray spectra, although this is likely to be sub-dominant~\cite{Tavecchio:2013fwa,Cerruti:2014iwa,Zdziarski:2015rsa}. Blazars may be further classified as either BL Lacertae (BL~Lacs) or Flat Spectrum Radio Quasars (FSRQs), which are characterized by the absence or presence of broad optical/ultraviolet emission lines, respectively.  

Before \emph{Fermi}, few blazars had been identified in gamma rays, and to estimate the size of this population, one had to extrapolate based on those observed at lower frequencies.  However, \emph{Fermi} brought the discovery of many more blazars in the gamma-ray band, making it possible to study their properties directly~\cite{Collaboration:2010gqa, Ajello:2011zi,Ajello:2013lka, DiMauro:2013zfa, Giommi:2015ela, Padovani:2014cha}.  Most recently, 403 blazars (with $|b| > 15^\circ$) from the First LAT AGN Catalog~\cite{Abdo:2010ge} were studied~\cite{Ajello:2015mfa}.  FSRQs and BL~Lacs were considered together in the same sample to improve statistics.  We use the best-fit luminosity and spectral energy distributions given in~\cite{Ajello:2015mfa} (specifically, the luminosity-dependent density evolution, or LDDE, scenario) to model the blazar component in our simulated data and refer to it as the ``Blazar--1'' model.  Alternatively, we also consider BL~Lacs and FSRQs separately, adding up their respective contributions using the LDDE1 model from~\cite{Ajello:2013lka} and the LDDE model from~\cite{Ajello:2011zi}, which we refer to as the ``Blazar--2'' model. This model predicts a much flatter source-count distribution below the \emph{Fermi} detection threshold, with more low-flux sources. The two source-count models approximately bracket the current theoretical uncertainty in the faint-end slope of blazars, and we use them to study the response of different blazar models to the NPTF, although this is meant to be purely illustrative and by no means exhaustive.

\begin{figure*}[!htbp] 
   \centering
   \includegraphics[width=0.9\textwidth]{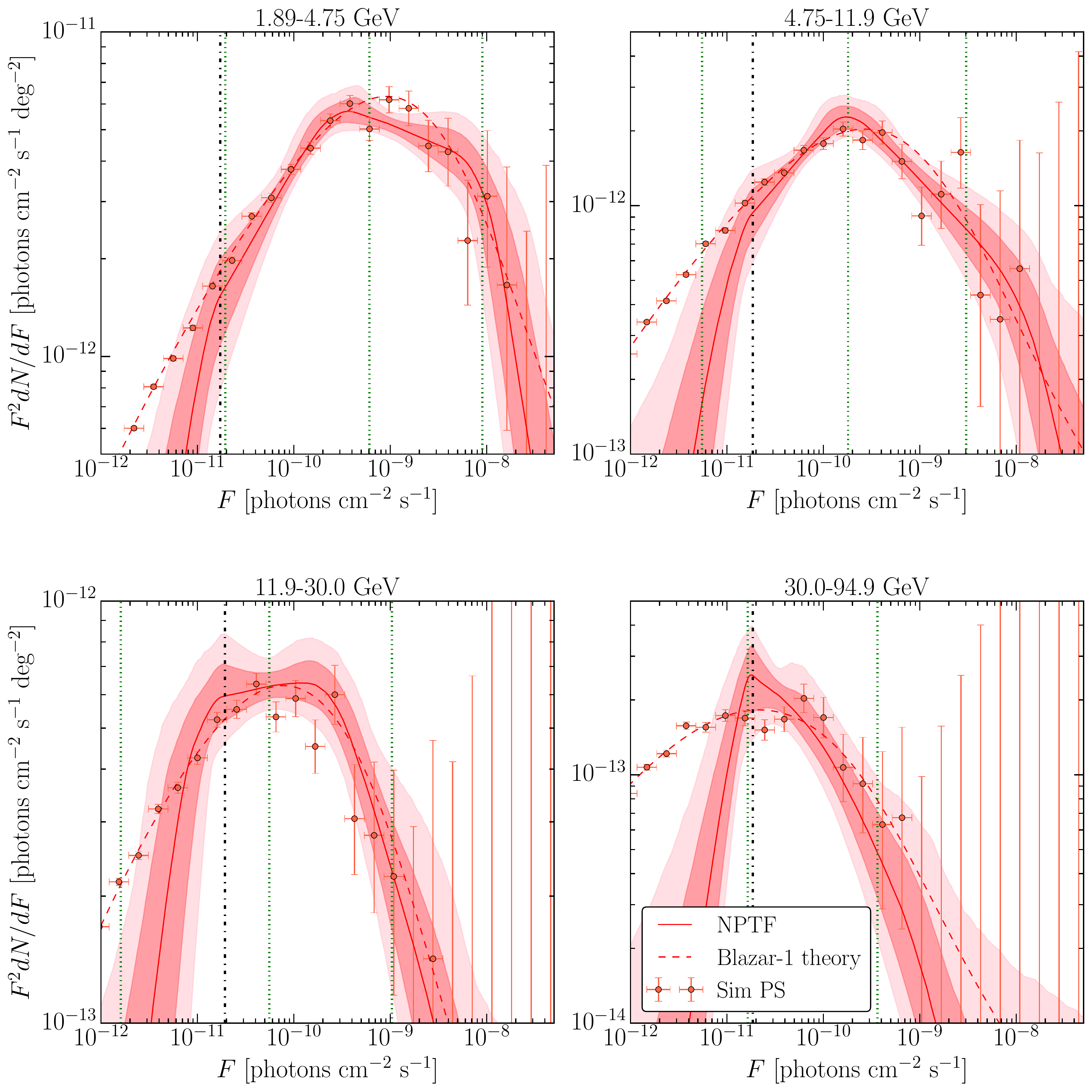} 
   \caption{The source-count distribution of the isotropic-PS population obtained by running the NPTF on simulated data in which the EGB arises from the Blazar--1 model~\cite{Ajello:2015mfa}.  Results are presented for the four energy bins considered.  The source-count distribution of the input blazar model (dashed red) matches the posterior for the isotropic PSs (68 and 95\% credible intervals, constructed pointwise, shaded in red) well at fluxes corresponding to counts above $\sim$1 photon (vertical, dot-dashed black).  The vertical dotted green lines indicate the fluxes at which 90\%, 50\%, and 10\% of the flux is accounted for, on average, by sources with larger flux (from left to right, respectively).  The red points show the histogram of the simulated PSs, with 68\% Poisson error bars (vertical).  Note that the NPTF loses sensitivity to sources contributing less than $\sim$1 photon; as a result, the NPTF result does not match the simulated data well below the dot-dashed black line.  }
   \label{fig:bl1dnds}
\end{figure*}

Figure~\ref{fig:bl1dnds} shows the best-fit source-count distributions recovered when the NPTF analysis is run on the Blazar--1 simulated data map, assuming the PSF3 instrument response function.  In each panel, the dark (light) red band is the 68\% (95\%) credible interval for the isotropic-PS source-count distribution as recovered from the posterior and  constructed pointwise in flux. The red line shows the median source-count distribution, constructed in the same way. The dashed red curve, on the other hand, indicates the source-count distribution of the blazar model used to generate the simulated data.  A flux histogram of the simulated PSs for the particular realization shown here is given by the red points, with vertical error bars indicating the 68\% credible interval associated with Poisson counting statistics on the number of sources in that bin.  Notice that these error bars become large at high fluxes because there are very few sources per flux bin. 

In general, the reconstructed source-count distribution is in good agreement with the input source-count distribution at intermediate fluxes, with uncertainties becoming large at low and high fluxes.  At high flux, this is due to the fact that it is unlikely to draw a bright source from the underlying source-count distribution.  At low fluxes, it is difficult to distinguish PS emission from genuinely isotropic emission.  To illustrate this point, we also mark the flux that corresponds to a single photon on average (in the particular energy range, region-of-interest, and event class) with the vertical dot-dashed black line.  At fluxes corresponding to counts near or below $\sim$1 photon, it is difficult to distinguish PS emission from smooth emission with the NPTF, as evidenced by the growing uncertainties.  In this low-flux regime, we do not expect that the NPTF will be able to fully recover the properties of the input source-count distribution.  

The vertical dotted green lines in Fig.~\ref{fig:bl1dnds} correspond to the fluxes above which 90\%, 50\%, and 10\% (from left to right) of the photon counts are accounted for, on average, by sources with larger flux.  Note that in the lowest energy bin, 90\% of the flux arises from sources that contribute more than one photon.  Moving towards higher energies, a larger fraction of the flux arises from sources that contribute less than one photon.  In all energy bins, more than 50\% of the flux is accounted for by sources that contribute more than a single photon each.

\begin{figure*}[!htbp] 
   \centering
  $ \begin{array}{cc}
   \scalebox{0.37}{\includegraphics{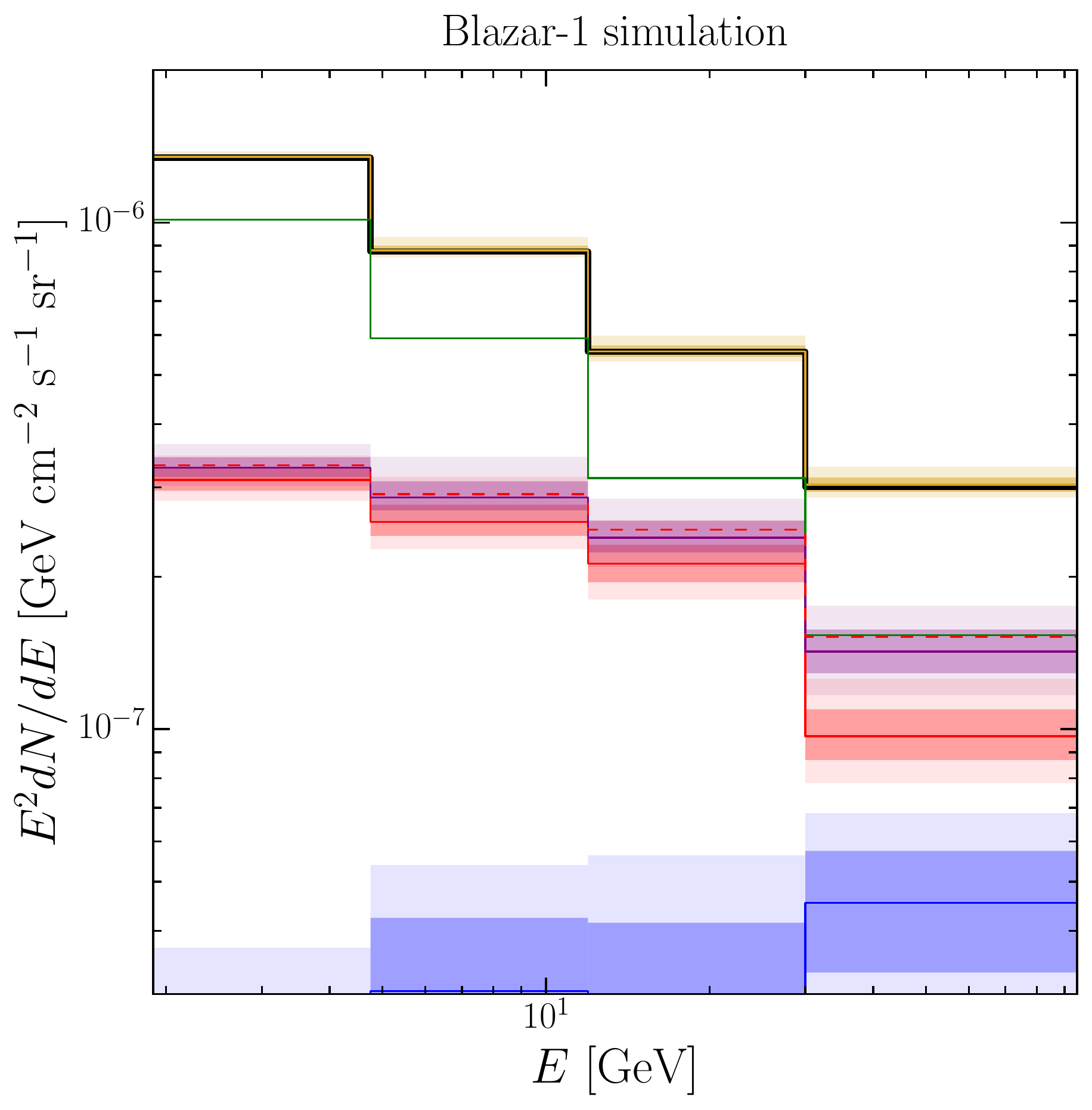}} &	\scalebox{0.37}{\includegraphics{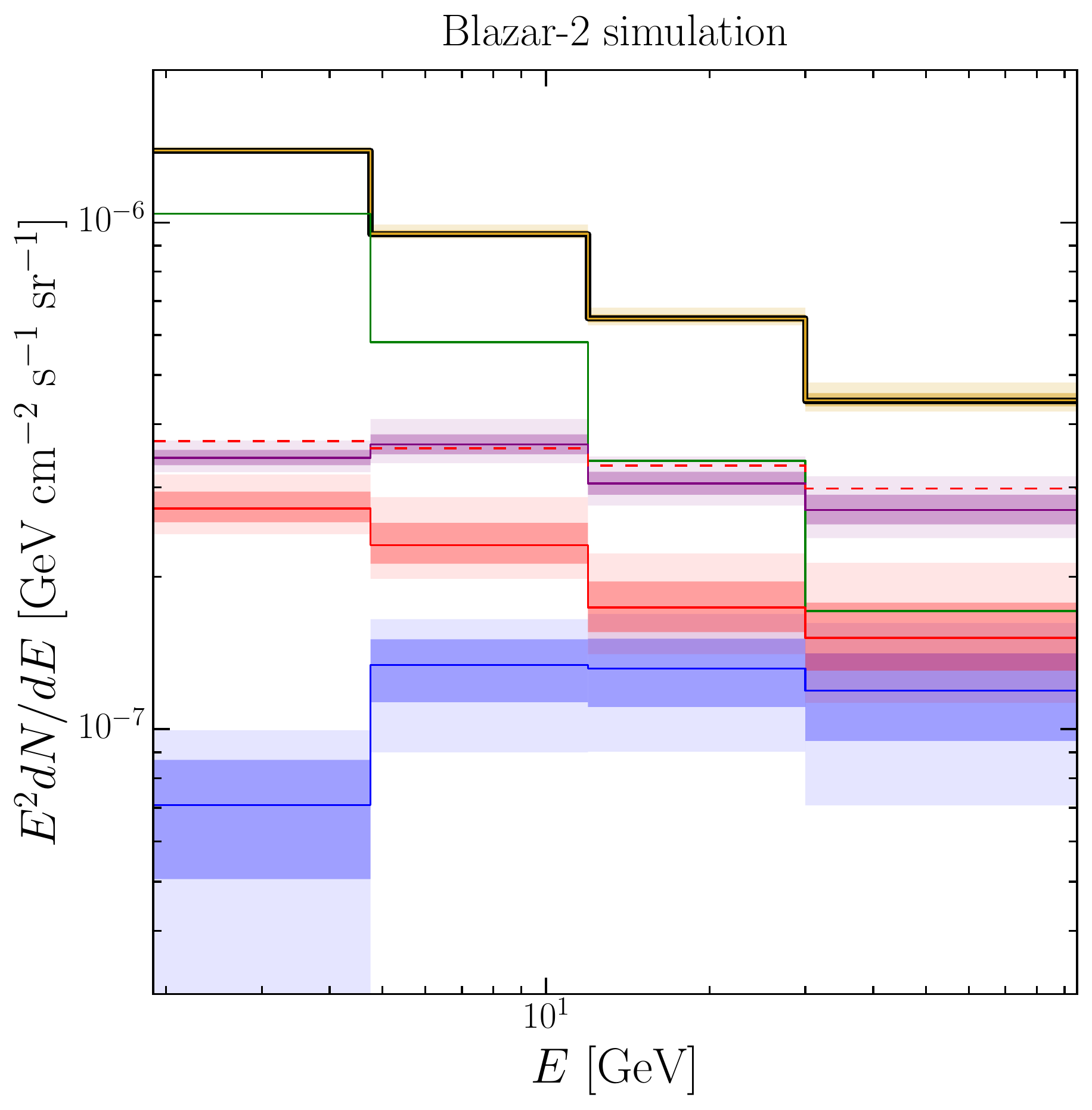}} \\
   \scalebox{0.37}{\includegraphics{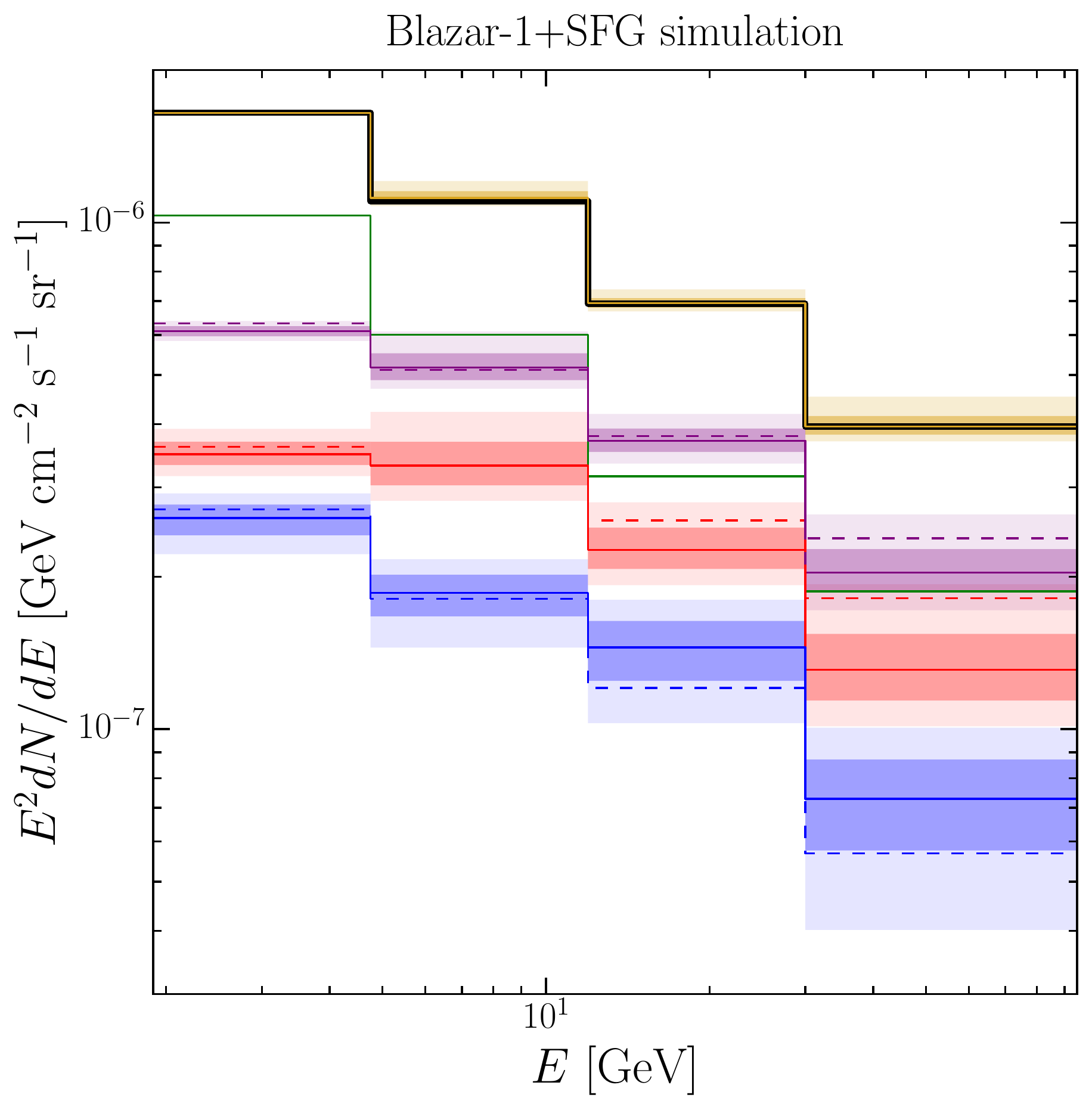}} &	\scalebox{0.37}{\includegraphics{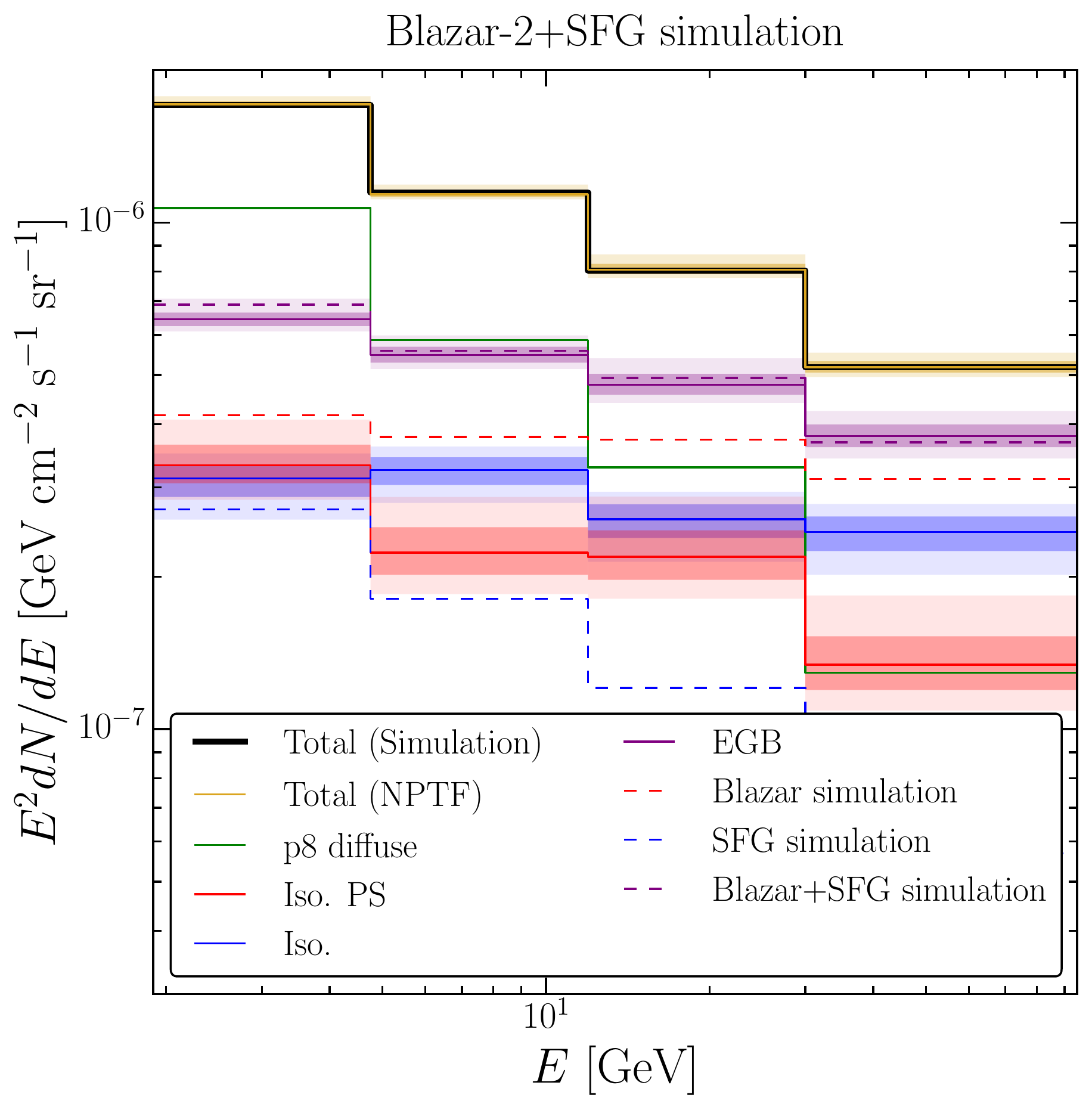}}
   \end{array}$
   \caption{The energy spectra for the isotropic and isotropic-PS templates in each energy bin considered; the 68 and 95\% credible intervals, constructed from the posterior distributions, are shown in blue and red, respectively.  The top row represents the results for simulated data, with {\it ultracleanveto} PSF3 instrument response function, in which the EGB consists of only Blazar--1 sources~\cite{Ajello:2015mfa} \textbf{(Top left)} or Blazar--2 sources~\cite{Ajello:2011zi, Ajello:2013lka} \textbf{(Top right)}.  The bottom row  shows the same results, except when SFGs~\cite{Tamborra:2014xia} are also included in the simulation.  The simulated spectrum for blazars (SFGs) is shown in dashed red (blue).  For the Blazar--1 model, the isotropic-PS template absorbs almost the entirety of the flux.   For the Blazar--2 model, both smooth and PS isotropic components absorb flux, but their sum (EGB, purple band) is consistent with the input.  When SFGs are also included, more emission is absorbed by the smooth isotropic template; however, the total emission absorbed by the smooth and PS isotropic templates is consistent with the expected total of SFG and blazar intensities.  The spectrum for Galactic diffuse emission is shown by the green line in each panel (median only).  The sum of all template emission (yellow band) agrees with the total spectrum of the simulated data.  Note that the energy spectrum of the bubbles template is not shown.}
   \label{fig:blESpec}
\end{figure*}

The corresponding energy spectra for the various templates are shown on the top left panel of Fig.~\ref{fig:blESpec}.  As is evident, these blazars show up as PSs under the NPTF; indeed, the smooth isotropic flux (blue) is sub-dominant in each energy bin.  
Overlaid in dashed red is the spectrum for the simulated Blazar--1 sources.  The sum of the smooth and PS isotropic components---which is simply the EGB intensity---is consistent with the simulated spectrum for the blazar model.  The green curve shows the median of the posterior for the galactic diffuse model spectrum.  The energy spectrum of the diffuse model is softer than that for blazars, so that the diffuse model dominates more at low energies than at high.  The sum of the components (yellow band) is consistent with the total flux in the simulated data (black lines) at 68\% confidence.

As a contrasting example, we also simulate the Blazar--2 model, which predicts more low-flux sources than the previous example we considered.  The best-fit source-count distributions for the Blazar--2 simulated maps are shown in Fig.~\ref{fig:bl2dnds}. Once again, we see good agreement between the input data and the recovered source-count distribution above the single-photon sensitivity threshold. In this case, however, the reference model predicts a larger fraction of flux coming from sources below this threshold.  For example, about 50\% of the flux comes from sub-single photon sources in the second energy bin, and this fraction only increases further at higher energies.  The corresponding energy spectrum is shown in the top right panel of Fig.~\ref{fig:blESpec}.  As expected, an increasing amount of flux is absorbed by the Poissonian isotropic template.  However, the EGB spectrum, shown by the purple band, is still consistent with the input spectrum for the Blazar--2 model.

\afterpage{
\begin{figure*}[!htbp] 
   \centering
   \includegraphics[width=0.9\textwidth]{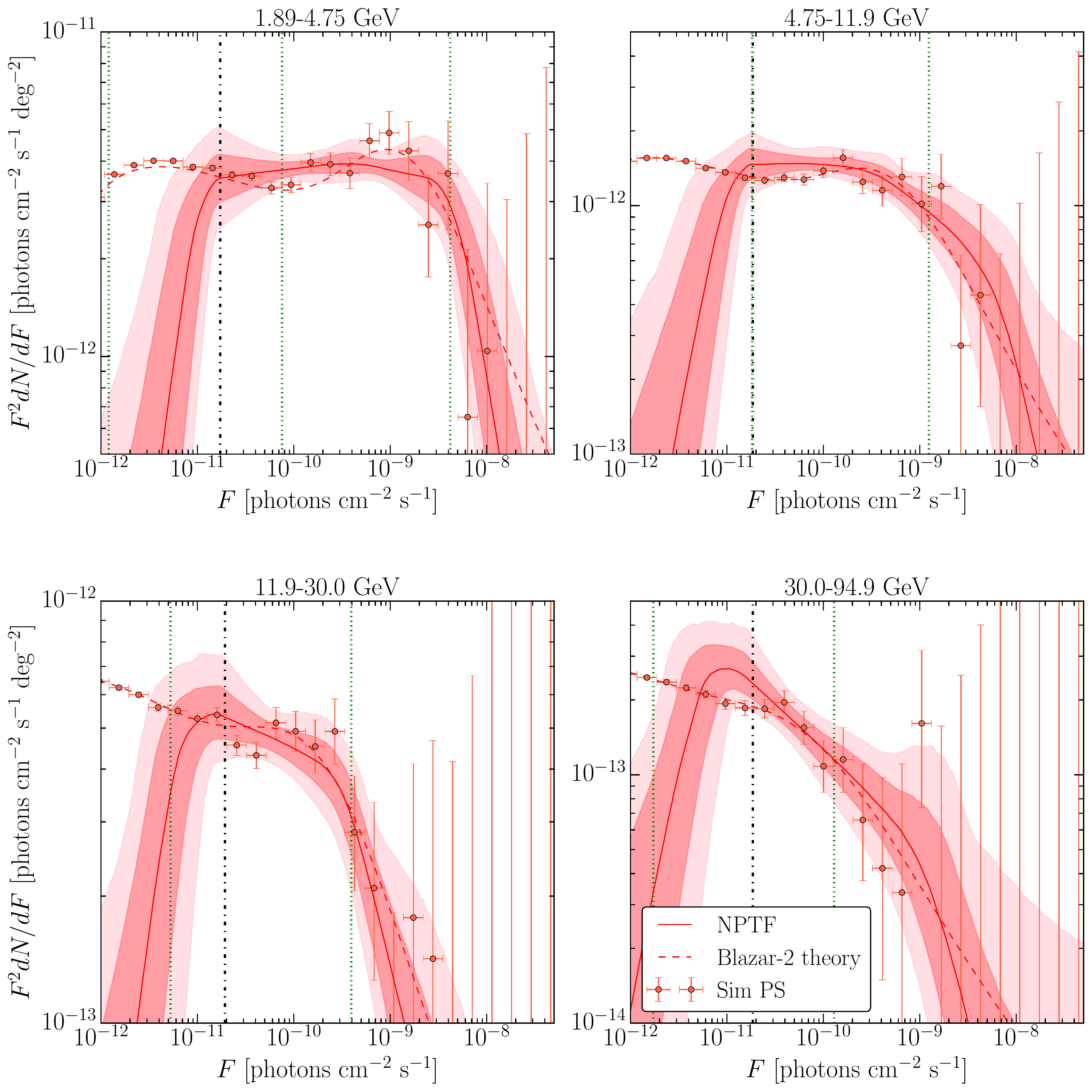} 
   \caption{Same as Fig.~\ref{fig:bl1dnds}, except for the Blazar--2 model~\cite{Ajello:2011zi, Ajello:2013lka}.}
   \label{fig:bl2dnds}
\end{figure*}
\clearpage
}

To further quantify the ability of the NPTF to reconstruct the blazar flux as PS emission, it is convenient to consider the ratios $I_\text{iso}^\text{PS} / I_\text{blazar-sim}$ in each energy bin, where $I_\text{iso}^\text{PS}$ is the PS intensity found by the NPTF and $ I_\text{blazar-sim}$ is the blazar intensity in the simulation.  For the Blazar--1 model, we find\footnote{Throughout this work, best-fit values indicate the 16$^\text{th}$, 50$^\text{th}$, and 84$^\text{th}$ percentiles of the appropriate posterior probability distributions. } 
\begin{equation}
\frac{I_\text{iso}^\text{PS}}{ I_\text{blazar-sim}} = [0.94_{-0.04}^{+0.05}, 0.88_{-0.05}^{+0.07}, 0.86_{-0.07}^{+0.08}, 0.64_{-0.07}^{+0.08}] \nonumber
\end{equation}
in each of the four respective energy bins, while for the Blazar--2 model, we find
\begin{equation}
\frac{I_\text{iso}^\text{PS}}{ I_\text{blazar-sim}} = [0.74_{-0.04}^{+0.06}, 0.64_{-0.05}^{+0.07}, 0.53_{-0.06}^{+0.07}, 0.51_{-0.07}^{+0.09}]\,, \nonumber
\end{equation}
 for the particular Monte Carlo realizations shown.\footnote{Different Monte Carlo realizations are found to induce variations consistent with the quoted statistical uncertainties, generally on the order of 5\%.}
For the Blazar--2 scenario, more flux goes into smooth isotropic emission, which is why the PS fractions are correspondingly smaller in each energy bin.  Note that, in both scenarios, the fraction of the blazar flux absorbed by the PS template decreases at higher energies, where the photon counts become less numerous and a higher fraction of the blazar flux is generated by sub-threshold sources.  As a result, the intensities $I_\text{iso}^\text{PS}$ should be interpreted as lower bounds on the blazar flux; this intuition is validated by the fact that the ratios  $I_\text{iso}^\text{PS} / I_\text{blazar-sim}$ tend to be less than unity.  

Next, we explore whether including more quartiles of the {\it ultracleanveto} data, as ranked by PSF, increases our ability to reconstruct the blazar flux as PSs under the NPTF.  When including more quartiles of data, there are two competing effects that determine our ability to constrain the PS flux: on the one hand,  we increase the effective area, but on the other hand, we worsen the angular resolution of the data set.  We investigate these effects by repeating the Monte Carlo tests described above using the PSF1--3 instrument response function
, and here we simply quote the fractions 
\begin{equation}
\frac{ I_\text{iso}^\text{PS}}{ I_\text{blazar-sim}} = [0.78_{-0.05}^{+0.06}, 0.81_{-0.06}^{+0.07}, 0.72_{-0.06}^{+0.06}, 0.57_{-0.05}^{+0.06}]\,  \nonumber
 \end{equation}
 for a generic realization of the Monte Carlo simulations for the Blazar--2 model.  The PSF1--3 event type increases our ability to distinguish between the blazar emission and smooth emission compared to the PSF3 event type.

\subsection{Star-Forming Galaxies}

Star-forming galaxies (SFGs) like our own Milky Way are individually fainter, though much more numerous, than blazars.  The modeling of SFGs in the gamma-ray band is highly uncertain, as \emph{Fermi} has only detected eight SFGs thus far~\cite{Fornasa:2015qua}.  However, SFGs could still contribute a sizable fraction of the total flux observed by \emph{Fermi}.  Even though SFGs are PSs, their flux is expected to be dominated by a large population of dim sources degenerate with smooth isotropic emission.  Under the NPTF, therefore, we expect that the majority of their  emission will be absorbed by the smooth isotropic template.  
To illustrate this point, we simulate SFGs using the luminosity function and energy spectrum from~\cite{Tamborra:2014xia}.  In that work, input from infrared wavelengths was used to construct a model for the infrared flux from SFGs.
Then, a scaling relation was used to convert from infrared to gamma-ray luminosities.  The contributions from quiescent and starburst SFGs were considered separately, along with SFGs that host an AGN.  Note, however, that other models predict less emission from SFGs than this particular case---see \emph{e.g.},~\cite{Makiya:2010zt,Inoue:2011bm,Ackermann:2012vca}.

We also performed tests using simulated SFGs. We find that while the NPTF does detect a small PS component in the first few energy bins, as the result of a few SFGs above the sensitivity threshold of the NPTF in those energy bins, by far most of the SFG emission is detected as smooth isotropic emission, with the ratio $I_\text{iso}^\text{PS} / I_\text{iso}\lesssim 1/100$ in all energy bins, where $I_\text{iso}$ is the intensity of smooth isotropic emission.  Moreover, the intensity $I_\text{iso}$ is consistent with the simulated EGB (SFG flux) in all energy bins, at 68\% confidence.

\subsection{Blazar and SFG combination}
\begin{figure*}[!htbp] 
   	\begin{center}$
	\begin{array}{cc}
	\scalebox{0.43}{\includegraphics[width=1\textwidth]{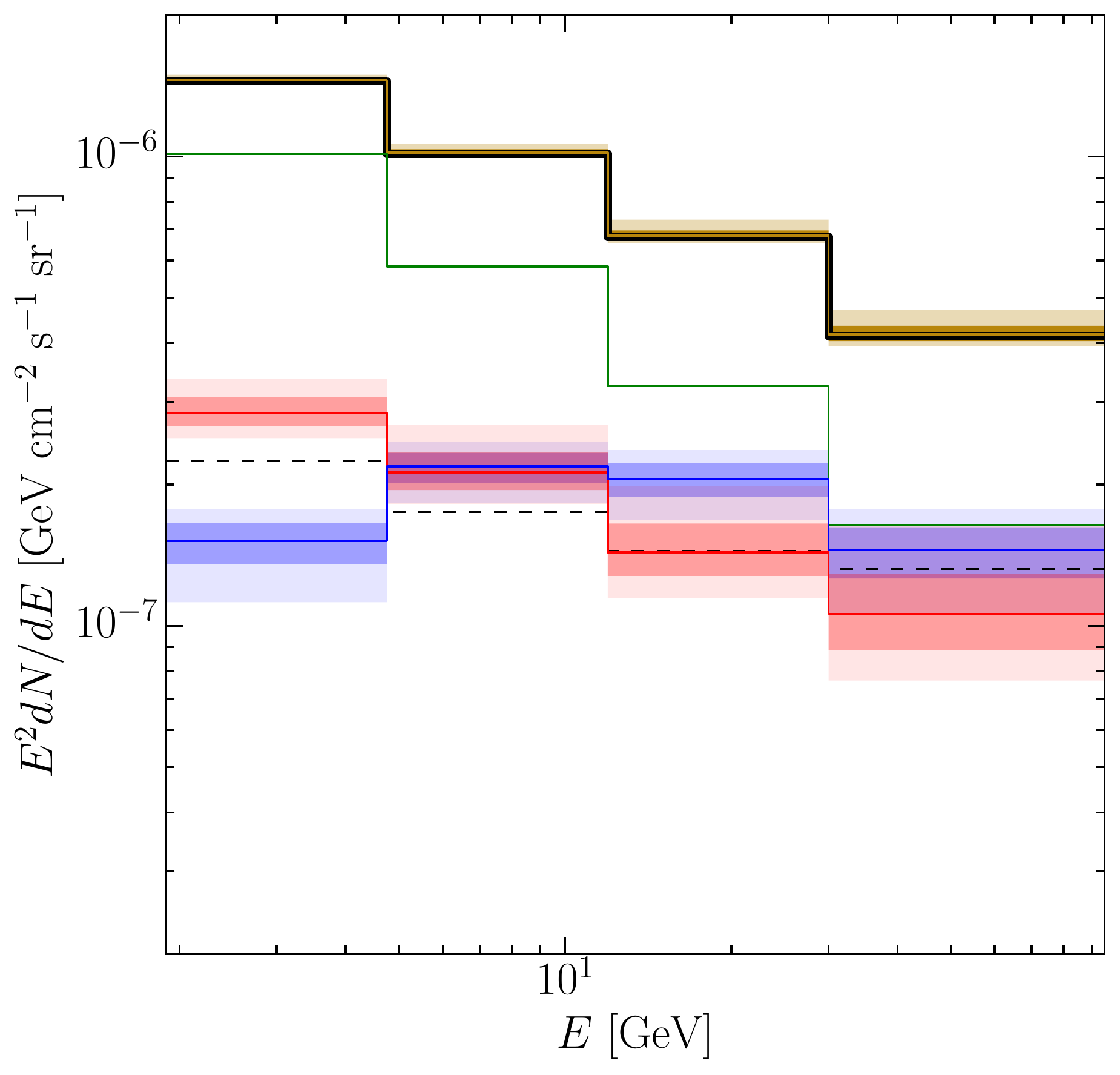}} &\scalebox{0.43}{\includegraphics[width=1\textwidth]{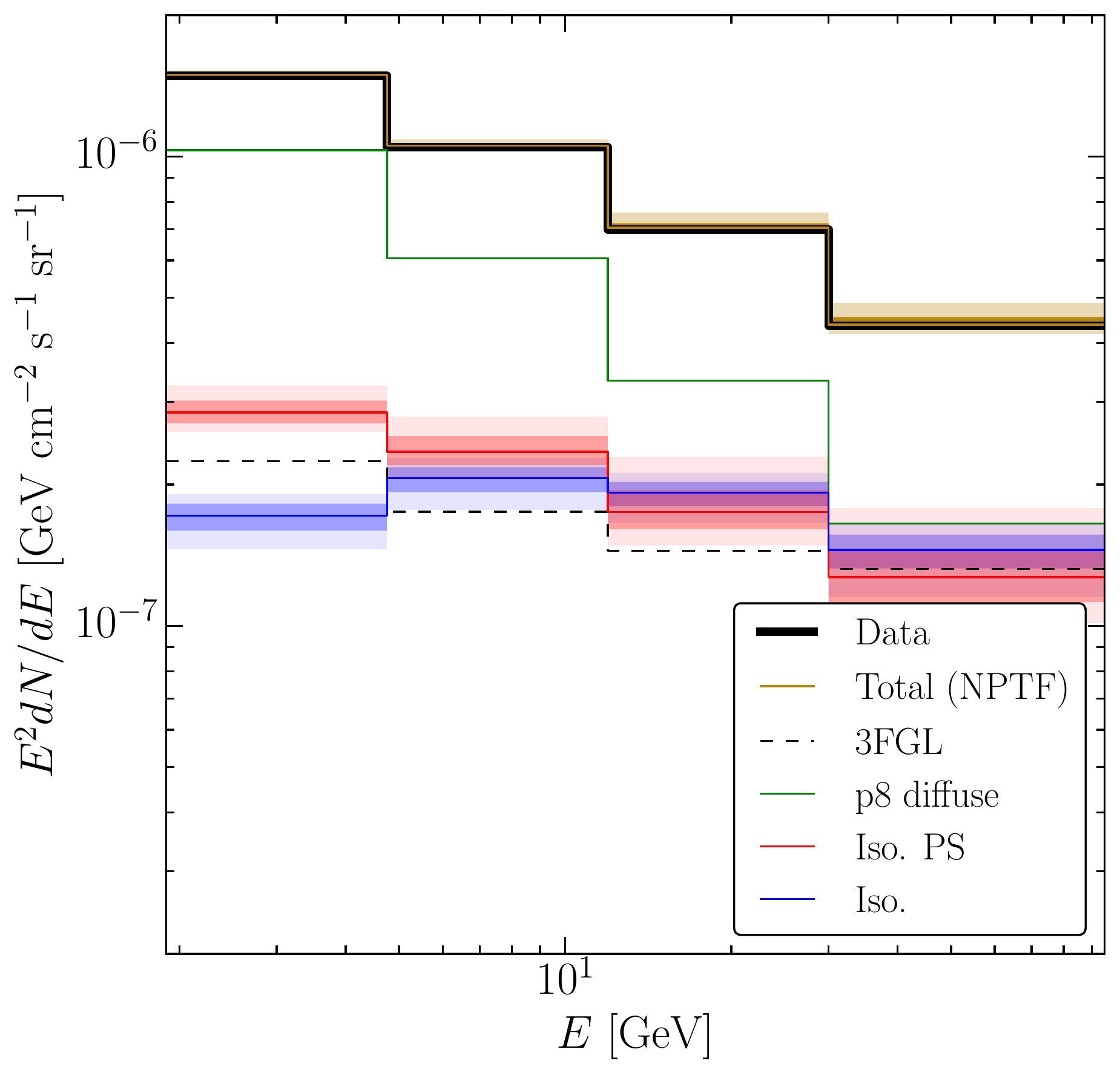}} 
	\end{array}$
	\end{center}
\caption{Best-fit energy spectra for the NPTF analysis using Pass~8 {\it ultracleanveto} data and the \texttt{p8r2} foreground model.  The left (right) panel shows the PSF3 (PSF1--3) results.  The 68 and 95\% credible intervals, constructed from the posterior distributions in each energy bin, are shown for the isotropic-PS and smooth isotropic templates in red and blue, respectively.  The median intensity for the foreground model is also shown (green).  The sum of all the components (yellow band) agrees with the total spectrum of the \emph{Fermi} data (black). The \emph{Fermi} bubbles contribution is subdominant (averaged over the full region of interest) and is thus not plotted.  For comparison, the spectrum of the 3FGL sources is shown in dashed black.  We caution the reader that, at higher energies, the 3FGL spectra are driven by extrapolations from low energies where the statistics are better.  The systematic uncertainties associated with this extrapolation are difficult to quantify and are not shown here.}
   \label{fig:nptfdata} 
\end{figure*}

A perhaps more realistic scenario for testing the NPTF is to consider a scenario where both SFGs and blazars contribute to the EGB.  Therefore, we create simulated maps that include both components and test them on the NPTF.  The recovered energy spectra for the SFG + Blazar--1 (Blazar--2) example is shown in the bottom left (right) panel of Fig.~\ref{fig:blESpec}.  In both cases, the PS spectrum is consistent with that found in the blazar-only simulations, which are shown in the top panels in that figure.  
The reconstructed source-count distributions for these examples are not shown, as they are consistent with those found in the blazar-only cases.

In the case of of the Blazar--1 model, the spectra of the smooth isotropic emission and the PS emission trace the spectra of the input SFG population and blazar population, respectively. In the case of the Blazar--2 model, the PS flux is further below the input blazar spectrum, as was found in the blazar-only simulations.  However, the smooth isotropic emission is further above the simulated SFG spectrum.  In both cases, the sum of the smooth isotropic emission and PS emission (EGB) is consistent with the simulated blazar plus SFG flux.  

There is, in fact, a subtle difference between the PS distribution recovered with and without the addition of a SFG population.  The difference becomes noticeable when comparing the fractions 
\begin{equation}
\frac{I_\text{iso}^\text{PS}}{I_\text{blazar-sim}} = [0.97_{-0.05}^{+0.06}, 1.00_{-0.09}^{+0.11}, 0.87_{-0.07}^{+0.09}, 0.72_{-0.09}^{+0.12}] \nonumber
\end{equation}
for SFG + Blazar--1 and 
\begin{equation}
\frac{I_\text{iso}^\text{PS}}{ I_\text{blazar-sim} }= [0.80_{-0.06}^{+0.08}, 0.59_{-0.06}^{+0.07}, 0.59_{-0.06}^{+0.08}, 0.43_{-0.05}^{+0.06}] \nonumber
\end{equation} 
for SFG + Blazar--2 to the corresponding values for the blazar-only simulations.  In the simulations with SFGs, the fractions  $I_\text{iso}^\text{PS} / I_\text{blazar-sim}$ are generally higher and have larger uncertainties.  The reason for this is that the SFG emission is degenerate with an enhanced sub-threshold component to the PS source-count distribution.

Simulating data with the PSF1--3 instrument response function, we find that the ratios $I_\text{iso}^\text{PS} / I_\text{blazar-sim}$ are somewhat closer to unity than in the PSF3 case.  In particular, for the SFG  +  Blazar--2 model simulations,
\begin{equation}
\frac{I_\text{iso}^\text{PS}}{ I_\text{blazar-sim}} = [1.03_{-0.13}^{+0.20}, 0.73_{-0.05}^{+0.06}, 0.66_{-0.06}^{+0.07}, 0.57_{-0.06}^{+0.07}] \,.  \nonumber
\end{equation}
The improved exposure allows the NPTF to probe lower fluxes and to therefore recover a larger fraction of the isotropic-PS emission.

\section{Low-Energy Analysis: 1.89--94.9 GeV} 
\label{sec:lowenergy}

The findings from the previous section illustrate that the NPTF procedure is able to set strong constraints on the PS (\emph{e.g.}, blazar) and smooth Poissonian (\emph{e.g.}, SFGs, mAGN) contributions to the EGB.
 In this section, we focus on the energy range from 1.89--94.9~GeV, and begin by presenting the results of our benchmark analysis on the real \emph{Fermi} data.  This is followed by a detailed discussion of potential systematic uncertainties and their effects on the conclusions.  

\subsection{ Pass~8~{\it ultracleanveto} Data}
\label{sec:benchmark}

\subsubsection{Top PSF Quartile}

We begin by analyzing the {\it ultracleanveto} PSF3 data for $\abs{b} \geq 30^\circ$, using the \texttt{p8r2} foreground model.  This is referred to as the ``benchmark analysis'' throughout the text.  Table~\ref{tab:bestfit} provides the best-fit intensities for each spectral component, as a function of energy, and the best-fit spectra are plotted in the left panel of Fig.~\ref{fig:nptfdata}.   The \texttt{p8r2} diffuse model is shown in green (median only), while the smooth isotropic and isotropic-PS posteriors are shown by the blue and red bands, respectively.  The best-fit spectrum for PSs with $|b| > 30^\circ$ in the 3FGL catalog~\cite{Acero:2015hja} is shown by the dashed black line in Fig.~\ref{fig:nptfdata}; the spectrum as plotted should be treated with care as systematic uncertainties are not properly accounted for.  In particular, the 3FGL catalog includes sources between 0.1--300~GeV.  At the high end of this range, the spectrum is driven to a large extent by extrapolations from lower energies, where the statistics are better.  The potential errors associated with such extrapolations are difficult to quantify and are not shown in Fig.~\ref{fig:nptfdata}.    
As a result, a direct comparison between the 3FGL spectrum and our results is difficult to make, especially in the highest energy bins.  For this reason, we have a dedicated NPTF study for energies greater than 50~GeV in Sec.~\ref{sec:highenergy}.  Those results are compared to the \emph{Fermi} 2FHL catalog~\cite{TheFermi-LAT:2015ykq}, which is explicitly constructed at higher energies and is likely a more faithful representation of above-threshold PSs in this regime.    

\begin{table*}[phtb]
\renewcommand{\arraystretch}{1.4}
\setlength{\tabcolsep}{5pt}
\begin{center}
\resizebox{\textwidth}{!}{%

\begin{tabular}{ c  | c  c  c c  c   }
\toprule
 Energy & $I_\text{EGB}$&$I_\text{iso}^\text{PS}$ & $I_\text{iso}$ & $I_\text{diff}$ & $I_\text{bub}$   \\
$[\text{GeV}]$ &  \multicolumn{5}{c}{$\left[\text{cm}^{-2}\text{ s}^{-1}\text{ sr}^{-1}\right]$}    \\%
\midrule
1.89--4.75  
&  $1.38_{-0.04}^{+0.05} \times 10^{-7}$ & $9.00_{-0.54}^{+0.66} \times 10^{-8}$ & $4.82_{-0.52}^{+0.43} \times 10^{-8}$ & $3.22_{-0.02}^{+0.02} \times 10^{-7}$ & $2.90_{-0.69}^{+0.67} \times 10^{-8}$\\
4.75--11.9 &    
$5.46_{-0.22}^{+0.24} \times 10^{-8}$ & $2.68_{-0.21}^{+0.26} \times 10^{-8}$ & $2.77_{-0.21}^{+0.18} \times 10^{-8}$ & $7.38_{-0.16}^{+0.15} \times 10^{-8}$ & $1.44_{-0.39}^{+0.39} \times 10^{-8}$   \\
11.9--30.0  & 
$1.76_{-0.09}^{+0.10} \times 10^{-8}$ & $7.17_{-0.76}^{+0.99} \times 10^{-9}$ & $1.04_{-0.08}^{+0.08} \times 10^{-8}$ & $1.63_{-0.07}^{+0.07} \times 10^{-8}$ & $5.18_{-2.23}^{+2.35} \times 10^{-9}$  \\  
30.0-94.9  &  
$5.74_{-0.41}^{+0.46} \times 10^{-9}$ & $2.40_{-0.38}^{+0.48} \times 10^{-9}$ & $3.30_{-0.42}^{+0.39} \times 10^{-9}$ & $3.73_{-0.33}^{+0.31} \times 10^{-9}$ & $1.46_{-0.92}^{+1.25} \times 10^{-9}$  \\
\bottomrule
\end{tabular}}
\end{center}
\caption{Best-fit intensities for all templates used in the NPTF analysis of Pass~8 {\it ultracleanveto} PSF3 data and the \texttt{p8r2} foreground model.  Note that the \emph{Fermi} bubbles template intensity is defined relative to the interior of the bubbles, while the intensities of the other templates are computed with respect to the region $\abs{b} \geq 30^\circ$.  The best-fit EGB intensity, which is the sum of the smooth and PS isotropic contributions, is also shown.  
}
\label{tab:bestfit}
\end{table*}
\begin{table*}[phtb]
\renewcommand{\arraystretch}{1.3}
\setlength{\tabcolsep}{3pt}
\begin{center}
\resizebox{\textwidth}{!}{%

\begin{tabular}{ c  | c  c  c c |  c c c   }
\toprule
 Energy & $n_1$ & $n_2$ & $n_3$ & $n_4$ & $F_{b,3}$ & $F_{b,2}$ & $F_{b,1}$   \\
$[\text{GeV}]$ &  & & & & \multicolumn{3}{c}{$\left[\text{cm}^{-2}\text{ s}^{-1}\right]$}    \\
\midrule
1.89--4.75 &  
$3.96_{-0.80}^{+0.68}$ & $2.04_{-0.05}^{+0.05}$ & $1.74_{-0.37}^{+0.19}$ & $-0.40_{-1.05}^{+1.18}$ & $1.13_{-0.52}^{+0.39} \times 10^{-11}$ & $1.22_{-0.56}^{+2.00} \times 10^{-10}$ & $1.43_{-0.46}^{+0.51} \times 10^{-8}$
   \\
4.75--11.9 &    
$3.84_{-0.86}^{+0.78}$ & $2.13_{-0.13}^{+0.15}$ & $1.91_{-0.12}^{+0.09}$ & $-0.44_{-1.03}^{+1.21}$ & $1.16_{-0.51}^{+0.47} \times 10^{-11}$ & $2.95_{-1.79}^{+1.80} \times 10^{-10}$ & $5.52_{-2.06}^{+2.66} \times 10^{-9}$\\
11.9--30.0  & 
$3.54_{-0.91}^{+0.96}$ & $2.42_{-0.32}^{+0.41}$ & $1.97_{-0.13}^{+0.11}$ & $-0.14_{-1.15}^{+1.13}$ & $1.11_{-0.50}^{+0.52} \times 10^{-11}$ & $3.47_{-1.76}^{+1.56} \times 10^{-10}$ & $2.83_{-1.34}^{+1.34} \times 10^{-9}$    \\  
30.0-94.9  &  
$3.63_{-0.98}^{+0.89}$ & $1.83_{-0.47}^{+0.52}$ & $2.51_{-0.21}^{+0.29}$ & $-0.20_{-1.16}^{+1.15}$ & $1.02_{-0.46}^{+0.47} \times 10^{-11}$ & $2.48_{-1.36}^{+1.86} \times 10^{-10}$ & $1.68_{-0.65}^{+0.68} \times 10^{-9}$   \\
\bottomrule
\end{tabular}}
\end{center}
\caption{Best-fit parameters for the source-count distributions recovered for each energy bin; the flux breaks  $F_{b,i}$ and indices $n_i$ are labeled from highest to lowest ($F_{b,i} > F_{b,i+1}$).  These values correspond to the NPTF analysis of Pass~8 {\it ultracleanveto} PSF3 data with the \texttt{p8r2} foreground model. The median and 68\% credible  intervals are recovered from the posterior distributions.
}
\label{tab:bestfit_dndf}
\end{table*}

\begin{figure*}[!htbp] 
   \centering
   \includegraphics[width=0.9\textwidth]{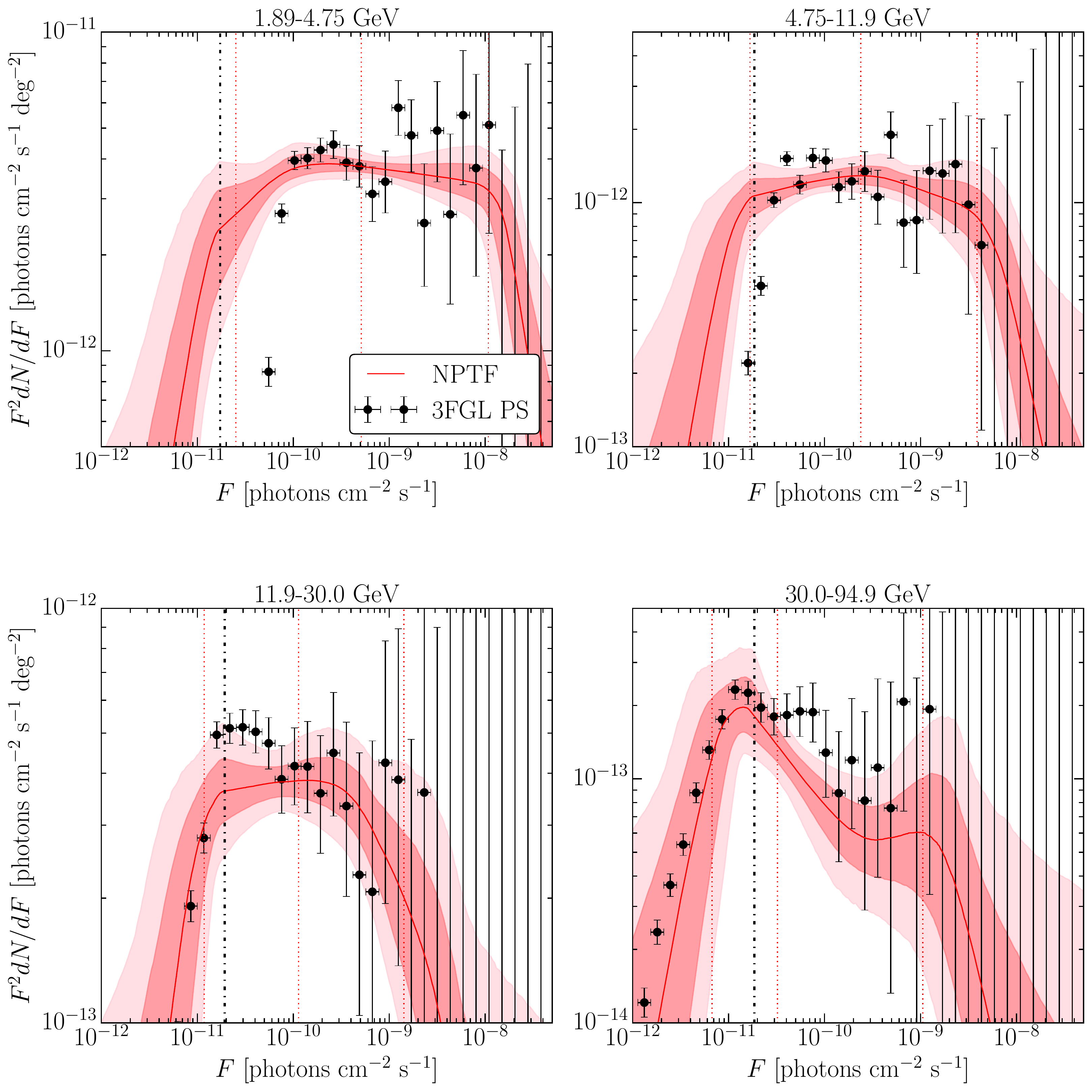} 
   \caption{The best-fit source-count distribution, as a function of energy, for the isotropic-PS population obtained by the NPTF analysis of Pass~8 {\it ultracleanveto} PSF3 data with the \texttt{p8r2} foreground model.  The median (red line) and 68 and 95\% credible intervals (shaded red bands) are shown.  The vertical dot-dashed black line denotes the $\sim$1 photon boundary, below which the NPTF begins to lose sensitivity.  The vertical dotted red lines indicate the fluxes at which 90\%, 50\%, and 10\% of the flux is accounted for, on average, by sources of larger flux (from left to right, respectively). The black points correspond to the \emph{Fermi} 3FGL sources, with 68\% statistical error bars (vertical).  The NPTF is expected to be sensitive down to the $\sim$1 photon limit, extending the reach to sources below the 3FGL detection threshold.  This is most apparent in the lowest energy bin, where the apparent 3FGL flux threshold is  $\sim$10 times higher than that for the NPTF.   
 We caution the reader that, at higher energies, the 3FGL spectra are driven by extrapolations from low energies where the statistics are better.  The systematic uncertainties associated with this extrapolation are difficult to quantify and are not included in the source counts shown here.}
   \label{fig:dndsdata}
\end{figure*}

The source-count distributions reconstructed from the NPTF are shown in Fig.~\ref{fig:dndsdata}, with best-fit parameters provided in Tab.~\ref{tab:bestfit_dndf}.  For comparison, the binned 3FGL source-count distributions are also plotted; the vertical error bars represent 68\% statistical uncertainties  and do not account for systematic uncertainties.  
A few trends are clearly visible.  First, each flux break tends to have large uncertainties.  This may be a reflection of the fact that the real source-count distribution is not a simple triply-broken power law, but rather a more complicated function, as in the blazar simulations of Sec.~\ref{sec:simulations}.  Therefore, the best-fit values for each of these parameters, when viewed independently, may be somewhat deceptive.  As is evident in Fig.~\ref{fig:dndsdata}, the posteriors for the breaks and indices are distributed in such a way as to describe a smooth concave function for $F^2 dN/dF$.

At very high and very low flux, the uncertainties on the indices ($n_1$ and $n_4$, respectively) become large.  At high flux, this is simply due to the fact that there are very few sources, so the source-count distribution falls off rapidly.  At low flux, the large uncertainties on $n_4$ arise from the difficulty in distinguishing the isotropic-PS contribution from its smooth counterpart.  Indeed, below the single-photon boundary (dot-dashed black line), the NPTF analysis starts to lose sensitivity.  The posterior distributions for the slopes above (below) the highest (lowest) break are highly dependent on the priors and so the quoted values in Tab.~\ref{tab:bestfit_dndf} should be treated with care.

The presence of any distinctive breaks encodes information about the number of source populations as well as their evolutionary properties.  In all energy bins, we see that the NPTF places the lowest break, $F_{b,3}$, close to the one-photon sensitivity threshold and the highest break, $F_{b,1}$, in the vicinity of the highest-flux 3FGL source (see Tab.~\ref{tab:bestfit_dndf} for the exact values).  The evidence for an additional break, $F_{b,2}$, at intermediate fluxes  varies depending on the energy bin.  From 1.89--4.75~GeV, there is strong indication for a break at fluxes $\sim$$10^{-10}$ ph cm$^{-2}$ s$^{-1}$, with the index $n_2 \approx 2.04$ above the break hardening to $n_3 \approx 1.74$ below the break.  In the two subsequent energy bins, up to $\sim$$30$ GeV, we also find evidence that the source-count distribution hardens as we move from high fluxes to below the second break, with the index $n_3$ below the second break $\sim$1.9-2.0 in both cases.
  In the last bin, the uncertainties are too large to determine if the source-count distribution changes slope at any flux above the lowest break $F_{b,1}$.

\subsubsection{Top Three PSF Quartiles}
\label{sec:benchmark_top3}
\begin{figure*}[!htbp] 
   \centering
   \includegraphics[width=0.9\textwidth]{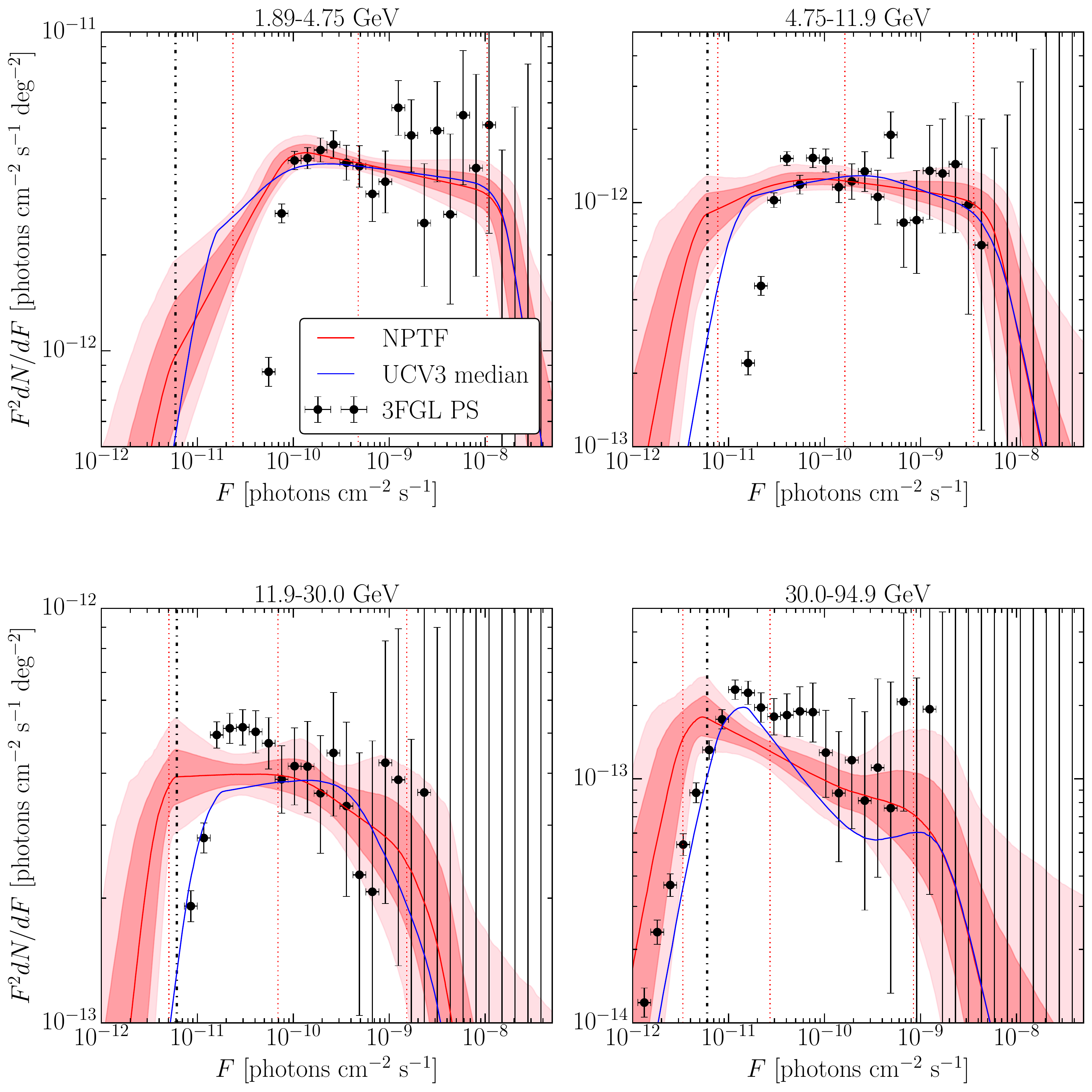} 
   \caption{The same as Fig.~\ref{fig:dndsdata}, except using the top three quartiles (PSF1--3) of the Pass 8 {\it ultracleanveto} data.  The median source-count distribution for the PSF3 analysis is shown in blue.  
   }
   \label{fig:dndsdata_top3}
\end{figure*}

The benchmark analysis described in the previous section used only the top quartile (PSF3) of the Pass 8 {\it ultracleanveto} data set.  This restriction selects events with the best angular resolution, but at the price of reducing the total photon count.  In Sec.~\ref{sec:simulations}, we showed that  including the top three quartiles of the Pass 8 {\it ultracleanveto} data may help constrain the source-count distribution at low fluxes.  With that in mind, we now investigate how the results of the benchmark analysis change when using the PSF1--3 {\it ultracleanveto} data set.

In general, the best-fit intensities for the individual spectral components are consistent within uncertainties with those obtained using only the top quartile of data.  
The PS flux does increase slightly  in going from PSF3 to PSF1--3 in the upper energy bins due to the increased exposure.  More specifically, the ratios of the median PS intensities measured with {\it ultracleanveto} PSF1--3 data to those measured with PSF3 data are $[1.00,1.06,1.19,1.19]$ in the four increasing energy bins. 
This can also be seen in the associated spectral intensity plot (right panel of Fig.~\ref{fig:nptfdata}), where the red bands are further above the 3FGL line in the last energy bins than in the corresponding plot for the PSF3 analysis (left panel).   
The intensity of the EGB is seen to increase slightly, in all energy bins, when going from PSF3 to PSF1--3 data, potentially suggesting additional cosmic-ray contamination with the looser photon-quality cuts, though the increases in EGB intensities are within statistical uncertainties.     

The best-fit source-count distributions recovered by the NPTF with PSF1--3 data are shown in Fig.~\ref{fig:dndsdata_top3}.  For reference, the blue curve shows the best-fit for the PSF3--only analysis.  The most important difference between the PSF3 and PSF1--3 results is that the source-count distributions extend to lower flux with PSF1--3 data.  This is due to the fact that the exposure in each energy bin, averaged over the region of interest, is larger for the top three quartiles compared to the top quartile alone.  As a result, the flux corresponding to single-photon detection is lower (compare the vertical dot-dashed line in Fig.~\ref{fig:dndsdata_top3} with that in Fig.~\ref{fig:dndsdata}), which improves the NPTF reach.  Thus, the PSF1--3 analysis is sensitive to more sub-threshold sources.  
Note that the same trend was observed in the simulation tests in Sec.~\ref{sec:simulations} in going from PSF3 to PSF1--3 data sets.

Other than the location of the lowest break, which is lower due to the increased exposure, all other source-count distribution parameters are consistent, within uncertainties, between analyses.  At the lowest energy, the break at \mbox{$F_{b,2} \sim 10^{-10}$} photons~cm$^{-2}$~s$^{-1}$ is even more pronounced, with an index $n_2 \sim 2.10$ above the break and $n_3 \sim 1.75$ below the break.  In the highest energy bin, the structure observed in the source-count distribution for the benchmark analysis has smoothed out.

\subsection{Systematic Tests}
\label{sec:systematictests}

The previous subsection illustrated how the results of the NPTF change when additional {\it ultracleanveto} PSF quartiles are included in the analysis. We also tested the stability of our analysis to variations in the region of interest, \emph{Fermi} event class,  foreground modeling, \emph{Fermi} bubbles, PSF modeling, and choice of priors.  

Figure~\ref{fig:systematicsplot} briefly summarizes the results.  The EGB intensity as measured by \emph{Fermi} is shown by the gray band.  To obtain this band, we use the best-fit power-law spectrum with exponential cut-off provided in~\cite{Ackermann:2014usa}; the width of the gray band is found by varying between best-fit values for the three foreground models considered in that paper (Models A/B/C) and does not include statistical uncertainties, which become increasingly important at high energies.  The smooth isotropic intensity, and thus the intensity of the EGB, is subject to large systematic uncertainties.   As expected, the variation in smooth isotropic intensity is most pronounced when using the {\it source} event class, which contains more cosmic-ray contamination.    However, the spectrum of emission from PSs as captured by the NPTF appears robust to all the systematic effects considered here.  This is the primary conclusion of this subsection.  We now describe in detail the systematic tests that were conducted for the low-energy analysis.

\subsubsection{Region of Interest}

As a first cross-check on the stability of the results presented in Sec.~\ref{sec:benchmark}, we explore the effects of altering the region of interest.  While we previously defined the region of interest with $|b| \geq 30^\circ$, we now loosen this constraint and consider the case $|b| \geq 10^\circ$.  Extending the region of interest closer to the Galactic disk increases the amount of data being analyzed, but at the cost of potentially more contamination from diffuse foreground emission and local PSs.  As shown in Fig.~\ref{fig:systematicsplot}, the best-fit intensities for the isotropic and isotropic-PS components are equivalent, within errors, to their counterparts in the benchmark analysis. 

We also ran the NPTF on the Northern ($b > 30^\circ$) and Southern ($b < -30^\circ$) hemispheres separately.  The intensities for the EGB, IGRB, and PS components are systematically lower (higher) for the Northern (Southern) analysis than for the benchmark case.  

\subsubsection{Event class}

We explored the implications of broadening the {\it ultracleanveto} data set to include the top three quartiles in Sec. \ref{sec:benchmark_top3}.   Now, we consider the implications of repeating the NPTF analysis on the {\it source} data with PSF1--3.  This event class has looser photon-quality cuts, which leads to larger overall exposure, but significantly more cosmic-ray contamination.  In general, it is not recommended to use {\it source} data for IGRB studies; for our purposes, however, it will be intriguing to see how the increased photon statistics affect the recovered source-count distribution for the PS component.  As shown in Fig.~\ref{fig:systematicsplot}, the EGB intensity is far larger than that recovered by the benchmark analysis and overpredicts \emph{Fermi}'s EGB result in most energy bins.  The sharp rise in the EGB intensity can be traced to a substantial fraction of smooth isotropic emission, which is expected for this event class at most energies.  Most importantly, the intensity of the isotropic-PS component is consistent, within uncertainties, with that found in the benchmark analysis.\footnote{The recovered PS intensity is slightly larger with {\it source} PSF1--3 data as compared to {\it ultracleanveto} PSF3 data, which is likely due to the increased exposure in the {\it source} PSF1--3 data set. 
}   This is a  confirmation that the NPTF is able to successfully constrain the source-count distribution even in a data set with significantly more smooth isotropic flux.  


\subsubsection{Foreground Model}
\label{sec:foreground}

A potentially significant source of systematic uncertainty in the NPTF analysis is due to mis-modeling of high-energy gamma-rays produced in cosmic-ray propagation in the Milky Way~\cite{Ackermann:2012pya}.  These high-energy photons arise from  bremsstrahlung of electrons off the interstellar medium, boosted pion decay, and inverse Compton (IC)  emission off the interstellar radiation field.  Our benchmark analysis uses 
the associated foreground model for the Pass~8 data set (\emph{gll\_iem\_v06.fits}), denoted here as \texttt{p8r2}.  The total diffuse emission in \texttt{p8r2} is modeled as a linear combination of several sources, some of which are traced by maps of gas column densities, which serve as templates for the pion and bremsstrahlung emission.  The IC component is modeled using the \texttt{GALPROP} package~\cite{Strong:2007nh}.\footnote{\url{http://galprop.stanford.edu/}}  These individual templates are fit to the data, and used to identify `extended emission excesses' that are identified directly and then added back into the model~\cite{Acero-2016}.

To better assess the uncertainties due to the foreground modeling, we repeat the NPTF analysis using several other foreground models made available by \emph{Fermi}.  In particular, we use the \emph{gll\_iem\_v02\_P6\_V11\_DIFFUSE.fits} diffuse emission model, denoted as \texttt{p6v11}, which was initially developed for the Pass~6 data set.\footnote{\url{http://fermi.gsfc.nasa.gov/ssc/data/access/lat/ring_for_FSSC_final4.pdf}}  \texttt{p6v11} is distinct from \texttt{p8r2} in that it uses older gas and IC maps and does not include templates for large-scale structure or extended emission excesses.  The Pass~7 model \emph{gal\_2yearp7v6\_v0.fits}, denoted as \texttt{p7v6},\footnote{\url{http://fermi.gsfc.nasa.gov/ssc/data/access/lat/Model_details/Pass7_galactic.html}} is a compromise as it uses updated gas and IC maps and includes some large-scale extended structures, such as Loop~1 and the \emph{Fermi} bubbles.

The NPTF results using the \texttt{p6v11} and \texttt{p7v6} foreground models are summarized in Fig.~\ref{fig:systematicsplot}.
In general, we observe that the intensity of the PS components is consistent with that for the benchmark analysis in all energy bins.  However, variations occur in the smooth isotropic intensity.  Typically, more IGRB intensity is recovered with \texttt{p6v11} and \texttt{p7v6}, versus \texttt{p8r2}.  The differences are particularly dramatic in the first two energy bins and are more severe for \texttt{p6v11}.  The net consequence is that the EGB intensity is higher than the expected range from \emph{Fermi}.  The enhancement in the isotropic component may arise from the fact that each foreground model incorporates large-scale diffuse structures differently---with \texttt{p6v11} being the least inclusive and \texttt{p8r2} being the most inclusive.  We note, however, that the fit to data with the \texttt{p8r2} foreground model, from the point of view of the Bayesian evidence, is much better than the analogous fit with the  \texttt{p6v11} model; the fit with the \texttt{p7v6} model is intermediate. 

\subsubsection{The Bubbles Template}

To better understand how dependent the analysis is on the details of the \emph{Fermi} bubbles template, we simply removed the template from the analysis.  This has indiscernible effects on the final results.  We see in Fig.~\ref{fig:systematicsplot} that the EGB, IGRB, and PS intensities are consistent, within uncertainties, to the corresponding values in the benchmark study.  

\subsubsection{Point Spread Function}

The PSF can affect the photon-count distribution because it can redistribute photons between pixels, and must therefore be properly accounted for in the calculation of the photon-count probability distributions.  For the primary analyses presented in this work, the PSF is modeled using a King function.  However, to test the sensitivity of the results to mis-modeling of the PSF, we have also repeated the NPTF analysis using a two-dimensional Gaussian in the calculation of the photon-count probability distributions, with a width set to give the correct 68\% containment radius.  As shown in Fig.~\ref{fig:systematicsplot}, the NPTF results remain unchanged with this substitution.  

\subsubsection{Priors}

\begin{figure*}[!phtb] 
   \centering
   \includegraphics[width=\textwidth]{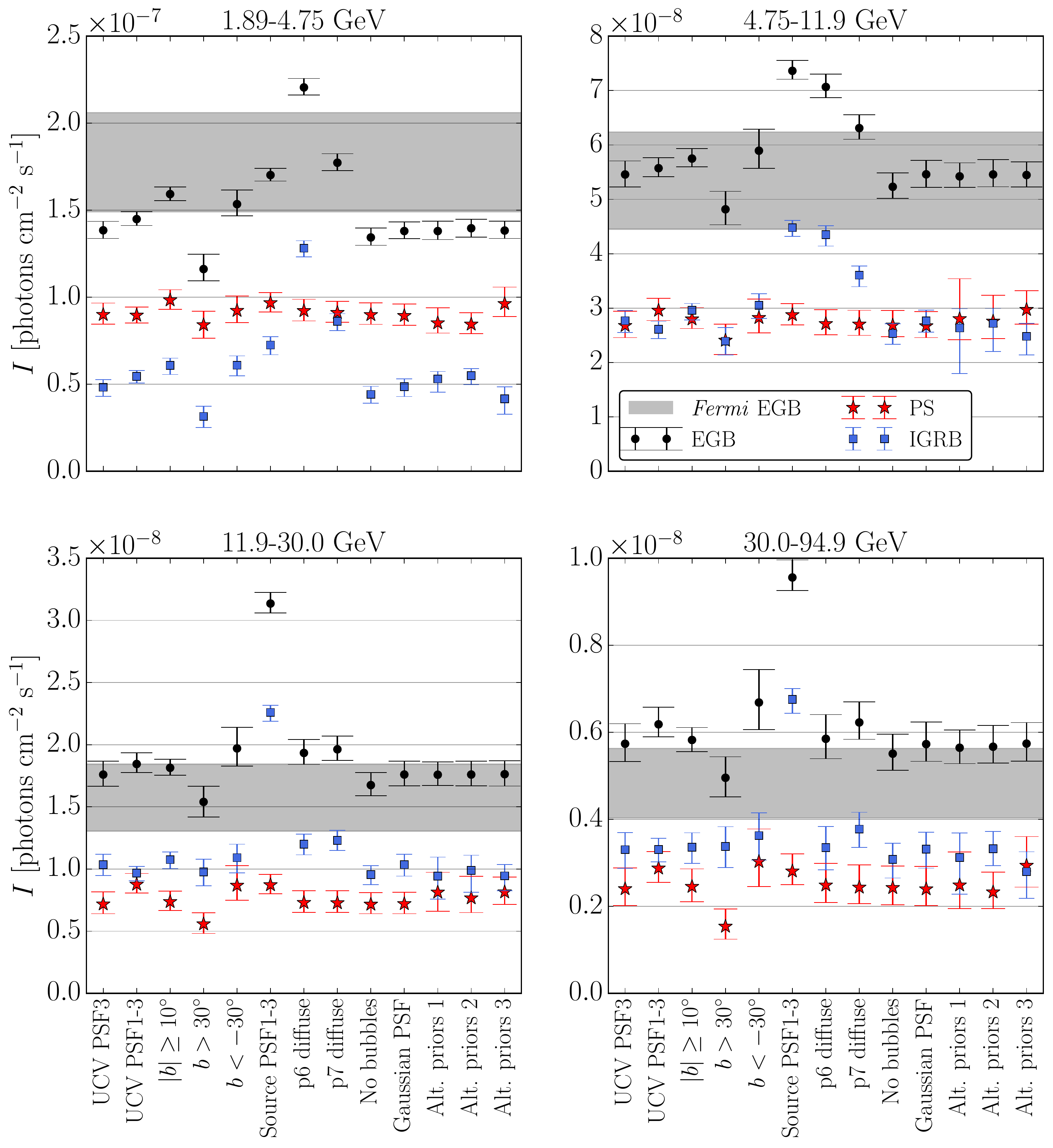} 
   \caption{Comparison of the EGB (black circles), IGRB (blue squares), and PS (red stars) intensities recovered by the NPTF for the various systematic tests described in Sec.~\ref{sec:systematictests}
   .  Note that `UCV' is shorthand for {\it ultracleanveto}.  The gray band is meant to indicate the systematic uncertainty associated with the measured \emph{Fermi} EGB~\cite{Ackermann:2014usa} (see text for more details).}
   \label{fig:systematicsplot}
\end{figure*}

Our choice of priors, given in Tab.~\ref{tab:priors}, is carefully chosen to both avoid biasing the posterior for the source-count distribution while at the same time allowing breaks at both high and low flux.  This is meant to properly account for the fact that the source-count distribution is not well constrained by the data at very high fluxes, where the mean expected number of sources over the full region is much less than unity, and at very low fluxes, where the mean photon-count per source is much less than unity.  Our choice of priors is further justified by the simulated data studies, presented in Sec.~\ref{sec:simulations}, which show that the NPTF can successfully constrain the emission from blazar models.  However, one may still be concerned that these particular choice of priors might bias the recovered source-count distribution in a particular way.  For that reason, we have tried many variations to the priors shown in Tab.~\ref{tab:priors}, three of which (labeled `Alt. priors 1--3')  are described below and shown in Fig.~\ref{fig:systematicsplot}:
\begin{itemize}
\item {\it Alternate prior 1}:  All priors are the same as in Tab.~\ref{tab:priors}, except for those on the breaks, which are changed to $[0.1,10]$, $[10,40]$, and $[40, 2 \times S_\text{b,max}]$ ph for $S_{b,1}$, $S_{b,2}$, and $S_{b,3}$, respectively. 
\item {\it Alternate prior 2}:  As above, except changing the priors for the breaks to $[1,20]$, $[20,S_\text{b,max}/2]$, and $[S_\text{b,max}/2, 2 \times S_\text{b,max}]$ ph, respectively.
\item {\it Alternate prior 3}:  All priors are the same as in Tab.~\ref{tab:priors}, except for that of $n_4$, which is changed to $[1,1.99]$.
\end{itemize}

The first two examples address the possibility that the break priors might artificially sculpt the source-count distribution and the recovered PS intensity, while the third example addresses how the source-count distribution is dealt with at fluxes below the lowest break, where the distribution is not well constrained by the data.  In many classes of blazar models, such as those considered in Sec.~\ref{sec:simulations}, the index below the lowest break ($n_4$) is greater than unity, so that the total number of PSs $\sim$$\int_{F_\text{min}} dF \, dN / dF$ diverges as the minimum flux cut-off $F_\text{min}$ is taken to zero.

It is useful to know if the recovered PS intensity, $I_\text{iso}^\text{PS}$, tends to under or overshoot the simulated blazar intensity, $I_\text{blazar-sim}$, when using the alternate priors.  With that in mind, we run the NPTF on simulated maps, as in Sec.~\ref{sec:simulations}, constructed from both the SFG + Blazar--1 model as well as the SFG + Blazar--2 model.   For \emph{Alternate prior 1}, we find that
\begin{equation}
{I_\text{iso}^\text{PS} \over I_\text{blazar-sim}} = [0.87_{-0.04}^{+0.05},0.93_{-0.08}^{+0.17},0.92_{-0.15}^{+0.23},0.61_{-0.07}^{+0.11}]  \nonumber
\end{equation}
and
\begin{equation}
{I_\text{iso}^\text{PS} \over I_\text{blazar-sim}} = [0.68_{-0.05}^{+0.06},0.59_{-0.09}^{+0.15},0.52_{-0.05}^{+0.07},0.37_{-0.03}^{+0.05}]  \nonumber
\end{equation} 
for the SFG + Blazar--1 and SFG + Blazar--2 models, respectively, with {\it ultracleanveto} PSF3 instrument response function.
With {\it Alternate prior 1}, we see larger uncertainties, with the PS template capable of absorbing more flux in particular.  With \emph{Alternate prior 2}, on the other hand, we find more noticeable differences in the medians as well as in the uncertainties.  In particular, for the SFG + Blazar--1 and SFG + Blazar--2 models, we find
\begin{equation}
{I_\text{iso}^\text{PS} \over I_\text{blazar-sim}} = [1.01_{-0.10}^{+0.12}, 1.27_{-0.31}^{+0.16}, 1.25_{-0.15}^{+0.12}, 0.73_{-0.12}^{+0.21}]  \nonumber
\end{equation}
and
\begin{equation}
{I_\text{iso}^\text{PS} \over I_\text{blazar-sim} }= [0.74_{-0.06}^{+0.19}, 0.94_{-0.19}^{+0.20}, 0.61_{-0.10}^{+0.17}, 0.41_{-0.05}^{+0.09}] \,, \nonumber
\end{equation} 
respectively.  In the Blazar--1 model case, it is important to notice that at intermediate energies the NPTF tends to over-predict $I_\text{blazar-sim}$ at the $\sim$20\% level.  With \emph{Alternate prior 3}, the results are 
\begin{equation}
{I_\text{iso}^\text{PS} \over I_\text{blazar-sim}} = [1.06_{-0.09}^{+0.15}, 1.10_{-0.09}^{+0.14}, 1.00_{-0.10}^{+0.14}, 0.85_{-0.11}^{+0.15}]  \nonumber
\end{equation}
and
\begin{equation}
{I_\text{iso}^\text{PS} \over I_\text{blazar-sim}} = [0.92_{-0.09}^{+0.16}, 0.77_{-0.14}^{+0.39}, 0.69_{-0.08}^{+0.12}, 0.53_{-0.06}^{+0.10}] \,, \nonumber
\end{equation} 
for the Blazar--1 and Blazar--2 models.  The \emph{Alternate prior 3} results are consistently closer to unity than the first two alternate prior results. 

As may be seen in Fig.~\ref{fig:systematicsplot}, the median values for the PS intensities recovered from the NPTF analyses with alternate priors are generally consistent with those found in the baseline study.  The {\it Alternate prior 3} PS intensities are slightly enhanced in all energy bins compared to the baseline results---following our expectations from the simulation results presented above---though the two results are consistent within statistical uncertainties.  


\section{High-Energy Analysis: 50--2000 GeV}
\label{sec:highenergy}

We now consider the NPTF results at high energies from 50--2000~GeV.  The number of photons available decreases when moving to higher energies, so we loosen the restrictions on the PSF quartiles to maximize the sensitivity potential of the NPTF.  In this section, the majority of the analyses are done using all quartiles of the {\it ultracleanveto} data, though we also show results using all quartiles of {\it source} data.  For the same reason, we widen the ROI to $|b| > 10^\circ$ rather than $30^\circ$, although the results are not sensitive to this cut, as we will show.

\begin{figure*}[!phtb] 
   \centering
   \includegraphics[width=\textwidth]{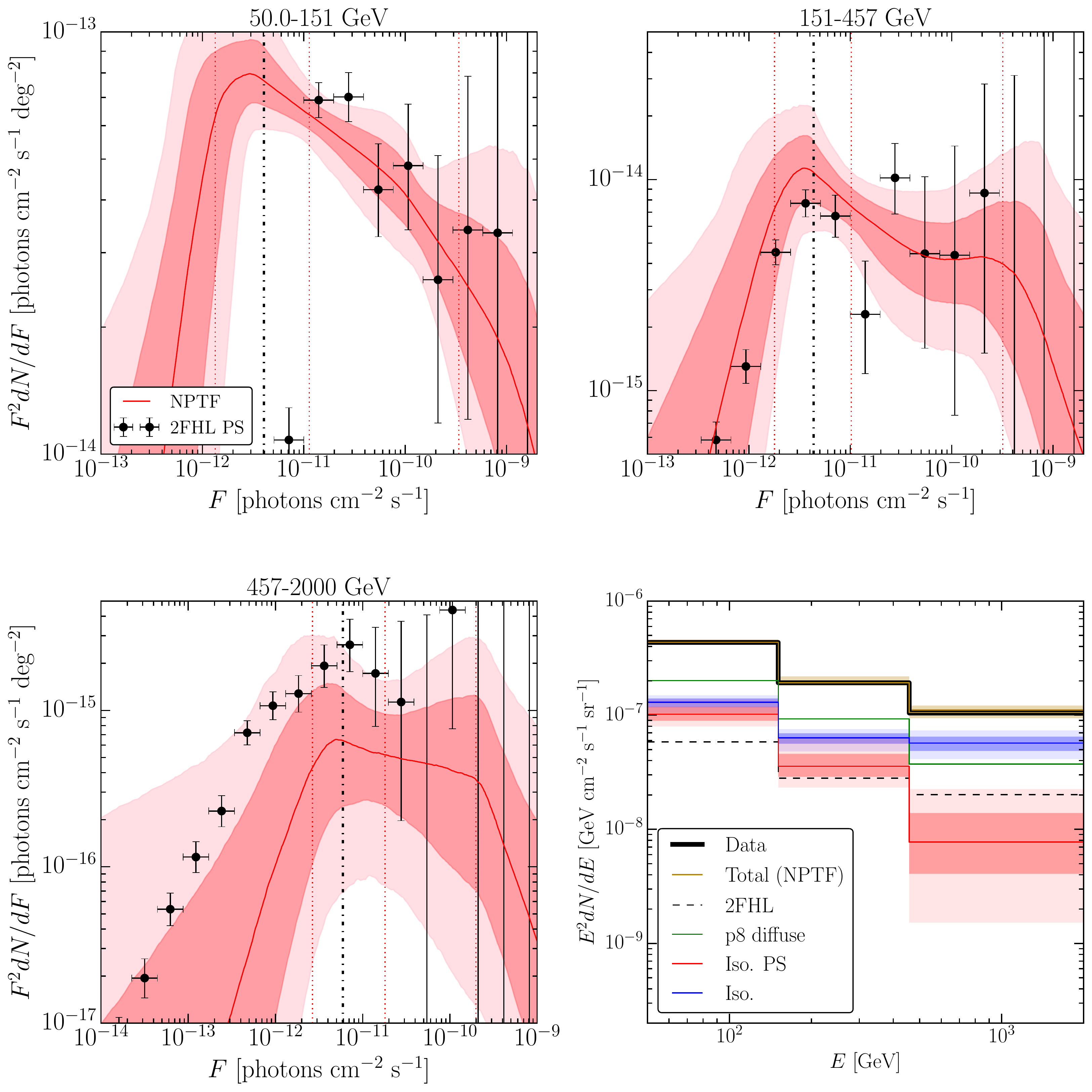} 
   \caption{NPTF results for the high-energy analysis of all quartiles of Pass8 {\it ultracleanveto} data. Top row and bottom left panel:  The best-fit source-count distribution for the isotropic-PS population, for each separate energy bin, is shown using the same format conventions as Fig.~\ref{fig:dndsdata}.  The black points correspond to the \emph{Fermi} 2FHL sources~\cite{Ackermann:2015uya}, with 68\% statistical error bars (vertical).  Bottom right panel: Best-fit energy spectrum.  The 68 and 95\% credible intervals are shown for the isotropic-PS and smooth isotropic templates in red and blue, respectively.  The median intensity for the foreground is also included (green).  The sum of all the components (yellow band) agrees with the total spectrum of the \emph{Fermi} data (black).  The spectrum of the 2FHL sources is provided in dashed black.  
   Note that, as for the 3FGL case, the spectra of 2FHL sources are driven at the high end by extrapolations from lower energies; the associated uncertainties are not shown here. }
   \label{fig:dndsdata_HE}
\end{figure*}

The best-fit energy spectra recovered by the NPTF analysis for the high-energy study of  {\it ultracleanveto} data is shown in the bottom right panel of Fig.~\ref{fig:dndsdata_HE}.
The fit results are compared with the best-fit  energy spectrum for sources in \emph{Fermi}'s 2FHL catalog~\cite{Ackermann:2015uya} (dashed black line).  This recently-published catalog is based on 80 months of data and focuses on hard sources in the range from 50--2000~GeV.  Statistical and systematic uncertainties are not accounted for in the determination of the 2FHL spectrum in Fig.~\ref{fig:dndsdata_HE}; these are likely non-negligible, especially at the highest energies.

The best-fit source-count distributions for the three energy bins are also shown in Fig.~\ref{fig:dndsdata_HE}, in the top row and bottom left panel.  The black points in those panels denote the 2FHL source-count distributions, with vertical error bars indicating 68\% Poisson errors.  The statistical errors on the 2FHL sources are large due to the fact that there are not many sources.  In all energy bins, the NPTF places the lowest break close to the single-photon sensitivity threshold (vertical dot-dashed line) and the highest break in the vicinity of the brightest 2FHL source, just as in the low-energy analysis.  Most notably in the 50--151~GeV bin, the NPTF probes unresolved sources with fluxes nearly an order-of-magnitude below the apparent 2FHL threshold.  We find no evidence for an additional intermediate-flux break in any of the energy bins, although it is difficult to make conclusive statements due to the large uncertainties in the individual source-count distributions. 

\begin{figure*}[!phtb] 
   \centering
   \includegraphics[width=\textwidth]{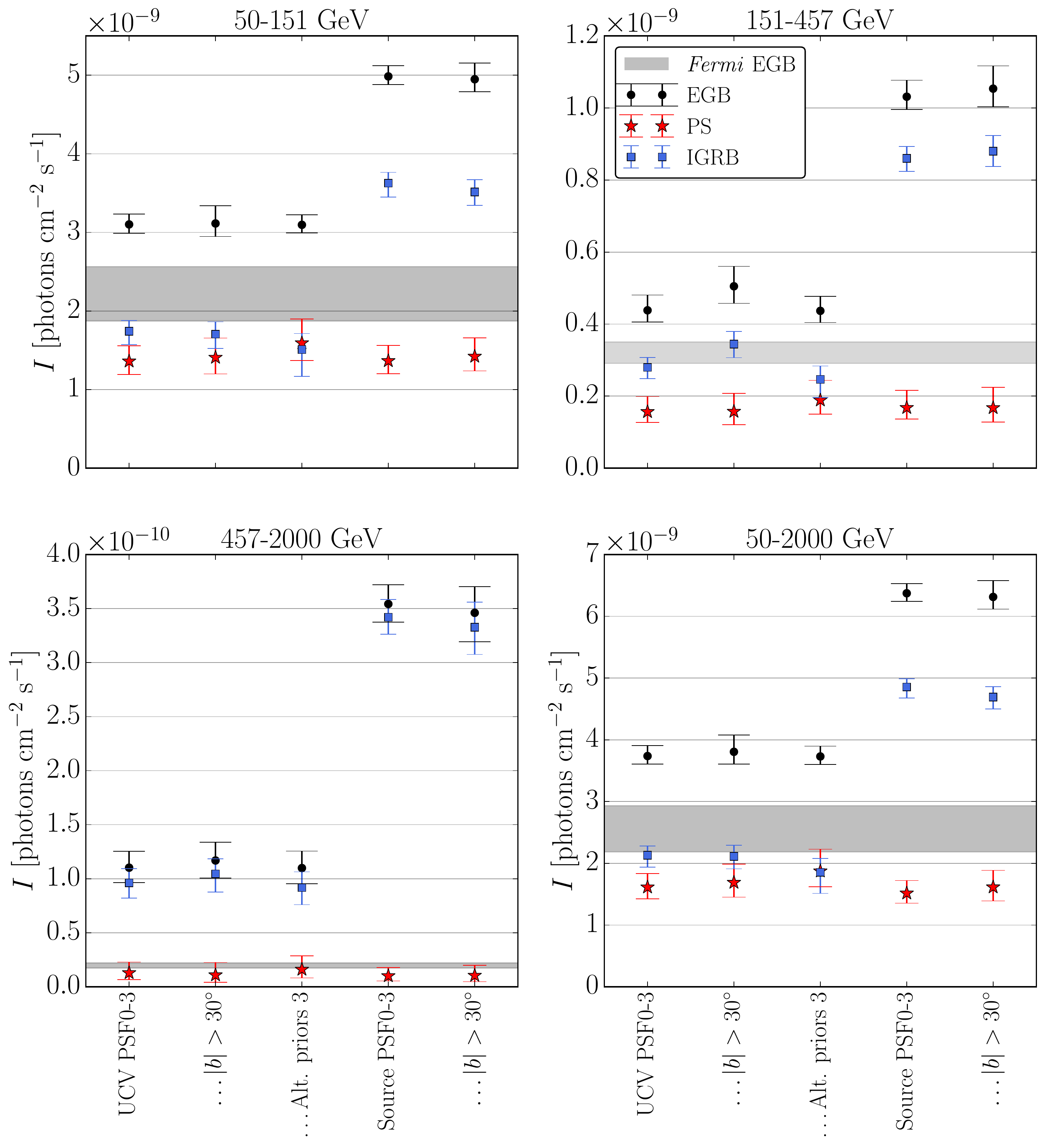} 
   \caption{Comparison of the EGB (black circles), IGRB (blue squares), and PS (red stars) intensities recovered by the NPTF for the various systematic tests specific to high energies.  The gray band indicates the systematic uncertainty associated with the measured \emph{Fermi} EGB~\cite{Ackermann:2014usa}.  
   }
   \label{fig:systematics_HE}
\end{figure*}

We have completed a number of systematic tests of the high-energy analyses that include looking at all quartiles of the {\it source} data, requiring $|b| > 30^\circ$ for both event classes, and using the third alternate prior choice, with $n_4 > 1$.   The results are summarized in Fig.~\ref{fig:systematics_HE}.
Importantly, the isotropic-PS intensity is consistent across all the tests.  However, the EGB intensities recovered by the NPTF are, in general, higher than those measured by \emph{Fermi}.  This discrepancy is likely due to increased cosmic-ray contamination above $\sim$100~GeV, as suggested by the high IGRB intensities recovered by the NPTF at these energies.  Indeed, the \emph{Fermi} EGB study on Pass~7 data~\cite{Ackermann:2014usa} used dedicated event classes with specific data cuts to minimize such contributions.  Such an analysis is beyond the scope of this study, as our primary focus is on the PS populations.  We simply caution the reader that the derived intensity for the smooth isotropic component in the high-energy analyses is subject to potentially large contamination.
\begin{figure*}[!htbp] 
   	\begin{center}$
	\begin{array}{cc}
	\scalebox{0.43}{\includegraphics{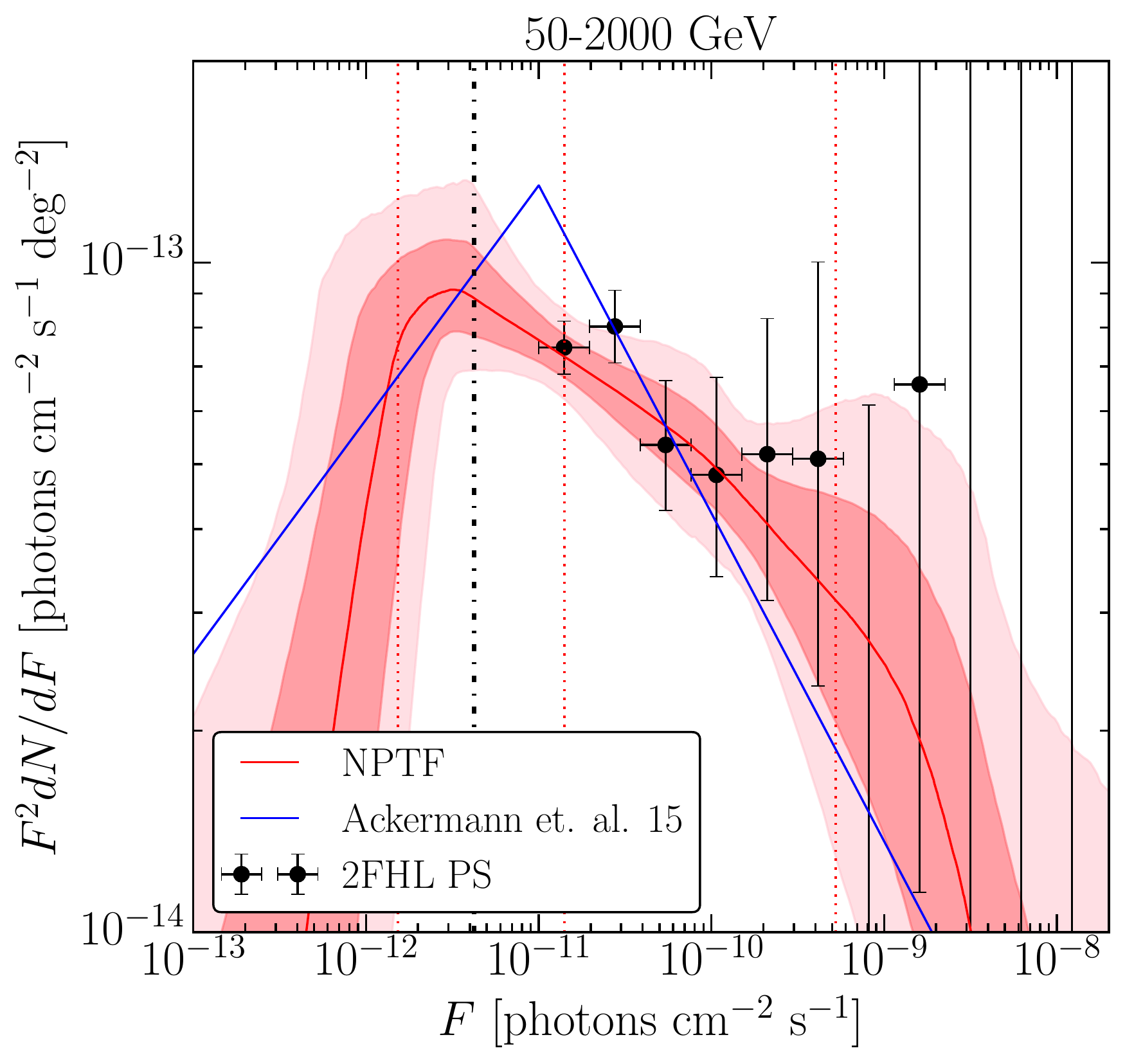}} &\scalebox{0.5}{\includegraphics{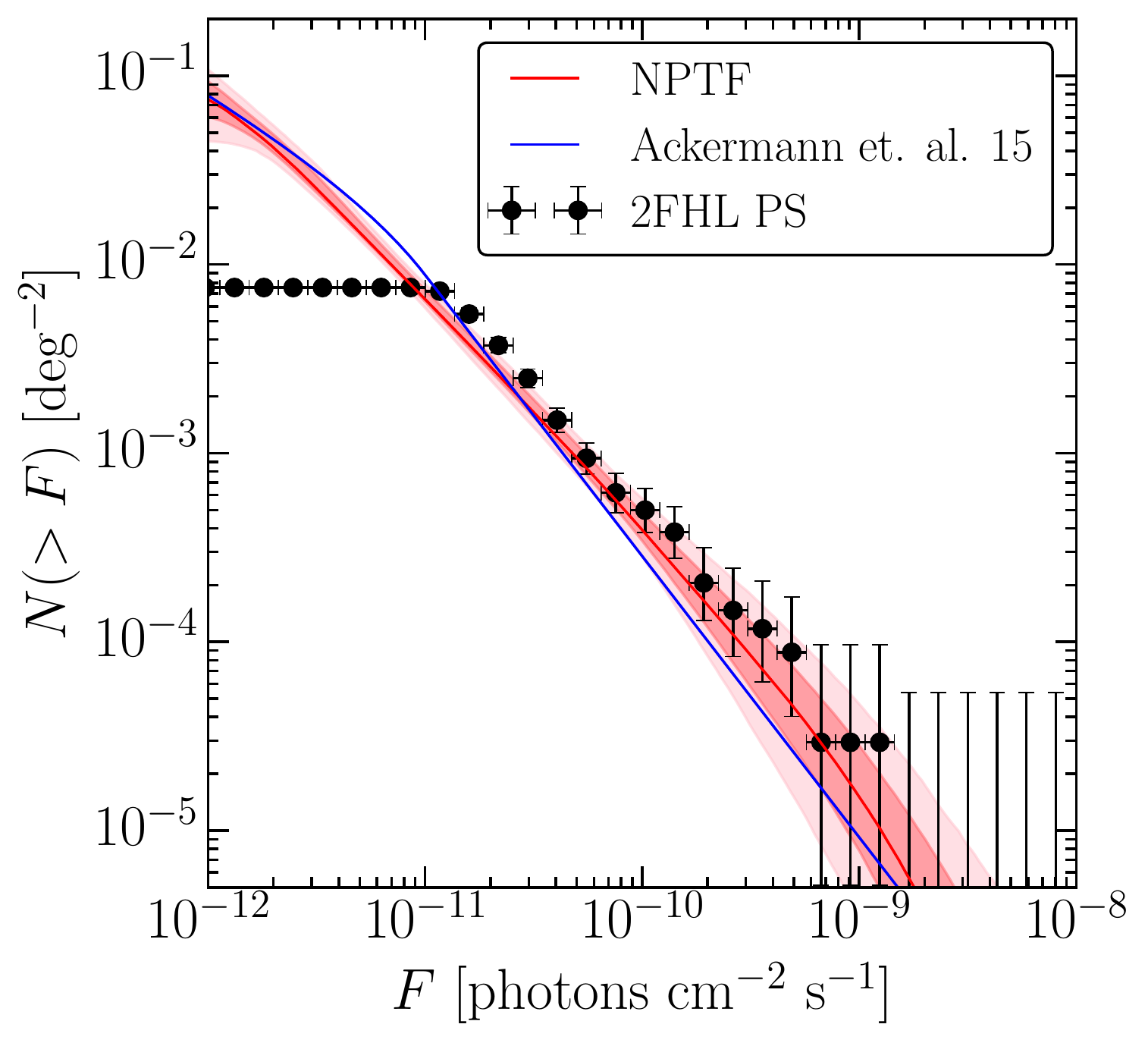}} 
	\end{array}$
	\end{center}
\caption{\textbf{(Left)}  Best-fit source-count distribution in the wide-energy bin from 50--2000~GeV using all quartiles of Pass~8 {\it ultracleanveto} data.  The black points indicate the 2FHL sources, and the blue line denotes the best-fit source-count from~\cite{TheFermi-LAT:2015ykq} that corresponds to the same energy bin. \textbf{(Right)}  A comparison of the cumulative source-count distribution for the same analysis.}
   \label{fig:dnds_0_10} 
\end{figure*}

It is possible to make stronger statements about the best-fit source-count distribution at high energies if we consider the wide-energy bin from 50--2000~GeV.  The results are shown in the left panel of Fig.~\ref{fig:dnds_0_10}.  Due to the improved statistics, the uncertainties on the source-count distribution are smaller than those for the three sub-bins.  Other than the low-flux sensitivity break, the NPTF finds no preference for  an additional break.  The intermediate-flux break, $F_{b,2}$, is essentially unconstrained as a result, and the power-law slope above (below) it are consistent within uncertainties: $n_2 = 2.28_{-0.22}^{+0.28}$ and $n_3 = 2.17_{-0.09}^{+0.12}$, respectively.  We compare this result to the best-fit source-count distribution (blue line) published by \emph{Fermi} for sources in this same energy range~\cite{TheFermi-LAT:2015ykq}.  There are important differences between the two analyses.  In the \emph{Fermi} study, simulated maps were created using several different source-count distributions, parametrized as singly broken power laws.  The histogram of the photon-count distribution for each of these maps, averaged over the full region of interest, was compared to the actual data, and a fit was done to select the simulated maps that most closely resembled the data.  
This method is related to but in many ways distinct from the NPTF.  The NPTF considers the difference between Poissonian and non-Poissonian photon probability distributions at the pixel-by-pixel level, instead of averaging the distributions over the full region.  Moreover, in our analysis we rely on semi-analytic techniques to calculate the photon-count probability distributions as we scan over the space of model parameters, instead of relying on Monte Carlo samples to numerically construct these distributions.  As a result, we are able to consider source-count distributions with additional degrees of freedom and also scan over the normalizations of all of the background templates, which tend to be well determined given the pixel-by-pixel nature of the fit.  In contrast, the intensity of all Poissonian models in~\cite{TheFermi-LAT:2015ykq}, including the smooth isotropic emission, was kept fixed while scanning over the source-count distribution degrees of freedom.

The cumulative source-count plot is provided in the right panel of Fig.~\ref{fig:dnds_0_10}.  Our result is in good agreement with the 2FHL sources above the catalog sensitivity threshold $\sim$$10^{-11}$~ph cm$^{-2}$~s$^{-1}$.  In the first few flux bins above this threshold, there appear to be more 2FHL sources than what is predicted by the NPTF, although the results are still consistent within uncertainties.  This may be due to the Eddington bias~\cite{Eddington} where extra sources are observed above threshold due to upward statistical fluctuations from sources immediately below.

Based on the results in Fig.~\ref{fig:dnds_0_10}, we can project the expected number of these sources that may be observed by the Cherenkov Telescope Array (CTA)~\cite{2011arXiv1111.2183C,Dubus:2012hm}.  For energies above 50~GeV, the CTA flux sensitivity is $\sim2.93\times10^{-12}$~cm$^{-2}$~s$^{-1}$ for 50 hours of observation per field-of-view (5$\sigma$ detection).\footnote{\url{https://portal.cta-observatory.org/Pages/CTA-Performance.aspx}}  For 250 hours total of observation time, this covers $\sim$190 deg$^2$ of sky, assuming a $7^\circ$ field-of-view.  As shown in Fig.~\ref{fig:dnds_0_10}, the NPTF predicts a density of $0.029^{+0.008}_{-0.005}$ deg$^{-2}$ for sources above this threshold.  This translates to $5.51_{-0.95}^{+1.52}$ detected sources, more than double what had previously been estimated for similar observing parameters~\cite{Dubus:2012hm}.  Relaxing the observing time per source and assuming, as in~\cite{TheFermi-LAT:2015ykq}, that a quarter of the sky is surveyed in 240 hours at 5mCrab sensitivity, then the NPTF predicts $161^{+30}_{-20}$ sources.  This is lower, and in slight tension, with the $200\pm45$ sources predicted by the \emph{Fermi} study using the blue source-count distribution illustrated in Fig.~\ref{fig:dnds_0_10}.

\section{Discussion and Conclusions}
\label{sec:conclusions_igrb}

The primary focus of this chapter is to characterize the properties of the PSs contributing to the EGB in a data-driven manner.  To achieve this, we use a novel analysis method, referred to as Non-Poissonian Template Fitting (NPTF), which takes advantage of photon-count statistics to distinguish diffuse and PS contributions to gamma-ray maps with non-trivial spatial variations.  We presented the NPTF results on \emph{Fermi} Pass~8 data at low (1.89--94.9~GeV, $|b| > 30^\circ$) and high (50--2000~GeV, $|b| > 10^\circ$) energies.  For the first time, the intensity and source-count distributions for the isotropic PSs have been  obtained as a function of energy, up to 2~TeV.  The best-fit source-count distributions probe fluxes below the current detection threshold for the \emph{Fermi} 3FGL and 2FHL catalogs, providing information on the unresolved populations. 

Through extensive studies of how the NPTF responds to simulated populations, we have shown that the analysis procedure reproduces the properties of input source classes.  Therefore, the features of the best-fit source-count distributions obtained from the data provide a potential wealth of information about the source populations of the EGB.  While a detailed interpretation of the source-count distributions in terms of particular theoretical models is beyond the scope of this study, several important trends were observed.  

In this chapter, the source-count distributions are parametrized as triply-broken power laws in the NPTF.  At all energies, a break is fit at low (high) fluxes, below (above) which the analysis method loses sensitivity.  Of particular interest is whether an additional break, $F_{b,2}$, is preferred at intermediate flux.  We find a break in the lowest energy bin (1.89--4.75~GeV) at $1.22_{-0.56}^{+2.00}\times10^{-10}$~cm$^{-2}$~s$^{-1}$ with slope $2.04_{-0.05}^{+0.05}$ above and $1.74_{-0.37}^{+0.19}$ below.  In the subsequent two energy bins, 4.75--11.9~GeV and 11.9--30.0~GeV, there is a mild indication that the source-count distribution hardens below the intermediate flux break, though the change in slope is not as robust and significant as in the lowest energy bin.  At higher energies, above $\sim$30 GeV, there is no indication that the source-count distribution changes slope at the intermediate break.  This trend is in line with the expectations from the blazar simulations in Sec.~\ref{sec:simulations}.  For example, in both Figs.~\ref{fig:bl1dnds} and~\ref{fig:bl2dnds}
, which show the results of the NPTF run on simulated data with the Blazar--1 and Blazar--2 models, we find evidence for curvature in the source-count distribution at intermediate fluxes in the lowest energy bins, while at higher energies the recovered source-count distribution appears as a single power law at fluxes above the sensitivity threshold of the NPTF.  In the energy bin from 50--2000~GeV the best-fit value for $F_{b,2}$ is essentially unconstrained and the slopes above and below it are consistent within uncertainties:  $2.28_{-0.22}^{+0.28}$ and $2.17_{-0.09}^{+0.12}$.

\begin{figure}[!htbp] 
   \centering
   \includegraphics[width=0.8\textwidth]{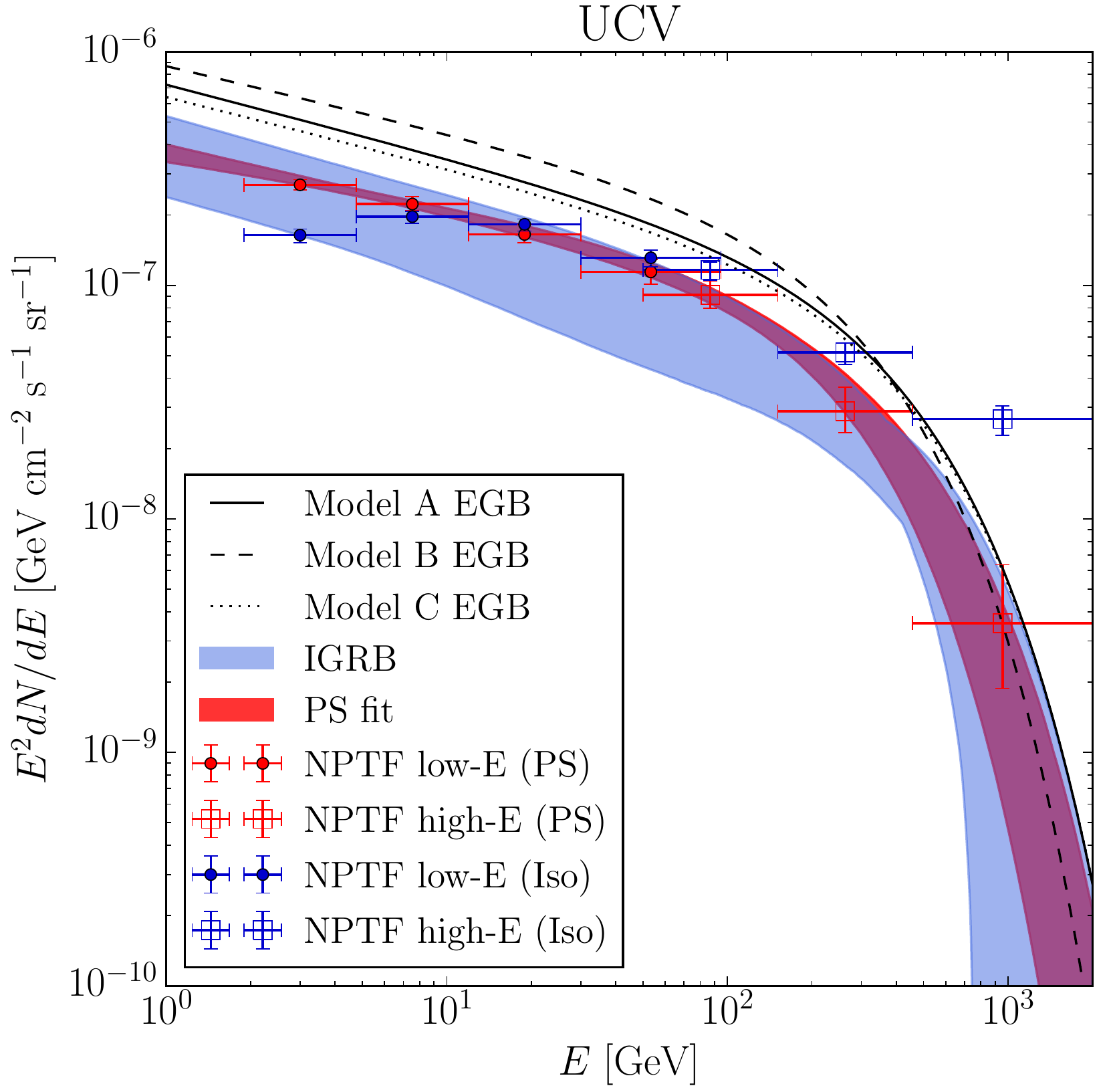} 
   \caption{Global fit to the PS intensity spectrum recovered by the NPTF. The results of the NPTF low-energy analysis on {\it ultracleanveto} PSF1--3 data and the high-energy analysis on {\it ultracleeanveto} PSF0--3 data are shown (filled red circles and open red boxes, respectively).  The red band indicates the best-fit (68\% credible interval) to a power law with exponential cutoff.  For comparison, the best-fit \emph{Fermi} EGB spectra from~\cite{Ackermann:2014usa} are shown for three different diffuse background models (Model A--C).  The blue band indicates the estimated IGRB spectrum, obtained by subtracting the PS spectrum from the \emph{Fermi} EGB; the spread includes the statistical uncertainty from the PS intensity as well as the systematic uncertainty on the EGB.  We also plot the best-fit smooth isotropic spectrum recovered by the NPTF (filled blue circles and open blue boxes).  The results are in good agreement with the estimated IGRB result (blue band) below $\sim$100 GeV, but overestimate the result at higher energies due to cosmic-ray contamination.  
    } 
   \label{fig:global}
\end{figure}

The NPTF also provides the best-fit intensities for the isotropic-PS populations as a function of energy.  Figure~\ref{fig:global} illustrates this spectrum for analyses done using the {\it ultracleanveto} event class.  The filled red circles (open red boxes) show the results for the dedicated low (high)-energy analysis, with PSF1--3 data used at low energies and PSF0--3 data at high energies.  For comparison, the \emph{Fermi} EGB spectrum is shown by the black line~\cite{Ackermann:2014usa}.  This corresponds to the best-fit intensity using the Model A diffuse background from that study.  To illustrate the systematic uncertainty on this curve, we also plot the spectra for diffuse models B and C (dashed and dotted, respectively).

\begin{table*}[t]
\renewcommand{\arraystretch}{1.5}
\setlength{\tabcolsep}{5.2pt}
\begin{center}
\resizebox{\textwidth}{!}{%
\begin{tabular}{  c c   c  c  c  c c c c}
\toprule
$I_\text{EGB}$ & \multicolumn{4}{c }{Low-Energy Analysis} & \multicolumn{4}{c}{High-Energy Analysis}  \Tstrut\Bstrut	\\   
& 1.89--4.75 & 4.75--11.9 & 11.9--30 & 30--94.9 & 50--151 & 151--457 & 457--2000 & 50--2000 \Tstrut\Bstrut \\
\midrule
Scenario A& $0.62^{+0.04}_{-0.02}$ & $0.53^{+0.03}_{-0.03}$ & $0.48^{+0.03}_{-0.03}$ &  $0.47^{+0.05}_{-0.04}$ &$0.44^{+0.06}_{-0.05}$ & $0.36^{+0.08}_{-0.06}$ & $0.12^{+0.09}_{-0.06}$ & $0.43^{+0.05}_{-0.04}$\Tstrut\Bstrut \\
Scenario B& $0.54^{+0.03}_{-0.03}$ & $0.60^{+0.04}_{-0.03}$ &  $0.61^{+0.06}_{-0.05}$ &  $0.66^{+0.09}_{-0.07}$ & $0.67^{+0.10}_{-0.09}$ & $0.51^{+0.13}_{-0.09}$ &  $0.58^{+0.45}_{-0.27}$ & $0.68^{+0.09}_{-0.08}$ \Tstrut\Bstrut  \\
\bottomrule 
\end{tabular}}
\end{center}
\caption{PS fractions ($I_\text{PS}/I_\text{EGB}$) for the low (PSF1--3) and high-energy (PSF0--3) analyses, using {\it ultracleanveto} data, with energy sub-bins in units of GeV.   The first row (`Scenario A') uses the EGB intensity obtained in this study using foreground model \texttt{p8r2}; however, this scenario likely overestimates the $I_\text{EGB}$ at energies above $\sim$100~GeV due to cosmic-ray contamination.  The second row shows the PS fractions calculated with respect to the \emph{Fermi} EGB intensity from~\cite{Ackermann:2014usa}, with foreground Model A (`Scenario B').  Although the \emph{Fermi} analysis uses a different foreground model, it takes advantage of a dedicated event selection above $\sim$100~GeV that mitigates effects of additional contamination.  }
\label{tab:fractions}
\end{table*}

The PS fraction, defined as $I_\text{PS}/I_\text{EGB}$, is provided in Tab.~\ref{tab:fractions} for each energy bin.  While using the EGB intensity derived in this study (`Scenario A') is the most self-consistent comparison, this may underestimate the PS contribution above $\sim$100~GeV, where the NPTF appears to recover too much smooth isotropic emission due to increased cosmic-ray contamination in the data sets used, as already discussed.  Therefore, we also show the PS fractions calculated relative to the \emph{Fermi} EGB intensity from~\cite{Ackermann:2014usa} for diffuse model A (`Scenario B').  The comparison to the EGB as measured in~\cite{Ackermann:2014usa} is not fully self consistent, since, for example, the foreground modeling and data sets in~\cite{Ackermann:2014usa} differ from those used in this study to measure $I_\text{PS}$.  However, the advantage of this comparison is that the \emph{Fermi} analysis uses special event-quality cuts to mitigate contamination, and thus their measure of $I_\text{EGB}$ is likely more faithful than that presented in this study.  These results are shown in the second row of Tab.~\ref{tab:fractions}.  For the low-energy analysis, the PS fractions are consistent, within uncertainties, when $I_\text{EGB}$ is taken from our study or \emph{Fermi}'s.\footnote{For `Scenario B', the quoted uncertainties only include those measured in this work for $I_\text{PS}$.  For $I_\text{EGB}$, we use the best-fit value given in~\cite{Ackermann:2014usa}.}  The substantial differences occur at high-energies, where our result is systematically lower than the fractions based on \emph{Fermi}'s EGB intensity.

In general, we find that approximately 50--70\% of the EGB consists of PSs in the energy ranges considered.  To interpret these results, we use the ratios $I_\text{iso}^\text{PS} / I_\text{blazar-sim}$ obtained in the simulation studies of Sec.~\ref{sec:simulations}.  In that section, we showed that the efficiency for the NPTF to recover the flux for the Blazar--2 model (with PSF1--3) is 
$\sim$100\% in the first energy bin and drops to $\sim$60\% in the fourth energy bin.  For the Blazar--1 model, the efficiencies are consistently higher than the Blazar--2 scenario.  These two blazar models are meant to illustrate extreme scenarios, with the Blazar--1 model having a significant fraction of the total flux arising from high-flux sources, while low-flux sources dominate instead in the Blazar--2 case. 
The high efficiency of the NPTF to recover the blazar component at low energies, combined with the PS fractions observed in the data (Tab.~\ref{tab:fractions}), clearly suggests that there is a substantial non-blazar component of the EGB up to energies $\sim$30 GeV.   The interpretation of the results in the energy bin from 30.0--94.9 GeV is less clear.  A proper interpretation of the results at higher energies in terms of evidence for or against a non-blazar component of the EGB requires dedicated blazar simulations, which we leave to future work.  
 
Our results tend to predict fewer PSs (and photons from PSs) where we do overlap with previous studies.  For example, a similar photon-count analysis was used by~\cite{Zechlin:2015wdz} to study 1--10~GeV energies in the Pass~7 Reprocessed data.  They found an $\sim$80\% PS fraction at these energies.  At the lowest energies that we probe---which admittedly do not extend down as low as $\sim$1~GeV---we only find a $\sim$54\% PS fraction (relative to Model A).  
  Systematic uncertainties, as shown in Fig.~\ref{fig:systematicsplot}, can affect the recovered PS intensities at the ${\cal O}(10\%)$ level, which can partially alleviate the tension between our results.  

Above 50~GeV, the NPTF procedure predicts that $0.68_{-0.08}^{+0.09}$ of the EGB consists of PSs, with systematic uncertainties estimated at approximately $\pm 10\%$.  This fraction is smaller, and in slight tension, with the predicted value $0.86_{-0.14}^{+0.16}$ obtained in previous work~\cite{TheFermi-LAT:2015ykq}.  The fact that our results suggest that there is more diffuse isotropic emission at high energies may help alleviate the tension between \cite{TheFermi-LAT:2015ykq} and the hadronuclear ($pp$) interpretation of IceCube's PeV neutrinos \cite{Murase:2013rfa}.  Some models suggest, for example, that these very-high-energy neutrinos are produced in hadronuclear interactions, along with high-energy gamma-rays that would contribute to the IGRB~\cite{Murase:2013rfa, Tamborra:2014xia,Ando:2015bva,Hooper:2016gjy}.  If the smooth isotropic gamma-ray spectrum (\emph{i.e.}, the non-blazar spectrum) is suppressed above 50~GeV in the \emph{Fermi} data, it could put such scenarios in tension with the data ~\cite{Bechtol:2015uqb,Murase:2015xka}; however, that does not necessarily appear to be the case  given the results of our analysis~\cite{Murase:2016gly}.  With that said, and as already mentioned, dedicated blazar simulations at high energies are needed to properly interpret our results at these energies.

The PS spectrum in Fig.~\ref{fig:global} is well-modeled (reduced \mbox{$\chi^2 = 1.18$}) as a power law with an exponential cut-off:
\es{powerexp}{
{d N \over dE} = C \left( {E \over 0.1 \, \, \text{GeV} } \right)^{-\gamma} \exp{ \left( - {E \over E_\text{cut}} \right) } \,,
}
where $C=6.91_{-1.29}^{+1.44} \times 10^{-5}$ GeV$^{-1}$cm$^{-2}$s$^{-1}$sr$^{-1}$,  $\gamma= 2.26_{-0.05}^{+0.05}$, and $E_\text{cut} = 289_{-86.3}^{+127}$~GeV are the best-fit parameters.\footnote{Repeating the fit using the results from the NPTF analyses with {\it source} data returns similar results, though the PS spectrum is slightly enhanced relative to the {\it ultracleanveto} result.  In particular, with {\it source} data, we find $C=7.98_{-1.40}^{+1.58} \times 10^{-5}$ GeV$^{-1}$cm$^{-2}$s$^{-1}$sr$^{-1}$,  $\gamma= 2.29_{-0.05}^{+0.04}$, and $E_\text{cut} = 325_{-78.1}^{+117}$~GeV, with reduced $\chi^2 = 0.93$. }  Note that the fit is done taking into account the uncertainties on the PS intensities in the energy sub-bins.  The global fit for the PS spectrum is shown in Fig.~\ref{fig:global} by the red band, which denotes the 68\% credible interval.  Interestingly, the index $\gamma$ and cut-off $E_\text{cut}$ that we extract from the fit are very similar to the values found in~\cite{Ackermann:2014usa}, which used the same functional form to fit the EGB spectrum.  Subtracting our PS spectrum from the EGB spectral fits gives the blue band in Fig.~\ref{fig:global}.  The band includes statistical uncertainties from our global fit as well as systematic uncertainties associated with varying between Models A-C.  The blue band is an estimate of the IGRB spectrum and we compare it to the smooth isotropic spectrum recovered by the NPTF (blue points).  Note that the two are consistent, within the large uncertainties, below $\sim$100~GeV; above this energy, our IGRB value is expectedly high. 

\begin{figure*}[!phtb] 
   	\begin{center}
	$\begin{array}{c}
\includegraphics[scale=0.6]{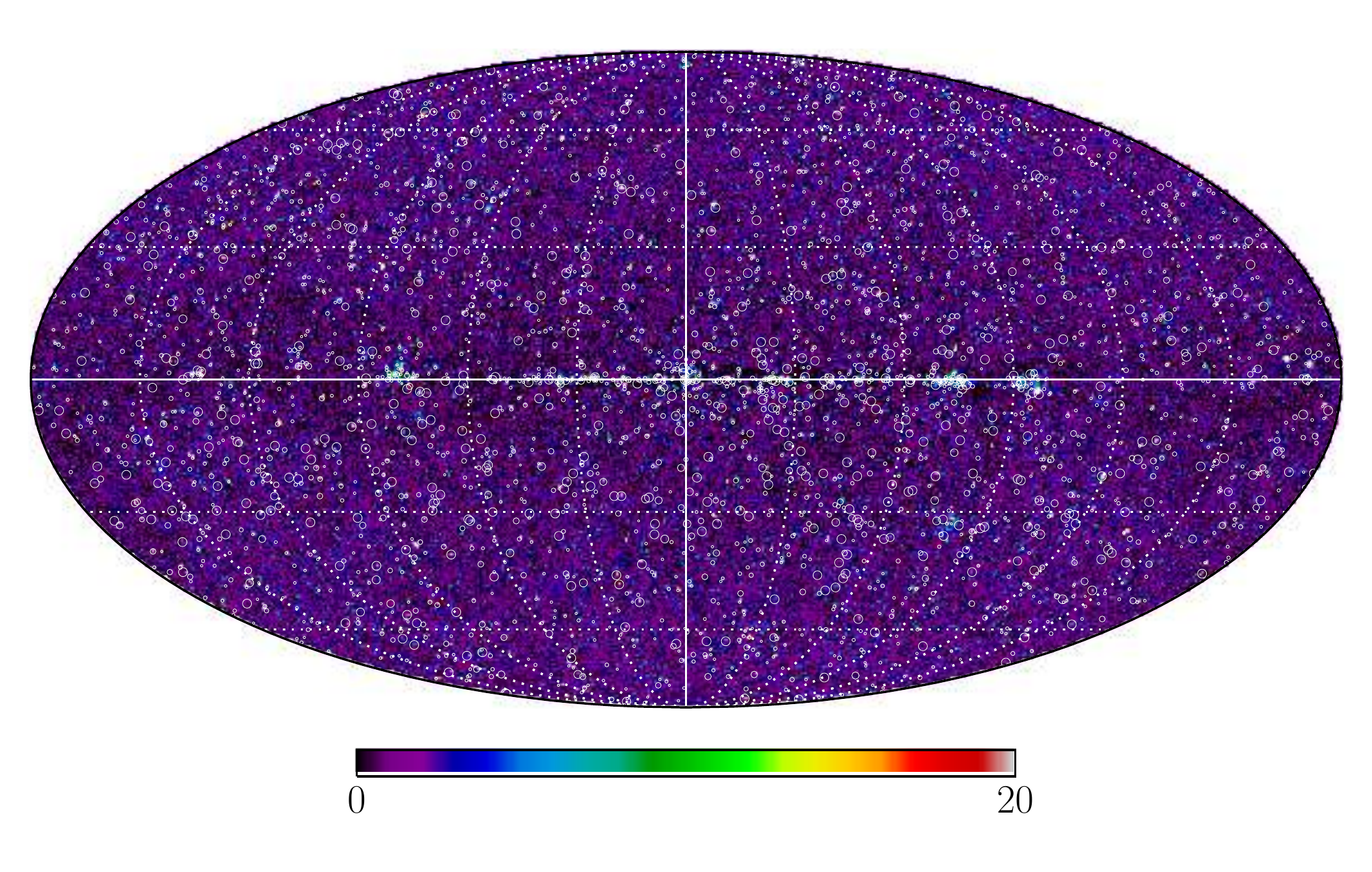} \\
\includegraphics[scale=0.6]{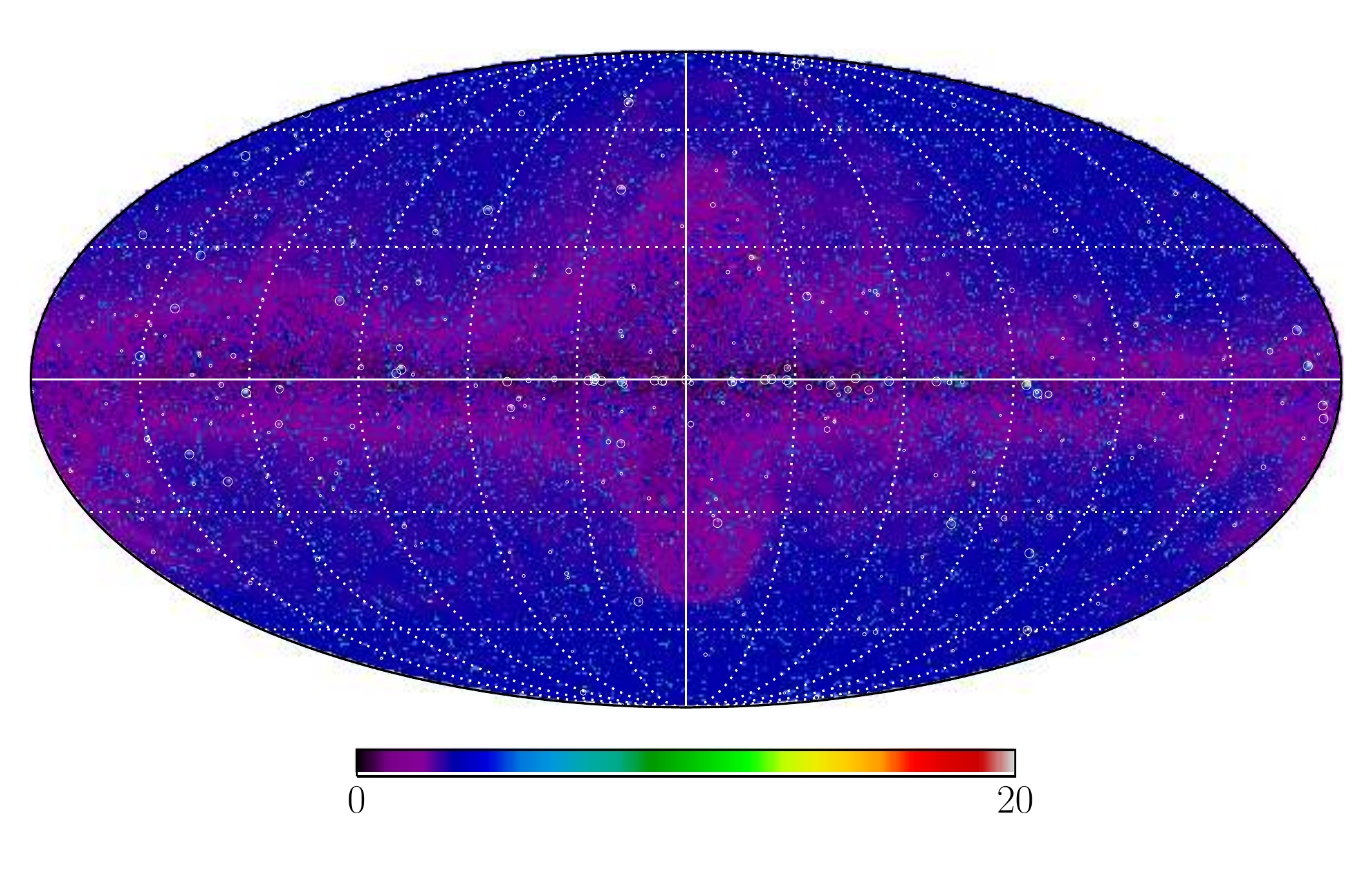}
	\end{array}$
	\end{center}
\caption{Full-sky maps showing the value (clipped at 20) of $-\log \epsilon_p$ in each pixel $p$.  The larger the value of $-\log \epsilon_p$, the more likely the pixel contains a point source.  \textbf{(Top)}  Results using {\it ultracleanveto} data (PSF3) for energies 1.89--94.9~GeV.  \emph{Fermi} 3FGL sources are indicated by the white circles, with radii weighted by the predicted number of photon counts for a given source.  \textbf{(Bottom)} Results using all quartiles of {\it ultracleanveto} data for 50--2000~GeV.  Circles now represent \emph{Fermi} 2FHL sources.  The data for this figure is available upon request. }
   \label{fig:hotspot} 
\end{figure*}

The NPTF allows us to make statistical statements about the properties of source populations contributing to the EGB, but at the expense of identifying the precise locations of these sources.  However, it is still possible to make probabilistic statements about these locations.  To do so, we compare the observed photon count in a given pixel, $n_p$, to the mean expected value, $\mu_p$, without accounting for PSs.  To determine $\mu_p$ we include the diffuse background, smooth isotropic emission, and the {\it Fermi} bubbles templates, with normalizations as determined from the NPTF.
The pixel-dependent survival function is defined as
\begin{equation}
\epsilon_p \equiv 1- \text{CDF}\left[ \mu_p, n_p \right] \, ,
\end{equation}
where CDF is the Poisson cumulative distribution function.  The smaller the value of $\epsilon_p$ (or, conversely, the larger the value of $-\log \epsilon_p$), the more probable it is that the pixel contains a PS.  Figure~\ref{fig:hotspot} shows full-sky maps of $-\log \epsilon_p$ for both low (1.89--94.9~GeV) and high (50--2000~GeV) energies.\footnote{ Digital versions of these maps are available upon request.}  The white circles indicate the presence of a 3FGL (2FHL) source for the low- \mbox{(high-)}energy map, with the radii proportional to the predicted photon counts for the sources.  There is good correspondence between the hottest pixels, as determined by $-\log \epsilon_p$,  and the brightest resolved sources.  Pixels that are correspondingly less ``hot" tend to be associated with less-bright 3FGL (or 2FHL) sources.  Of particular interest are the hot pixels not already identified by the published catalogs.  In the region $|b| \gtrsim 30^\circ$ ($|b| \gtrsim 10^\circ$) in the low- (high-)energy analysis, these are likely the sources lending the most weight to the NPTF below the catalog sensitivity thresholds.  While more sophisticated algorithms are needed to further refine the candidate source locations, Fig.~\ref{fig:hotspot} provides a starting point for identifying the spatial locations of potential new sources to help guide, for example, future TeV gamma-ray observations and cross-correlations with other data sets, such as the IceCube ultra-high-energy neutrinos.

Deciphering the constituents of the EGB remains an important goal in the study of high-energy gamma-ray astrophysics, with broad implications extending from the production of PeV neutrinos to signals of dark matter annihilation or decay. The \emph{Fermi} LAT has already played an important role in the discovery of many new sources in the GeV sky.  By taking advantage of the statistical properties of unresolved populations, our results provide a glimpse at the aggregate properties of the sources that lie below the detection threshold of these published catalogs and suggest a wealth of detections for future observatories.

\subsection{Implication for Dark Matter Annihilation Searches}

Pinning down the origin of 50-70\% of the extragalactic gamma-ray sky as being of point source origin narrows down the potential contribution of more exotic sources such as the integrated emission of annihilating dark matter in halos around far-away galaxies and clusters. This would lead to an improvement in constraints on annihilating DM obtained by studying their contribution to the isotropic gamma-ray background (IGRB), such as those presented in~\cite{Ackermann:2015tah,Ajello:2015mfa}, potentially by a factor of a few. 

There are a few drawbacks to this approach, however. The contribution of relatively nearby halos to an annihilation signal is expected to dominate due to the late-time clustering of matter (which boosts the annihilation signal) as well as our favored location in the Local Group where we are surrounded by halos and clusters of a larger size than those around a randomly chosen place in the Universe. This fact is not optimally taken into account in IGRB analyses. Secondly, IGRB analyses for dark matter annihilation cannot conclusively discover a DM signal due to the irreducible isotropic background of astrophysical origin -- only constraints on its properties are possible. 

In the next part of this thesis, we will systematically build up the best way to search for extragalactic dark matter annihilation, focusing on emission from nearby galaxies and clusters. We will develop a framework to characterize the distribution of nearby extragalactic dark matter halos (Ch.~\ref{ch:groups_sim}) and look for this structure in \emph{Fermi} data (Ch.~\ref{ch:groups_data}).

\sectionline

\chapter{Mapping Extragalactic Dark Matter Annihilation with Galaxy Surveys}
\label{ch:groups_sim}

This chapter is based on an edited version of \emph{Mapping Extragalactic Dark Matter Annihilation with Galaxy Surveys: A Systematic Study of Stacked Group Searches},  \href{https://journals.aps.org/prd/abstract/10.1103/PhysRevD.97.063005}{Phys.Rev. \textbf{D97} (2018) 063005} \href{https://arxiv.org/abs/1709.00416}{[arXiv:1709.00416]} with Mariangela Lisanti, Nicholas Rodd, Benjamin Safdi and Risa Wechsler~\cite{Lisanti:2017qoz}. The results of this chapter have been presented at the following conferences and workshops: \emph{TeV Particle Astrophysics (TeVPA) 2017} in Columbus, OH (August 2017), \emph{Dark Matter, Neutrinos and their Connection (DA$\nu$CO)} in Odense, Denmark (August 2017), \emph{Workshop on Statistical Challenges in the Search for Dark Matter} in Banff, Canada (February 2018) and \emph{Recontres de Blois 2018} in Blois, France (June 2018).

\ifdefined\printmode
\else
\clearpage
\fi

\section{Introduction}

\lettrine[lines=3]{D}{ark} matter (DM) annihilation into visible final states remains one of the most promising avenues for discovering non-gravitational interactions in the dark sector.  While an individual annihilation event is rare, the probability of observing it can be maximized by searching for excess photons in regions of high dark matter density.  The center of the Milky Way is potentially one of the brightest regions of DM annihilation as seen from Earth, but the astrophysical uncertainties associated with the baryonic physics at the heart of our Galaxy motivate exploring other targets.  Gamma-ray studies of DM-dominated dwarf galaxies in the Local Group currently provide some of the most robust constraints on the annihilation cross section~\cite{Fermi-LAT:2016uux, Ackermann:2015zua}.  However, many more potential targets are available beyond the Local Group.  This chapter proposes a new analysis strategy to search for DM emission from hundreds more DM halos identified in galaxy group catalogs.  

A variety of methods have been used to study gamma-ray signatures of extragalactic DM annihilation, including modeling potential contributions to the Isotropic Gamma-Ray Background~\cite{Bengtsson:1990xf,Bergstrom:2001jj,Ullio:2002pj,Bottino:2004qi,Bertone:2004pz,Bringmann:2012ez,Ajello:2015mfa, DiMauro:2015tfa, Ackermann:2015tah, Feng:2016fkl}, measuring the \emph{Fermi} auto-correlation power spectrum~\cite{Ackermann:2012uf,Fornasa:2016ohl,Ando:2006cr,Ando:2013ff}, and cross-correlating the \emph{Fermi} data with galaxy counts~\cite{Branchini:2016glc, Xia:2011ax,Ando:2014aoa,Ando:2013xwa,Xia:2015wka,Regis:2015zka,Cuoco:2015rfa,Ando:2016ang}, cosmic shear~\cite{Camera:2014rja,Troster:2016sgf,Choi:2015mnp,Camera:2012cj,Shirasaki:2015nqp,Shirasaki:2014noa,Shirasaki:2016kol} and lensing of the Cosmic Microwave Background~\cite{Fornengo:2014cya, Feng:2016fkl}.  These methods typically rely on using a probabilistic distribution of the DM annihilation signal on the sky.  Our approach is more deterministic in nature.  In particular, we treat a collection of known galaxies as seeds for DM halos.  The properties of each galaxy---such as its luminosity and redshift---enable one to deduce the characteristics of its associated halo and the expected DM-induced gamma-ray flux from that particular direction in the sky.  In this way, we can build a map of the expected DM annihilation flux that traces  the observed distribution of galaxy groups. 

In certain ways, our approach resembles that used in previous studies  of DM annihilation from individual galaxy clusters.  For example, most recently the Andromeda galaxy~\cite{Ackermann:2017nya} and Virgo cluster~\cite{Ackermann:2015fdi} have been the subject of dedicated study by the \emph{Fermi} Collaboration.  Other work has inferred the properties of the DM halos associated with galaxy clusters detected in X-rays~\cite{Ackermann:2010rg, Ando:2012vu,Ackermann:2013iaq,Anderson:2015dpc,Rephaeli:2015nca,2016A&A...589A..33A,Liang:2016pvm}.  Most of these studies focused on a small number of galaxy clusters and obtained DM sensitivities weaker than those from dwarf galaxies.  

Recent advancements in the development of galaxy group catalogs allow us to now build a full-sky map of the nearby galaxies that should be the brightest DM gamma-ray emitters.  Catalogs based primarily on the 2MASS Redshift Survey (2MRS)~\cite{ Huchra:2011ii} provide an unprecedented amount of information regarding a group's constituents and halo properties~\cite{Tully:2015opa,2017ApJ...843...16K,Lu:2016vmu}.  This information allows us to build a list of the brightest extragalactic DM targets on the sky and to perform a stacked analysis for gamma-ray emission from them.  A gamma-ray line search using this methodology was recently performed by Ref.~\cite{Adams:2016alz}.  Our focus is on continuum DM signatures, which carry considerably more complications in terms of the treatment of astrophysical backgrounds.  

In the upcoming Chapter~\ref{ch:groups_data}, we present results implementing a stacked analysis of the group catalogs from Ref.~\cite{Tully:2015opa,2017ApJ...843...16K} on \emph{Fermi} data and show explicitly that this method yields competitive sensitivity to the dwarf searches.  Here, we present the full details of the analysis method and a thorough discussion of the systematic uncertainties involved in deducing the DM-induced flux associated with a given galaxy group.  To fully understand these uncertainties, we apply these methods on mock data where it is possible to compare the inferred DM properties to their true values.  For this purpose, we use the \texttt{DarkSky} cosmological $N$-body simulation~\cite{Skillman:2014qca,Lehmann:2015ioa} and an associated galaxy catalog from Ref.~\cite{Lehmann:2015ioa}.  We emphasize that, while we illustrate the analysis method on gamma-ray data, it can also be applied to other wavelengths and even other messengers, such as neutrinos.  

This chapter is organized as follows. In Sec.~\ref{sec:galaxyfilteringpipeline}, we describe how to build DM annihilation flux maps starting from a galaxy group catalog and discuss the associated systematic uncertainties.  Sec.~\ref{sec:stats} presents a detailed description of the statistical methods that we follow to implement the stacking. We show the results of applying the limit-setting and signal recovery procedures on mock data in Sec.~\ref{sec:smallrois} and conclude in Sec.~\ref{sec:conclusions}.  Appendix~\ref{app:JDrelations} provides a  detailed discussion of the $J$-factor expressions used in the main text.  

\section{Tracing Dark Matter Flux with Galaxy Surveys}
\label{sec:galaxyfilteringpipeline}

In this Section, we describe how to construct catalogs of extragalactic DM targets starting from a list of galaxy groups.  We begin by reviewing the properties of the galaxy group catalogs and then describe how to predict the DM signal from a given galaxy group and quantify the systematic uncertainties of this extrapolation.  

\subsection{Galaxy and Halo Catalogs}

The approach that we use throughout this work relies on galaxy surveys as an input.  Different galaxy catalogs span a range of redshifts and luminosities. Optimal catalogs for DM searches should cover as much of the sky as possible (to increase statistics) and sample low redshifts ($z\lesssim 0.1$).  The strength of the DM signal increases at lower redshifts due to accretion of mass at late times, affecting both the halo mass distribution and substructure~\cite{Ando:2014aoa}.  In contrast, the integrated gamma-ray flux of standard astrophysical sources, such as Active Galactic Nuclei and star-forming galaxies, is expected to peak at higher redshifts between $\sim$0.1 and $\sim$2 depending on the specific source class and model for its unresolved contribution~\cite{Ando:2014aoa, Xia:2015wka}.  

The Two Micron All-Sky Survey Extended Sources Catalog (2MASS XSC)~\cite{Bilicki:2013sza,Huchra:2011ii}  satisfies the criteria listed above and has been used extensively in past cross-correlation studies~\cite{Ando:2013xwa,Ando:2014aoa,Ando:2016ang,Cuoco:2015rfa,Regis:2015zka,Xia:2011ax,Xia:2015wka}. The XSC is an all-sky infrared survey that consists of approximately one million galaxies up to a limiting magnitude of $K = 13.5$~mag. Several redshift surveys based on the 2MASS XSC map the redshifts associated with these galaxies. The 2MRS~\cite{Huchra:2011ii}, for example, samples about 45,000 galaxies in the 2MASS XSC with redshifts to a limiting magnitude of $K=11.75$ mag. This corresponds to a nearly complete galaxy sample up to redshifts of $z=0.03$, which is ideal for DM studies.

Galaxies from large surveys such as 2MASS can be organized into group catalogs.  A group of gravitationally-bound galaxies shares a DM host halo.  The  brightest galaxy in the group is referred to as the central galaxy; the additional galaxies are bound satellites surrounded by their own subhalos.  As we will see, the total luminosity of the galaxies in the group is a good predictor of the mass of the DM host halo.  A variety of group finders have been developed and applied to the 2MASS data set~\cite{Tully:2015opa,Lu:2016vmu,2017ApJ...843...16K}, using the 2MRS which adds information in the redshift dimension.  The groups in these catalogs range from cluster scales with $\sim$190 members and associated halo masses of $\sim$10$^{15}$~M$_\odot$, down to much smaller systems with only a single member. Galaxy group catalogs are especially relevant for the present study, since (as will be shown) halo properties tend to be correlated with properties of galaxy groups rather than those of individual galaxies.

While in the upcoming Chapter~\ref{ch:groups_data} we use information from the 2MASS group catalogs in the analysis of \emph{Fermi} data, we focus on a catalog of simulated galaxies and halos here.  We use the \texttt{DarkSky-400} cosmological $N$-body simulation (version \texttt{ds14\_i})~\cite{Skillman:2014qca,Lehmann:2015ioa} and an associated $r$-band galaxy catalog.  Using the code \texttt{2hot}~\cite{Warren:2013vma}, \texttt{DarkSky-400}  follows the evolution of $4096^3$ particles (DM-only) of mass $7.63 \times 10^7$~M$_\odot$ in a box 400~Mpc$\,h^{-1}$ per side.  Initial perturbations are tracked from $z=93$ to today, assuming $(\Omega_M, n_s, \sigma_8, h) = (0.295, 0.968, 0.834, 0.688)$.   The halo catalog was generated using the \texttt{Rockstar} halo finder~\cite{Behroozi:2011ju, Lehmann:2015ioa}.  Crucially, the simulation covers the relevant redshift space for DM studies.\footnote{The snapshot of the simulation analyzed in this work is taken at $z = 0$, but we will refer to distance using redshift because that is the more appropriate language when applied to real data.}  In particular, an observer at the center of the simulation box has a complete sample of galaxies out to \mbox{$z \sim 0.045$}, with the furthest galaxies extending out to \mbox{$z \sim 0.067$}.  In our work, we only consider groups located within $z\lesssim 0.03$, which is the approximate redshift cutoff of the catalogs in Ref.~\cite{Tully:2015opa,2017ApJ...843...16K,Lu:2016vmu}. We include only well-resolved halos in our analysis by imposing a lower cut-off of $5\times10^{11}$~M$_\odot$ on the mass of included host halos. The associated galaxy catalog is generated using the abundance matching technique following Ref.~\cite{Behroozi:2010rx,Reddick:2012qy} with luminosity function and two-point correlation measurements from the Sloan Digital Sky Survey (SDSS).  Specifically, the $\alpha=0.5$ model from Ref.~\cite{Lehmann:2015ioa} is used, which was shown to provide the best fit to SDSS two-point clustering.  The \texttt{DarkSky} galaxy catalog contains the same information that would be found in, \emph{e.g.} the 2MASS galaxy catalog and associated group catalogs, such as individual galaxy luminosities and sky locations. 

\begin{figure*}[htbp]
   \centering
   \includegraphics[width=0.9\textwidth]{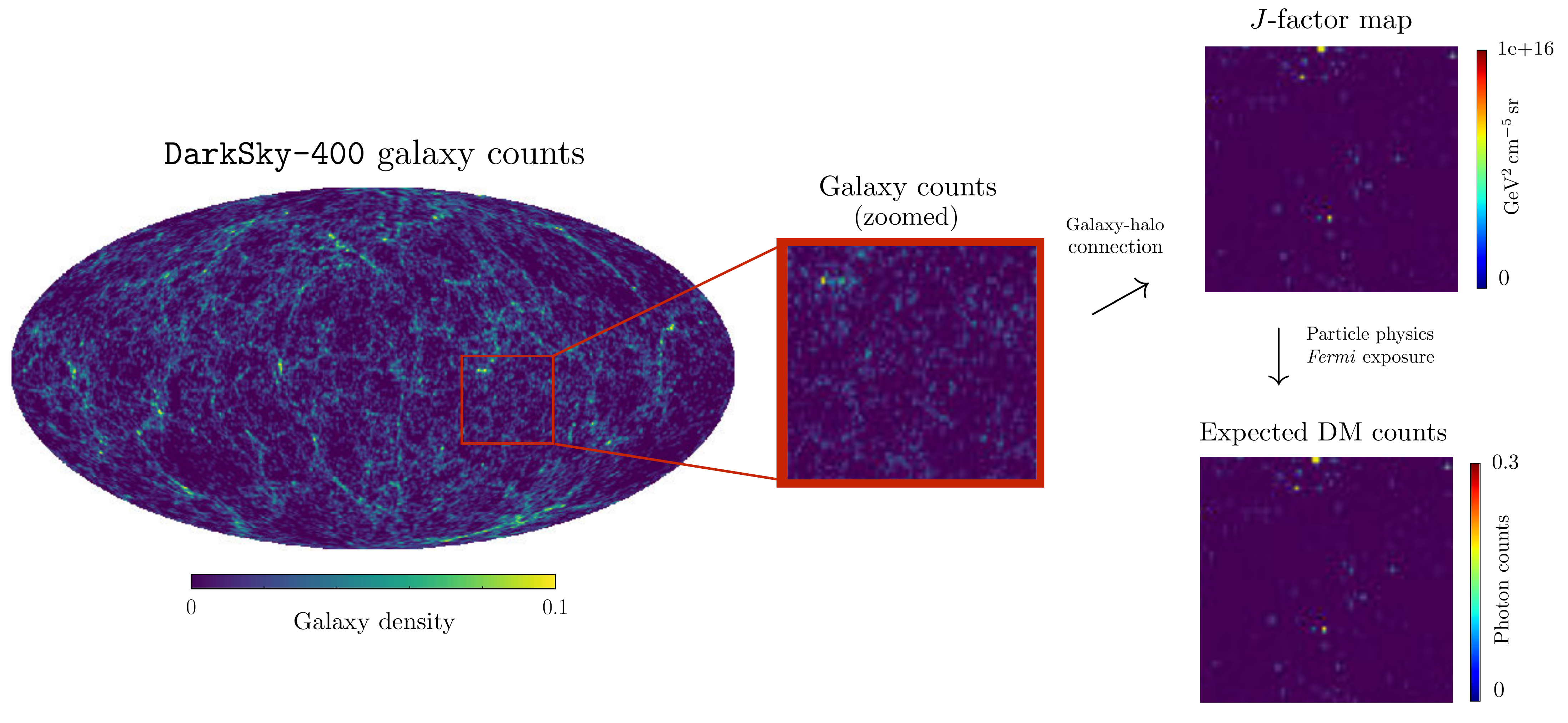}
   \caption{A schematic illustration of the analysis procedure as applied to \texttt{DarkSky}.   We begin with a sky map of galaxy counts (center left).  The \texttt{DarkSky} group catalog categorizes the galaxies into groups, which likely share a common DM halo.  From the \texttt{DarkSky} group catalog, we build a map of the $J$-factors for the host halos, as shown in the top right.  In reality, the properties of the halos surrounding each group of galaxies must be inferred from its total luminosity.  For a given DM model (here, a 100~GeV particle annihilating to $b\bar{b}$ with cross section $\langle \sigma v \rangle \approx 10^{-24}$ cm$^3$s$^{-1}$) and detector energy range (here, $\sim0.9-1.4$ GeV) the DM annihilation flux can be obtained (bottom right).  Going from the map of $J$-factors to that of DM counts also requires knowledge of the \emph{Fermi} exposure. Note that the full sky map has been subjected to 2$^{\circ}$ Gaussian smoothing.}
   \label{fig:DarkSkycounts}
\end{figure*}

Figure~\ref{fig:DarkSkycounts} shows a sky map of the galaxy counts in \texttt{DarkSky} up to $z=0.03$ for an observer at the center of the simulation box.  It is a \texttt{HEALPix}~\cite{Gorski:2004by} map with resolution \texttt{nside=128}. To first approximation, the galaxies are isotropically distributed throughout the sky.  However, regions of higher and lower galaxy density are clearly visible. Note that this is shown for a particular sky realization and placing the observer in different parts of the \texttt{DarkSky} box would change the regions of contrasting galaxy density.

\subsection{Dark Matter Annihilation Flux Map}
\label{sec:dmflux}

One can predict the DM annihilation flux associated with a halo that surrounds a given galaxy group.  This requires knowing the halo's properties, including its mass and concentration.  In this subsection, we discuss how to determine the flux when the halo's properties are known exactly.  Then, in the following subsection, we consider how to generalize the results to the more realistic scenario where the halo properties have to be inferred.

Each halo in \texttt{DarkSky} is fit by the \texttt{Rockstar} halo finder with 
a Navarro-Frenk-White (NFW) distribution~\cite{Navarro:1995iw} of the form
\begin{equation}
\rho_\text{NFW}(r)=\frac{\rho_{s}}{r/r_{s}\,(1+r/r_{s})^{2}}\, ,
\label{eq:NFW}
\end{equation}
where $r_s$ is the scale radius and $\rho_s$ is the normalization.  The NFW parameters are determined from the parameters that are provided for each DM halo---specifically, its redshift $z$, virial mass $M_\text{vir}$, virial radius $r_\text{vir}$, and virial concentration parameter $c_{\text{vir}}=r_\text{vir}/r_{s}$.

In the simplest scenarios, the annihilation flux factorizes as 
\begin{equation}
\frac{d\Phi}{dE_{\gamma}} = \frac{d\Phi_\text{pp}}{dE_{\gamma}}\times J \, ,
\label{eq:flux}
\end{equation}
where $E_\gamma$ is the photon energy and $\Phi_\text{pp}$  ($J$) encodes the particle physics (astrophysical) dependence.
The particle physics contribution is given by
\begin{equation}
\frac{d\Phi_\text{pp}}{dE_{\gamma}}=\frac{\langle\sigma v\rangle}{8\pi m_{\chi}^{2}}\sum_i \text{Br}_{i}\, \left. \frac{dN_{i}}{dE_{\gamma}'} \right|_{E'_{\gamma} = (1+z) E_{\gamma}},
\end{equation}
where $m_\chi$ is the DM mass, $\langle \sigma v\rangle $ is its annihilation cross section, $\text{Br}_{i}$ is its branching fraction to the $i^\text{th}$ annihilation channel, $ dN_{i}/dE_{\gamma}$ is the photon energy distribution in this channel, which is modeled using  PPPC4DMID~\cite{Cirelli:2010xx}, and $z$ is the redshift.  We consider the case of annihilation into the $b \bar{b}$ channel as a generic example of a continuum spectrum. Of course, the exact limits will vary for different spectra, and one should  consider a range of final states when applying the method to data, or use model independent-approaches (see, {\it e.g.}, Ref.~\cite{Elor:2015tva,Elor:2015bho}).
\begin{figure}[t]
   \centering
     \includegraphics[width=0.8\textwidth]{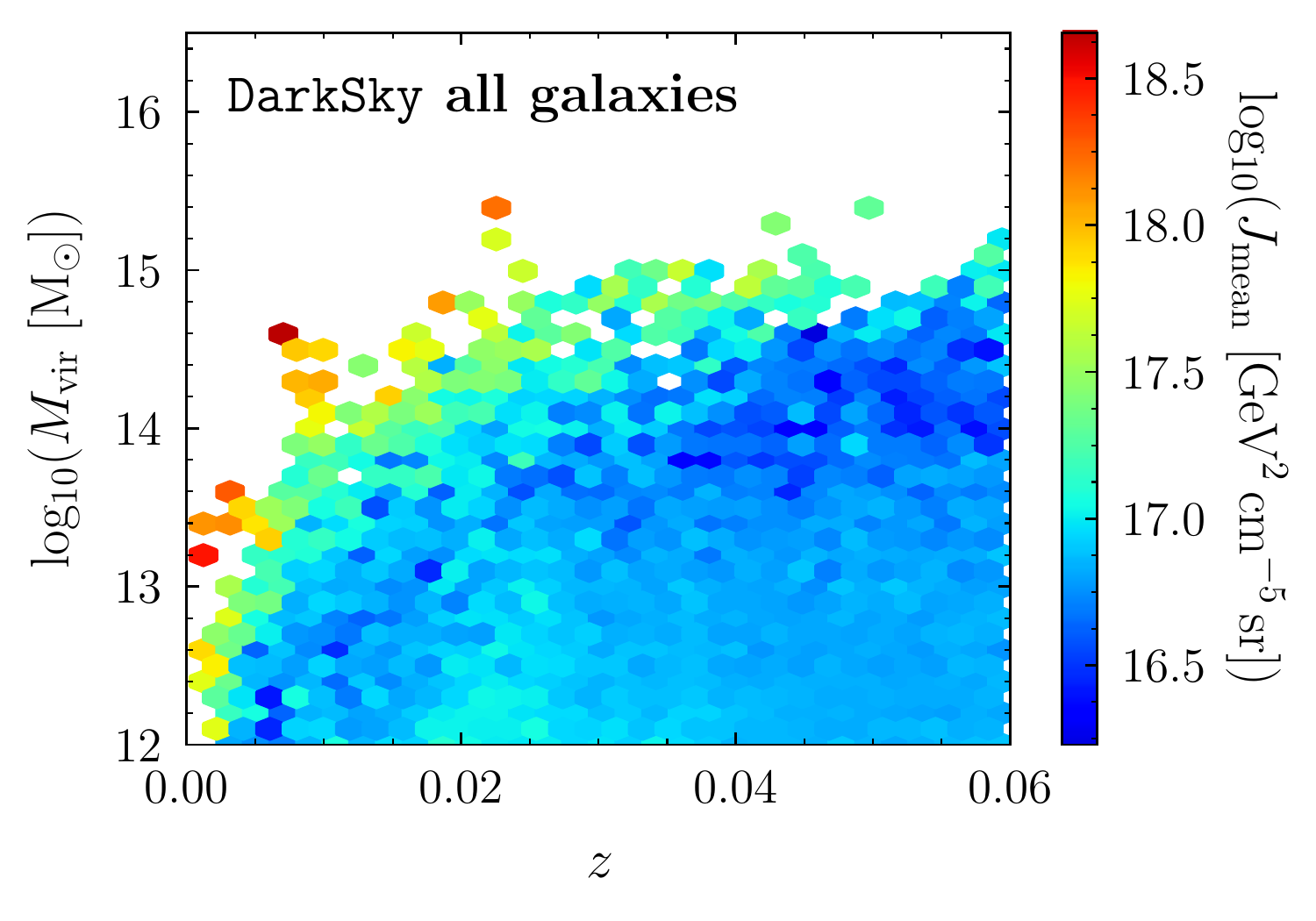}
   \caption{Heatmap of $J$-factors for the halos associated with all the galaxy groups in \texttt{DarkSky}, as a function of redshift and virial mass.  For this example, the observer is placed in the center of the simulation box. }
   \label{fig:heatmap}
\end{figure}

The $J$-factor is defined as the integral along the line-of-sight of the squared DM density of the observed object:\footnote{As defined, the $J$-factor has units of [${\rm GeV}^2 \cdot {\rm cm}^{-5} \cdot {\rm sr}$]. This definition is convenient for extragalactic objects, but beware because another common definition of the $J$-factor involves dividing out by a solid angle to remove the units of $[{\rm sr}]$. A detailed discussion of the units is provided in Appendix~\ref{app:JDrelations}.}
\begin{equation}
J = \left(1+b_\text{sh}[M_\text{vir}]\right)\,  \int ds \, d\Omega\,\rho^{2}_\text{NFW}(s,\Omega)\,,
\label{eq:jfactor}
\end{equation}
where $s$ is the line-of-sight distance and $b_\text{sh}[M_\text{vir}]$ is the so-called boost factor.  The boost factor accounts for the enhancement in the flux due to the annihilation in DM substructure (subhalos, subhalos within subhalos and so on\ldots), and is usually the dominant source of systematic uncertainty in extragalactic DM annihilation studies.   For the case of extragalactic objects, one can obtain a closed form solution that is an excellent approximation to the integral in Eq.~\ref{eq:jfactor}, which is proportional to
\begin{equation}
J \propto \left(1+b_\text{sh}[M_\text{vir}]\right) \frac{M_{\rm vir} \, c_{\rm vir}^3\, \rho_c}{d_c^2[z]}\,,
\label{eq:jscale}
\end{equation}
where $d_c$ is the comoving distance (a function of redshift, $z$), $\rho_c$ is the critical density, and $c_\text{vir}$ is the concentration.  In our analysis, we calculate the $J$-factor exactly, but the scaling illustrated in Eq.~\ref{eq:jscale} is useful for understanding the dependence of $J$ on the halo mass and concentration. The derivation of the $J$-factor expression is reviewed in detail in Appendix~\ref{app:JDrelations}, where we also show the result for the Burkert profile.

Figure~\ref{fig:DarkSkycounts} illustrates the truth $J$-factor map associated with \texttt{DarkSky}, obtained by putting the observer in the center of the simulation box.  This map is constructed by applying Eq.~\ref{eq:jfactor} to all host halos in the \texttt{DarkSky} catalog and using the boost model from Ref.~\cite{Bartels:2015uba} to describe the contribution from substructure.  Once the $J$-factors are known, the expected photon counts per pixel can be determined using Eq.~\ref{eq:flux} and \emph{Fermi}'s exposure map.  This is also shown in Fig.~\ref{fig:DarkSkycounts}, assuming a DM particle with $m_\chi = 100$~GeV that annihilates to $b \bar{b}$ with $\langle \sigma v \rangle \approx 10^{-24}$~cm$^3\,$s$^{-1}$.  Not all the pixels that contain one or more galaxies correspond to significant regions of DM annihilation.  The DM annihilation flux is largest for the most massive, concentrated, and/or closest galaxy groups.  

Note that when constructing Fig.~\ref{fig:DarkSkycounts}, we perform the angular integrals in Eq.~\ref{eq:jfactor} as a function of angular extent, $\Omega$.  In doing so, we implicitly assume that the boost factor is simply a multiplicative factor.  In reality, the boost factor likely broadens the angular profile, because the subhalo annihilation should extend further away from the halo center. However, since the angular extent of the annihilation in most halos is small compared to the instrument point-spread function (PSF), we do not model this extension here. Some nearby halos may have significantly larger angular extent, as would be expected for the Andromeda galaxy. Nevertheless, such considerations need to be made case by case and are discussed in detail in the next chapter, where we choose to exclude Andromeda due to its size.

Figure~\ref{fig:heatmap} is a heatmap representing the average $J$-factor, for a given $M_\text{vir}$ and $z$, of the \texttt{DarkSky} halos in the above configuration.  The halos span a wide range of masses and redshifts, with $J$-factors averaging over several orders of magnitude from $\sim$~10$^{16.5-18.5}~{\rm GeV}^2\,{\rm cm}^{-5}\,{\rm sr}$.     
The largest $J$-factors are observed for the most massive, cluster-sized halos at $z \sim $~0.01--0.02, as well as for less-massive halos at smaller redshifts ($z \lesssim 0.01$). 

\subsection{Uncertainties in Halo Modeling}
\label{sec:uncertainties}

Now, we consider more carefully the systematic uncertainties associated with modeling the halo properties.  
A halo with an NFW density profile has a $J$-factor dictated by its parameters as given in Eq.~\ref{eq:jscale}.  In addition to the distance, the $J$-factor also depends on the virial mass and concentration.\footnote{Note that uncertainties on the halo redshift also feed into the $J$-factor.  However, we consider this uncertainty to be subdominant for spectroscopically determined redshifts. For nearby halos, where the relation between distance and redshift is nontrivial, the uncertainty on the distance can be noticeably larger, and as high as $\sim$5\%~\cite{Tully:2016ppz}. Nonetheless, even such uncertainties are considerably smaller than those associated with the mass and concentration, and so we do not consider them.}  
  Therefore, any uncertainty in the determination of these halo properties is propagated through to the uncertainty on the DM annihilation flux.  Up until now, we have taken the halo mass and concentration directly from \texttt{DarkSky}, but in practice these parameters need to be inferred from properties of the observed galaxy groups.
 
\begin{figure*}
   \centering
   \includegraphics[width=0.45\textwidth]{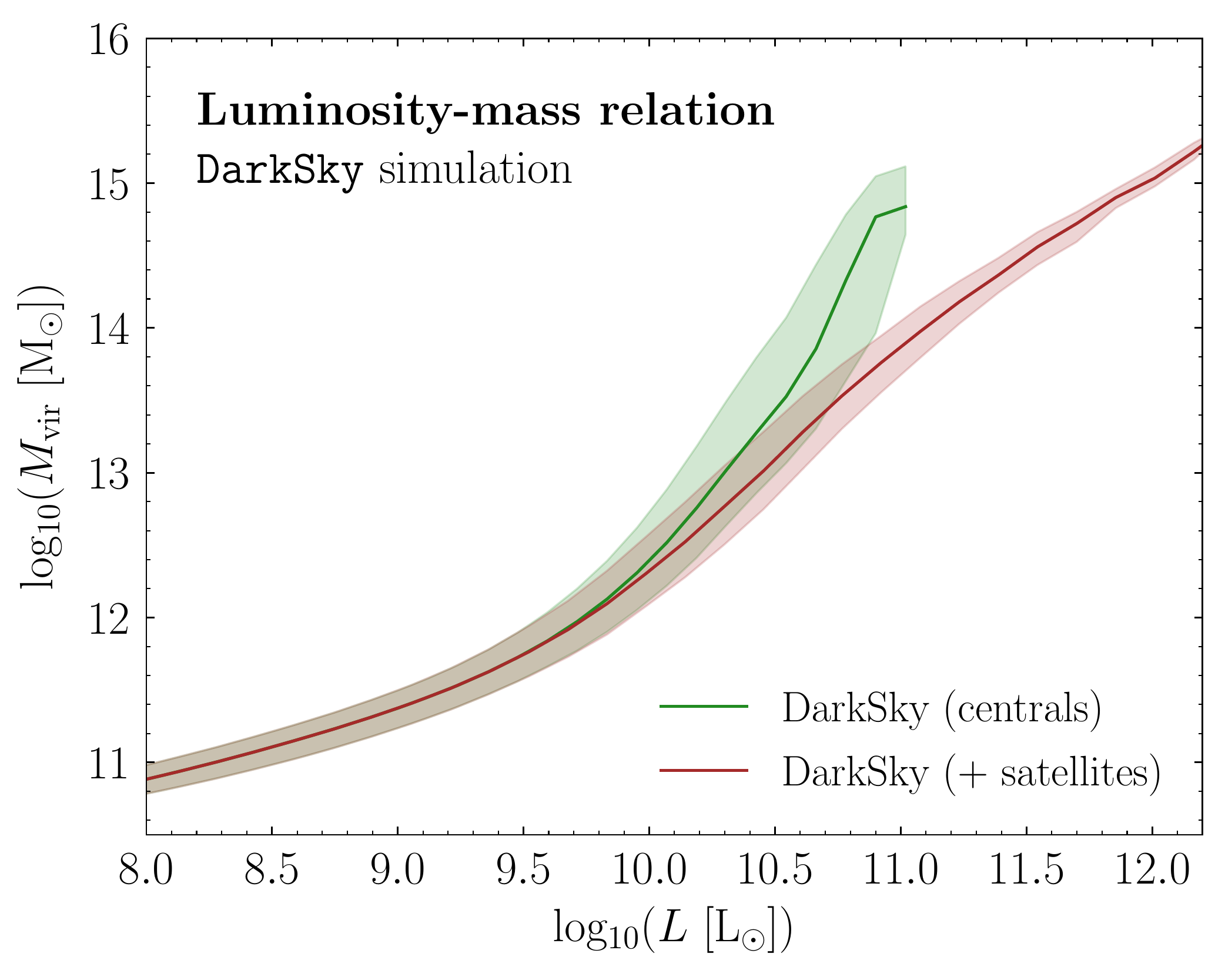}
   \includegraphics[width=0.45\textwidth]{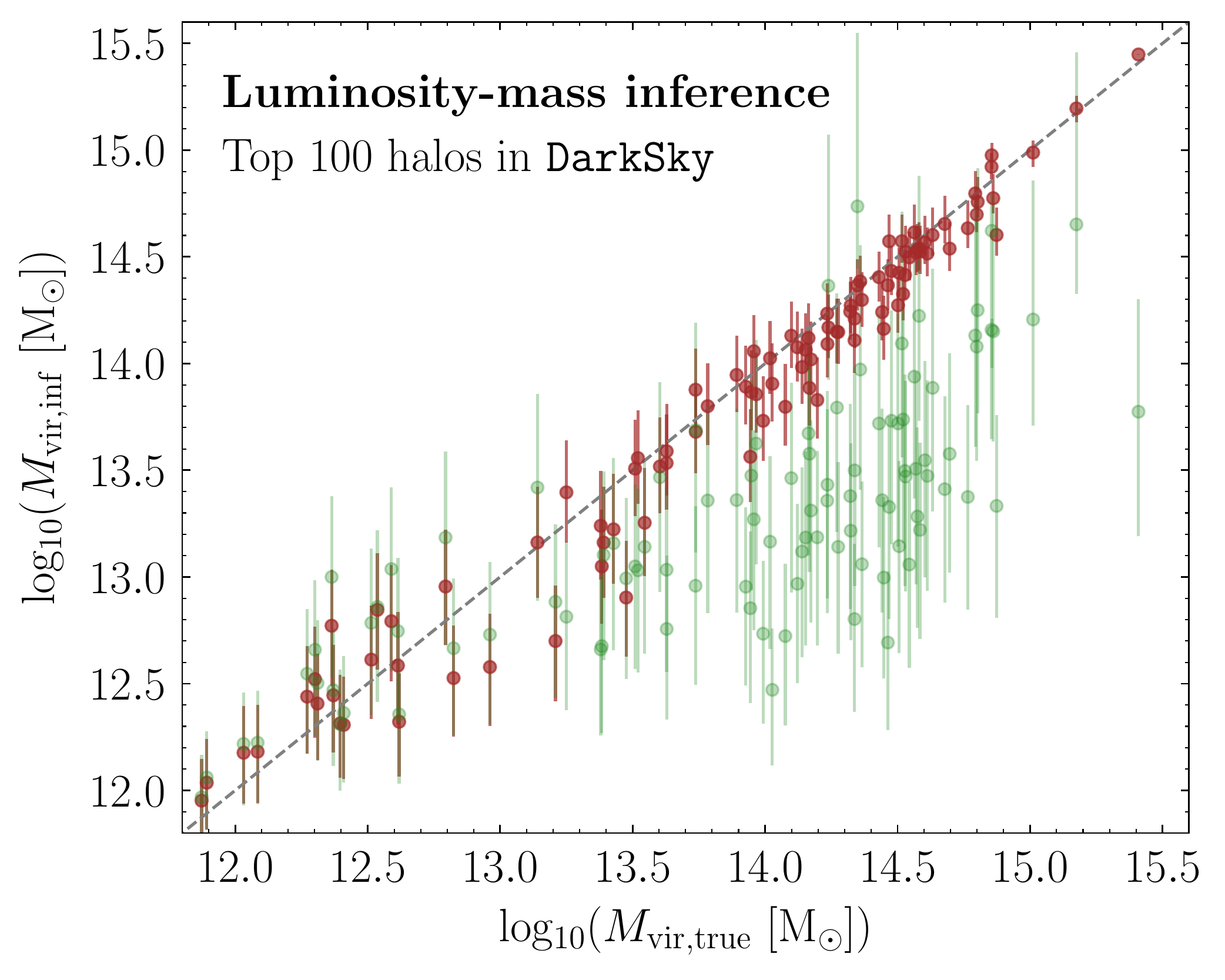}
   \caption{\textbf{(Left)} From \texttt{DarkSky}, we obtain the host halo mass as a function of absolute luminosity.  The green line represents the best-fit $M(L)$ relation when the central galaxy luminosity $(L_\mathrm{cen})$ is used to infer the host halo mass, while the red line uses the total luminosity $L_\mathrm{tot}$ (central + satellite).  The shaded region denotes the 68\% containment region in each case. \textbf{(Right)} Halo masses and uncertainties, inferred using the $M(L_\mathrm{cen})$ relation (green) and the $M(L_\mathrm{tot})$ relation (red).  The inclusion of the satellite luminosity allows one to better recover the halo mass.}
   \label{fig:ltom}
\end{figure*}

Within \texttt{DarkSky}, the halo mass can be inferred from the absolute luminosity of its associated galaxy group.  We obtain a deterministic $M(L)$ relation following a procedure similar to that in Ref.~\cite{Vale:2005mw}, which derived a phenomenological relation between the $K$-band galaxy luminosity and the mass of its DM halo.  The left panel of Fig.~\ref{fig:ltom} shows the true masses for the \texttt{DarkSky} halos, as a function of central galaxy luminosity (green) or the total luminosity, which includes the luminosity of the satellite galaxies (red).  The \texttt{DarkSky} catalog provides the associations for all galaxies, central and satellite, so we include all satellites that are associated to the group when calculating the total absolute luminosity. This is similar to what is done in published group catalogs~\cite{Tully:2015opa,2017ApJ...843...16K,Lu:2016vmu}, where they account for the loss in luminosity of satellite galaxies that are farther away.  

From Fig.~\ref{fig:ltom}, we see that the spread in the associated halo mass increases above $\sim10^{10}$~L$_\odot$, up to the brightest galaxy at $\sim10^{11}$~L$_\odot$, when the central galaxy luminosity is used.  In contrast, the spread is significantly smaller when the total luminosity is used, making it a better predictor for the halo mass.  As demonstrated in the right panel of Fig.~\ref{fig:ltom}, including the satellite luminosities allows one to better reconstruct the halo mass.  Therefore, we use the median $M(L)$ relation thus obtained as our fiducial case to infer the central mass estimate, and we use the spread in the $M(L)$ relation to infer the uncertainty on the mass.  Note that the $M(L)$ relation shown in Fig.~\ref{fig:ltom} is constructed by binning the \texttt{DarkSky} data in luminosity and calculating the 16, 50, and 84 percentiles in $M_\text{vir}$; different results would be obtained by binning in $M_\text{vir}$ and then constructing the percentiles from the luminosity distributions.   This procedure is similar to that adopted by galaxy group catalogs to infer the halo mass~\cite{Tully:2015opa,Lu:2016vmu,2017ApJ...843...16K}. Using this $M(L)$ relation, we can infer the halo mass and uncertainty  for each galaxy-group host halo in \texttt{DarkSky}. 

DM halos of the same mass can have very different characteristics, usually reflecting their distinct formation history and environment.  One such characteristic is the halo's virial concentration $c_\text{vir} = r_\text{vir}/r_s$. 
The scale radius is the relevant quantity to compare to as it indicates an isothermal slope for the density profile, which is required for a flat rotation curve.  The virial radius corresponds to the spherical volume within which the mean density is $\Delta_{c}$ times the critical density of the Universe at that redshift. We use $\Delta_c(z) = 18\pi^2 +82x-39x^2$ with $x = \Omega_{m}(1+z)^3/[\Omega_{m}(1+z)^3 + \Omega_{\Lambda}]-1$ in accordance with Ref.~\cite{Bryan:1997dn}.  The cosmology associated with the \texttt{DarkSky} simulation is used throughout, with $\Omega_\Lambda = 0.705$,  $\Omega_m = 0.295$ and $h = 0.688$. 

\begin{figure*}[t]
   \centering
  \includegraphics[width=0.49\textwidth]{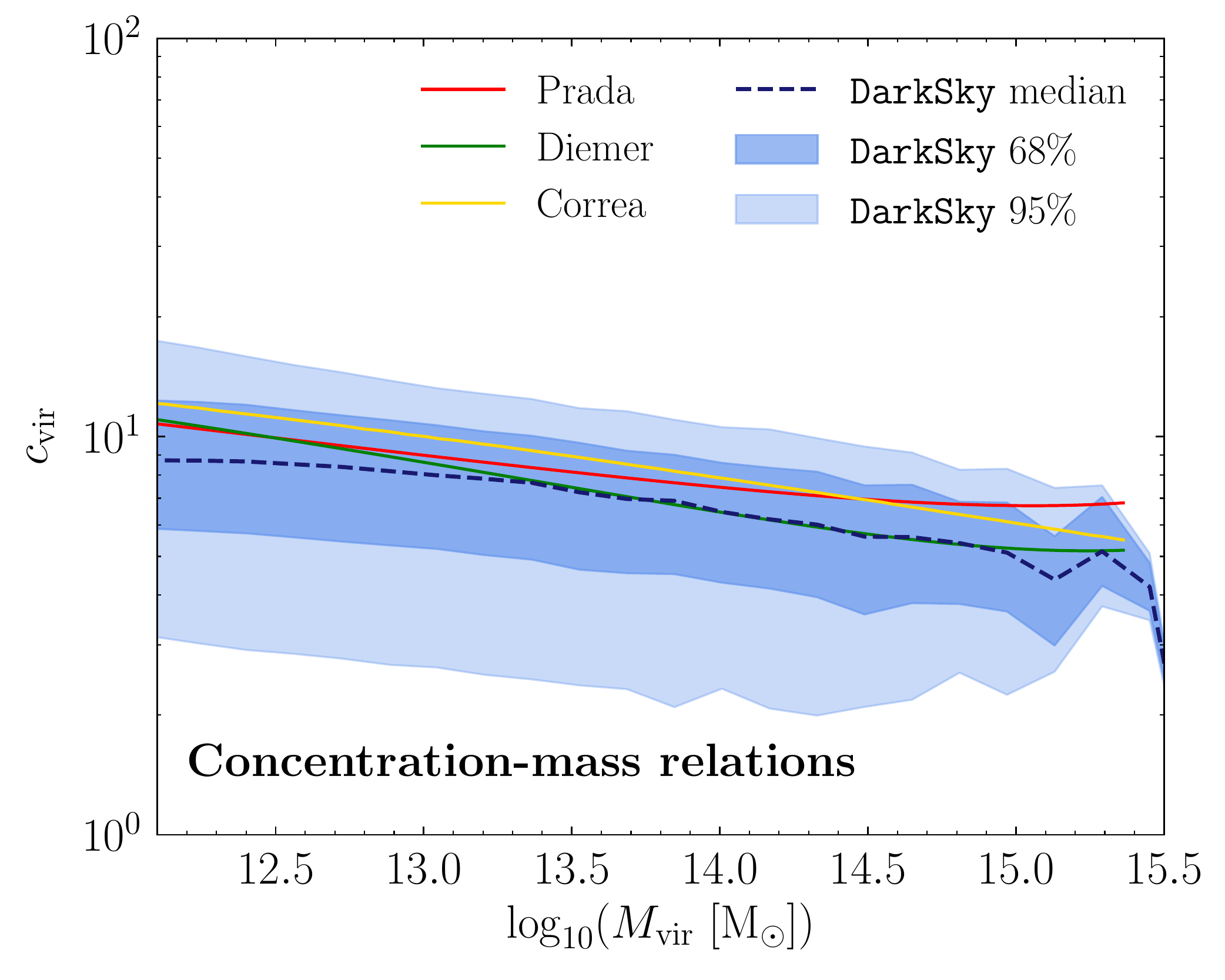} 
   \includegraphics[width=0.49\textwidth]{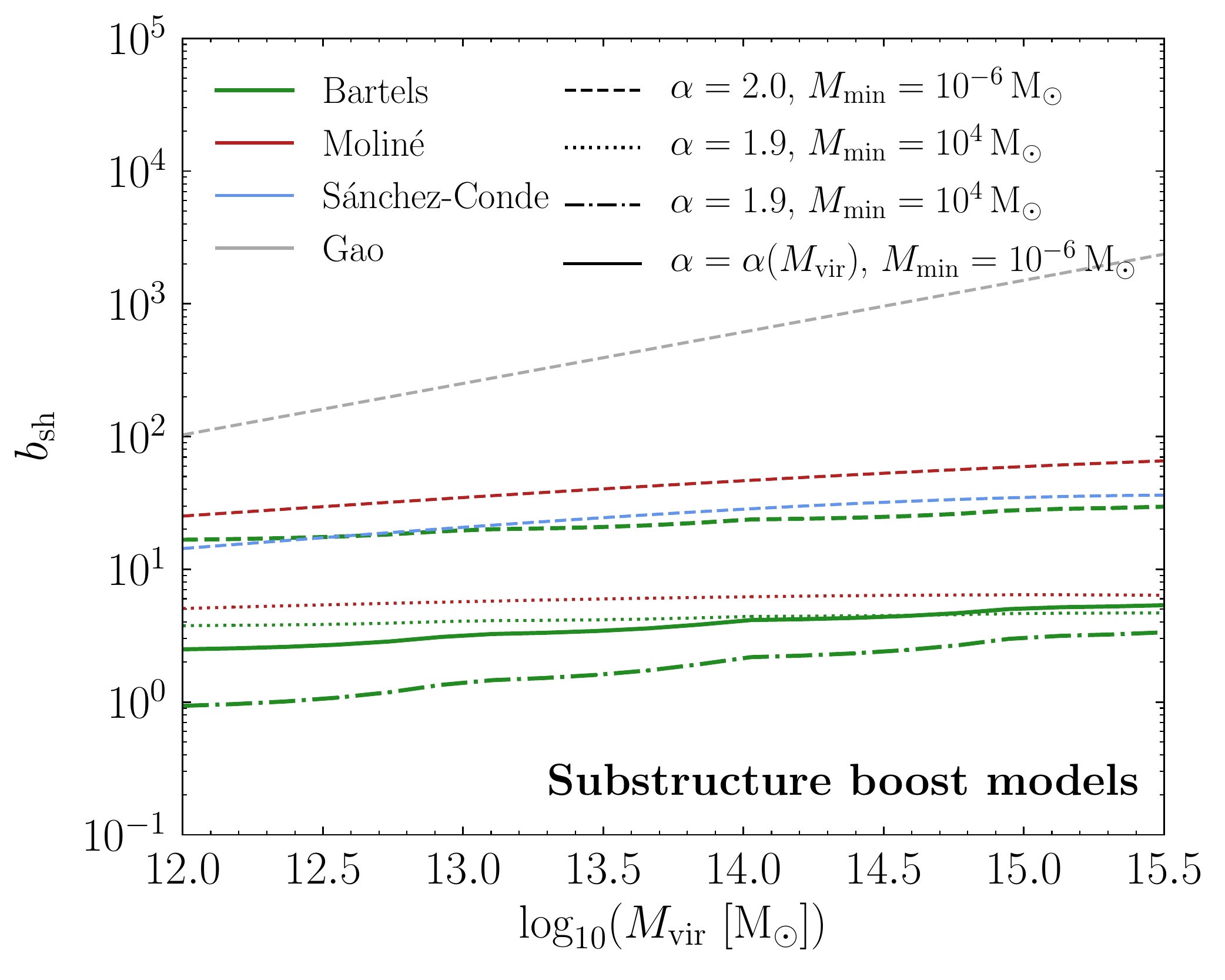} 
         \caption{\textbf{(Left)} The median concentration-mass relation in \texttt{DarkSky} (dashed black) along with the middle 68 and 95\% spread (blue regions) compared with models found in the literature.  For comparison, we also show the models of Correa \emph{et al.} (yellow)~\cite{Correa:2015dva}, Diemer and Kravtsov (green)~\cite{Diemer:2014gba}, and Prada \emph{et al.} (red) \cite{Prada:2011jf}.  All concentration models are evaluated for the \texttt{DarkSky} cosmology. \textbf{(Right)}  Boost models found in the literature as a function of host halo mass. As a conservative choice, we select the Bartels and Ando model~\cite{Bartels:2015uba} shown in thick solid green. In blue, red and gray, we compare this to the boost models of S{\'a}nchez-Conde \emph{et al.}~\cite{Sanchez-Conde:2013yxa}, Molin{\'e} \emph{et al.}~\cite{Moline:2016pbm}, and Gao \emph{et al.}~\cite{Gao:2011rf}, respectively. The line type (dashed, dotted, dot-dashed, and solid) denotes the assumption being made on the slope of the subhalo mass function, $\alpha$, and the mass cutoff, M$_\text{min}$.  }
    \label{fig:concentrations}
\end{figure*}

In general, the concentration correlates strongly with halo mass due to the dependence of halo formation time on mass---on average, lower mass halos tend to be more concentrated because they collapsed earlier, when the Universe was denser.  For the same reason, the concentration is sensitive to the cosmology, which determines how early halos start to assemble. 
The concentration of field halos has been extensively studied and several concentration-mass relations have been proposed in the literature, usually based on $N$-body simulations or physically motivated analytic approaches~\cite{Duffy:2008pz,Prada:2011jf,Sanchez-Conde:2013yxa,Diemer:2014gba,Correa:2015dva,Moline:2016pbm,Klypin:2014kpa,Dutton:2014xda}. In the left panel of Fig.~\ref{fig:concentrations}, we show the median value of the concentration-mass relation derived directly from the \texttt{DarkSky} simulation, as well as the middle 68 and 95\% spread. The middle 68\% scatter in the relation is typically in the range 0.14-0.19 across the halo mass range considered. For comparison, we also show several concentration models that are commonly used in the literature.  As is standard in the literature~\cite{Hutten:2016jko,Sanchez-Conde:2013yxa}, we model the uncertainty in the concentration, for a given virial mass, as a log-normal distribution around its median value.  

To summarize, it is possible to infer the halo mass from the luminosity of the galaxy group and to then obtain the concentration.  The final remaining property that is needed to solve for the $J$-factor in Eq.~\ref{eq:jfactor} is the boost factor, which depends on the distribution and minimum cutoff of the subhalos' mass.  The boost factor encapsulates the complicated dependence of the subhalo mass distribution on both the particle physics assumptions of the DM model as well as the dynamics of the host halo formation.  A variety of different boost models typically used in the literature are illustrated in the right panel of Fig.~\ref{fig:concentrations}.  As our fiducial case, we adopt the boost model of Ref.~\cite{Bartels:2015uba} (labeled as `Bartels Boost Model'), which self-consistently accounts for  the concentration-mass relation of subhalos (compared to field halos) as well as the effects of tidal stripping. Specifically, in the subhalo mass function $dn/dM_\text{sh}\propto M_\text{sh}^{-\alpha}$, we use a minimum subhalo mass cutoff of $M_\text{min}=10^{-6}$~M$_{\odot}$ and slope $\alpha$ that varies self-consistently with host halo mass while accounting for evolution effects (see Ref.~\cite{Bartels:2015uba} for details).

We have now built up a framework that allows us to determine the expected DM annihilation flux map associated with a catalog of galaxy groups.  Next, we show how to use this information to search for signals of DM  from hundreds of galaxy groups.  

\section{Statistical Methods}
\label{sec:stats}

In this work, we introduce and study a statistical procedure to search for gamma-ray signals from DM by stacking galaxy groups.  All analyses discussed here are run on mock data, which is based on the expected astrophysical contributions to the real \emph{Fermi} data set. When building this mock data set, we include contributions from (1) the diffuse  emission, for which we use the \textit{Fermi} Collaboration's \texttt{p7v6} model; (2) isotropic emission;  (3) emission from the \textit{Fermi} Bubbles \cite{Su:2010qj}; and (4) emission from point sources in the \textit{Fermi} 3FGL catalog \cite{Acero:2015hja}. The overall flux normalization  for each component must be known \emph{a priori} to create the mock data. To obtain this, we fit spatial maps of (1)--(4) above to the actual \emph{Fermi} data.  We use 413 weeks of UltracleanVeto (all PSF quartile) Pass 8 data collected between August 4, 2008 and July 7, 2016. We break the data into 40 equally log-spaced energy bins between 200~MeV and 2~TeV, applying the recommended data cuts: zenith angle $< 90^{\circ}$, \texttt{DATA\_QUAL} $> 0$, and \texttt{LAT\_CONFIG} $=1$.  To minimize the Galactic contamination in this initial fit, we mask the region $|b|<30^{\circ}$ as well as the 68\% containment radius for the 300 brightest and most variable sources in the 3FGL catalog. We emphasize that these masks are only used when creating the mock data and not in the stacked analysis. The fitting procedure described here provides the expected astrophysical background contribution from the real data.  Monte Carlo (MC) is then generated by summing up these contributions and taking a Poisson draw from the resulting map.  In the following discussion, we will show how results vary over different MC realizations of the mock data as a demonstration of Poisson fluctuations in the photon distribution. 

We now describe in detail the statistical procedure we employ to implement the stacking analysis on the mock data.  We perform a template-fitting profile likelihood analysis in a 10$^{\circ}$ region-of-interest (ROI) around each group.  Template studies amount to describing the sky by a series of spatial maps (called templates).  The normalization of each template is proportional to its relative gamma-ray flux.  We use five templates in our study.  The first four are associated with the known astrophysical sources (1)--(4) described above. Within $10^\circ$ of the halo center, we independently float the normalization of each 3FGL source.\footnote{The results do not change when floating all the point sources together as one combined template. This can potentially cause problems when implemented on data, however, because the 3FGL normalizations can be erroneous in certain energy bins.  Allowing the normalizations of the sources to float separately helps to mitigate this potential problem.}  Sources outside this region may potentially contribute within the ROI because of the tails of the {\it Fermi} PSF.  Therefore, between 10$^{\circ}$ and 18$^{\circ}$ of the halo center,  we float the sources as a single template. The fifth and final template that we include is associated with the expected DM annihilation flux for the halo, which is effectively a map of the $J$-factor and is described in Sec.~\ref{sec:galaxyfilteringpipeline}.  Note that all templates have been carefully smoothed using the \textit{Fermi} PSF. The diffuse model is smoothed with the \textit{Fermi} Science Tools, whereas other templates are smoothed according to the instrument response function using custom routines.  Mismodeling the smoothing of either the point sources or individual halos can potentially impact the  results.

A given mock data set, $d$, is divided into 40 log-spaced energy bins indexed by $i$. Each energy bin is then spatially binned using \texttt{HEALPix}~\cite{Gorski:2004by} with \texttt{nside}=128 and individual pixels indexed by $p$.  In this way, the full data set is reduced to a two-dimensional array of integers $n_i^p$ describing the number of photons in energy bin $i$ and pixel $p$.  For a given halo, indexed by $r$, only a subset of all the pixels in its vicinity are relevant.  In particular, the relevant pixels are those with centers within 10$^{\circ}$ of the object.  Restricting to these pixels leaves a subset of the data, which we denote by $n_i^{p,r}$. Template fitting dictates that this data is described with a set of spatial templates binned in the same way as the data, which we label as $T^{p,\ell}_i$, where $\ell$ indexes the different templates considered. The number of counts in a given pixel, energy bin, and region consists of a combination of these templates:
\begin{equation}
\mu_i^{p,r}({\boldsymbol \theta_i^r}) = \sum_{\ell} A^{r, \ell}_i\,T^{p,\ell}_i\,.
\end{equation}
Here, ${\boldsymbol \theta_i^r}$ represents the set of model parameters. For Poissonian template fitting, these are given by the normalizations of the templates $A^{r,\ell}_i$, \emph{i.e.}, ${\boldsymbol \theta_i^r} = \{ A^{r,\ell}_i \}$. Note that the template normalizations have an energy but not a spatial index, as the templates have an independent degree of freedom in each energy bin as written, but the spatial distribution of the model is fixed by the shapes of the templates themselves.  In principle, we could also remove this freedom in the relative emission across energy bins, because we have models for the spectra of the various background components, and in particular DM. Nevertheless, we still allow the template normalizations to float independently in each energy bin for the various backgrounds. This is more conservative than assuming a model for the background spectra, and in particular we can use the shape of the derived spectra as a check that the dominant background components are being correctly modeled. The spectral shape of the DM forms part of our model prediction, however, and once we pick a final state such as annihilation to two $b$-quarks, we fix the relative emission between the energy bins.

As we assume that the data comes from a Poisson draw of the model, the appropriate likelihood in energy bin $i$ and ROI $r$ is
\begin{equation}
\mathcal{L}_i^r(d_i^r | {\boldsymbol \theta}_i^r) = \prod_p \frac{\mu_i^{p,r}({\boldsymbol \theta}_i^r)^{n_i^{p,r}} e^{-\mu_i^{p,r}({\boldsymbol \theta}_i^r)}}{n_i^{p,r}!}\,.
\label{eq:pi}
\end{equation}
Of the templates that enter this likelihood, there are some we are more interested in than others. In particular, we care about the the DM annihilation intensity, which we denote as $\psi_i$.  We treat the normalizations of the templates associated with the known astrophysical emission as nuisance parameters, $\lambda_i^r$. Below, we will describe how to remove the nuisance parameters to reduce Eq.~\ref{eq:pi} to a likelihood profile that depends only on the DM annihilation intensity, but for now we have ${\boldsymbol \theta}_i^r = \{ \psi_i, \lambda_i^r \}$.

Importantly, the nuisance parameters have different values between ROIs, but the DM parameters do not.  This is because the DM parameters, such as the DM mass, annihilation rate, and set of final states, are universal, while the parameters that describe the astrophysical emission can vary from region to region.
We do, however, profile over the $J$-factor uncertainty in each ROI.
Explicitly, each halo is given a model parameter $J^r$, which is described by a
log-normal distribution around the central value $\log_{10} J_{\rm c}^r$ with width $\sigma_r = \log_{10} J_{\rm err}^r$, both of which depend on the object and hence ROI considered.  The $J$-factor error, $J_{\rm err}^r$, is determined by propagating the errors associated with the mass and concentration of a given halo. To account for this, we append the following addition onto our likelihood as follows:
\begin{equation}\begin{aligned}
&\mathcal{L}_i^r \left( d_i^r | \boldsymbol{\theta}_i^r \right) \to \mathcal{L}_i^r \left( d_i^r | \boldsymbol{\theta}_i^r \right) \\
&\times \frac{1}{\ln(10) J_{\rm c}^r\sqrt{2\pi} \sigma_r} \exp \left[ - \frac{(\log_{10} J^r - \log_{10} J_{\rm c}^r)^2}{2\sigma_r^2} \right]\,.
\label{eq:Jlognormal}
\end{aligned}\end{equation}
Note that this procedure does not account for any systematic uncertainties that can bias the determination of the $J$-factor.

The nuisance parameter $J^r$ can now be eliminated via the profile likelihood---see Ref.~\cite{Rolke:2004mj} for a review. Unlike for the other nuisance parameters, the value of $J^r$ does not depend on energy and so we eliminate the energy-dependent parameters first:
\begin{equation}
\mathcal{L}_i^r(d_i^r | \psi_i) = \max_{\{\lambda_i^r\}}\,\mathcal{L}_i^r \left( d_i^r | \boldsymbol{\theta}_i^r \right)\,.
\label{eq:firstprofile}
\end{equation}
The full implementation of the profile likelihood method as suggested by this equation requires determining the maximum likelihood for the $\lambda_i^r$ template coefficients, for every value of $\psi_i$.  Nevertheless, an excellent approximation to the profile likelihood, which is computationally more tractable, is simply to set the nuisance parameters to their maximum value obtained in an initial scan where all templates are floated.\footnote{The DM template is only included for energy bins above 1 GeV.  At lower energies, the large \textit{Fermi} PSF leads to confusion between the DM, isotropic and point source templates, which can introduce a spurious preference for the DM template.} 

Using this approach to determine the likelihood in Eq.~\ref{eq:firstprofile}, we can build a total likelihood by combining the energy bins. Once this is done, the likelihood  depends on the full set of DM intensities $\psi_i$, which are specified by a DM model $\mathcal{M}$, cross section $\langle \sigma v \rangle$, mass $m_{\chi}$, and $J$-factor via Eq.~\ref{eq:flux}. Explicitly:
\begin{equation}
\mathcal{L}^r(d_r | \mathcal{M}, \langle \sigma v \rangle, m_\chi, J^r) = \prod_i \mathcal{L}_i^r(d_i^r | \psi_i)\,,
\end{equation}
and recall that unlike the other parameters on the left hand side, the $J$-factor not only determines the $\psi_i$, but also enters the likelihood through the expression in Eq.~\ref{eq:Jlognormal}. We emphasize that in this equation, the DM model and mass specify the spectra, and thereby the relative weightings of the $\psi_i$, whereas the cross section and $J$-factor set the overall scale of the emission.

The remaining step to get the complete likelihood for a given halo $r$ is to remove $J^r$, again using profile-likelihood:
\begin{equation}
\mathcal{L}^r(d_r | \mathcal{M}, \langle \sigma v \rangle, m_\chi) = \max_{J^r} \mathcal{L}^r(d_r | \mathcal{M}, \langle \sigma v \rangle, m_\chi, J^r)\,.
\end{equation}
This provides the full likelihood for this object as a function of the DM model parameters.  The likelihood for the full stacked catalog is then simply a product over the individual likelihoods:
\begin{equation}
\mathcal{L}(d | \mathcal{M}, \langle \sigma v \rangle, m_\chi) = \prod_r \mathcal{L}^r(d_r | \mathcal{M}, \langle \sigma v \rangle, m_\chi)\,.
\label{eq:likelihoodobjprod}
\end{equation}
Using this likelihood, we define a test statistic (TS) profile as follows:
\begin{equation}\begin{aligned}
{\rm TS}(\mathcal{M}, \langle\sigma v\rangle, m_\chi) \equiv 2 &\left[ \log \mathcal{L}(d | \mathcal{M}, \langle\sigma v\rangle, m_\chi ) \right.\\
&\left.- \log \mathcal{L}(d | \mathcal{M}, \widehat{\langle\sigma v\rangle}, m_\chi ) \right]\,,
\label{eq:TSdef_darksky}
\end{aligned}\end{equation}
where $\widehat{\langle\sigma v\rangle}$ is the cross section that maximizes the likelihood for that DM model and mass. From here, we can use this TS, which is always nonpositive by definition, to set a threshold for limits on the cross-section.  When searching for evidence for a signal, we use an alternate definition of the test statistic defined as 
\es{eq:maxTS_darksky}{
{\rm TS}_\text{max}(\mathcal{M}, m_\chi) \equiv &2 \left[ \log \mathcal{L}(d | \mathcal{M}, \widehat{\langle\sigma v\rangle}, m_\chi ) \right. \\
 &\left.- \log \mathcal{L}(d | \mathcal{M}, \langle\sigma v\rangle =0, m_\chi ) \right] \,.
}

We implement template fitting with the package \texttt{NPTFit} \cite{Mishra-Sharma:2016gis}, which uses \texttt{MultiNest}~\cite{Feroz:2008xx,Buchner:2014nha} by default, but we have employed \texttt{Minuit}~\cite{James:1975dr} in our analysis.

\section{Analysis Results}
\label{sec:smallrois}

In this Section, we present the results of our analysis on mock data using the \texttt{DarkSky} galaxy catalog. We begin by describing the sensitivity estimates associated with this study, commenting on the impact of statistical as well as systematic uncertainties and studying the effect of stacking a progressively larger number of halos.  Then, we justify the halo selection criteria that are used by showing that we can recover injected signals on mock data. 

\subsection{Halo Selection and Limits}
\label{sec:limits}

We now discuss the results obtained by applying the halo inference pipeline described in Sec.~\ref{sec:galaxyfilteringpipeline} and the  statistical analysis described in Sec.~\ref{sec:stats} to  mock gamma-ray data.  We focus on the top 1000 galaxy groups in the \texttt{DarkSky} catalog, as ranked by the inferred $J$-factors of their associated halos, placing ourselves at the center of the simulation box. In addition, we mask regions of the sky associated with seven large-scale structures that are challenging to model accurately: the Large and Small Magellanic Clouds, the Orion molecular clouds, the galaxy NGC5090, the blazar 3C454.3, and the pulsars J1836+5925 and Geminga. This is done here for simulated data in order to closely track the analysis that will subsequently be performed on real \emph{Fermi} data.

While we start from an initial list of 1000 galaxy groups, we do not include all of them in the stacking procedure.  A galaxy group is excluded if: 
\begin{enumerate}
\item it is located within $|b| \leq 20^\circ$;
\item it is located less than $2^\circ$ from the center of another brighter group in the catalog;
\item it has TS$_\text{max}$ $> 9$ and $\langle \sigma v \rangle_\text{best}  > 10.0 \times \langle \sigma v \rangle_\text{lim}^*$\, ,
\end{enumerate}
where $\langle \sigma v \rangle_\text{best}$ is the best-fit cross section at any mass and $\langle \sigma v \rangle^*_\text{lim}$ is the best-fit limit set by \emph{any} halo at the specified DM mass.  Note that the second requirement is applied sequentially to the ranked list of halos, ordered by $J$-factor.  
We now explain the motivation for each of these requirements separately.  The first requirement listed above removes groups that are located close to the Galactic plane to reduce contamination from regions of high diffuse emission and the associated uncertainties in modeling these. The second requirement demands that the halos be reasonably well-separated, which avoids issues having to do with overlapping halos and accounting for multiple DM parameters in the same ROI. The non-overlap criterion of 2$^\circ$ is chosen based on the \emph{Fermi} PSF containment in the lowest energy bins used and on the largest spatial extent of gamma-ray emission associated with the extended halos, which collectively drive the possible overlap between nearby halos.

\begin{figure*}[t]
   \centering
   \includegraphics[width=0.45\textwidth]{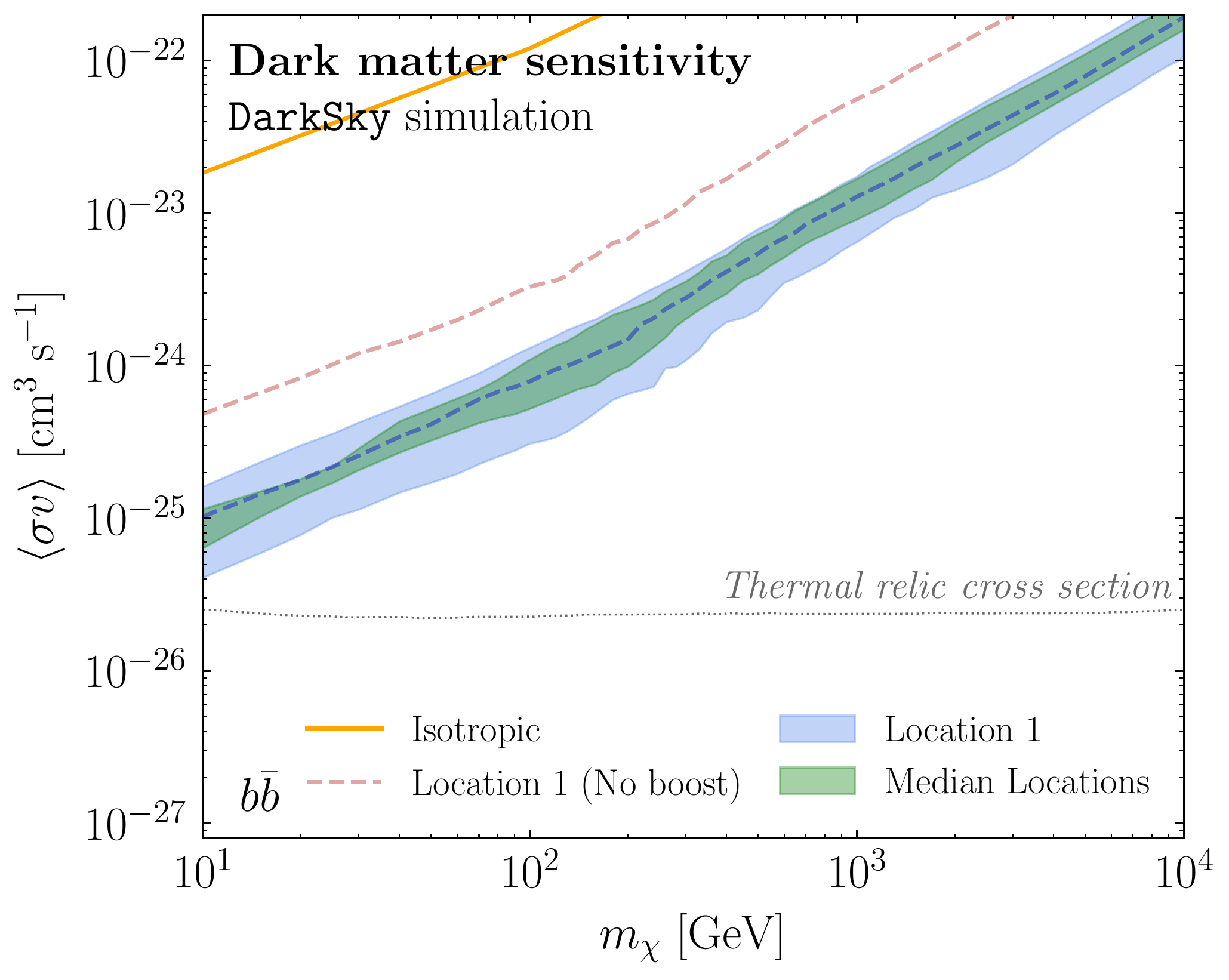} 
   \includegraphics[width=0.45\textwidth]{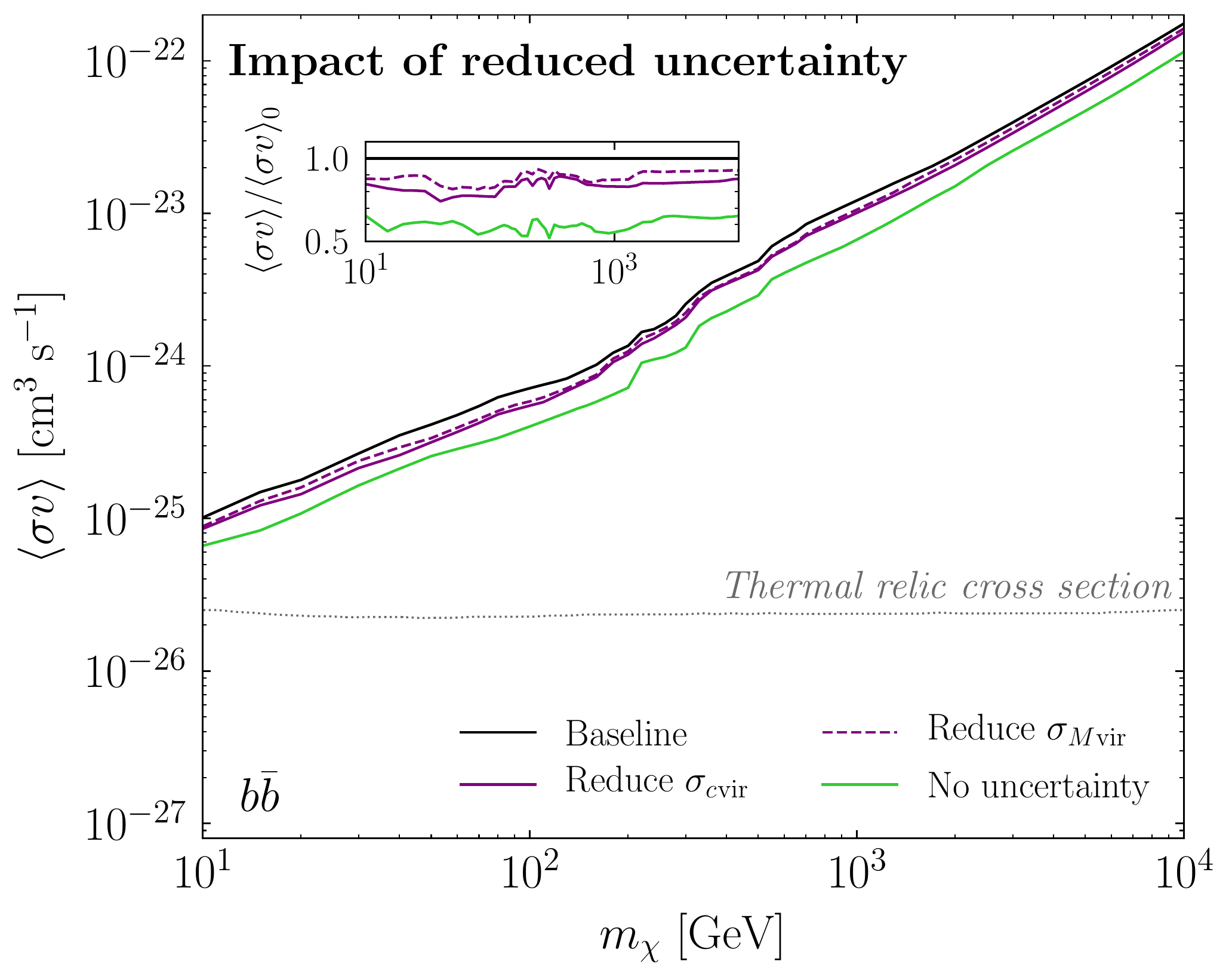} 
   \caption{\textbf{(Left)} The 95\% confidence limit on the DM annihilation cross section, $\langle \sigma v \rangle$, as a function of the DM mass, $m_\chi$, for the $b\bar{b}$ final state, assuming the fiducial boost factor model from Ref.~\cite{Bartels:2015uba} (dashed blue); the corresponding result with no boost factor is shown in dashed red.  These limits correspond to the default position where the observer is placed in the center of the \texttt{DarkSky} simulation box (`Location 1').  The blue band shows the middle 68\% spread in the median limits obtained from 100 Monte Carlo realizations of the mock data.  The green band shows the same spread on the median limits obtained from nine random observer locations within the \texttt{DarkSky} simulation box.  The orange line shows the limit obtained by requiring that DM emission not overproduce the observed isotropic gamma-ray intensity and highlights how the sensitivity improves when one resolves the DM structure.  The thermal relic cross section for a generic weakly interacting massive particle~\cite{Steigman:2012nb} is indicated by the thin dotted line. \textbf{(Right)} The effect of reducing the uncertainty on virial mass, $M_\text{vir}$, and concentration, $c_\text{vir}$, in the stacking analysis.  The case where no uncertainty on the $J$-factor is assumed (green) is compared with the baseline analysis (black). We also show the impact of individually reducing the uncertainty on the concentration (solid purple) or mass (dashed purple) by 50\% for each halo. The inset shows the ratio of the improved cross section limit to the baseline case.}
   \label{fig:DSLimits}
\end{figure*}

The final requirement excludes a galaxy group if it has an excess of at least 3$\sigma$ significance associated with the DM template that is simultaneously excluded by the other galaxy groups in the sample.  This selection is necessary because we expect that some galaxy groups will have true cosmic-ray-induced gamma-ray emission from conventional astrophysics in the real data, unrelated to DM.  To identify these groups, we take advantage of the fact that we are starting from a large population of halos that are all expected to be bright DM sources in the presence of a signal.  Thus, if one halo sets a strong limit on the annihilation rate and another halo, at the same time, has a large excess that is severely in conflict with the limit, then most likely the large excess is not due to DM.  The worry here is that we could have mis-constructed the $J$-factor of the halo that gave the strong limit, so that the real limit is not as strong as we think it is.  However, with the TS$_\text{max}$ and $\langle \sigma  v \rangle$ criteria outlined above, this does not appear to be the case.  In particular, we find that the criteria very rarely rejects halos due to statistical fluctuations.  For example, over 50 MC iterations of the mock data, $966\pm8$ halos (out of 1000) remain after applying the TS$_\text{max}$ and cross section cuts alone, and the excluded halos tend to have lower $J$-factors, since there the $\langle \sigma v\rangle$ requirement is more readily satisfied.  

We expect that this selection criteria will be very important on real data, however, where real excesses can abound.  In addition, as we will describe in the next subsection, injected signals are not excluded when the analysis pipeline is run on mock data.  In an ideal scenario, we would attempt to understand the origin of these excesses by correlating their emission to known astrophysics either individually or statistically. In the present analysis, however, we take the conservative approach of removing halos that are robustly inconsistent with a DM signal and leave a deeper understanding of the underlying astrophysics to future work.

 \begin{figure*}[t]
   \centering
   \includegraphics[width=1.0\textwidth]{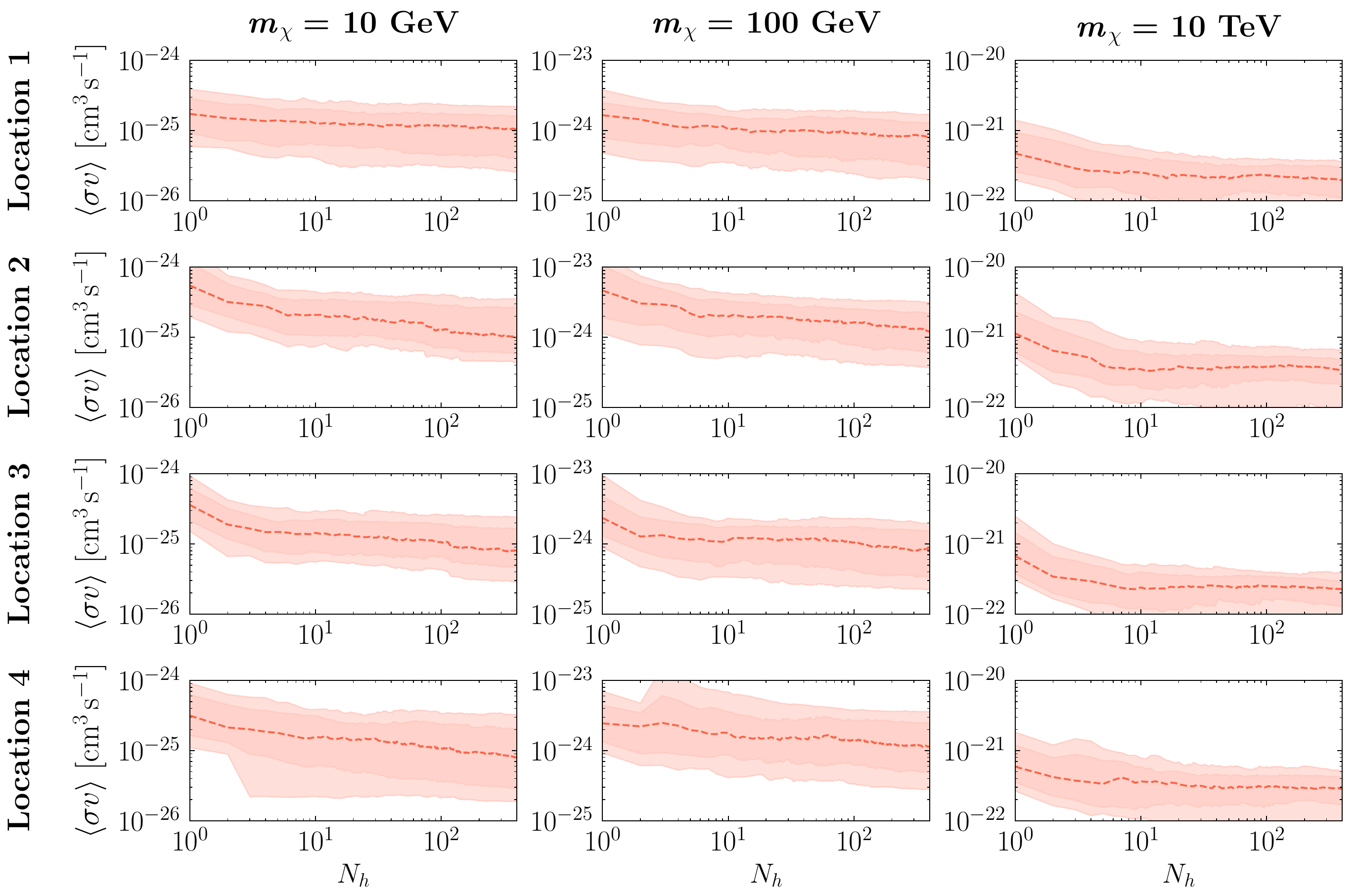}
   \caption{Variation of the limits as the number of galaxy groups (ranked by $J$-factor) included in the stacking, $N_h$, increases. The left, center, and right columns correspond to masses of 10~GeV, 100~GeV, and 10~TeV, respectively. Note that the scale of the y-axis varies between masses.  The four rows show how the limits vary for four different observer locations within the \texttt{DarkSky} simulation box.  }
   \label{fig:DSelephants}
\end{figure*}
We apply the procedure outlined in Secs.~\ref{sec:galaxyfilteringpipeline} and~\ref{sec:stats} to  the mock data to infer the 95\% confidence limit on the DM annihilation cross section.  The resulting sensitivity is shown by the blue dashed line in the left panel of Fig.~\ref{fig:DSLimits}, which uses the boost factor from Ref.~\cite{Bartels:2015uba}.  For comparison, we also show the limit assuming no boost factor (red dashed line); note that the boost factor model that we use provides a modest $\mathcal{O}(1)$ improvement to the limit.  Because the limit can potentially vary over different MC realizations of the mock data, we repeat the procedure for 100 MCs (associated with different Poisson realizations of the map); the blue band indicates the middle 68\% spread in the limit associated with this statistical variation.

To see how the limit depends on the observer's location within the \texttt{DarkSky} simulation box, we repeat the procedure described above over nine different locations.\footnote{The nine locations we used are at the following coordinates $(x,y,z)$ Mpc/h in \texttt{DarkSky}: $(200,200,200)$, $(100,100,100)$, $(100,100,300)$, $(100,300,100)$, $(300,100,100)$, $(300,300,100)$, $(100,300,300)$, $(300,100,300)$, $(300,300,300)$. The first listed location is our default position, and any time we use more than one location they are selected in order from this list.}  At each location, we perform 20 MCs and obtain the median DM limit.  The green band in the left panel of Fig.~\ref{fig:DSLimits} denotes the middle 68\% spread on the median bounds for each of the different sky locations. 
In general, we find that the results obtained by an observer at the center of the \texttt{DarkSky} box are fairly representative, compared to random locations.  Note, however, that this bound does not necessarily reflect the sensitivity reach one would expect to get with actual \emph{Fermi} data. The reason for this is that the locations probed in \texttt{DarkSky} do not resemble that of the Local Group in detail.
We will come back to this point below, when we compare the $J$-factors of the \texttt{DarkSky} halos to those from galaxy catalogs that map the local Universe.

The orange line in the left panel of Fig.~\ref{fig:DSLimits} shows the limit obtained by requiring that the DM emission from the groups not overproduce the measured isotropic gamma-ray component~\cite{Ackermann:2014usa}. This should not be compared to the published DM bounds obtained with the \emph{Fermi} Isotropic Gamma-Ray Background~\cite{Ackermann:2015tah} because that study accounts for the integrated effect of the DM annihilation flux from halos much deeper than those we consider here.  The inclusion of these halos results in a total flux that can be greater than those from our sample by over an order of magnitude. Nevertheless, this gives an idea of how much we gain by resolving the spatial structure of the local DM population and knowing the locations of the individual galaxy groups.

The right panel of Fig.~\ref{fig:DSLimits} shows the effect of propagating uncertainties associated with inferring the halo properties. The green line indicates how the limit improves when no uncertainties are assumed, \emph{i.e.}, we can perfectly reconstruct the virial mass and concentration of the halos. The sensitivity reach improves by roughly a factor of two in this case. We further show the effect of individually reducing the error on $M_\text{vir}$ (dashed purple line) and $c_\text{vir}$ (purple line) by 50\%. The reductions in the uncertainties provide only marginal improvements to the overall sensitivity, still far below the level of systematic uncertainty associated with extragalactic analyses in general.

\begin{figure}[t]
   \centering
   \includegraphics[width=0.8\textwidth]{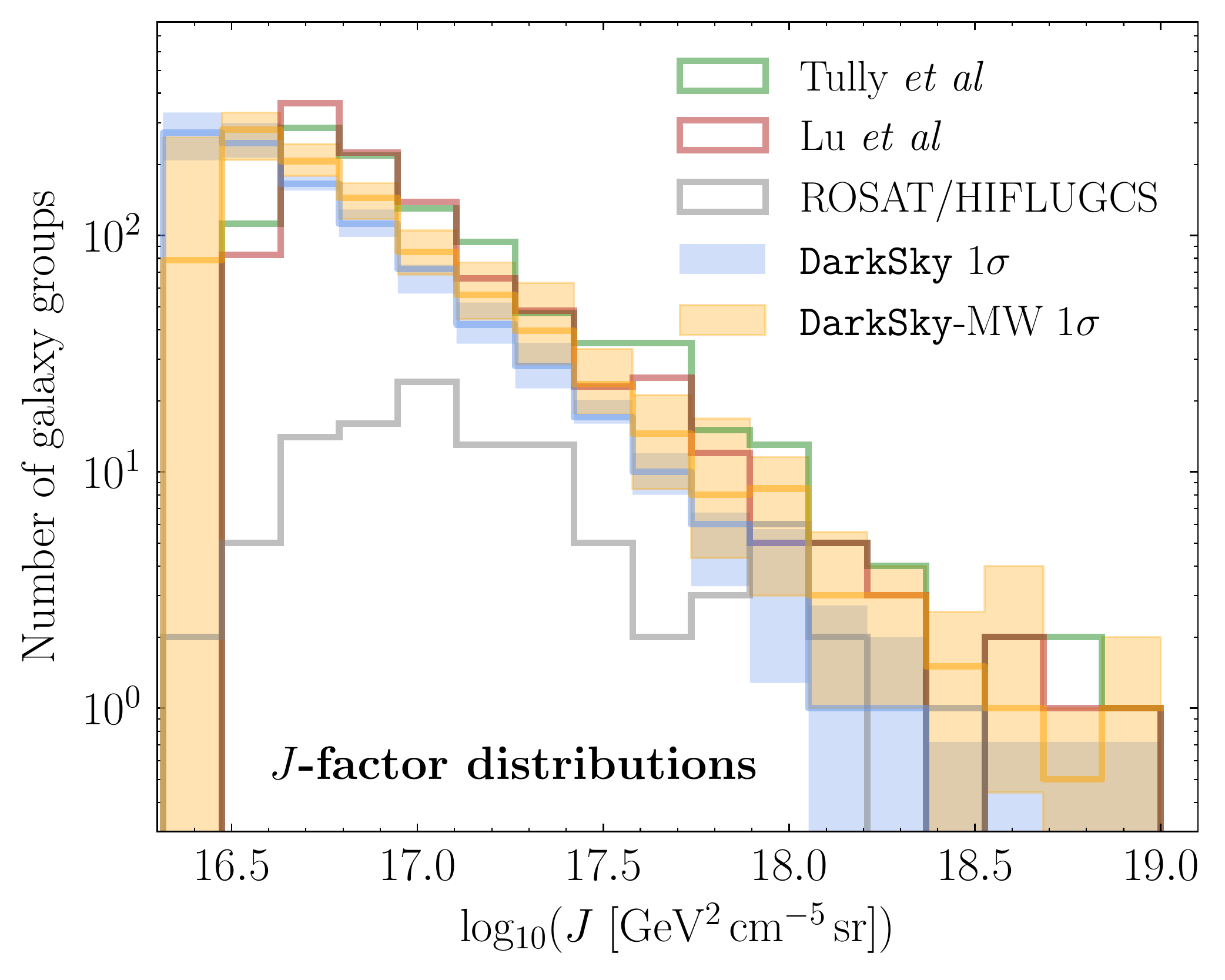} 
   \caption{Distribution of the top 1000 $J$-factors from the \texttt{DarkSky} catalog; the blue line  indicates the median distribution over nine random observer locations within the simulation box, with the blue band denoting the 68\% containment.  The orange line and band are the same, except for observers placed at ten random Milky~Way--like halos of mass  $\sim10^{12}$~M$_\odot$ in the box.  The distributions for the top 1000 $J$-factors in 2MRS galaxy-group catalogs are also shown; the green and red lines correspond to the Tully \emph{et al.}~\cite{Tully:2015opa,2017ApJ...843...16K} and the Lu \emph{et al.}~\cite{Lu:2016vmu} catalogs, respectively.  We also show the distribution (gray line) for the 106 galaxy clusters from the extended HIFGLUGCS catalog~\cite{Reiprich:2001zv,Chen:2007sz}, which is based on X-ray observations.  
   The $J$-factors for the real-world catalogs use the concentration model from Ref.~\cite{Correa:2015dva} and assume the \emph{Planck} 2015 cosmology~\cite{Ade:2015xua}, which is very similar to that used in \texttt{DarkSky}.
   }
   \label{fig:GroupCats}
\end{figure}

\begin{figure*}[htb]
   \centering
   \includegraphics[width=0.45\textwidth]{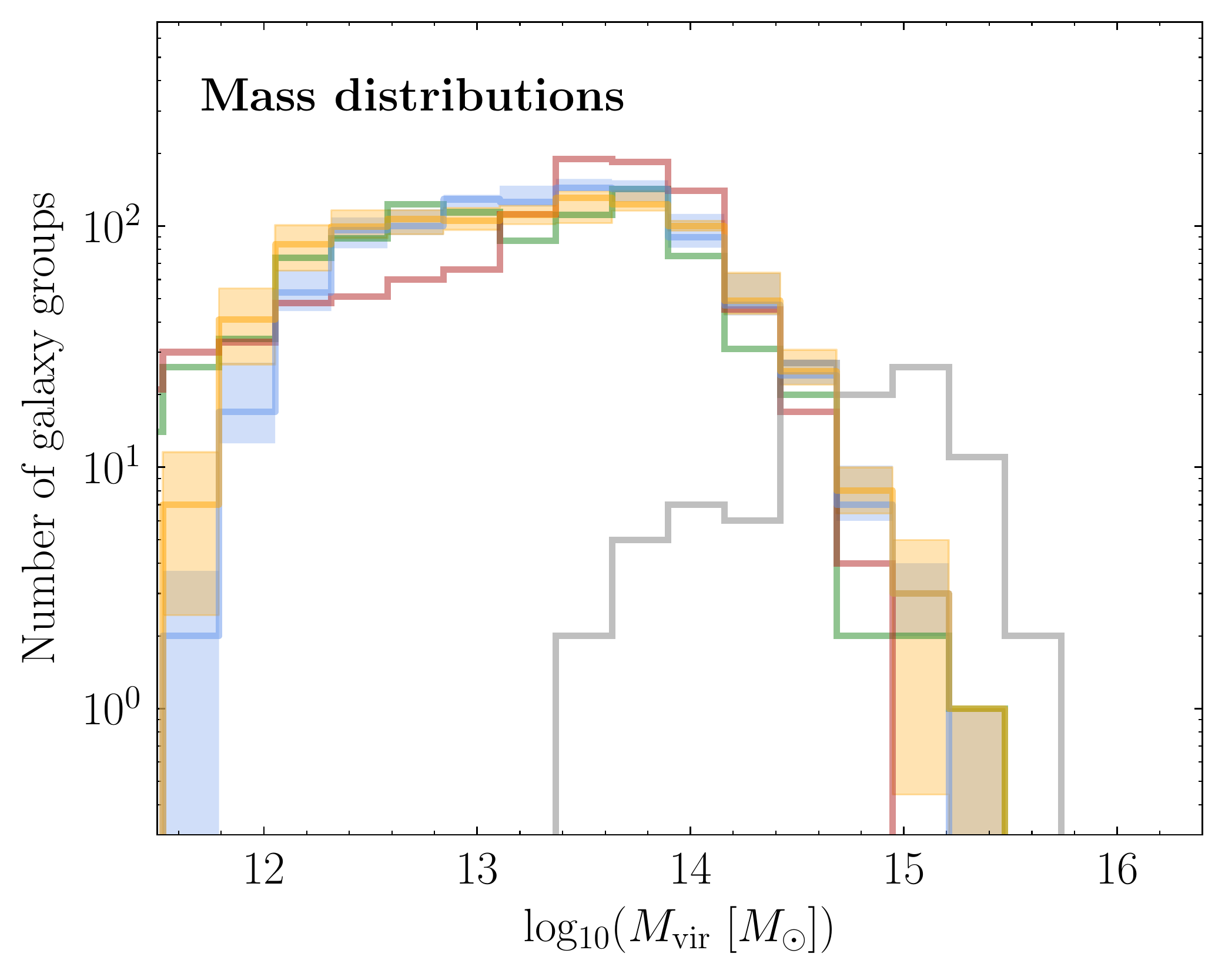}
     \includegraphics[width=0.45\textwidth]{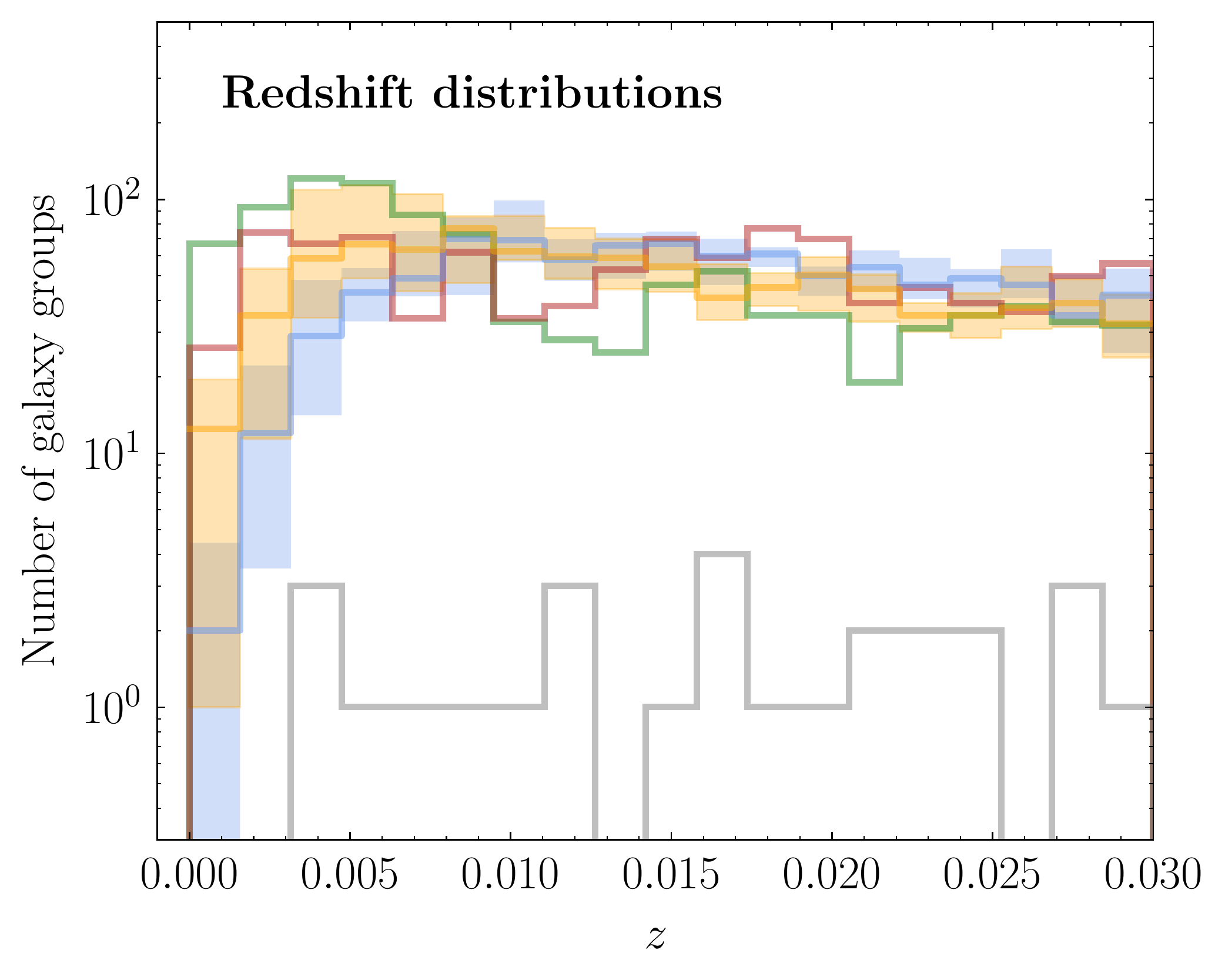}    
   \caption{Same as Fig.~\ref{fig:GroupCats}, except showing the mass function \textbf{(left)} and redshift distribution \textbf{(right)}.  Note that the redshift distribution for the HIFLUGCS clusters extends above $z\sim0.03$, even though these are not shown in the right panel.     }
   \label{fig:GroupCatsMZ}
\end{figure*}
  
It is interesting to study how the limit scales with the number of halos, $N_h$, included in the stacking procedure.  This result is shown in Fig.~\ref{fig:DSelephants} for $m_\chi = 10,$ $100,$ and $10^4$~GeV, for four different observer locations in the simulation box.  The dashed red line indicates the median 95\% confidence limit. The red bands are the 2.5, 16, 84 and 97.5 percentiles on the limit, obtained from 100 MC realizations of the mock data.  We observe that the limit typically improves continuously for the first $\sim$10 halos.  As more halos are included in the stacking, the gains diminish.  For some sky locations, the limit simply remains flat; for others we see some marginal improvements in the bounds.  These results are consistent, within uncertainties, between the DM masses and the different sky locations of the observer. 

We emphasize that the scaling on $N_h$ can be very different on applicaton to real data, because the distribution of $J$-factors in the random \texttt{DarkSky} locations is not representative of our own environment in the Local Group and also some halos can have residuals that are not related to DM but rather to mismodeling or real cosmic-ray--induced emission from the galaxy groups.  The former point is demonstrated in Fig.~\ref{fig:GroupCats}, where we histogram the top 1000 $J$-factors associated with the baseline \texttt{DarkSky} analysis (blue line/band).  For  comparison, we also show the distributions corresponding to 2MRS  galaxy group catalogs, specifically the Tully~\emph{et al.}~\cite{Tully:2015opa,2017ApJ...843...16K} (green line) and the Lu 
~\emph{et al.}~\cite{Lu:2016vmu} (red line) catalogs.  We see that the distribution of $J$-factors for the 2MRS catalogs is skewed towards higher values compared to that from \texttt{DarkSky}.  (Note that the cut-off at low $J$-factors is artificial and is simply a result of including 1000 halos for each catalog.)

The differences in the $J$-factor distributions can be traced to the redshift distribution of the galaxy groups, as illustrated in Fig.~\ref{fig:GroupCatsMZ}.  We see specifically that the mass function of the top 1000 \texttt{DarkSky} halos in each of the random sky locations sampled is roughly consistent with that observed in the 2MRS catalogs.  In contrast, the actual catalogs have more groups at lower $z$ than observed in the random \texttt{DarkSky} locations.  

While a random location in the \texttt{DarkSky} box does not resemble our own Local Group, we can try to find specific locations in the simulation box that do.  Therefore, we place the observer at ten random Milky Way--like halos in the simulation box, which have a mass $\sim10^{12}$~M$_\odot$.  More specifically, we select halos with mass $\log_{10}(M/\mathrm{M}_\odot) \in [11.8,12.2]$ and at least 100 Mpc\,$h^{-1}$ from the box boundaries.  The distribution of the top 1000 $J$-factors is indicated by the orange line/band in Fig.~\ref{fig:GroupCats}, while the corresponding mass and redshift distributions are shown in Fig.~\ref{fig:GroupCatsMZ}.  We see that the redshift---and, consequently, $J$-factor---distributions approach the observations, though the correspondence is still not exact.  A more thorough study could be done assessing the likelihood that an observer in \texttt{DarkSky} is located at a position that closely resembles the Local Group.  However, as our primary goal here is to outline an analysis procedure that we can apply to actual data, we simply conclude that our own local Universe appears to be a richer environment compared to a random location within the \texttt{DarkSky} simulation box, which bodes well for studying the actual \emph{Fermi} data.

\subsection{Signal Recovery Tests}
\label{sec:siginj}
 
\begin{figure*}[t]
   \centering
   \includegraphics[width=0.35\textwidth]{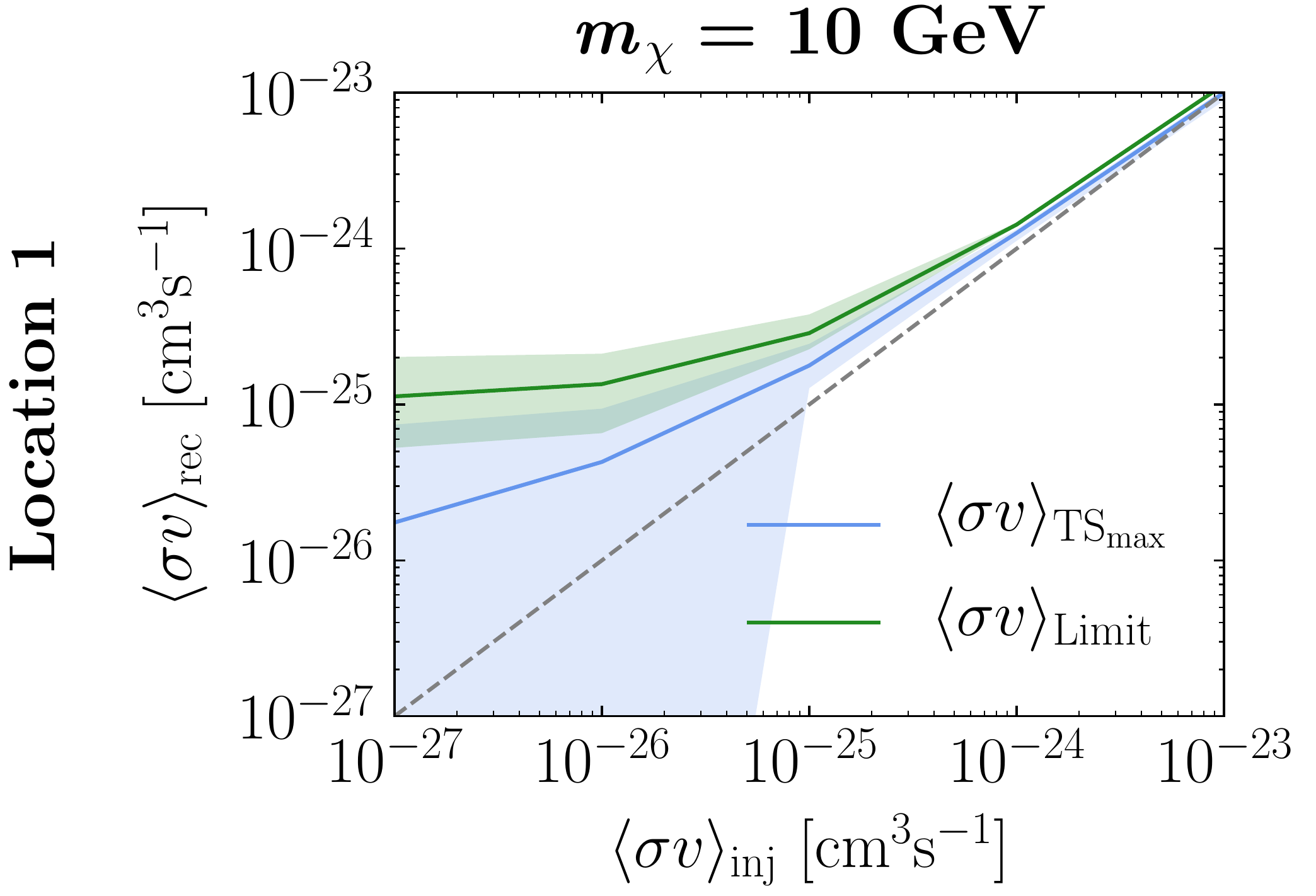} \hspace{0.01cm}
   \includegraphics[width=0.30\textwidth]{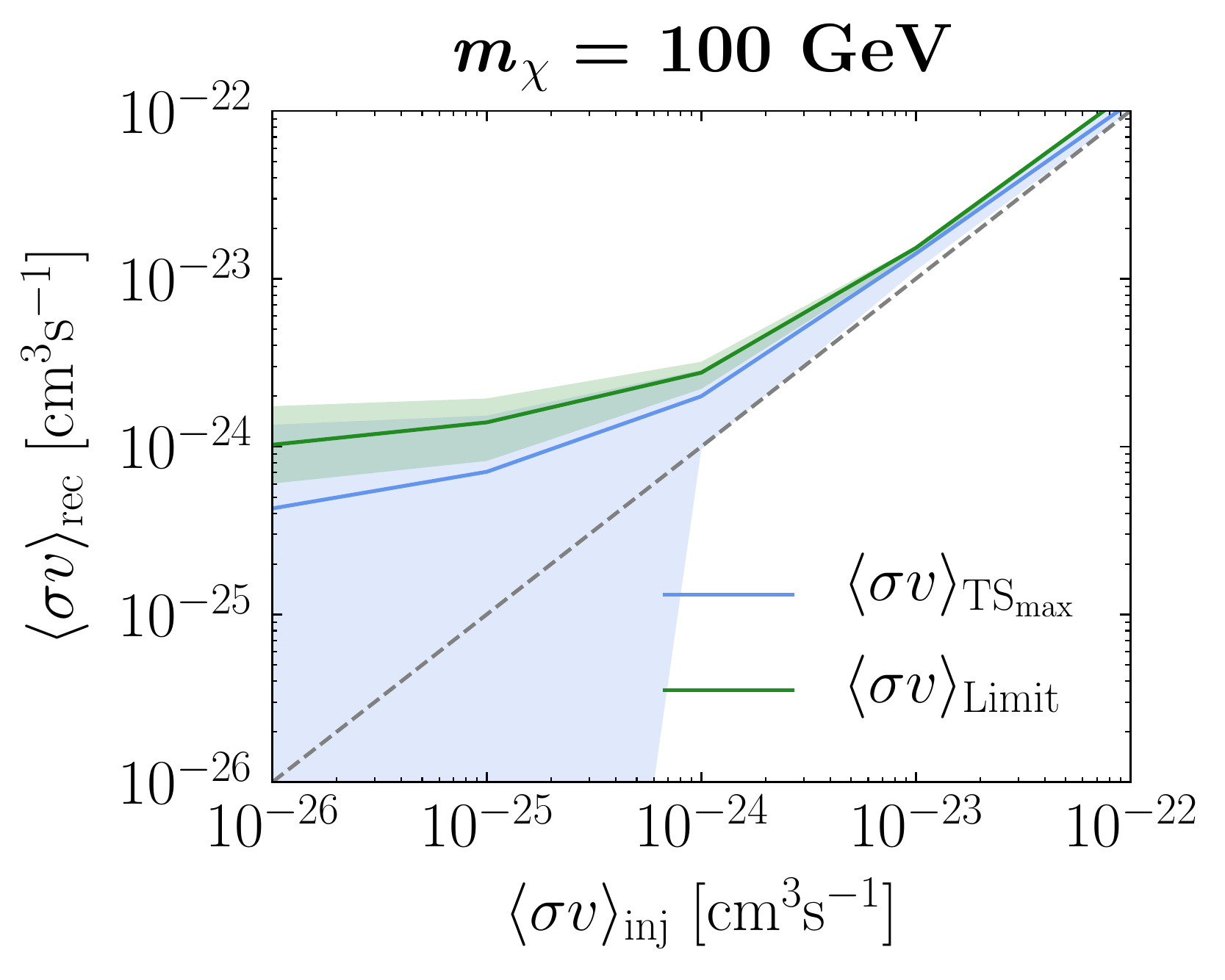} \hspace{0.01cm} 
   \includegraphics[width=0.30\textwidth]{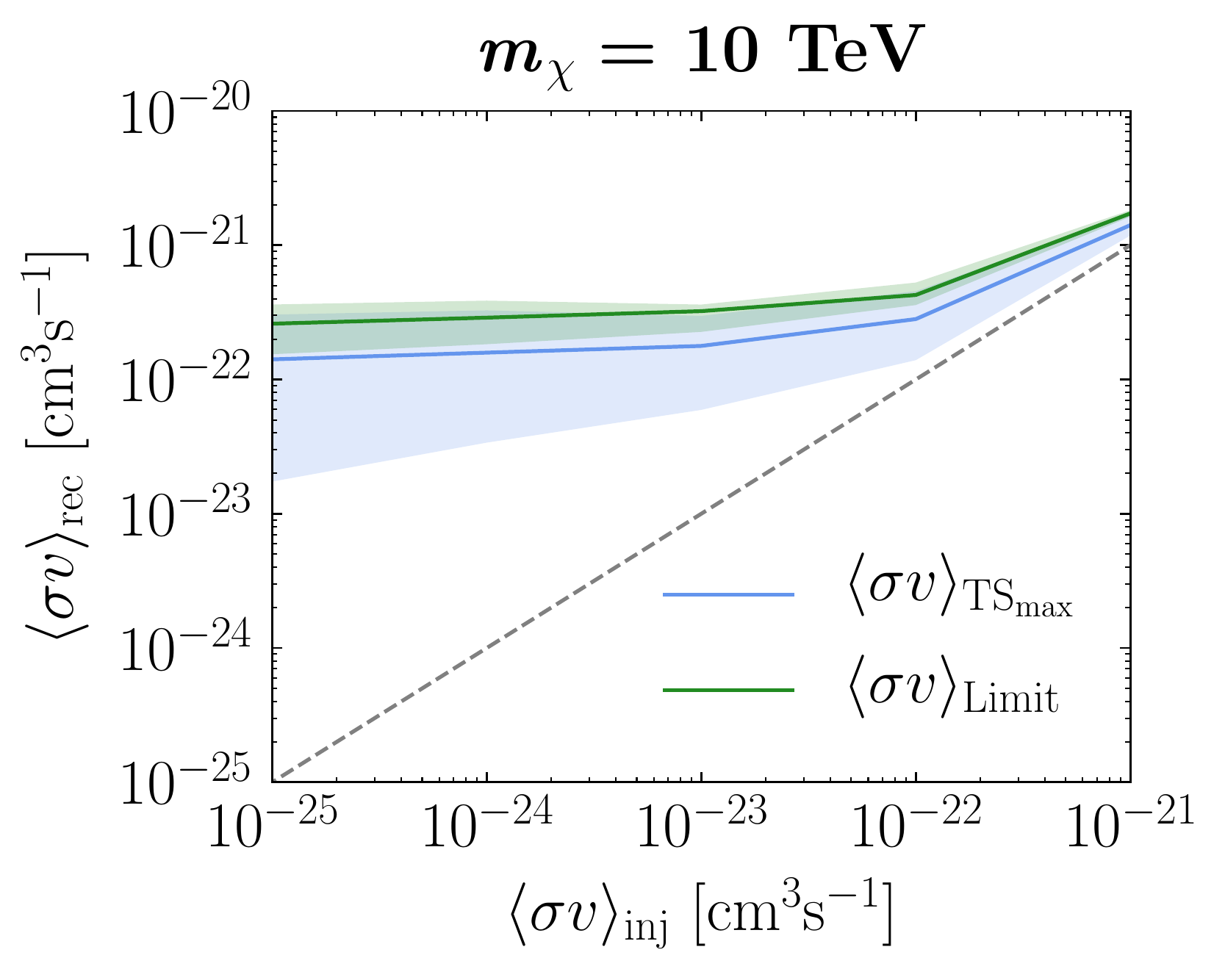} \\ \vspace{0.01cm}  
   \includegraphics[width=0.34\textwidth]{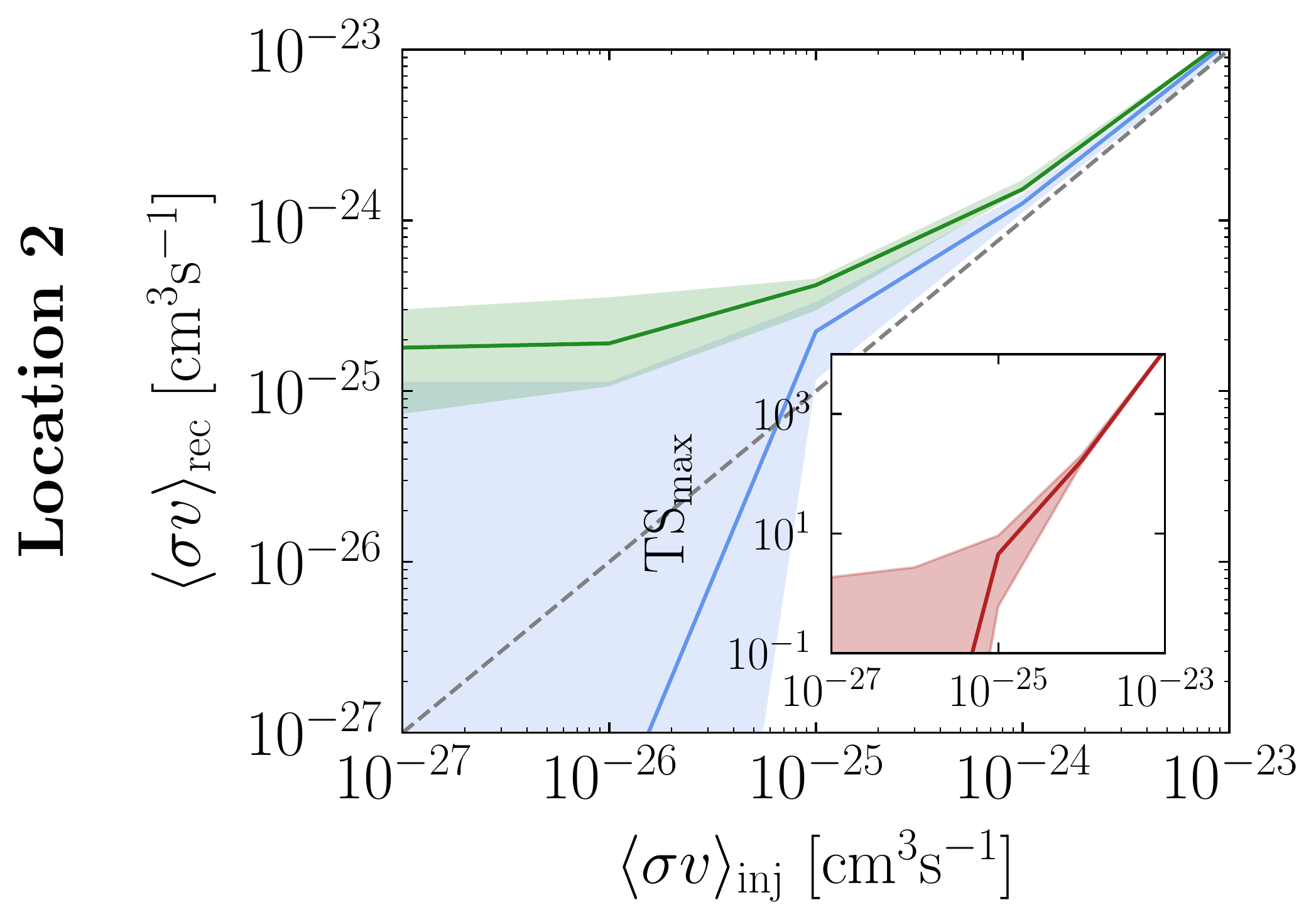} \hspace{0.01cm}
   \includegraphics[width=0.30\textwidth]{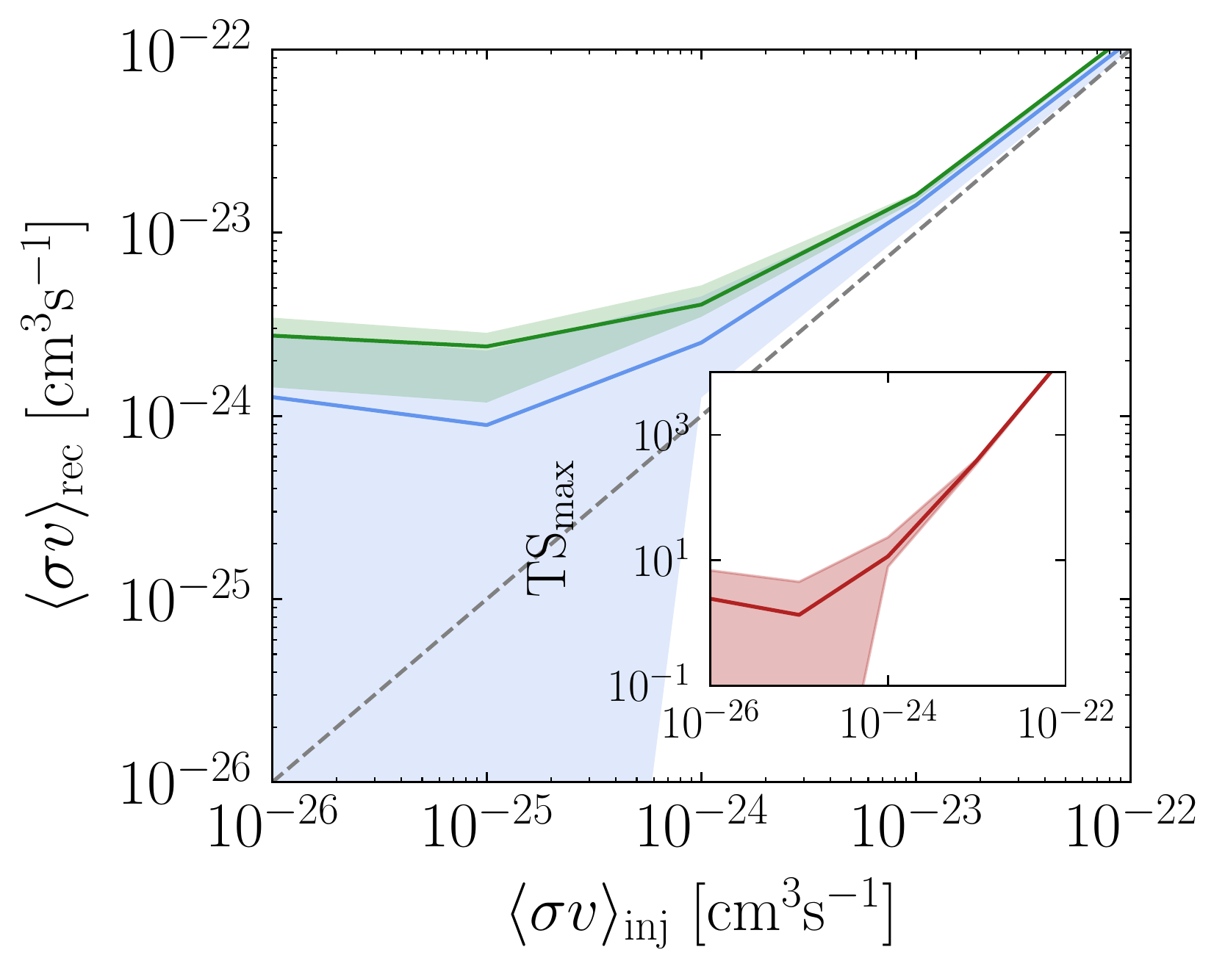} \hspace{0.01cm} 
   \includegraphics[width=0.30\textwidth]{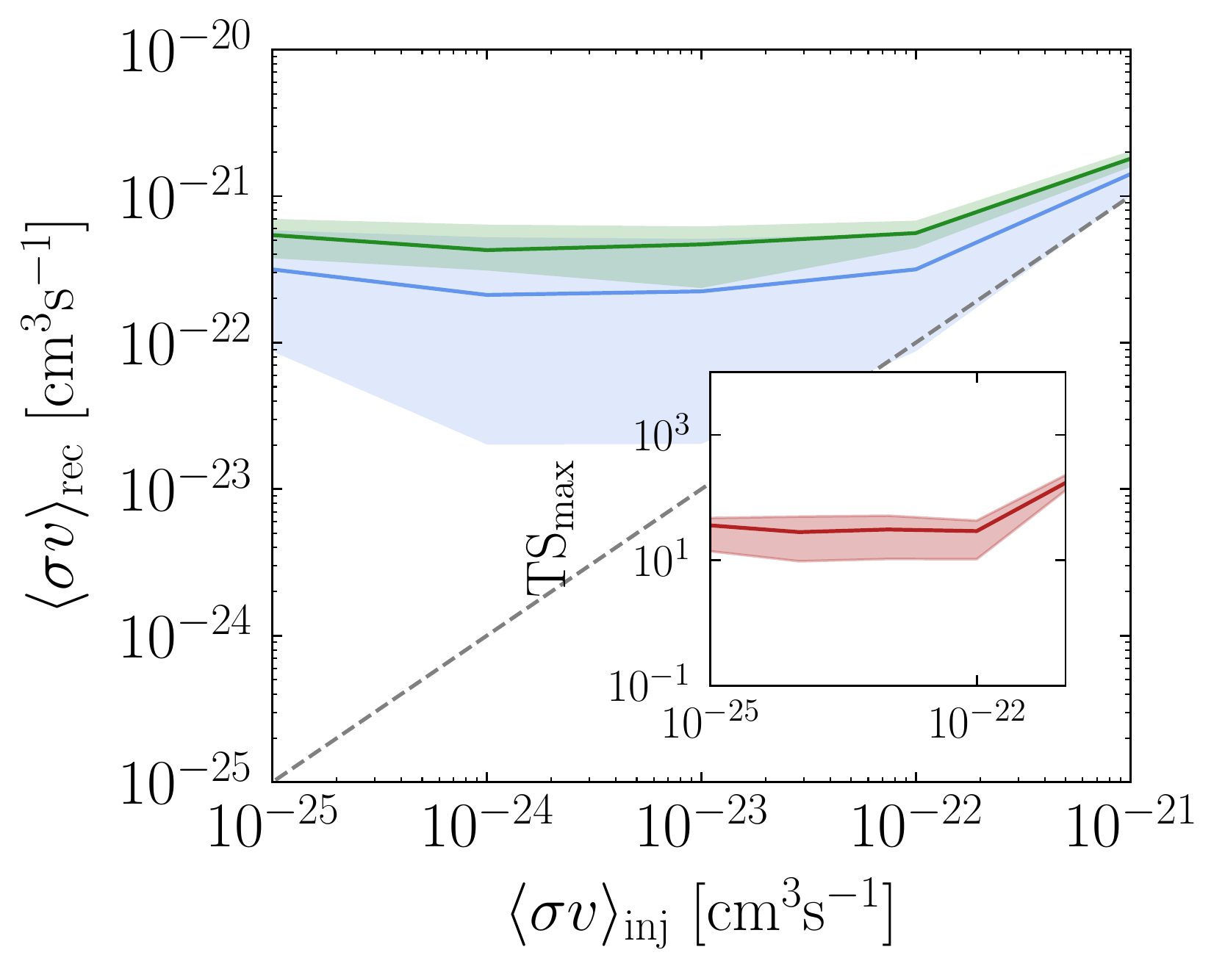}
   \caption{The results of injecting a DM signal with cross section $\langle \sigma v \rangle_\text{inj}$ into the mock data and studying the recovered cross section, $\langle \sigma v \rangle_\text{rec}$.  Each column shows the result for a different DM mass ($m_\chi = 10, 100, 10^4$~GeV), while each row shows a different observer location within the \texttt{DarkSky} simulation box.  The green line shows the 95\% confidence limit, with the green band denoting the 68\% containment region over twenty different Monte Carlo (MC) realizations of the mock data.  Critically, the limit never rules out an injected signal.  The blue line shows the median value of $\langle \sigma v\rangle_{\text{TS}_\text{max}}$, the cross section associated with the maximum test statistic (TS$_\text{max}$), over twenty MCs of the data. The blue band spans the median cross sections associated with TS$_\text{max} \pm 1$.  The maximum test statistic for each mass (with the band denoting the 68\% spread over MC realizations) is shown as an inset for each mass.  
   }
   \label{fig:DSinjsiglocs}
\end{figure*}

 It is critical that the halo selection criteria described in the previous section do not exclude a potential DM signal if one were present. To verify this, we have conducted extensive tests where we inject a signal into the mock data, pass it through the analysis pipeline and test our ability to accurately recover its cross section in the presence of the selection cuts.  Figure~\ref{fig:DSinjsiglocs} summarizes the results of the signal injection tests for two different observer locations in the \texttt{DarkSky} simulation box (top and bottom rows, respectively).  We inject a signal in the mock data that is associated with $b\bar{b}$ annihilation for three different masses ($m_\chi = 10, 100, 10^4$~GeV) that traces the DM annihilation flux map associated with \texttt{DarkSky}. The dashed line in each panel delineates where the injected cross section, $\langle \sigma v \rangle_\text{inj}$, matches the recovered cross section, $\langle \sigma v \rangle_\text{rec}$.  

The green line shows the 95\% one-sided limit on the cross section $\langle \sigma v \rangle_\text{rec}$ found using Eq.~\ref{eq:TSdef_darksky}, with a TS threshold corresponding to $\text{TS} = -2.71$.  The green band shows the 68\% containment region on this limit, constructed from twenty different MC realizations of the mock data set.  Importantly, the limit on $\langle \sigma v \rangle_\text{rec}$ roughly follows---but is slightly weaker than---the injected signal, up until the maximum sensitivity is reached and smaller cross sections can no longer be probed.  This behavior is generally consistent between the three DM masses tested and both sky locations.  We clearly see that the limit obtained by the statistical procedure never excludes an injected signal over the entire cross section range. 

Next, we consider the recovered cross section that is associated with the maximum test statistic, TS$_\text{max}$, in the total likelihood. The blue line in each panel of Fig.~\ref{fig:DSinjsiglocs} shows the median value of $\langle \sigma v\rangle_{\text{TS}_\text{max}}$ over 20 MCs of the mock data.  The blue band spans the median cross sections associated with $\text{TS}_\text{max}\pm1$.  The inset show the median and 68\% containment region for TS$_\text{max}$ as a function of the injected cross section.  The maximum test statistic is an indicator for the significance of the DM model and as such the $\langle \sigma v\rangle_{\text{TS}_\text{max}}$ distributions are only influenced by the data at high injected cross sections where TS$_\text{max}$ has begun to increase.  At lower injected cross sections, the distributions for $\langle \sigma v\rangle_{\text{TS}_\text{max}}$ are not meaningful.  

Two issues are visible in Fig.~\ref{fig:DSinjsiglocs}: (i) at high injected cross sections, the best-fit recovered cross sections are systematically around 1$\sigma$ too high, and (ii) at high DM masses and near-zero injected cross sections, the distribution of TS$_\text{max}$ deviates from the chi-square distribution (which can be seen based on the fact that the TS$_\text{max}$ flattens out with a non-zero median value).  The first issue stems from the way we model the $J$-factor contribution to the likelihood, while the second arises from the approximations we make to perform the profile likelihood in a computationally efficient manner.  

\section{Conclusions}
\label{sec:conclusions}

In this chapter, we introduced a procedure to build a full-sky map of extragalactic DM targets based on galaxy surveys and demonstrated this methodology using the \texttt{DarkSky} cosmological simulation.   Starting from the galaxies in the \texttt{DarkSky} catalog, we inferred the properties of their respective DM halos using the galaxy-halo connection.  In so doing, we identified the halos that are the brightest sources of extragalactic DM annihilation and which act as the best annihilation targets.  This procedure allows us to account for the fact that not all galaxy groups are expected to be bright DM emitters; the most massive, concentrated, and/or most nearby galaxies dominate the signals.  By building a map of extragalactic DM targets, we can focus our search for DM annihilation on the most relevant regions of sky.  This philosophy contrasts with that of cross-correlation studies, which treat all galaxies as equally good targets for DM.   

With a list of extragalactic DM halos in hand, as well as their inferred $J$-factors, we performed a stacked analysis to search for gamma-ray signatures of DM annihilation in mock data.  We described the likelihood procedure for the stacking analysis in detail.  There are two clear advantages to this approach over, say,  a full-sky template study.
First, focusing on smaller regions around each halo significantly reduces the sensitivity to mis-modeling of the foregrounds.  Second, uncertainties on the predicted DM annihilation flux can be straightforwardly included in the likelihood function.  In particular, we outlined how uncertainties in the $J$-factors, which arise from the determination of the virial mass and concentration, are marginalized over in the analysis. 

We presented limits on the DM annihilation cross section for mock data and, most importantly, demonstrated that the analysis procedure robustly recovers injected signals.  We found that the sensitivity improves by nearly two orders of magnitude when the structure of extragalactic DM emission on the sky is accounted for, rather than simply assuming an isotropic distribution.  Typically, the limit is dominated by the brightest $\mathcal{O}(10)$ halos in the stacking, though this varies depending on the location in the simulation box.  The $J$-factor distribution of nearby groups in our own Galaxy differs from the random locations sampled in the \texttt{DarkSky} box, which can change the number of halos that dominate the limit.  In actuality, one would want to continue adding halos to the analysis---ranked starting from the brightest $J$-factors---until the gains in the limit are observed to level off.

One advantage of using the \texttt{DarkSky} simulation in this initial study is that the truth information for all the halos is known.  We can therefore study how the DM limits improve when the virial mass and  concentration of the halos are known precisely.  For this ideal scenario, we find that that the limits improve by roughly 50\% over those obtained by marginalizing over uncertainties.  This suggests that a concrete way to improve the bounds on DM annihilation is to reduce the uncertainties on $M_\text{vir}$ and $c_\text{vir}$ for the brightest halos in the catalog.

The substructure boost factor remains one of the most difficult systematics to handle.  In this work, we use recent boost factor models that account for tidal stripping of subhalos.  This boost factor changes the limit by an $\mathcal{O}(1)$ factor, which is more conservative than other models sometimes used in extragalactic DM studies.  While the boost factor enhancement is fairly modest, it is still the dominant systematic uncertainty over the halo mass and  concentration.   

The analysis outlined in this chapter can be repeated on \emph{Fermi} data using published galaxy group catalogs.  In particular, the Tully \emph{et al.} catalogs~\cite{Tully:2015opa,2017ApJ...843...16K} and the Lu \emph{et al.} catalog~\cite{Lu:2016vmu} provide a map of the galaxy groups in the local Universe within $z\lesssim0.03$.  Both catalogs are based primarily on 2MRS, but use different clustering algorithms and halo mass determinations.  
Taken together, they provide a way to estimate the systematic uncertainties associated with the galaxy to halo mapping procedure.  Previous cluster studies on $\emph{Fermi}$ data~\cite{Ackermann:2010rg, Ando:2012vu,Ackermann:2013iaq,Anderson:2015dpc,Liang:2016pvm} used the extended HIghest X-ray FLUx Galaxy Cluster Sample (HIFLUGCS)~\cite{Reiprich:2001zv,Chen:2007sz}, which includes 106 of the brightest clusters observed in X-ray with the ROSAT all-sky survey.  These clusters cover redshifts from $0.0037 \lesssim z\lesssim 0.2$; the distribution of their $J$-factors, masses, and redshifts are shown in Fig.~\ref{fig:GroupCats} and~\ref{fig:GroupCatsMZ}.  In general, the 2MRS catalogs provide a larger number of groups that should be brighter in DM annihilation flux, so we expect a corresponding improvement in the sensitivity to annihilation signatures.  

The recent advancement of galaxy catalogs based on 2MRS and other nearby group catalogs allows us for the first time to map out the most important extragalactic DM targets in the nearby Universe.  This, in turn, enables us to perform a search that focuses on regions of sky where we expect the DM signals to be the brightest outside the Local Group.  We present the complete results of such an analysis, as applied to data, in the next chapter.

\sectionline

\chapter{A Search for Dark Matter Annihilation in Galaxy Groups}
\label{ch:groups_data}

This chapter is based on an edited version of \emph{A Search for Dark Matter Annihilation in Galaxy Groups},  \href{https://journals.aps.org/prl/abstract/10.1103/PhysRevLett.120.101101}{Phys.Rev.Lett. \textbf{120} (2018) 101101} \href{https://arxiv.org/abs/1708.09385}{[arXiv:1708.09385]} with Mariangela Lisanti, Nicholas Rodd and Benjamin Safdi~\cite{Lisanti:2017qlb}. The results of this chapter have been presented at the following conferences and workshops: \emph{TeV Particle Astrophysics (TeVPA) 2017} in Columbus, OH (August 2017), \emph{Dark Matter, Neutrinos and their Connection (DA$\nu$CO)} in Odense, Denmark (August 2017), \emph{Workshop on Statistical Challenges in the Search for Dark Matter} in Banff, Canada (February 2018) and \emph{Recontres de Blois 2018} in Blois, France (June 2018).

\section{Introduction}

\lettrine[lines=3]{W}{eakly}-interacting massive particles, which acquire their cosmological abundance through thermal freeze-out in the early Universe, are leading candidates for dark matter (DM).  Such particles can annihilate into Standard Model states in the late Universe, leading to striking gamma-ray signatures that can be detected with observatories such as the {\it Fermi} Large Area Telescope.  
Some of the strongest limits on the annihilation cross section have been set by searching for excess gamma-rays in the Milky Way's dwarf spheroidal satellite galaxies (dSphs)~\cite{Ackermann:2015zua,Fermi-LAT:2016uux}.  In this chapter, we present competitive constraints that are obtained using hundreds of galaxy groups within $z\lesssim0.03$. 

Chapter~\ref{ch:groups_sim} describes the procedure for utilizing  galaxy group catalogs in searches for extragalactic DM.  Previous attempts to search for DM outside the Local Group were broad in scope, but yielded weaker constraints than the dSph studies.  For example, limits on the annihilation rate were set by requiring that the DM-induced flux not overproduce the isotropic gamma-ray background~\cite{Ackermann:2015tah}.  These bounds could be improved by further resolving the contribution of sub-threshold point sources to the isotropic background~\cite{Zechlin:2016pme,Lisanti:2016jub}, or by  
looking at the auto-correlation spectrum~\cite{Ackermann:2012uf, Ackermann:2012uf,Ando:2006cr,Ando:2013ff}.  A separate approach involves cross-correlating~\cite{Xia:2011ax,Ando:2014aoa,Ando:2013xwa,Xia:2015wka,Regis:2015zka,Cuoco:2015rfa,Ando:2016ang} the {\it Fermi} data with galaxy-count maps constructed from, \emph{e.g.}, the Two Micron All-Sky Survey (2MASS)~\cite{Jarrett:2000me,Bilicki:2013sza}.  A positive cross-correlation was detected with 2MASS galaxy counts~\cite{Xia:2015wka}, which could arise from annihilating DM with mass $\sim$$10$--$100$~GeV and a near-thermal annihilation rate~\cite{Regis:2015zka}.  However, other source classes, such as misaligned Active Galactic Nuclei, could also explain the signal~\cite{Cuoco:2015rfa}.

\begin{table*}[htb]
\resizebox{\textwidth}{!}{%
\begin{tabular}{C{3.0cm}C{2.1cm}C{1.8cm}C{1.8cm}C{1.8cm}C{1.5cm}C{1.6cm}C{1.6cm}C{1.6cm}C{1.6m}}
\toprule
Name &   $\log_{10} J$  &  $\log_{10} M_\text{vir}$ &          $z \times 10^{3}$&        $\ell$ &        $b$ &  $\log_{10} c_\text{vir}$ &  $\theta_\text{s}$  &  $b_\text{sh}$   \\
 & {[GeV$^2$\,cm$^{-5}$\,sr]}& [$M_\odot$] &  & [deg] & [deg] & & [deg] &\\
\midrule

            NGC4472/Virgo &  19.11$\pm$0.35 &  14.6$\pm$0.14 &   3.58 &  283.94 &  74.52 &  0.80$\pm$0.18 &     1.15 &  4.53 \\
                  NGC0253 &  18.76$\pm$0.37 &  12.7$\pm$0.12 &   0.79 &   98.24 & -87.89 &  1.00$\pm$0.17 &     0.77 &  2.90 \\
                  NGC3031 &  18.58$\pm$0.36 &  12.6$\pm$0.12 &   0.83 &  141.88 &  40.87 &  1.02$\pm$0.17 &     0.64 &  2.76 \\
        NGC4696/Cen. &  18.33$\pm$0.35 &  14.6$\pm$0.14 &   8.44 &  302.22 &  21.65 &  0.80$\pm$0.18 &     0.47 &  4.50 \\
                  NGC1399 &  18.30$\pm$0.37 &  13.8$\pm$0.13 &   4.11 &  236.62 & -53.88 &  0.89$\pm$0.17 &     0.45 &  3.87 \\
\bottomrule
\end{tabular}}
\caption{The top five halos included in the analysis, as ranked by inferred $J$-factor, including the boost factor.  For each group, we show the brightest central galaxy and the common name, if one exists, as well as the virial mass, cosmological redshift, Galactic longitude $\ell$, Galactic latitude $b$, inferred virial concentration~\cite{Correa:2015dva}, angular extent, and boost factor~\cite{Bartels:2015uba}.  The angular extent is defined as $\theta_\text{s} \equiv \tan^{-1} (r_\text{s} / d_c[z])$, where $d_c[z]$ is the comoving distance and $r_\text{s}$ is the NFW scale radius.  A complete table of the galaxy groups used in this analysis, as well as their associated properties, are provided at \url{https://github.com/bsafdi/DMCat}.
}
\label{Jtab}
\end{table*}

An alternative to studying the full-sky imprint of extragalactic DM annihilation is to use individual galaxy clusters~\cite{Ackermann:2010rg, Ando:2012vu,Ackermann:2013iaq,Ackermann:2015fdi,Anderson:2015dpc,Rephaeli:2015nca,2016A&A...589A..33A,Liang:2016pvm,Adams:2016alz,Huang:2011xr}. Previous analyses along these lines have looked at a small number of $\sim$$10^{14}$--$10^{15}$~M$_\odot$ clusters whose properties were inferred from X-ray measurements~\cite{Reiprich:2001zv,Chen:2007sz}. Like the dSph searches, the cluster studies have the advantage that the expected signal is localized in the sky, which reduces the systematic uncertainties associated with modeling the foregrounds and unresolved extragalactic sources.  As we will show, however, the sensitivity to DM annihilation is enhanced---and is more robust---when a larger number of targets are included compared to previous studies.

 Our work aims to combine the best attributes of the cross-correlation and cluster studies to improve the search for extragalactic DM annihilation.  We use the galaxy group catalogs in Refs.~\cite{Tully:2015opa} and~\cite{2017ApJ...843...16K} (hereby T15 and T17, respectively), which contain accurate mass estimates for halos with mass greater than $\sim$$10^{12}$~M$_\odot$ and $z \lesssim 0.03$, to systematically determine the galaxy groups that are expected to yield the best limits on the annihilation rate.  The T15 catalog provides reliable redshift estimates in the range $0.01 \lesssim z \lesssim 0.03$, while the T17 catalog provides measured distances for nearby galaxies, $z \lesssim 0.01$, based on Ref.~\cite{Tully:2016ppz}. The T15 catalog was previously used for a gamma-ray line search~\cite{Adams:2016alz}, but our focus here is on the broader, and more challenging, class of continuum signatures.  We search for gamma-ray flux from these galaxy groups and interpret the null results as bounds on the annihilation cross section.   
 
 \section{Galaxy Group Selection}

The observed gamma-ray flux from DM annihilation in an extragalactic halo is proportional to both the particle physics properties of the DM, as well as its astrophysical distribution:
\es{particle}{
\frac{d\Phi}{dE_{\gamma}} &= \left.  J \, \times \frac{\langle\sigma v\rangle}{8 \pi m_{\chi}^{2}} \, \, \sum_i \text{Br}_{i}\, \frac{dN_{i}}{dE'_{\gamma}} \right|_{E_{\gamma}' = (1 +z) E_{\gamma}}   \,,
}
with units of $[{\rm counts} \,\,{\rm cm}^{-2} \, {\rm s}^{-1} \, {\rm GeV}^{-1}]$.  Here, $E_\gamma$ is the gamma-ray energy, $\langle \sigma v \rangle$ is the annihilation cross section, $m_\chi$ is the DM mass, $\text{Br}_{i}$ is the branching fraction to the $i^\text{th}$ annihilation channel, and $z$ is the cosmological redshift.  The energy spectrum for each channel is described the function $dN_{i}/dE_{\gamma}$, which is modeled using PPPC4DMID~\cite{Cirelli:2010xx}.  The $J$-factor that appears in~Eq.~\ref{particle} encodes the astrophysical properties of the halo.  It is proportional to the line-of-sight integral of the squared DM density distribution, $\rho_\text{DM}$, and is written in full as 
\begin{equation}
J = \left(1+b_\text{sh}[M_\text{vir}] \right)  \int ds\,d \Omega \,\rho^{2}_\text{DM}(s,\Omega) \, ,
\label{eq:Jfactor}
\end{equation}
where $b_\text{sh}[M_\text{vir}]$ is the boost factor, which accounts for the enhancement due to substructure.  For an extragalactic halo, where the comoving distance $d_c[z]$ is much greater than the virial radius $r_\text{vir}$, the integral in Eq.~\ref{eq:Jfactor} scales as $M_{\rm vir} c_{\rm vir}^3\rho_c/d_c^2[z]$ for the Navarro-Frenk-White (NFW) density profile~\cite{Navarro:1996gj}.  Here, $M_\text{vir}$ is the virial mass, $\rho_c$ is the critical density, and $c_\text{vir}=r_\text{vir}/r_s$ is the virial concentration, with $r_s$ the scale radius.  We infer $c_\text{vir}$ using the concentration-mass relation from Ref.~\cite{Correa:2015dva}, which we update with the \emph{Planck} 2015 cosmology~\cite{Ade:2015xua}.  
For a given mass and redshift, the concentration is modeled as a log-normal distribution with mean given by the concentration-mass relation.  We estimate the dispersion by matching to that observed in the \texttt{DarkSky-400} simulation for an equivalent $M_\text{vir}$~\cite{Lehmann:2015ioa}.  Typical dispersions range from $\sim$$0.14$--$0.19$ over the halo masses considered. 

The halo mass and redshift also determine the boost factor enhancement that arises from annihilation in DM substructure.  Accurately modeling the boost factor is challenging as it involves extrapolating the halo-mass function and concentration to masses smaller than can be resolved with current simulations.  Some previous analyses of extragalactic DM annihilation have estimated boost factors $\sim$$10^2$--$10^3$ for cluster-size halos (see, for example, Ref.~\cite{Gao:2011rf}) based on phenomenological extrapolations of the subhalo mass and concentration relations.  However, more recent studies indicate that the concentration-mass relation likely flattens at low masses~\cite{Anderhalden:2013wd,Ludlow:2013vxa,Correa:2015dva}, suppressing the enhancement. We use the model of Ref.~\cite{Bartels:2015uba}---specifically, the ``self-consistent" model with $M_\text{min} = 10^{-6}$~M$_\odot$---which accounts for tidal stripping of bound subhalos and yields a modest boost $\sim$$5$ for $\sim$$10^{15}$~M$_\odot$ halos. Additionally, we model the boost factor as a multiplicative enhancement to the rate in our main analysis, though we consider the effect of possible spatial extension from the subhalo annihilation in App.~\ref{supp:clusters}. In particular, we find that modeling the boost component of the signal as tracing a subhalo population distributed as $\rho_\text{NFW}$ rather than $\rho^{2}_\text{NFW}$ degrades the upper limits obtained by almost an order of magnitude at higher masses $m_\chi \gtrsim 500$ GeV while strengthening the limit by a small $\mathcal O(1)$ factor at lower masses $m_\chi \lesssim 200$ GeV. This is arguably a more plausible scenario, since the spatial distribution of subhalos  is expected to follow the overall shape of the dark matter halo rather than the annihilation profile (modulo baryonic effects).

The halo masses and redshifts are taken from the galaxy group catalog T15~\cite{Tully:2015opa}, which is based on the 2MASS Redshift Survey (2MRS)~\cite{Crook:2006sw}, and T17~\cite{2017ApJ...843...16K}, which compiles an inventory of nearby galaxies and distances from several sources.  The catalogs provide group associations for these galaxies as well as mass estimates and uncertainties of the host halos, constructed from a luminosity-to-mass relation. The mass distribution is assumed to follow a log-normal distribution with uncertainty fixed at 1\% in log-space (see Ch.~\ref{ch:groups_sim}), which translates to typical absolute uncertainties of 25-40\%.\footnote{To translate, approximately, between log- and linear-space uncertainties for the mass, we may write $x = \log_{10} M_\text{vir}$, which implies that the linear-space fractional uncertainties are $\delta M_\text{vir} / M_\text{vir} \sim (\delta x / x) \log M_\text{vir}$. } This is conservative compared to the 20\% uncertainty estimate given in T15 due to their inference procedure. The halo centers are assumed to coincide with the locations of the brightest galaxy in the group.  We infer the $J$-factor using Eq.~\ref{eq:Jfactor} and calculate its uncertainty by propagating the errors on $M_\text{vir}$ and $c_\text{vir}$, which we take to be uncorrelated.  Note that we neglect the distance uncertainties, which are expected to be $\sim$5\%~\cite{Tully:2016ppz,2017ApJ...843...16K}, as they are subdominant compared to the uncertainties on mass and concentration.  We compile an initial list of nearby targets using the T17 catalog, supplementing these with the T15 catalog.  We exclude from T15 all groups with Local Sheet velocity $V_\text{LS} < 3000$~km s$^{-1}$ ($z \lesssim 0.01$) and $V_\text{LS} > 10,000$~km s$^{-1}$ ($z \gtrsim 0.03$), the former because of peculiar velocity contamination and the latter because of large uncertainties in halo mass estimation due to less luminous satellites.  When groups overlap between the two catalogs, we preferentially choose distance and mass measurements from T17.

The galaxy groups are ranked by their inferred $J$-factors, excluding any groups that lie within $|b| \leq 20^\circ$ to mitigate contamination from Galactic diffuse emission.  We require that halos do not overlap to within $2^\circ$ of each other, which is approximately the scale radius of the largest halos.  The exclusion procedure is applied sequentially starting with a halo list ranked by $J$-factor.  We manually exclude Andromeda, the brightest halo in the catalog, because its large angular size is not ideally suited to our analysis pipeline and requires careful individual study~\cite{Ackermann:2017nya}.  
As discussed later in this chapter, halos are also excluded if they show large residuals that are inconsistent with DM annihilation in the other groups in the sample.  Starting with the top 1000 halos, we end up with 495 halos that pass all these requirements.  Of the excluded halos, 276 are removed because they fall too close to the Galactic plane, 134 are removed by the $2^\circ$ proximity requirement, and 95 are removed because of the cut on large residuals. Other than the manual exclusion of Andromeda, these selection criteria are identical to those introduced and tested in Ch.~\ref{ch:groups_sim} in the context of simulations.

Table~\ref{Jtab} lists the top five galaxy groups included in the analysis, labeled by their central galaxy or common name, if one exists.  We provide the inferred $J$-factor including the boost factor, the halo mass, redshift, position in Galactic coordinates, inferred concentration, and boost factor.  Additionally, we show $\theta_\text{s} \equiv \tan^{-1} (r_\text{s} / d_c[z])$ to indicate the spatial extension of the halo.  We find that $\theta_\text{s}$ is typically between the 68\% and 95\% containment radius for emission associated with annihilation in the halos, without accounting for spread from the point-spread function (PSF).  For reference, Andromeda has $\theta_\text{s} \sim 2.57^\circ$. A complete list of the analyzed galaxy groups is provided as Supplementary Data at \url{https://github.com/bsafdi/DMCat}.

\section{Data Analysis}

We analyze 413 weeks of Pass 8 {\it Fermi} data in the UltracleanVeto event class, from August 4, 2008 through July 7, 2016.  The data is binned in 26 logarithmically-spaced energy bins between 502~MeV and 251~GeV and spatially with a HEALPix pixelation~\cite{Gorski:2004by} with \texttt{nside}=128.\footnote{Our energy binning is constructed by taking 40 log-spaced bins between 200~MeV and 2~TeV and then removing the lowest four and highest ten bins, for reasons discussed in Ch.~\ref{ch:groups_sim} }  The recommended set of quality cuts are applied to the data corresponding to zenith angle less than $90^\circ$, $\texttt{LAT\_CONFIG}=1$, and $\texttt{DATA\_QUAL}>0$.\footnote{\url{https://fermi.gsfc.nasa.gov/ssc/data/analysis/documentation/Cicerone/Cicerone_Data_Exploration/Data_preparation.html}.}  We also mask known large-scale structures (see Ch.~\ref{ch:groups_sim}).

The template analysis that we perform using \texttt{NPTFit}~\cite{Mishra-Sharma:2016gis} is similar to that of previous dSph studies~\cite{Ackermann:2015zua,Fermi-LAT:2016uux}  and is detailed in Ch.~\ref{ch:groups_sim}.  We summarize the relevant points here.  Each region-of-interest (ROI), defined as the $10^\circ$ area surrounding each halo center, has its own likelihood.  In each energy bin, this  likelihood is the product, over all pixels, of the Poisson probability for the observed photon counts per pixel.  This probability depends on the mean expected counts per pixel, which depends on contributions from known astrophysical emission as well as a potential DM signal. Note that the likelihood is also multiplied by the appropriate log-normal distribution for $J$, which we treat as a single nuisance parameter for each halo and account for through the profile likelihood method.  

\begin{figure*}[t]
\centering
\includegraphics[width=0.8\textwidth]{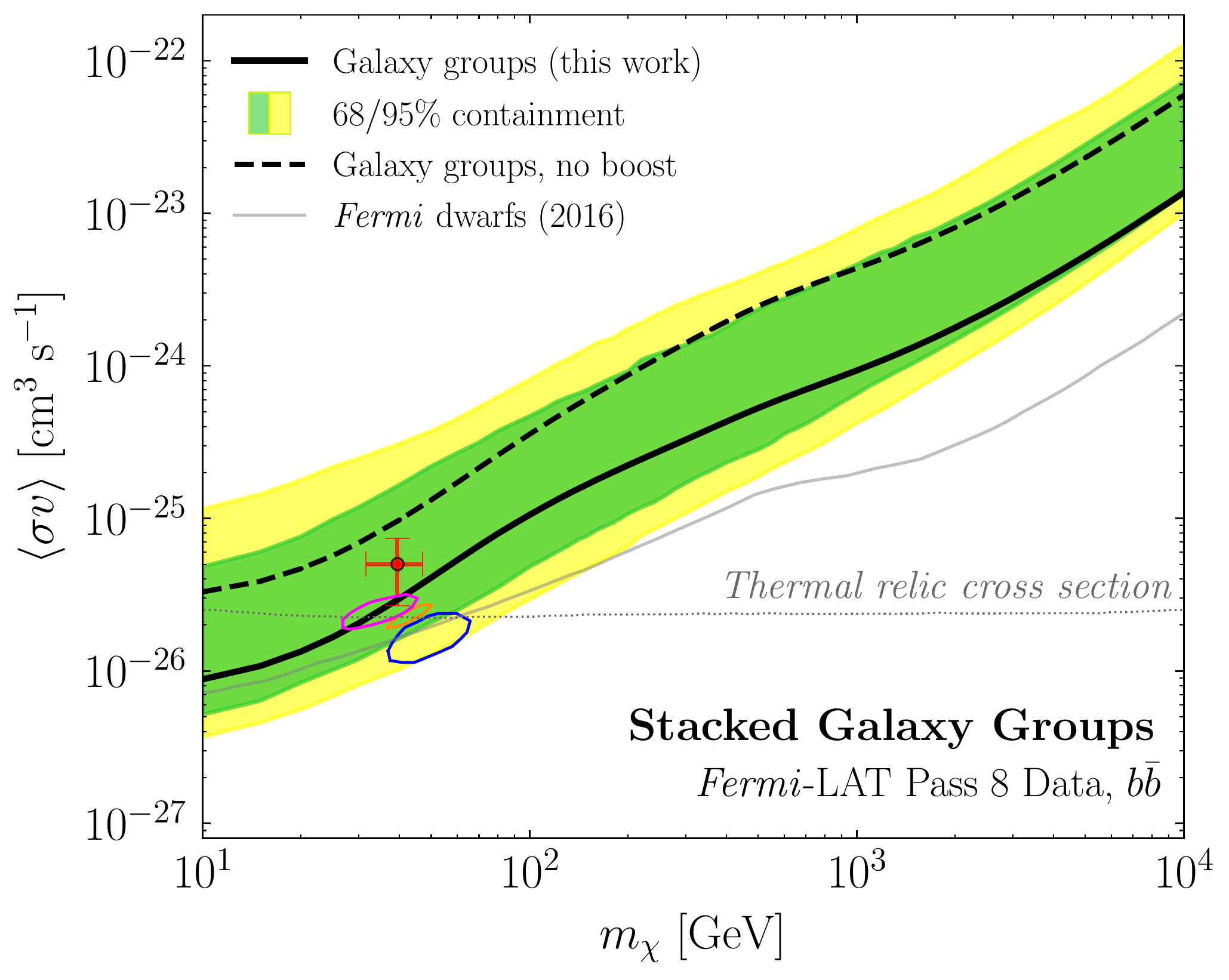} \hspace{4mm}
\caption{The solid black line shows the 95\% confidence limit on the DM annihilation cross section, $\langle \sigma v \rangle$, as a function of the DM mass, $m_\chi$, for the $b \bar b$ final state, assuming the fiducial boost factor~\cite{Bartels:2015uba}. The containment regions are computed by performing the data analysis multiple times for random sky locations of the halos.  For comparison, the dashed black line shows the limit assuming no boost factor.  The {\it Fermi} dwarf limit is also shown, as well as the $2$$\sigma$ regions where DM may contribute to the Galactic Center Excess (see text for details).  The thermal relic cross section for a generic weakly interacting massive particle~\cite{Steigman:2012nb} is indicated by the thin dotted line. Variations on the analysis (including results for final states other than $b \bar b$) and effects of systematics are presented in App.~\ref{supp:clusters}.}
\label{fig:bounds1}
\end{figure*}

\begin{figure*}[t]
\centering
\includegraphics[width=0.8\textwidth]{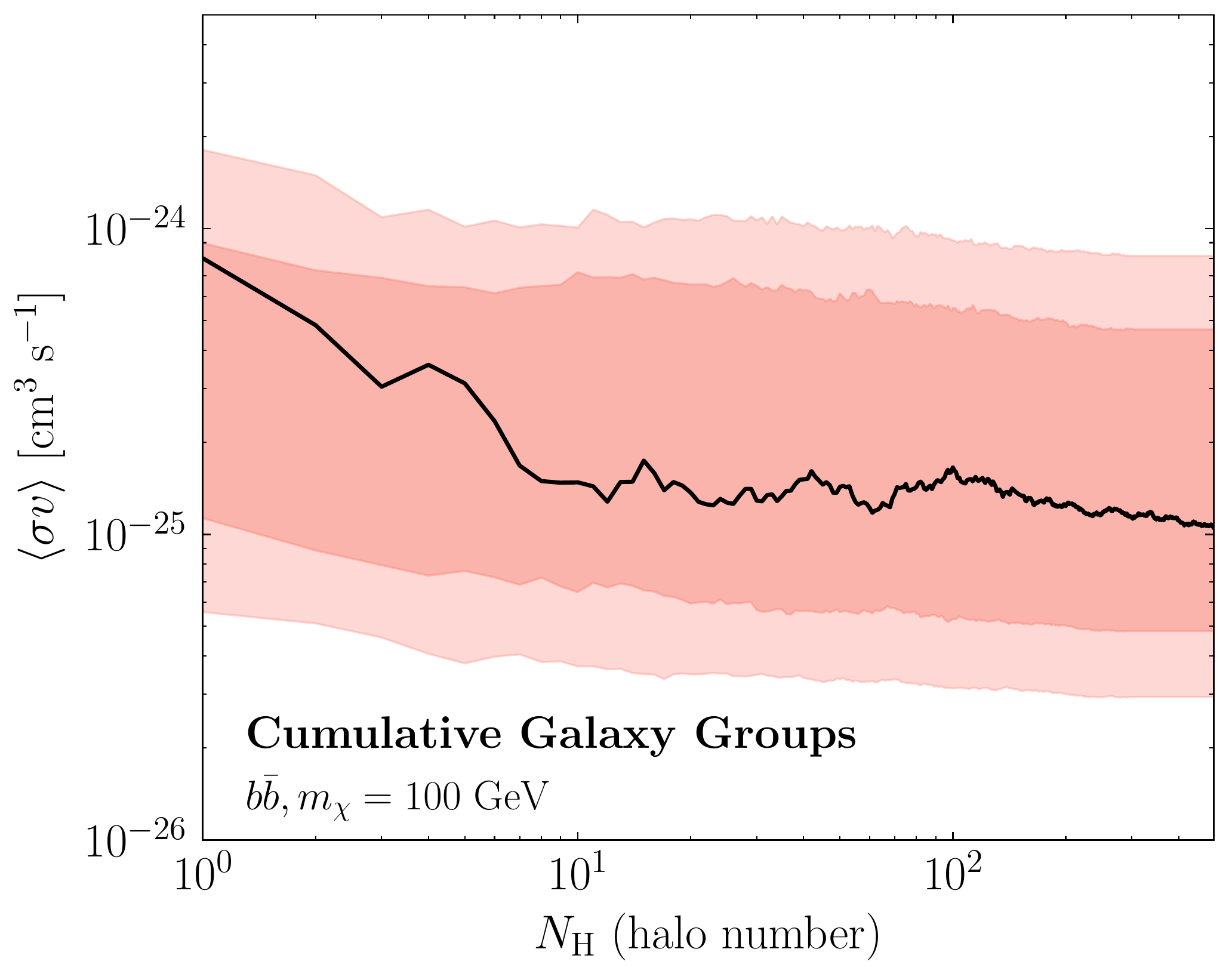}
\caption{The change in the limit for $m_\chi = 100$~GeV as a function of the number of halos that are included in the analysis, which are ranked in order of largest $J$-factor.  The result is compared to the expectation from random sky locations; the 68 and 95\% expectations from 200 random sky locations are indicated by the red bands.}
\label{fig:bounds2}
\end{figure*}

To model the expected counts per pixel, we include several templates in the analysis that trace the emission associated with: (i) the projected NFW-squared profile modeling the putative DM signal, (ii) the diffuse background, as described by the  {\it Fermi} \texttt{gll\_iem\_v06 (p8r2)} model, (iii) isotropic emission, (iv) the {\it Fermi} bubbles~\cite{Su:2010qj}, (v) 3FGL sources within $10^\circ$ to $18^\circ$ of the halo center, floated together after fixing their individual fluxes to the values predicted by the 3FGL catalog~\cite{Acero:2015hja}, and (vi) all individual 3FGL point sources within $10^{\circ}$ of the halo center.  Note that we do not model the contributions from annihilation in the smooth Milky Way halo because the brightest groups have peak flux significantly (approximately an order of magnitude for the groups in Tab.~\ref{Jtab}) over the foreground emission from Galactic annihilation and because we expect Galactic annihilation to be subsumed by the isotropic component.   

We assume that the best-fit normalizations (\emph{i.e.}, profiled values) of the astrophysical components, which we treat as nuisance parameters, do not vary appreciably with DM template normalization. This allows us to obtain the likelihood profile in a given ROI and energy bin by profiling over them in the presence of the DM template, then fixing the normalizations of the background components to the best-fit values and scanning over the DM intensity. We then obtain the total likelihood by taking the product of the individual likelihoods from each energy bin. In order to avoid degeneracies at low energies due to the large PSF, we only include the DM template when obtaining the best-fit background normalizations at energies above $\sim$$1$~GeV. At the end of this procedure, the likelihood is only a function of the DM template intensity, which can then be mapped onto a mass and cross section for a given annihilation channel. We emphasize that the assumptions described above have been thoroughly vetted in Ch.~\ref{ch:groups_sim}, where we show that this procedure is robust in the presence of a potential signal.

The final step of the analysis involves stacking the likelihoods from each ROI. The stacked log-likelihood, $\log \mathcal{L}$, is simply the sum of the log-likelihoods for each ROI.  It follows that the test statistic for data $d$ is defined as
 \begin{equation}\begin{aligned}
{\rm TS}(\mathcal{M}, \langle\sigma v\rangle, m_\chi) \equiv 2 &\left[ \log \mathcal{L}(d | \mathcal{M}, \langle\sigma v\rangle, m_\chi ) \right.\\
&\left.- \log \mathcal{L}(d | \mathcal{M}, \widehat{\langle\sigma v\rangle}, m_\chi ) \right]\,,
\label{eq:TSdef}
\end{aligned}\end{equation}
where $\widehat{\langle\sigma v\rangle}$ is the cross section that maximizes the likelihood for DM model $\mathcal{M}$.    The 95\% upper limit on the annihilation cross section is given by the value of $\langle\sigma v\rangle > \widehat{\langle \sigma v\rangle}$ where $\text{TS}=-2.71$.

Galaxy groups are expected to emit gamma-rays from standard cosmic-ray processes.  Using group catalogs to study gamma-ray emission from cosmic rays in these objects is an interesting study in its own right (see, {\it e.g.}, Ref.~\cite{Jeltema:2008vu,Huber:2013cia,Ackermann:2015fdi,Rephaeli:2015nca}), which we leave to future work.  For the purpose of the present analysis, however, we would like a way to remove groups with large residuals, likely arising from standard astrophysical processes in the clusters, to maintain maximum sensitivity to DM annihilation.  This requires care, however, as we must guarantee  that the procedure for removing halos does not remove a real signal, if one were present.  

We adopt the following algorithm to remove halos with large residuals that are inconsistent with DM annihilation in the other groups in the sample. A group is excluded if it meets two conditions. First, to ensure it is a statistically significant excess, we require twice the difference between the maximum log likelihood and the log likelihood with $\langle \sigma v \rangle = 0$ to be greater than 9 at any DM mass. This selects sources with large residuals at a given DM mass.  Second, the residuals must be strongly inconsistent with limits set by other galaxy groups. Specifically, the halo  must satisfy $\langle\sigma v\rangle_\text{best} > 10 \times \langle\sigma v\rangle^*_\text{lim}$, where $\langle\sigma v\rangle_\text{best}$ is the halo's best-fit cross section at \emph{any} mass and $\langle\sigma v\rangle^*_\text{lim}$ is the strongest limit out of all halos at the specified $m_\chi$. These conditions are designed to exclude galaxy groups where the gamma-ray emission is inconsistent with a DM origin.  This prescription has been extensively tested on mock data and, crucially, does not exclude injected signals (see Ch.~\ref{ch:groups_sim}).

\section{Results}

Figure~\ref{fig:bounds1} illustrates the main results of the stacked analysis.  The solid black line represents the limit obtained for DM annihilating to a $b \bar b$ final state using the fiducial boost factor model~\cite{Bartels:2015uba}, while the dashed line  shows the limit without the boost factor enhancement (results for final states other than $b\bar b$ are presented in App.~\ref{supp:clusters}).  To estimate the expected limit under the null hypothesis, we repeat the analysis by randomizing the locations of the halos on the sky 200 times, though still requiring they pass the selection cuts described above.  
The colored bands indicate the 68 and 95\% containment regions for the expected limit.  
The limit is consistent with the expectation under the null hypothesis.

Figure~\ref{fig:bounds2} illustrates how the limits evolve for the $b \bar b$ final state with $m_\chi = 100$~GeV as an increasing number of halos are stacked.  We also show the expected 68\% and 95\% containment regions, which are obtained from the random sky locations.  As can be seen, no single halo dominates the bounds.  For example, removing Virgo, the brightest halo in the catalog, from the stacking has no significant effect on the limit.  Indeed, the inclusion of all 495 halos buys one an additional order of magnitude in the sensitivity reach.

\begin{figure*}[htbp]
 \centering
  \includegraphics[width=0.9\textwidth]{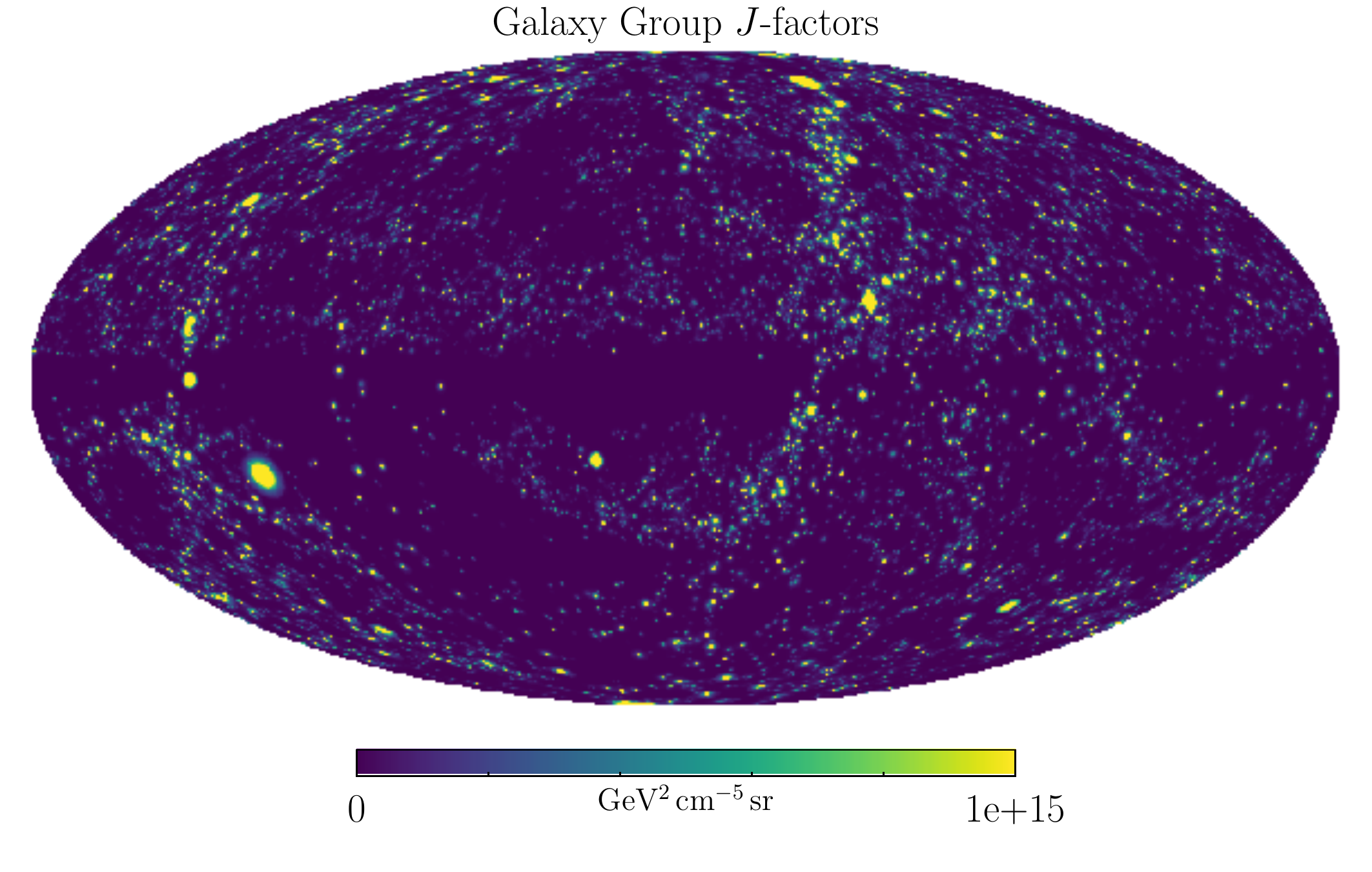}
 \caption{Mollweide projection of all the $J$-factors inferred using the T15 and T17 catalogs, smoothed at $2^\circ$ with a Gaussian kernel. If we could see beyond conventional astrophysics to an extragalactic DM signal, this is how it would appear on the sky.}
  \label{fig:jfactor_maps}
\end{figure*}

Fig.~\ref{fig:jfactor_maps} shows a Mollweide projection of all the $J$-factors inferred using the T15 and T17 catalogs,  smoothed at $2^\circ$ with a Gaussian kernel. The map is shown in Galactic coordinates with the Galactic Center at the origin. Looking beyond astrophysical sources, this is how an extragalactic DM signal might show up in the sky. Although this map has no masks added to it, a clear extinction is still visible along the Galactic plane. This originates from the incompleteness of the catalogs along the Galactic plane. 

The limit derived in this work is complementary to the published dSph bound~\cite{Ackermann:2015zua,Fermi-LAT:2016uux}, shown as the solid gray line in Fig.~\ref{fig:bounds1}. Given the large systematic uncertainties associated with the dwarf analyses (see~\emph{e.g.}, Ref.~\cite{Geringer-Sameth:2014qqa}), we stress the importance of using complementary targets and detection strategies to probe the same region of parameter space. Our limit also probes the parameter space that may explain the Galactic Center excess (GCE); the best-fit models are marked by the orange cross~\cite{Abazajian:2014fta}, blue~\cite{Calore:2014xka}, red~\cite{Gordon:2013vta}, and orange~\cite{Daylan:2014rsa} $2$$\sigma$ regions.  The GCE is a spherically symmetric excess of $\sim$GeV gamma-rays observed to arise from the center of the Milky Way~\cite{Goodenough:2009gk,Hooper:2010mq,TheFermi-LAT:2015kwa,Karwin:2016tsw}.  The GCE has received a considerable amount of attention because it can be explained by annihilating DM.  However, it can also be explained by more standard astrophysical sources; indeed, recent analyses have shown that the distribution of photons in this region of sky is more consistent with a population of unresolved point sources, such as millisecond pulsars, compared to smooth emission from DM~\cite{Lee:2015fea, Bartels:2015aea,Linden:2016rcf, FermiLAT:2017yoi}.  Because systematic uncertainties can be significant and hard to quantify in indirect searches for DM, it is crucial to have independent probes of the parameter space where DM can explain the GCE.  While our null findings do not exclude the DM interpretation of the GCE, their consistency with the dwarf bounds (which also cut into the GCE region) put it further in tension.  This does not, however, account for the fact that the systematics on the modeling of the Milky Way's density distribution can potentially alleviate the tension by changing the best-fit cross section for the GCE.  

\section{Conclusions}

This chapter presents the results of the first systematic search for annihilating DM in nearby galaxy groups.  We introduced and validated  a prescription to infer properties of DM halos associated with  these groups, thereby allowing us to build a map of DM annihilation in the local Universe.  Using this map, we performed a stacked analysis of several hundred galaxy groups and obtained bounds that exclude thermal cross sections for DM  annihilating to $b \bar b$ with mass below $\sim$$30$~GeV, assuming a conservative boost factor model.  These limits are competitive with those obtained from the \emph{Fermi} dSph analyses and are in tension with the range of parameter space that can explain the GCE.  Moving forward, we plan to investigate the objects with gamma-ray excesses to see if they can be interpreted in the context of astrophysical emission.  In so doing, we can also develop more refined metrics for selecting the optimal galaxy groups for DM studies.   

We include additional results in App.~\ref{supp:clusters} that further extends the results presented here.  There, we show limits for additional annihilation final states and the brightest individual halos.  We also show how the limits are affected by several analysis choices, such as the inclusion of Andromeda and Virgo, as well as a variety of systematic uncertainties.  A complete table of the galaxy groups used in this analysis, as well as their associated properties, are provided in Supplementary Data, which can be accessed at \url{https://github.com/bsafdi/DMCat}.  The catalog includes decay factors for all of the groups in addition to the annihilation $J$-factors.  We emphasize that the supplementary catalog is separate from the {\it Fermi} analysis presented here and may be used to search for extragalactic DM annihilation and decay into neutral cosmic rays, regardless of wavelength, messenger, and instrument.

\sectionline 

\appendix 

\chapter{$J$- and $D$-factors for Extragalactic Sources}
\label{app:JDrelations}

\lettrine[lines=3]{I}{n} this Appendix, we derive the $J$-factor relations used in the main text. We also derive the corresponding $D$-factor relations, which apply to the case of decaying DM.  Although we do not make use of the decay results in the main text, we include these results for completeness because much of our main analysis can be extended to the decaying case.
This Appendix is broken into three subsections. In the first of these, we detail the units and conventions used in our definition of the $J$- and $D$-factors.  After this, we derive an approximate form of the astrophysics factors for different DM density profiles and discuss the accuracy of the approximations made.  We conclude with a discussion of error propagation in the $J$-factors.  Note that several of the details presented in these appendices have been discussed elsewhere, see \emph{e.g.}, Ref.~\cite{Abdo:2010ex,Charbonnier:2011ft,Charbonnier:2012gf,Evans:2016xwx}. 

\section{Units and Conventions}

\subsection{Dark Matter Flux}

We begin by carefully outlining the units associated with the $J$- and $D$-factors. 
The flux, $\Phi$, associated with either DM annihilation or decay factorizes into two parts:
\begin{equation}\begin{aligned}
\frac{d\Phi^{\rm ann.}}{dE_{\gamma}} &= \frac{d\Phi_\text{pp}^{\rm ann.}}{dE_{\gamma}}\times J \, , \\
\frac{d\Phi^{\rm dec.}}{dE_{\gamma}} &= \frac{d\Phi_\text{pp}^{\rm dec.}}{dE_{\gamma}}\times D \, ,
\end{aligned}\end{equation}
where $E_\gamma$ is the photon energy and the `ann.'~(`dec.') superscripts denote annihilation~(decay).  
The particle physics factors are given by:
\begin{equation}\begin{aligned}
\frac{d\Phi_\text{pp}^{\rm ann.}}{dE_{\gamma}} &=\frac{\langle\sigma v\rangle}{8\pi m_{\chi}^{2}}\sum_i \text{Br}_{i}\, \frac{dN_{i}}{dE_{\gamma}}\,, \\
\frac{d\Phi_\text{pp}^{\rm dec.}}{dE_{\gamma}} &=\frac{1}{4\pi m_{\chi} \tau}\sum_i \text{Br}_{i}\, \frac{dN_{i}}{dE_{\gamma}}\,,
\end{aligned}\end{equation}
where $\langle \sigma v \rangle$ is the velocity-averaged annihilation cross section, $m_\chi$ is the DM mass, Br$_i$ is the branching fraction into the $i^\text{th}$ channel, $dN_i/dE_\gamma$ is the photon energy distribution associated with this channel, and $\tau$ is the DM lifetime.  The annihilation factor assumes that the DM is its own antiparticle; if this were not the case, and assuming no asymmetry in the dark sector, then the factor would be half as large.  The particle physics factors carry the following dimensions:
\begin{equation}\begin{aligned}
\left[ \frac{d\Phi_\text{pp}^{\rm ann.}}{dE_{\gamma}} \right] &= {\rm counts} \cdot {\rm cm}^3 \cdot {\rm s}^{-1} \cdot {\rm GeV}^{-3} \cdot {\rm sr}^{-1} \,, \\
\left[ \frac{d\Phi_\text{pp}^{\rm dec.}}{dE_{\gamma}} \right] &= {\rm counts} \cdot {\rm s}^{-1} \cdot {\rm GeV}^{-2} \cdot {\rm sr}^{-1}\,,
\label{eq:ppunits}
\end{aligned}\end{equation}
where `counts' refers to the number of gamma-rays produced in the interaction and the ${\rm sr}^{-1}$ is associated with the $1/4\pi$ in the particle physics factors.  Note that some references include this $4\pi$ in the definition of the $J$- or $D$-factors, but this is not the convention that we follow here.

The $J$- and $D$-factors are defined as follows:
\begin{equation}\begin{aligned}
J &= \left(1+b_\text{sh}[M_\text{vir}]\right)\,\int ds \, d\Omega \,\rho_{\rm DM}^2(s,\Omega) \,, \\
D &= \int ds\, d\Omega\, \rho_{\rm DM}(s,\Omega)\,,
\label{eq:JDdef}
\end{aligned}\end{equation}
where $b_\text{sh}[M_\text{vir}]$ is the subhalo boost factor.  The $J$- and $D$-factors carry the following units:
\begin{equation}\begin{aligned}
\left[ J \right] &= {\rm GeV}^2 \cdot {\rm cm}^{-5} \cdot {\rm sr}\,, \\
\left[ D \right] &= {\rm GeV} \cdot {\rm cm}^{-2} \cdot {\rm sr}\,.
\end{aligned}\end{equation}
Combining these with Eq.~\ref{eq:ppunits}, we find that 
\begin{equation}
\left[ \frac{d\Phi}{dE_{\gamma}} \right] = {\rm counts} \cdot {\rm cm}^{-2} \cdot {\rm s}^{-1} \cdot {\rm GeV}^{-1}\,
\end{equation}
for both the annihilation and decay case.  This means that $\Phi$ is given in units of counts per experimental effective area [${\rm cm}^2$] per experimental run time [${\rm s}$].  
In this work, we study extragalactic objects with small angular extent.  So long as each object is centered on the region-of-interest (ROI), we expect that all of its flux will be contained within the ROI as well.  This means that the photon counts obtained by integrating Eq.~\ref{eq:JDdef} over the entire sky corresponds to the total counts expected from that object in the ROI.  The situation is different when treating objects with a large angular extent that exceeds the size of the ROI---\emph{e.g.}, when looking for emission from the halo of the Milky Way.  In such cases, it is more common to divide the $J$- and $D$-factors by the solid angle of the ROI ($\Delta \Omega$) such that both they, and consequently $\Phi$, are averages rather than totals.  

\subsection{Halo Mass and Concentration}

We briefly comment here on different mass and concentration definitions (virial and 200) as relevant to our analysis.   Boost-factor models, concentration-mass relations, and masses are often specified in terms of 200 quantities, which must be converted to virial ones. In order to do this, we use the fact that
\begin{equation}\begin{aligned}
\frac{\rho_\text{s}}{\rho_c} \equiv \delta_\mathrm{c} = \frac{\Delta_\text{c}}{3}\frac{c^3}{\log{(1+c)}-c/(1+c)}
\label{eq:CritOverdens}
\end{aligned}\end{equation}
for the NFW profile~\cite{Navarro:1995iw}, where $\rho_s$ is the normalization of the density profile, $\rho_c$ is the critical density, $c$ is the concentration parameter, and $\delta_\mathrm{c}$ is the critical overdensity.  For virial quantities, $\Delta_c(z) = 18\pi^2 +82x-39x^2$ with $x = \Omega_{m}(1+z)^3/[\Omega_{m}(1+z)^3 + \Omega_{\Lambda}]-1$ in accordance with Ref.~\cite{Bryan:1997dn}, while for 200 quantities, $\Delta_c = 200$.  Therefore, Eq.~\ref{eq:CritOverdens} can be equated between the 200 and virial quantities and solved numerically to convert between definitions of the concentration.

For different mass definitions, we have 
\begin{equation}\begin{aligned}
\frac{M_{200}}{M_\text{vir}} = \left(\frac{c_{200}[M_{200}]}{c_\text{vir}[M_\text{vir}]}\right)^3\frac{200}{\Delta_\text{c}} \, ,
\label{eq:MassConvert}
\end{aligned}\end{equation}
where the concentration definitions on the right-hand side depend on $M_{200}$ and $M_\text{vir}$ and may have to be converted between each other and we have suppressed the redshift dependence for clarity.  Solving this numerically, we can convert between the two mass definitions.

\section{Approximate $J$- and $D$-factors}

For an extragalactic DM halo, the astrophysical factors in Eq.~\ref{eq:JDdef} can be approximated as:
\begin{equation}\begin{aligned}
J &\approx \left(1+b_\text{sh}[M_\text{vir}]\right)\, \frac{1}{d_c^2[z]} \int_V dV' \rho_{\rm DM}^2(r') \,, \\
D &\approx \frac{1}{d_c^2[z]} \int_V dV' \rho_{\rm DM}(r')\,,
\vspace{0.5in}
\label{eq:JDdapprox}
\end{aligned}\end{equation}
where the integrals are performed in a coordinate system centered on the halo, and $d_c[z]$ is the comoving distance, which is a function of redshift for a given cosmology.  The aim of this subsection is to derive Eq.~\ref{eq:JDdapprox} from Eq.~\ref{eq:JDdef} and to quantify the error associated with this approximation. 

To handle the $J$- and $D$-factors simultaneously, we consider the following integral over all space:
\begin{equation}
\int ds \, d\Omega \,\rho_{\rm DM}^n(s,\Omega)\,,
\end{equation}
with $n \geq 1$. Here, $s$ is playing the role of a radius in a spherical coordinate system centered on the Earth.  Therefore, we can rewrite the measure as
\begin{equation}
\int s^2\, ds \, d\Omega \,\, \frac{\rho_{\rm DM}^n(s,\Omega)}{s^2} = \int dV\, \frac{\rho_{\rm DM}^n(s,\Omega)}{s^2}\,.
\end{equation}

Next, we transform to a coordinate system (denoted by primed quantities) that is centered at the origin of the halo described by $\rho_{\rm DM}$.  Because this change of coordinates is only a linear translation, it does not induce a Jacobian and $dV = dV'$.  Assuming that the Earth is located at a position $\mathbf{r}$ from the halo center and the DM interaction occurs at position $\mathbf{r'}$, then $s = | \mathbf{r} - \mathbf{r}'| $ and
\begin{equation}
\int dV\, \frac{\rho_{\rm DM}^n(s,\Omega)}{s^2} = \int dV'\, \frac{\rho_{\rm DM}^n(r',\Omega')}{r^{\prime 2} - 2 d_c r' \cos \theta' + d_c^2}\,,
\label{eq:haloframe}
\end{equation}
where we take $|\mathbf{r}| = d_c$ and $\mathbf{r} \cdot \mathbf{r}' = d_c\, r' \, \cos\theta'$.

Eq.~\ref{eq:haloframe} can be simplified by taking advantage of several properties of the halo density.  First, it is spherically symmetric about the origin of the primed coordinate system.  Second, it only has finite support in $r'$.  In particular, it does not make sense to integrate the object beyond the virial radius, $r_{\rm vir}$. This allows us to rewrite the integral as follows:
\vspace{0.1in}

\begin{eqnarray}
\label{eq:multiline}
\int dV'\, \frac{\rho_{\rm DM}^n(r',\Omega')}{r^{\prime 2} - 2 d_c r' \cos \theta' + d_c^2}
&=& \int_0^{r_{\rm vir}} dr'\, \int d\Omega'\, \frac{\rho_{\rm DM}^n(r')}{r^{\prime 2} - 2 d_c r' \cos \theta' + d_c^2} \\
&=& \frac{2 \pi}{d_c^2} \, \int_0^{r_{\rm vir}} dr'\, \rho_{\rm DM}^n(r')  \int_0^{\pi} d\theta'\, \frac{\sin \theta'}{1 - 2 (r'/d_c) \cos \theta' + (r'/d_c)^2} \notag \\
&=& \frac{2\pi}{d_c^2} \, \int_0^{r_{\rm vir}} dr'\, \frac{ \rho_{\rm DM}^n(r')}{2\,(r'/d_c)} \ln \left[ \frac{((r'/d_c)+1)^2}{((r'/d_c)-1)^2} \right]\,. \notag
\end{eqnarray}

For extragalactic objects, $d_c \gg r_{\rm vir} \geq r'$.  As a result, we can take advantage of the following  expansion:
\begin{equation}
\frac{1}{2x} \ln \left[ \frac{(x+1)^2}{(x-1)^2} \right]  = 2 \left[ 1 + \frac{1}{3} x^2 + \mathcal{O} \left(x^4\right) \right] \, ,
\label{eq:logexpand}
\end{equation}
where $x= r'/d_c$.  It follows that the leading-order approximation to Eq.~\ref{eq:multiline} is \begin{equation}\begin{aligned}
\int ds \, d\Omega \,\rho_{\rm DM}^n(s,\Omega) = \frac{1}{d_c^2} \int dV' \rho_{\rm DM}^n(r') \,,
\end{aligned}\end{equation}
which when inserted into Eq.~\ref{eq:JDdef} gives Eq.~\ref{eq:JDdapprox}, as claimed. 

We can calculate the size of the neglected terms in Eq.~\ref{eq:logexpand} to quantify the accuracy of this approximation.  We take the parameters of the halo with the largest $J$-factor in the catalog to estimate the largest error possible amongst the \texttt{DarkSky} halos.  For this halo, the fractional correction to the $J$-factor of the first neglected term in the expansion is $\mathcal{O}(10^{-5})$ for either an NFW or Burkert profile (described below), whilst for the $D$-factor it is $\mathcal{O}(10^{-4})$. These values are significantly smaller than the other sources of uncertainty present in estimating these quantities and so we conclude that the approximations in Eq.~\ref{eq:JDdapprox} are sufficient for our purposes.

\section{Analytic Relations}

Starting from the approximate forms given in Eq.~\ref{eq:JDdapprox} and specifying a DM density profile $\rho_{\rm DM}$, the $J$- and $D$-factors can often be determined exactly.   We will now demonstrate that the final results only depend on the distance, mass, and concentration of the halo---for a given substructure boost model and cosmology.

As a starting point, consider the NFW profile:
\begin{equation}
\rho_{\rm NFW}(r) = \frac{\rho_s}{r/r_s(1+r/r_s)^2}\,.
\end{equation}
The parameter $r_s$ is the scale radius and dictates how sharply peaked the core of the DM distribution is.
Starting from this distribution, the volume integral in the $J$-factor evaluates to
\begin{equation}\begin{aligned}
\int dV'\, \rho_{\rm NFW}^2(r') &= 4\pi \rho_s^2 r_s^2 \int_0^{r_{\rm vir}} \frac{dr'}{(1+r'/r_s)^4} \\
&= \frac{4\pi}{3} \frac{\rho_s^2 r_{\rm vir}^3}{c_{\rm vir}^3} \left[ 1 - \frac{1}{(1+c_{\rm vir})^3} \right]\,,
\label{eq:NFWVolumeInt}
\end{aligned}\end{equation}
where $c_{\rm vir} = r_{\rm vir}/r_s$ is the virial concentration.  To remove the normalization factor $\rho_s$ from this equation, we can write the virial mass of the halo as
\begin{equation}\begin{aligned}
M_{\rm vir} &\equiv \int dV'\, \rho_{\rm NFW}(r') \\
&= 4\pi \rho_s \frac{r_{\rm vir}^3}{c_{\rm vir}^3} \left[ \ln \left( 1 + c_{\rm vir} \right) - \frac{c_{\rm vir}}{1+c_{\rm vir}} \right]\,,
\end{aligned}\end{equation}
which, when combined with Eq.~\ref{eq:NFWVolumeInt}, gives
\begin{equation}\begin{aligned}
\int dV'\, \rho_{\rm NFW}^2(r)
=\,&\frac{M_{\rm vir}^2 c_{\rm vir}^3}{12\pi r_{\rm vir}^3} \left[ 1 - \frac{1}{(1+c_{\rm vir})^3} \right] \\
\times &\left[ \ln \left( 1 + c_{\rm vir} \right) - \frac{c_{\rm vir}}{1+c_{\rm vir}} \right]^{-2}\,.
\end{aligned}\end{equation}
Stopping here, we would conclude that the $J$-factor scales as $M_{\rm vir}^2$.  However, for a given $M_{\rm vir}$ and cosmology, $r_{\rm vir}$ is not an independent parameter. Using the results of Ref.~\cite{Bryan:1997dn}, we can write:
\begin{equation}
\frac{3M_{\rm vir}}{4\pi r_{\rm vir}^3} = \rho_c \Delta_c[z]\,,
\end{equation}
where $\rho_c$ is the critical density and 
\begin{equation}\begin{aligned}
\Delta_c[z] &\equiv  18\pi^2 + 82\,x[z] - 39\,x[z]^2\,, \\
x[z] &\equiv \frac{\Omega_m \left( 1 + z \right)^3}{\Omega_m \left( 1 + z \right)^3 + \Omega_{\Lambda}} - 1\,.
\end{aligned}\end{equation}
This relation can then be used to remove $M_{\rm vir}/r_{\rm vir}^3$ from the volume integral and we conclude that
\begin{align}
J_{\rm NFW} \approx &\left(1+b_\text{sh}[M_\text{vir}]\right) \frac{M_{\rm vir} c_{\rm vir}^3 \rho_c \Delta_c[z]}{9 d_c^2[z]} \label{eq:JNFWfull} \\
\times &\left[ 1 - \frac{1}{(1+c_{\rm vir})^3} \right] \left[ \ln \left( 1 + c_{\rm vir} \right) - \frac{c_{\rm vir}}{1+c_{\rm vir}} \right]^{-2}\,. \notag
\end{align}
We see the additional mass dimension required from the fact this scales as $M_{\rm vir}$ not $M_{\rm vir}^2$ is carried by $\rho_c$. The $c_{\rm vir}^3$ dependence highlights that the annihilation flux is critically dependent upon how sharply peaked the halo is.   To summarize, Eq.~\ref{eq:JNFWfull} demonstrates that the $J$-factor is fully specified by three halo parameters for a given substructure boost model and cosmology: the redshift $z$, mass $M_{\rm vir}$, and concentration $c_{\rm vir}$.


The basic scalings and dependence shown above are not peculiar to the NFW profile, but are in fact more generic. To demonstrate this, we can repeat the above exercise for the cored Burkert profile~\cite{Burkert:1995yz}:
\begin{equation}
\rho_{\rm Burkert}(r) = \frac{\rho_B}{(1+r/r_B)(1+(r/r_B)^2)}\,,
\end{equation}
which is manifestly non-singular as $r \to 0$ unlike the NFW profile.  Here, $\rho_B$ and $r_B$ are the Burkert analogues of $\rho_s$ and $r_s$ in the NFW case, but they are not exactly the same.  Indeed, following \emph{e.g.}, Ref.~\cite{Bartels:2015uba}, by calculating physically measurable properties of halos such as the radius of maximum rotational velocity for both the NFW and Burkert cases and setting them equal, we find
\begin{equation}
r_B \simeq 0.7 r_s\,.
\end{equation}
We will replace $r_B$ with a concentration parameter $c_B = r_{\rm vir}/r_B$.  Following the same steps as for the NFW profile, we arrive at:
\begin{align}
J_{\rm Burkert} \approx &\left(1+b_\text{sh}[M_\text{vir}]\right) \frac{4M_{\rm vir} c_B^3 \rho_c \Delta_c[z]}{3 d_c^2[z]} \\
\times &\left[ \frac{c_B(1+c_B+2c_B^2)}{(1+c_B)(1+c_B^2)} - \arctan(c_B) \right] \notag \\
\times &\left[ \ln \left[ (1+c_B)^2 (1+c_B^2) \right] - 2 \arctan(c_B) \right]^{-2}\,, \notag
\end{align}
from which we see that $J \sim (1+b_\text{sh}) M_\text{vir} c_B^3\rho_c/d_c^2[z]$.

For the case of decaying DM, the approximate integral given in Eq.~\ref{eq:JDdapprox} can be evaluated independent of any choice for the halo profile.   Specifically:
\begin{equation}\begin{aligned}
D &\approx \frac{1}{d_c^2[z]} \int_V dV' \rho_{\rm DM}(r) = \frac{M_{\rm vir}}{d_c^2[z]}\,,
\label{eq:Dfactor}
\end{aligned}\end{equation}
where the second equality follows from the fact that the volume integral gives the virial mass exactly.
For DM decays in relatively nearby halos, the emission can be quite extended, as the flux is not as concentrated towards the center of the halo as in the annihilation case.  As such, it is often useful to have a version of the extragalactic $D$-factor where one only integrates out to some angle $\theta$ on the sky from the center of the halo, or equivalently to a distance $R = \theta \cdot d_c(z) < r_{\rm vir}$. In this case:\begin{align}
D &\approx \frac{M_{\rm vir}}{d_c^2(z)} \\
&\times \left[ \ln \left( 1 + \frac{c_{\rm vir} R}{r_{\rm vir}[M_{\rm vir}]} \right) - \frac{c_{\rm vir}}{r_{\rm vir}[M_{\rm vir}]/R + c_{\rm vir}} \right] \notag \\
&\times \left[ \ln (1 + c_{\rm vir}) - \frac{c_{\rm vir}}{1+c_{\rm vir}} \right]^{-1}\,, \notag
\end{align}
for the NFW profile, where we have made explicit the fact that $r_{\rm vir}$ is a function of $M_{\rm vir}$.  When $R=r_{\rm vir}$, this reduces to the simple result in Eq.~\ref{eq:Dfactor}.

\sectionline
\chapter{Supplementary Material on Cluster Searches}
\label{supp:clusters}

\section{Extended results}
\label{sec:extended}

\begin{figure*}[b]
  \centering
	\includegraphics[width=.45\textwidth]{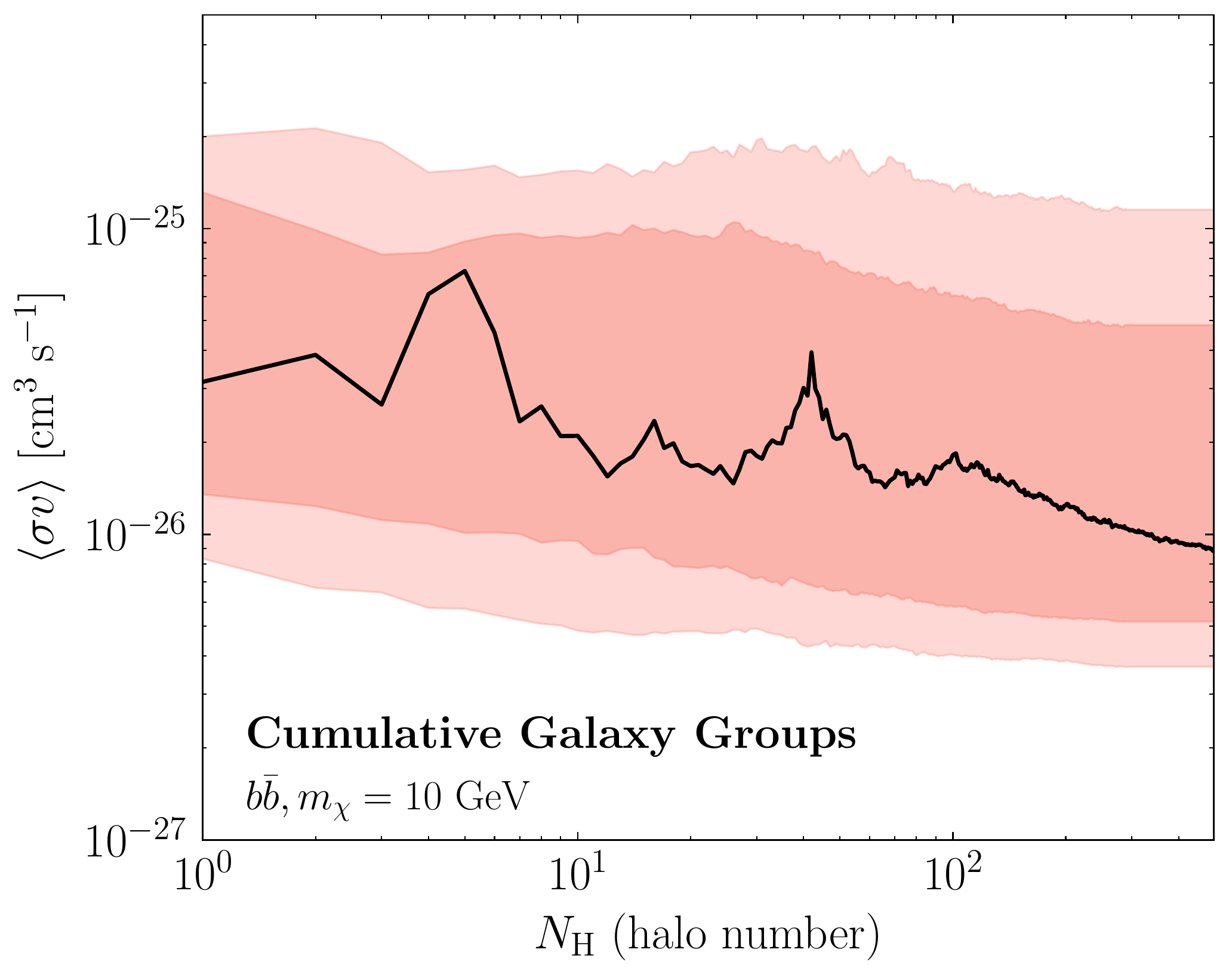} 
	\includegraphics[width=.45\textwidth]{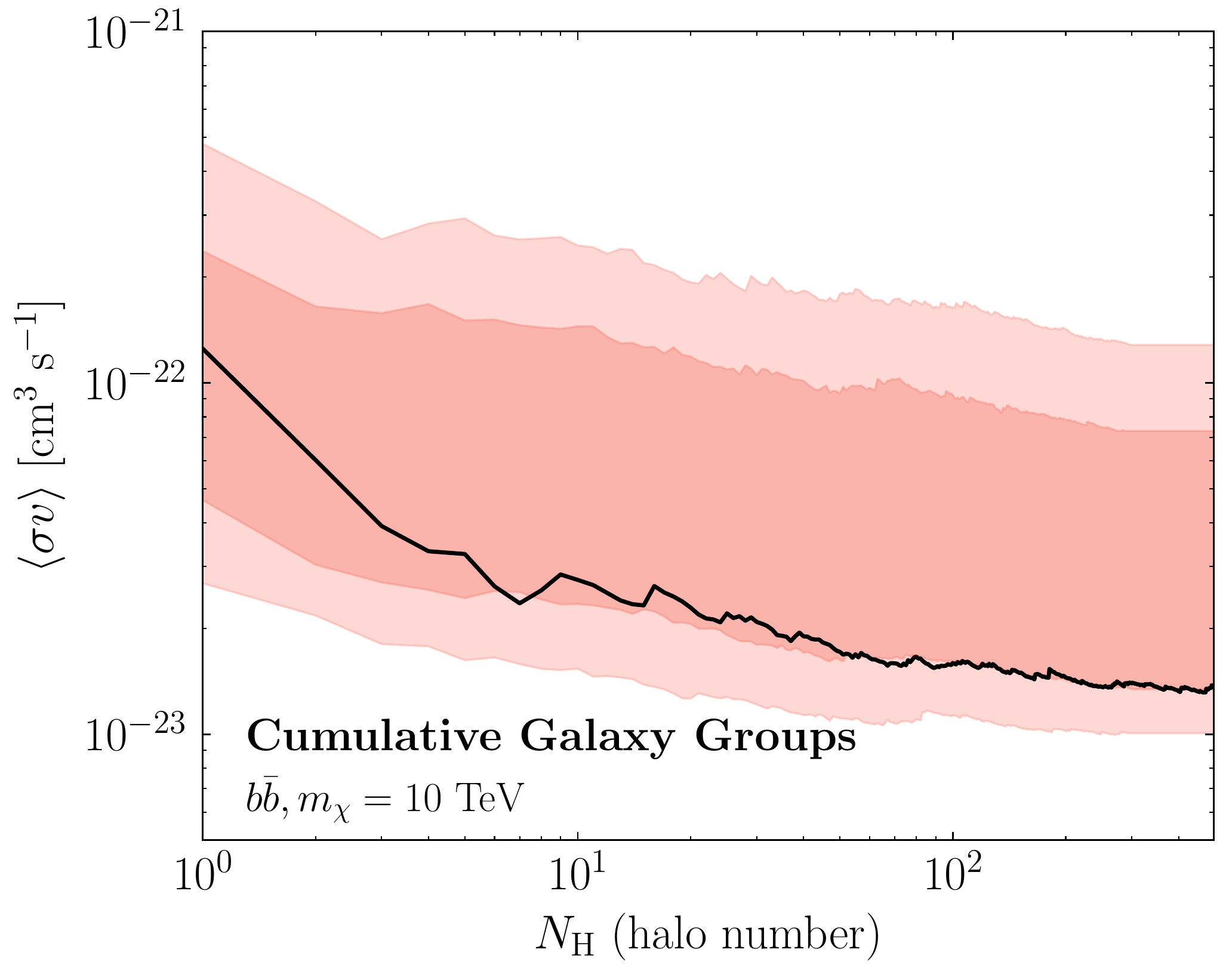} 
  \caption{The change in the limit on the $b\bar{b}$ annihilation channel as a function of the number of halos included in the stacking, for $m_\chi= 10$~GeV \textbf{(left)} and 10~TeV \textbf{(right)}. The 68 and 95\% expectations from 200 random sky locations are indicated by the red bands.}
  \label{fig:moreelephants}
\end{figure*}

\lettrine[lines=3]{I}{n} the main analysis, Fig.~\ref{fig:bounds2}  demonstrates how the limit on the  $b\bar b$ annihilation cross section  depends on the number of halos included in the stacking, for the case where  $m_\chi = 100$~GeV. In Fig.~\ref{fig:moreelephants}, we show the corresponding plot for $m_\chi = 10$~GeV (left) and $10$~TeV (right).  As in the 100~GeV case, we see that no single halo dominates the bound and that stacking a large number of halos considerably improves the sensitivity.

\begin{figure*}[t]
  \centering
   \includegraphics[width=0.45\textwidth]{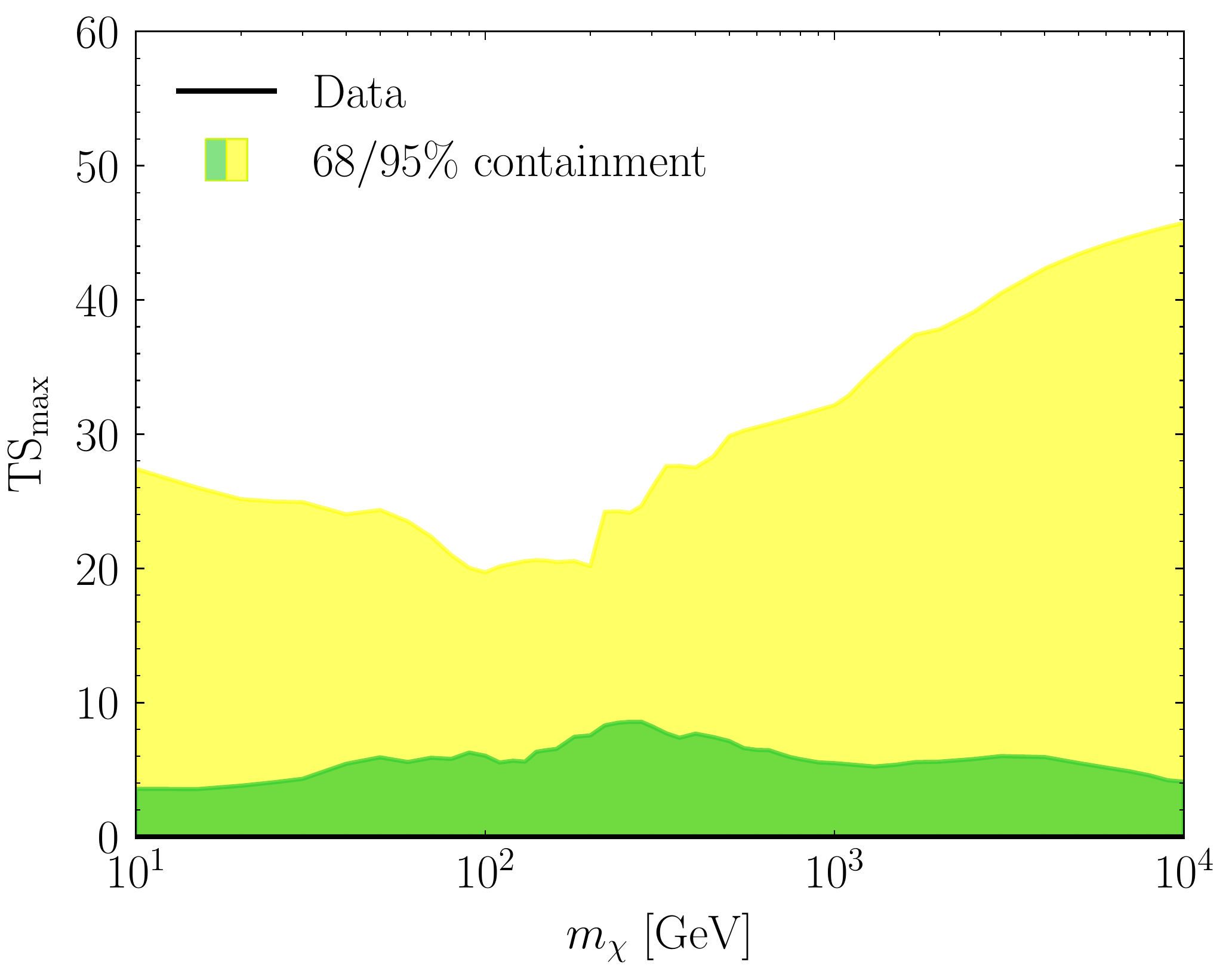}
  \includegraphics[width=.45\textwidth]{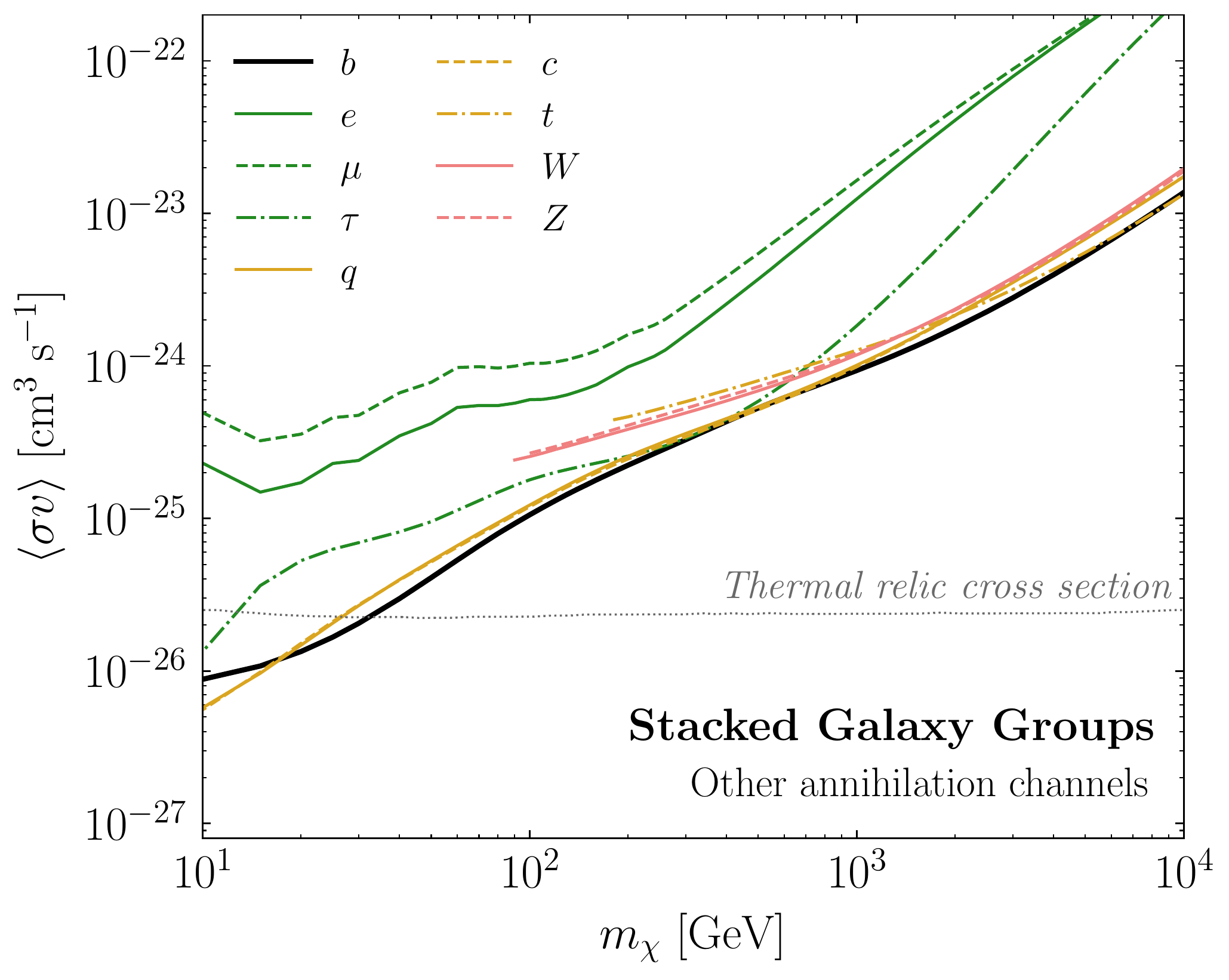}
  \caption{\textbf{(Left)} Maximum test statistic, TS$_\text{max}$, for the stacked analysis comparing the model with and without DM annihilating to $b \bar b$.  The green~(yellow) bands show the 68\%~(95\%) containment over multiple random sky locations.  \textbf{(Right)} The 95\% confidence limits on the DM annihilation cross section, as a function of the DM mass, for the Standard Model final states indicated in the legend.  These limits assume the fiducial boost factor taken from Ref.~\cite{Bartels:2015uba}.  Note that we neglect Inverse Compton emission and electromagnetic cascades, which can be relevant for the leptonic decay channels at high energies.}
  \label{fig:other_lims}
\end{figure*}

The left panel of Fig.~\ref{fig:other_lims} shows the maximum test statistic, TS$_\text{max}$, recovered for the stacked analysis in the $b\bar{b}$ channel.  For a given data set $d$, we define the maximum test-statistic in preference for the DM model, relative to the null hypothesis without DM, as 
\es{maxTS}{
{\rm TS}_\text{max}(\mathcal{M}, m_{\rm \chi}) \equiv & \,2 \left[ \log \mathcal{L}(d | \mathcal{M}, \widehat{\langle\sigma v\rangle}, m_\chi ) - \log \mathcal{L}(d | \mathcal{M}, \langle\sigma v\rangle =0, m_\chi ) \right] \, ,
}
where $\widehat{\langle\sigma v\rangle}$ is the cross section that maximizes the likelihood for DM model $\mathcal{M}$.  The observed TS$_\text{max}$ is negligible at all masses and well-within the null expectation (green/yellow bands), consistent with the conclusion that we find no evidence for DM annihilation.  \vspace{0.1in}

\noindent  {\bf Other Annihilation Channels.}  
In general, DM may annihilate to a variety of Standard Model final states.  Figure~\ref{fig:other_lims} (right) interprets the results of the analysis in terms of limits on additional final states that also lead to continuum gamma-ray emission.  Final states that predominantly decay hadronically  ($W^+ W^-$, $ZZ$, $q \bar{q}$, $c \bar c$, $b \bar b$, $t \bar t$) give similar limits because their energy spectra are mostly set by boosted pion decay.  The leptonic channels ($e^+ e^-$, $\mu^+ \mu^-$) give weaker limits because gamma-rays predominantly arise from final-state radiation or, in the case of the muon, radiative decays.  The $\tau^+ \tau^-$ limit is intermediate because roughly 35\% of the $\tau$ decays are leptonic, while the remaining are hadronic.   Of course, the DM could annihilate into even more complicated final states than the two-body cases considered here and the results can be extended to these cases~\cite{Elor:2015tva,Elor:2015bho}.  Note that the limits we present for the leptonic final states are conservative, as they neglect Inverse Compton (IC) emission and electromagnetic cascades, which are likely important at high DM masses---see~\emph{e.g.}, Ref.~\cite{Cirelli:2009dv, Murase:2012xs}.  A more careful treatment of these final states requires modeling the magnetic field strength and energy loss mechanisms within the galaxy groups. 
\vspace{0.1in}

\begin{figure*}[t]
  \centering
	\includegraphics[width=.32\textwidth]{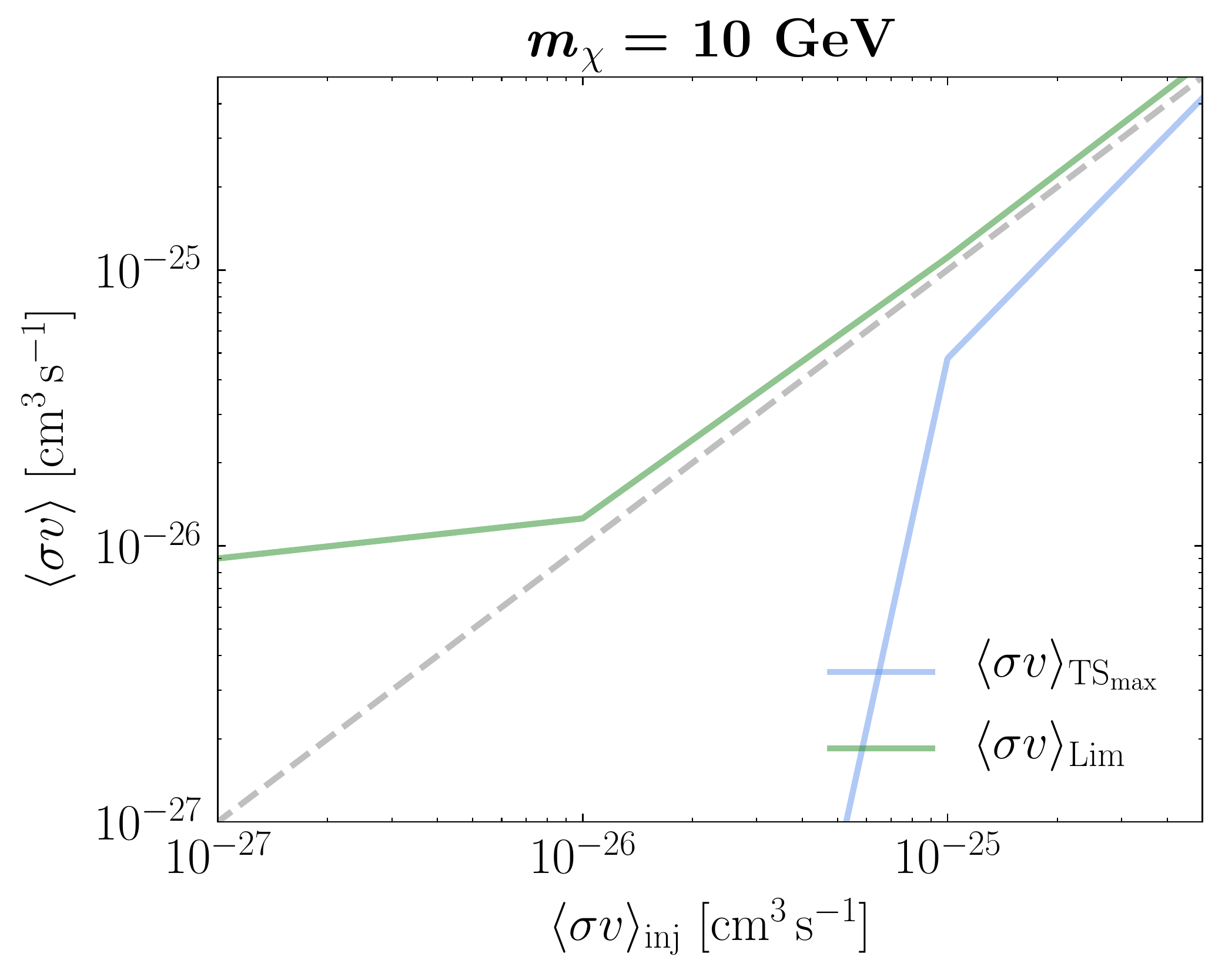}
	\includegraphics[width=.32\textwidth]{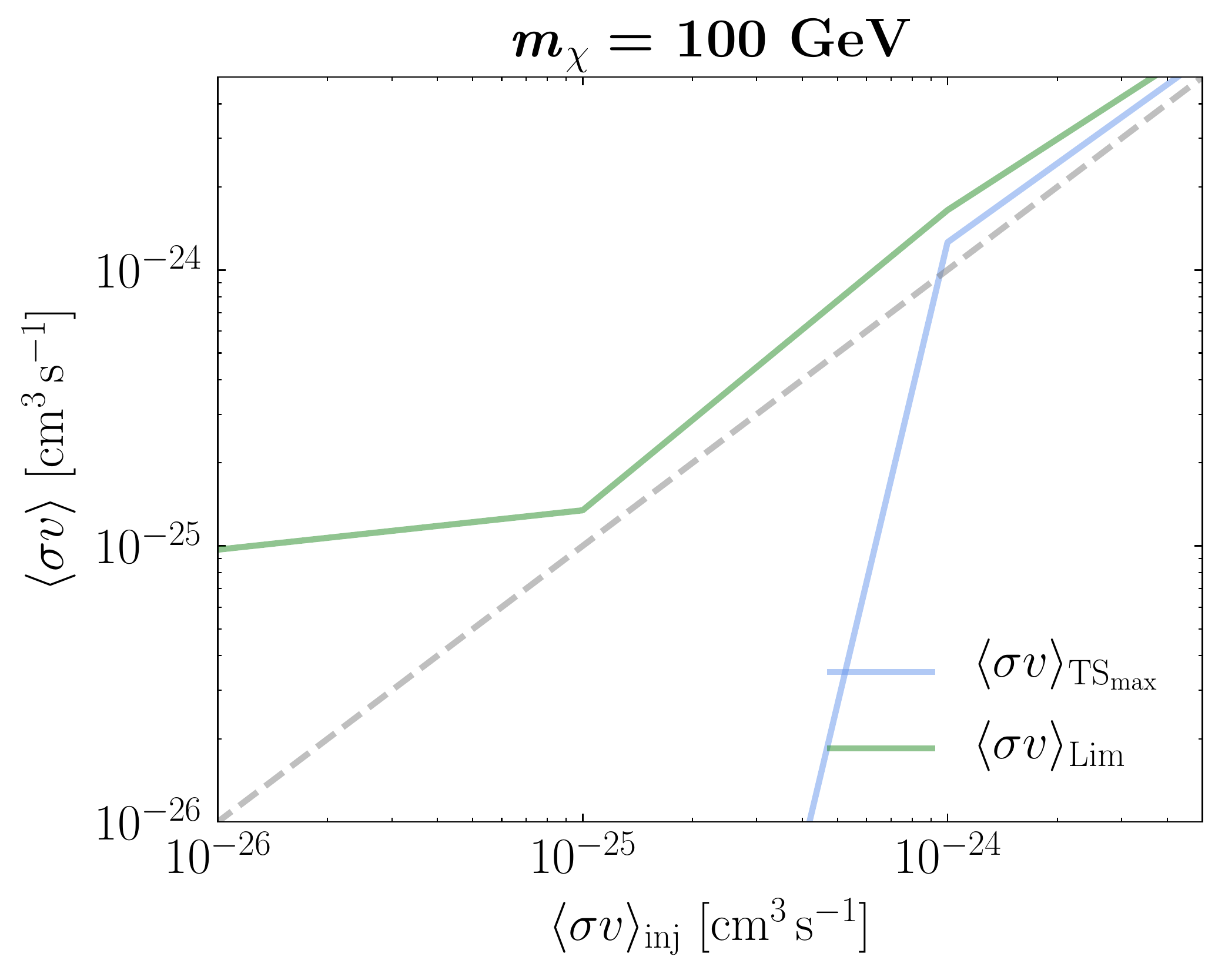}
	\includegraphics[width=.32\textwidth]{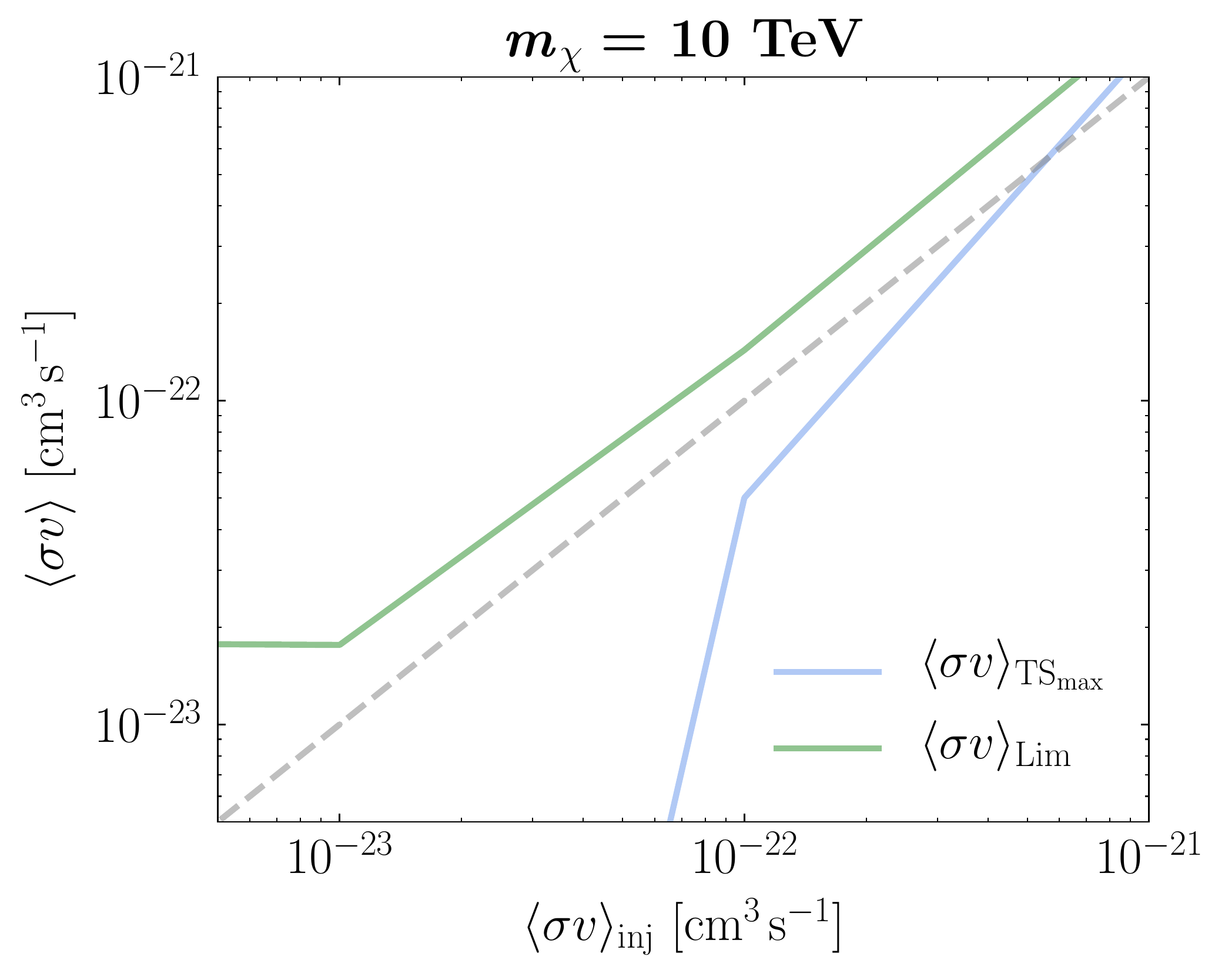} \\
	\includegraphics[width=.32\textwidth]{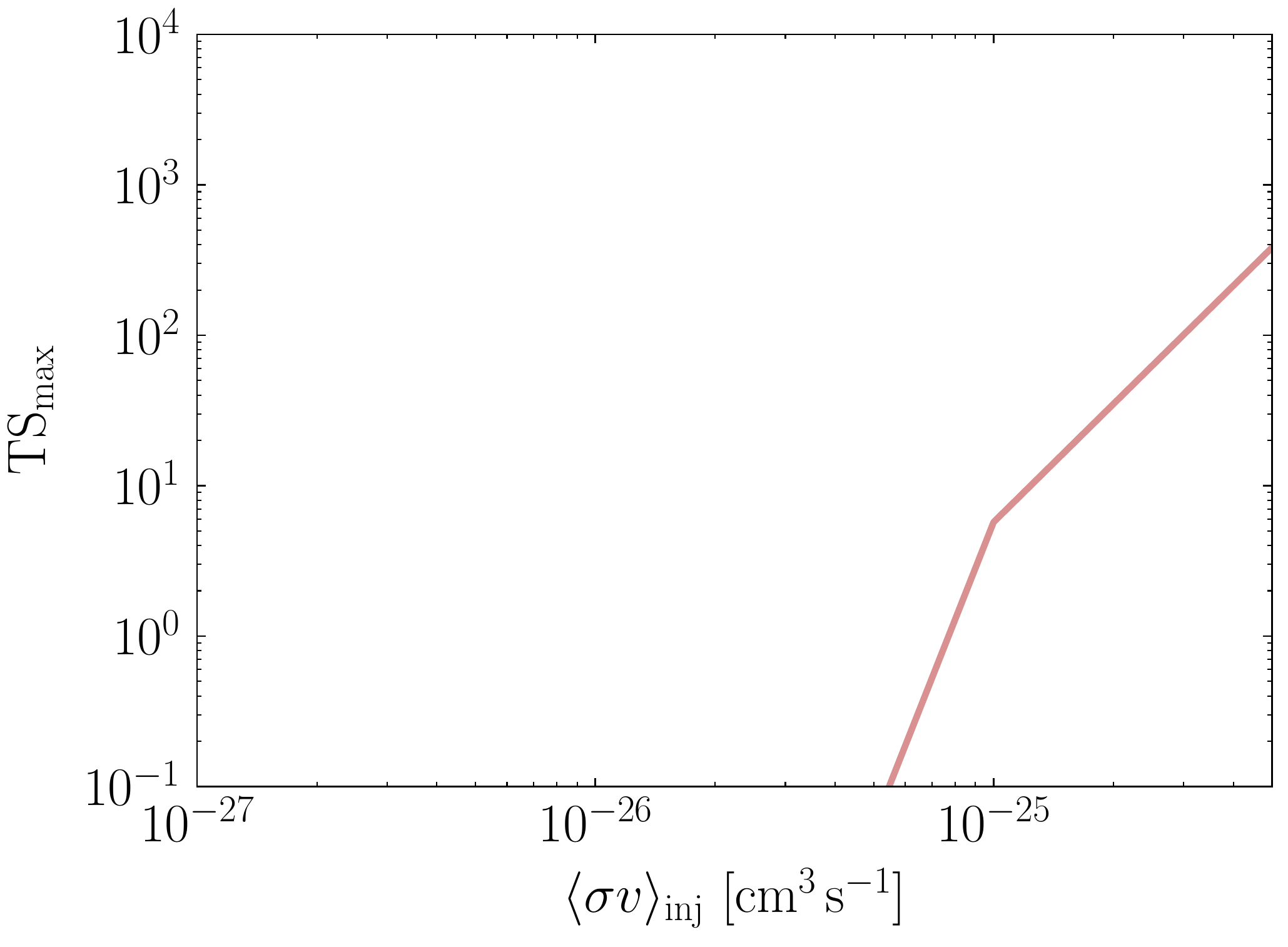}
	\includegraphics[width=.32\textwidth]{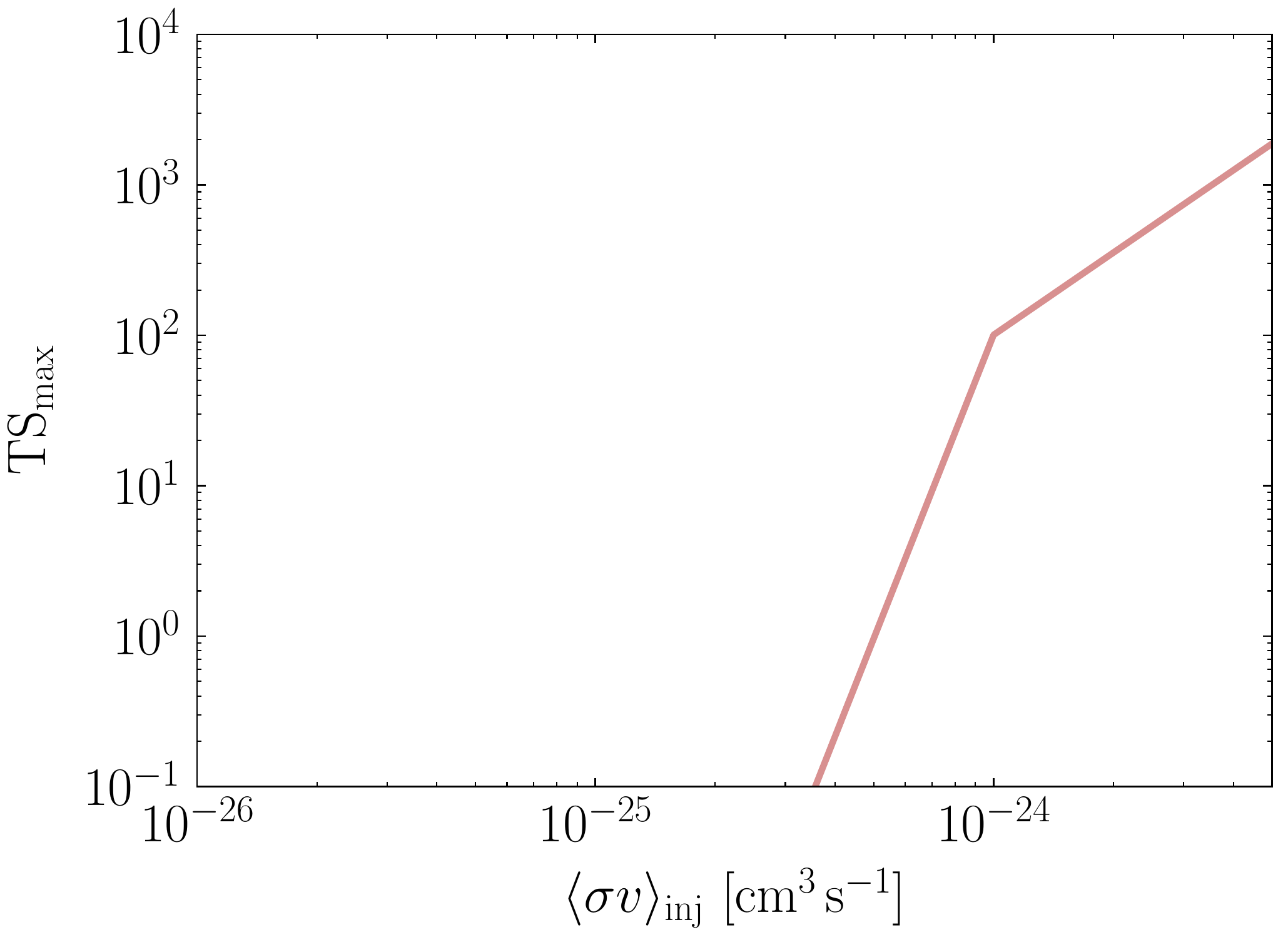}
	\includegraphics[width=.32\textwidth]{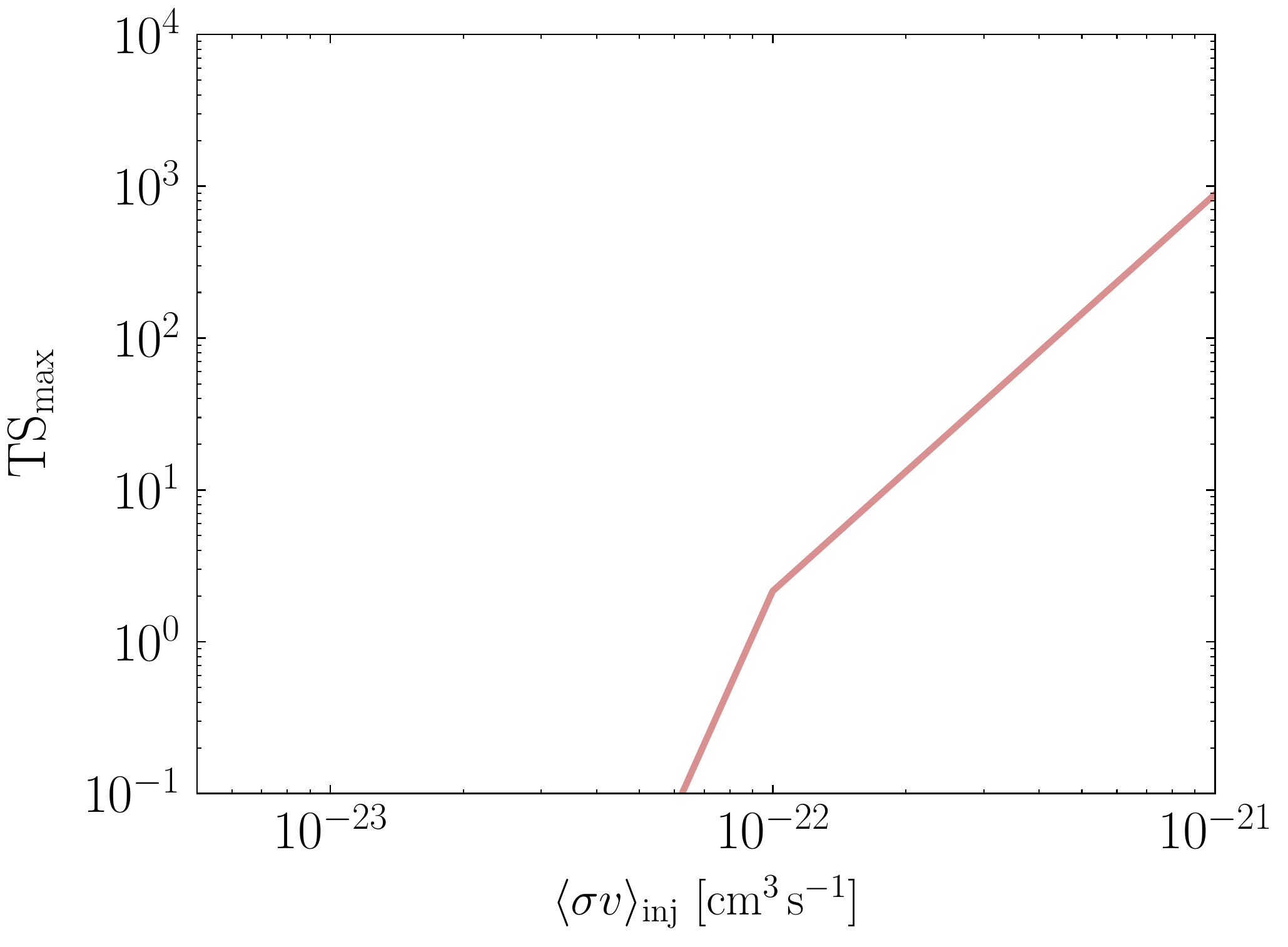}
  \caption{\textbf{(Top)} Recovered cross section at maxiumum test statistic, TS$_\text{max}$, (blue line) and limit (green line) obtained for various signals injected on top of the data. \textbf{(Bottom)} The maximum test statistic obtained at various injected cross section values. }
  \label{fig:injsig}
\end{figure*}

\noindent  {\bf Injected Signal.} An important consistency requirement is to ensure that the limit-setting procedure does not exclude a putative DM signal. The likelihood procedure employed here was extensively vetted in Ch.~\ref{ch:groups_sim}, where we demonstrated that the limit never excludes an injected signal.  In Fig.~\ref{fig:injsig}, we demonstrate a data-driven version of this test. In detail, we inject a DM signal on top of the actual data set used in the main analysis, focusing on the case of DM annihilation to $b \bar{b}$ for a variety of cross sections and masses. We then apply the analysis pipeline to these maps.  The top panel of Fig.~\ref{fig:injsig} shows the recovered cross sections, as a function of the injected values.  The green line corresponds to the 95\% cross section limit, while the blue line shows the best-fit cross section.  Note that statistical uncertainties arising from DM annihilation photon counts are not significant here, as the dominant source of counts arises from the data itself. 
The columns correspond to 10, 100, and 10$^4$~GeV DM annihilating to $b \bar b$ (left, center, right, respectively).  The bottom row shows the maximum test statistic in favor of the model with DM as a function of the injected cross section.  The best-fit cross sections are only meaningful when the maximum test statistic is $\gtrsim 1$, implying evidence for DM annihilation.     
We see that across all masses, the cross section limit  (green line) is always weaker than the injected value.  Additionally, the recovered cross section (blue line) closely approaches that of the injected signal as the significance of the DM excess  increases.   
\vspace{0.1in}

\noindent  {\bf Results for Individual Halos.}  Here, we explore the properties of the individual galaxy groups that are included in the stacked analysis.  These galaxy groups are taken from the catalogs in Ref.~\cite{Tully:2015opa} and~\cite{2017ApJ...843...16K}, which we refer to as T15 and T17, respectively.  Table~\ref{tab:tully_extended} lists the top 25 galaxy groups, ordered by the relative brightness of their inferred $J$-factor.  If a group in the table is not labeled with a checkmark, then it is not included in the stacking because one of the following conditions is met:
\es{eq:selection}{
\begin{cases} 
   |b| \leq 20^\circ \, , \\
   \text{overlaps another halo to within}~2^\circ~\text{of its center} \, ,\\
   \text{TS}_\text{max} > 9\, \text{ and } (\sigma v)_\text{best} > 10 \times (\sigma v)^*_\text{lim} \, .
\end{cases}
}
Note that the overlap criteria is applied sequentially in order of increasing $J$-factor.
These selection criteria have been extensively studied on mock data in Ch.~\ref{ch:groups_sim} and have been verified to not exclude a potential DM signal, even on data as discussed above. Of the five halos with the largest $J$-factors that are excluded, Andromeda is removed because of its large angular extent, and the rest fail the latitude cut. 

The exclusion of Andromeda is not a result of the criteria in Eq.~\ref{eq:selection}, so some more justification is warranted. As can be seen in Table~\ref{tab:tully_extended}, the angular extent of Andromeda's  scale radius, $\theta_{s}$, is significantly larger than that of any other halo.  To justify $\theta_{s}$ as a proxy for angular extent of the emission, we calculate the 68\% (95\%) containment angle of the expected DM annihilation flux, without accounting for the PSF, and find 1.2$^{\circ}$ (4.4$^{\circ}$). This can be contrasted with the equivalent numbers for the next most important halo, Virgo, where the corresponding  68\% (95\%) containment angles are 0.5$^{\circ}$ (2.0$^{\circ}$). 
Because Andromeda is noticeably more extended beyond the \textit{Fermi} PSF, one must carefully model the spatial distribution of both the smooth DM component and the substructure.  Such a dedicated analysis of Andromeda was recently performed by the \emph{Fermi} collaboration~\cite{Ackermann:2017nya}.  Out of an abundance of caution, we remove Andromeda from the main joint analysis, but we do show how the limits change when Andromeda is included further below.

Figure~\ref{fig:individual_lims} shows the individual limits on the $b\bar{b}$ annihilation cross section for the top ten halos that pass the selection cuts and Fig.~\ref{fig:individual_maxts}  shows the maximum test statistic (TS$_\text{max}$), as a function of $m_\chi$, for these same halos. The green and yellow bands in Fig.~\ref{fig:individual_lims} and~\ref{fig:individual_maxts} represent the 68\% and 95\% containment regions obtained by randomly changing the sky location of each individual halo 200 times (subject to the selection criteria listed above). 
As is evident, the individual limits for the halos  are consistent with expectation under the null hypothesis---\emph{i.e.}, the black line falls within the green/yellow bands for each of these halos.  Some of these groups have been analyzed in previous cluster studies.  For example, the \emph{Fermi} Collaboration provided DM bounds for Virgo~\cite{Ackermann:2015fdi}; our limit is roughly consistent with theirs, and possibly a bit stronger, though an exact comparison is difficult to make due to differences in the data set and DM model assumptions.\footnote{Note that the $J$-factor in Ref.~\cite{Ackermann:2015fdi} is a factor of $4\pi$ too small.}

Figure~\ref{fig:individual_flux} provides the 95\% upper limits on the gamma-ray flux associated with the DM template for each of the top ten halos.  The upper limits are provided for 26 energy bins and compared to the expectations under the null hypothesis.  The upper limits are generally consistent with the expectations under the null hypothesis, though small systematic discrepancies do exist for a few halos, such as NGC3031, at high energies.  This could be due to subtle differences in the sky locations and angular extents between the objects of interest and the set of representative halos used to create the null hypothesis expectations. 

To demonstrate the case of a galaxy group with an excess, we show the TS$_\text{max}$ distribution and the limit for NGC6822 in Fig.~\ref{fig:maxTSoneobject}.  This object fails the selection criteria because it is too close to the Galactic plane. However, it also exhibits a TS$_\text{max}$ excess and, as expected, the limit is weaker than the expectation under the null hypothesis. \vspace{0.1in}

\noindent  {\bf Sky maps.} In Fig.~\ref{fig:individual_skyrois}, we show the counts map in $20^\circ \times 20^\circ$ square regions around each of the top nine halos that pass the selection cuts.  For each map, we show all photons with energies above $\sim$500 MeV, indicate all {\it Fermi} 3FGL point sources with orange stars, and show the extent of $\theta_s$ with a dashed orange circle.  Given a DM signal, we would expect to see emission extend out to $\theta_s$ at the center of these images.

\begin{table*}[htb]
\footnotesize
\resizebox{\textwidth}{!}{%
\begin{tabular}{C{3cm}C{2.1cm}C{1.5cm}C{1.5cm}C{1.2cm}C{1.2cm}C{1.3cm}C{1.2cm}C{1.2cm}C{1.2cm}C{0.8cm}}
\toprule
Name &   $\log_{10} J$  &  $\log_{10} M_\text{vir}$ &          $z \times 10^{3}$&        $\ell$ &        $b$ &  $\log_{10} c_\text{vir}$  & $\theta_\text{s}$ &   $b_\text{sh}$ & TS$_\text{max}$ &  Incl. \\
& {[GeV$^2$ cm$^{-5}$ sr]} & [$M_\odot$] &  & [deg] & [deg] & & [deg] & &\\
\midrule
             Andromeda &  19.79$\pm$0.36 &  12.4$\pm$0.12 &   0.17 &  121.51 & -21.79 &  1.04$\pm$0.17 &     2.57 &  2.64 &   2.92 &             \\
         NGC4472/Virgo &  19.11$\pm$0.35 &  14.6$\pm$0.14 &   3.58 &  283.94 &  74.52 &  0.80$\pm$0.18 &     1.15 &  4.53 &   1.04 &  \checkmark \\
               NGC5128 &  18.89$\pm$0.37 &  12.9$\pm$0.12 &   0.82 &  307.88 &  17.08 &  0.99$\pm$0.17 &     0.88 &  3.14 &   0.00 &             \\
               NGC0253 &  18.76$\pm$0.37 &  12.7$\pm$0.12 &   0.79 &   98.24 & -87.89 &  1.00$\pm$0.17 &     0.77 &  2.90 &   0.63 &  \checkmark \\
              Maffei 1 &  18.68$\pm$0.37 &  12.6$\pm$0.12 &   0.78 &  136.23 &  -0.44 &  1.01$\pm$0.17 &     0.71 &  2.81 &   7.26 &             \\
              NGC6822 &  18.59$\pm$0.37 &  10.7$\pm$0.10 &   0.11 &   25.34 & -18.40 &  1.17$\pm$0.17 &     0.77 &  1.70 &  16.65 &             \\
               NGC3031 &  18.58$\pm$0.36 &  12.6$\pm$0.12 &   0.83 &  141.88 &  40.87 &  1.02$\pm$0.17 &     0.64 &  2.76 &   0.00 &  \checkmark \\
     NGC4696/Centaurus &  18.33$\pm$0.35 &  14.6$\pm$0.14 &   8.44 &  302.22 &  21.65 &  0.80$\pm$0.18 &     0.47 &  4.50 &   6.60 &  \checkmark \\
               NGC1399 &  18.30$\pm$0.37 &  13.8$\pm$0.13 &   4.11 &  236.62 & -53.88 &  0.89$\pm$0.17 &     0.45 &  3.87 &   0.72 &  \checkmark \\
                IC0356 &  18.26$\pm$0.36 &  13.5$\pm$0.13 &   3.14 &  138.06 &  12.70 &  0.92$\pm$0.17 &     0.43 &  3.51 &   0.02 &             \\
               NGC4594 &  18.26$\pm$0.35 &  13.3$\pm$0.13 &   2.56 &  299.01 &  51.30 &  0.94$\pm$0.17 &     0.43 &  3.36 &   0.00 &  \checkmark \\
               IC1613 &  18.17$\pm$0.37 &  10.6$\pm$0.10 &   0.17 &  129.74 & -60.58 &  1.18$\pm$0.17 &     0.48 &  1.67 &   1.72 &             \\
  Norma &  18.16$\pm$0.33 &  15.1$\pm$0.15 &  17.07 &  325.29 &  -7.21 &  0.74$\pm$0.18 &     0.39 &  5.17 &   0.00 &  \checkmark \\
               NGC4736 &  18.12$\pm$0.36 &  12.2$\pm$0.12 &   1.00 &  124.83 &  75.76 &  1.05$\pm$0.17 &     0.38 &  2.58 &   0.00 &             \\
      NGC1275/Perseus &  18.12$\pm$0.33 &  15.0$\pm$0.15 &  17.62 &  150.58 & -13.26 &  0.75$\pm$0.18 &     0.37 &  5.16 &   0.93 &  \checkmark \\
               NGC3627 &  18.11$\pm$0.35 &  13.0$\pm$0.13 &   2.20 &  241.46 &  64.36 &  0.98$\pm$0.17 &     0.35 &  3.23 &  27.24 &             \\
        NGC1316/Fornax &  18.01$\pm$0.36 &  13.5$\pm$0.13 &   4.17 &  239.98 & -56.68 &  0.92$\pm$0.17 &     0.32 &  3.49 &   2.33 &             \\
               NGC5236 &  18.01$\pm$0.36 &  12.2$\pm$0.12 &   1.09 &  314.58 &  31.98 &  1.05$\pm$0.17 &     0.33 &  2.56 &  22.08 &             \\
                IC0342 &  18.00$\pm$0.37 &  11.8$\pm$0.11 &   0.73 &  138.52 &  10.69 &  1.09$\pm$0.17 &     0.34 &  2.33 &   1.92 &             \\
               NGC4565 &  17.97$\pm$0.35 &  13.1$\pm$0.13 &   2.98 &  229.92 &  86.07 &  0.96$\pm$0.17 &     0.30 &  3.28 &  41.15 &             \\
 Coma &  17.96$\pm$0.33 &  15.2$\pm$0.15 &  24.45 &   57.20 &  87.89 &  0.73$\pm$0.18 &     0.31 &  5.21 &   2.35 &  \checkmark \\
        NGC1553/Dorado &  17.94$\pm$0.36 &  13.4$\pm$0.13 &   4.02 &  265.56 & -43.51 &  0.94$\pm$0.17 &     0.30 &  3.41 &   0.08 &  \checkmark \\
         NGC3311/Hydra &  17.94$\pm$0.34 &  14.4$\pm$0.14 &  10.87 &  269.55 &  26.41 &  0.82$\pm$0.17 &     0.30 &  4.32 &   0.04 &  \checkmark \\
               NGC3379 &  17.93$\pm$0.37 &  12.9$\pm$0.12 &   2.42 &  233.64 &  57.77 &  0.99$\pm$0.17 &     0.29 &  3.11 &   0.00 &  \checkmark \\
               NGC5194 &  17.93$\pm$0.37 &  12.6$\pm$0.12 &   1.84 &  104.86 &  68.53 &  1.01$\pm$0.17 &     0.30 &  2.81 &   4.94 &  \checkmark \\
        \bottomrule
\end{tabular}}
\caption{The top 25 halos included from the T15~\cite{Tully:2015opa} and T17~\cite{2017ApJ...843...16K} catalogs, as ranked by inferred $J$-factor, which includes the boost factor.  For each group, we show the brightest central galaxy and the common name, if one exists, as well as the virial mass, cosmological redshift, Galactic coordinates, inferred concentration using Ref.~\cite{Correa:2015dva}, angular extension, boost factor using the fiducial model from Ref.~\cite{Bartels:2015uba}, and the maximum test statistic (TS$_\text{max}$) over all $m_\chi$ between the model with and without DM annihilating to $b \bar b$. A checkmark indicates that the halo satisfies the selection criteria and is included in the stacking analysis.  A complete listing of all the halos used in this study is provided as Supplementary Data.
}
\label{tab:tully_extended}
\end{table*}

\afterpage{
\begin{figure*}[p]
  \centering
  \includegraphics[width=0.9\textwidth]{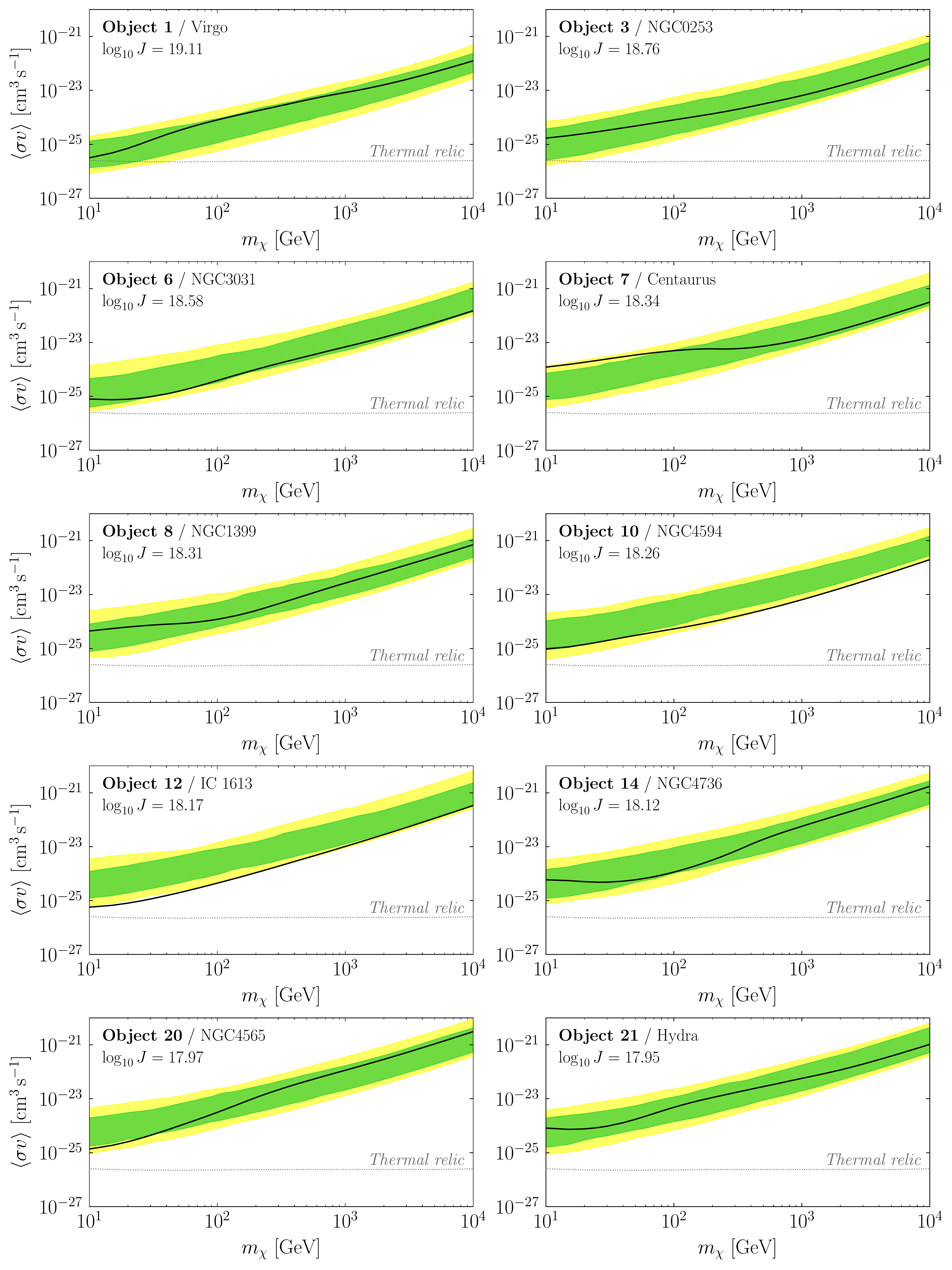}
  \caption{The 95\% confidence limit on the DM annihilation cross section to the $b \bar b$ final state for each of the top ten halos listed in Tab.~\ref{tab:tully_extended} that pass the selection cuts. For each halo, we show the 68\% and 95\% containment regions (green and yellow, respectively), which are obtained by placing the halo at 200 random sky locations.  The inferred ${J}$-factors, assuming the fiducial boost factor model~\cite{Bartels:2015uba}, are provided for each object.}
  \label{fig:individual_lims}
\end{figure*}
\clearpage}

\afterpage{
\begin{figure*}[htbp]
 \centering
  \includegraphics[width=0.9\textwidth]{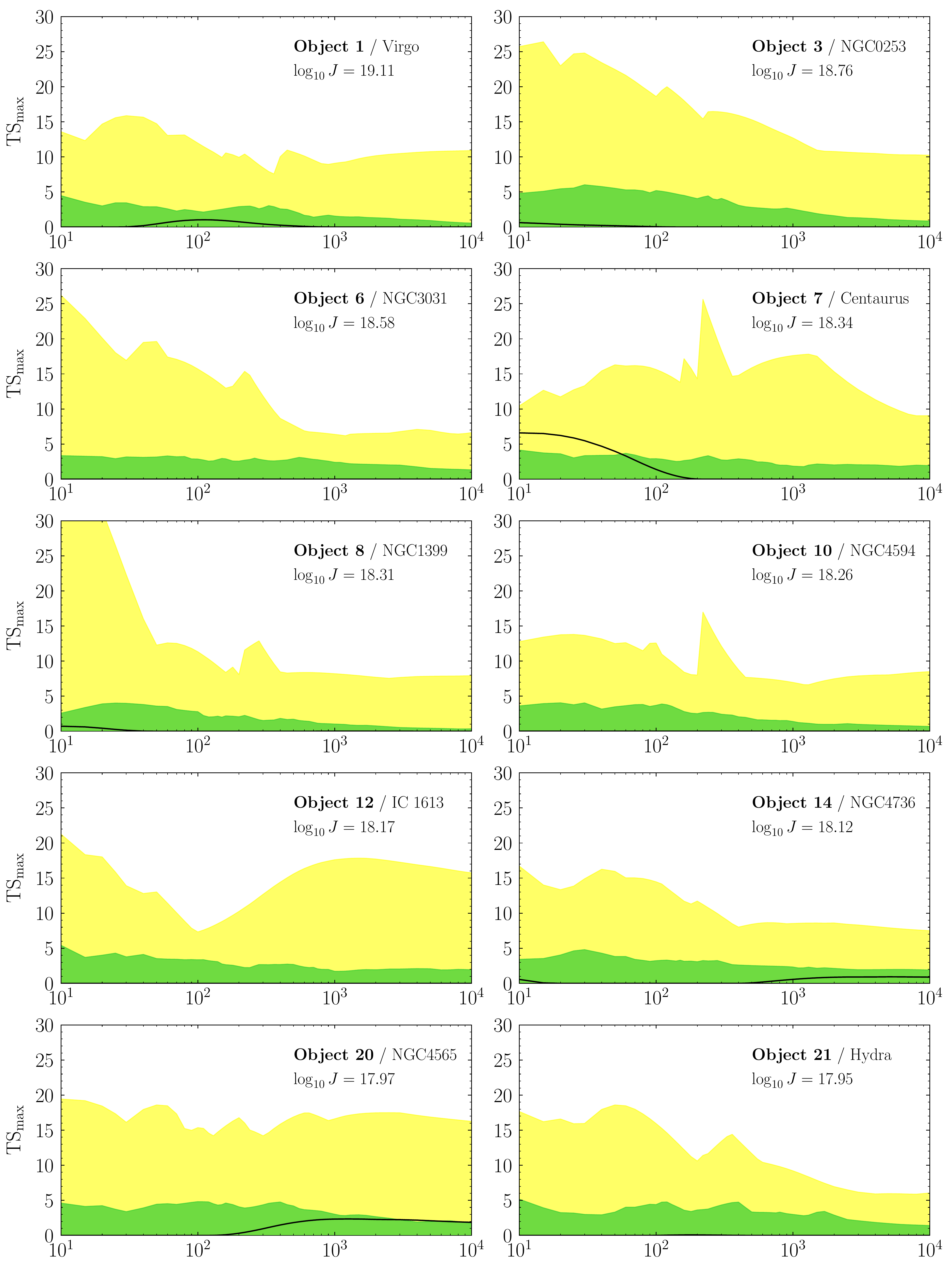}
 \caption{Same as Fig.~\ref{fig:individual_lims}, except showing the maximum test statistic (TS$_\text{max}$) for each individual halo, as a function of DM mass. These results correspond to the $b \bar b$ annihilation channel.}
  \label{fig:individual_maxts}
\end{figure*}
\clearpage}

\afterpage{
\begin{figure*}[htbp]
 \centering
  \includegraphics[width=0.9\textwidth]{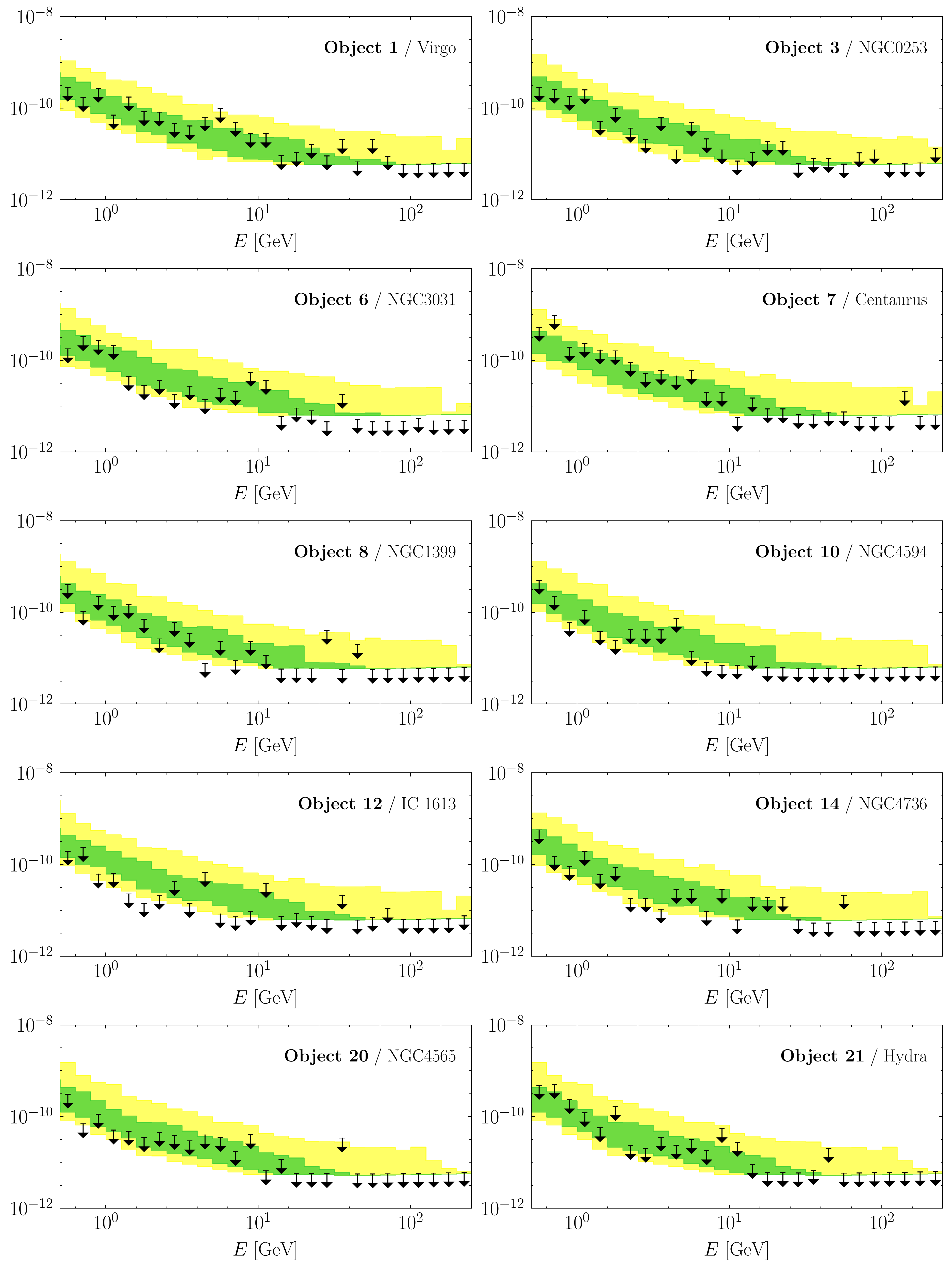}
 \caption{Same as Fig.~\ref{fig:individual_lims}, except showing the 95\% upper limit on the gamma-ray flux correlated with the DM annihilation profile in each halo.  We use 26 logarithmically spaced energy bins between 502~MeV and 251~GeV. 
 }
  \label{fig:individual_flux}
\end{figure*}
\clearpage}

\begin{figure*}[htbp]
  \centering
  \includegraphics[width=0.98\textwidth]{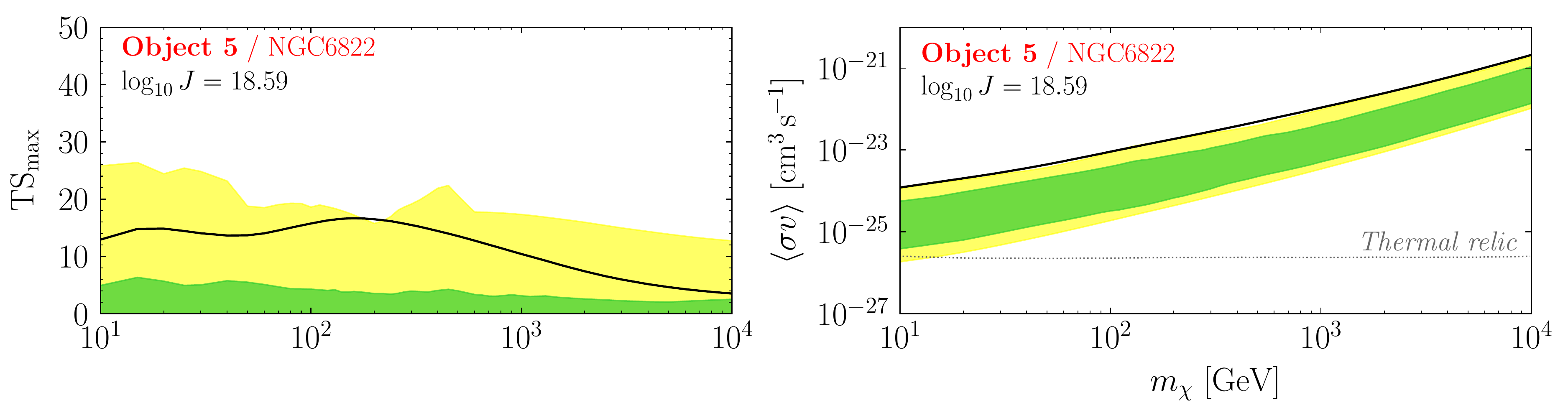}
  \caption{NGC6822 has one of the largest $J$-factors of the objects in the catalog, but it fails the selection requirements because of its proximity to the Galactic plane.  We show the analog of Fig.~\ref{fig:individual_maxts} \textbf{(left)} and Fig.~\ref{fig:individual_lims} \textbf{(right)}. We see that this object  has a broad TS$_\text{max}$ excess over many masses and a weaker limit than expected from random sky locations.}
  \label{fig:maxTSoneobject}
\end{figure*}

\afterpage{
\begin{figure*}[htbp]
  \centering
  \includegraphics[width=0.9\textwidth]{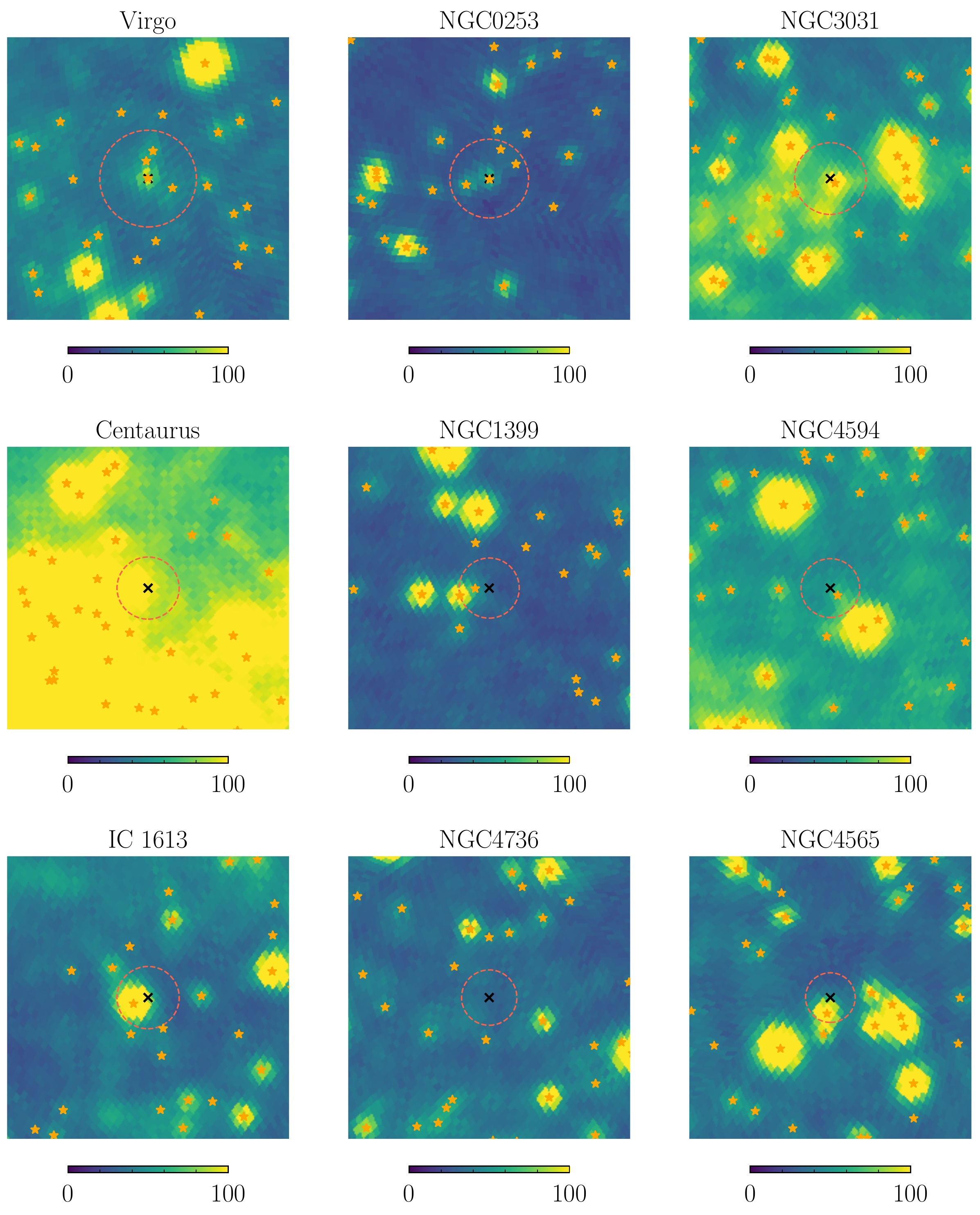}
  \caption{The \textit{Fermi}-LAT data centered on the top nine  halos that are included in the stacked sample. We show the photon counts (for the energies analyzed) within a $20^\circ$$\times$$20^\circ$ square centered on the region of interest. The dotted circle shows the scale radius $\theta_\mathrm{s}$, which is a proxy for the scale of DM annihilation, and the orange stars indicate the \emph{Fermi} 3FGL point sources.}
  \label{fig:individual_skyrois}
\end{figure*}
\clearpage}

\newpage

\section{Variations on the Analysis}
\label{sec:systematics}

We have performed a variety of systematic tests to understand the robustness of the results presented in the main body of the analysis.  Several of these uncertainties are discussed in detail in Ch.~\ref{ch:groups_sim}; here, we focus specifically on how they affect the results of the data analysis.  \vspace{0.1in}

\begin{figure*}[b]
  \centering
	\includegraphics[width=.45\textwidth]{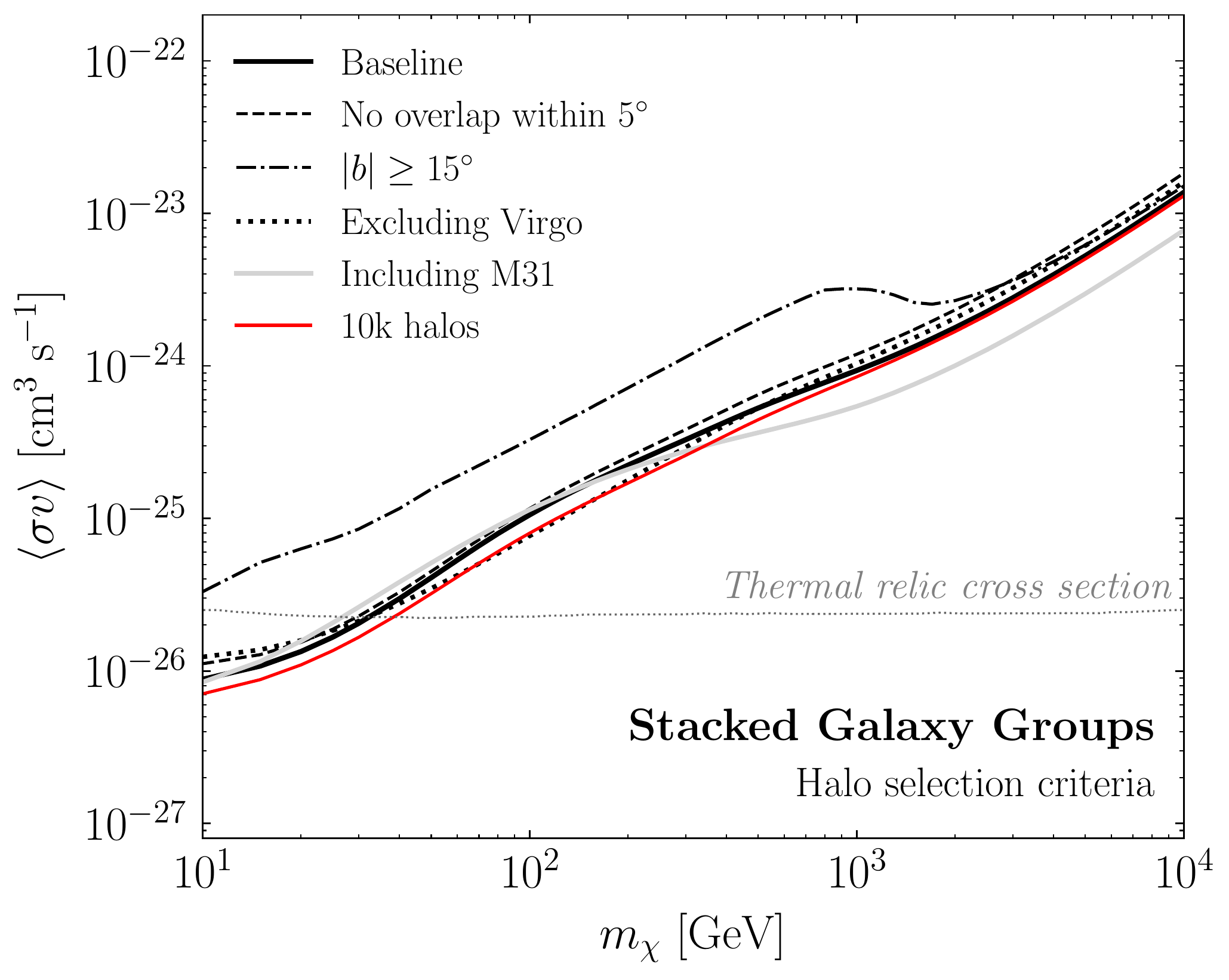} 
	\includegraphics[width=.45\textwidth]{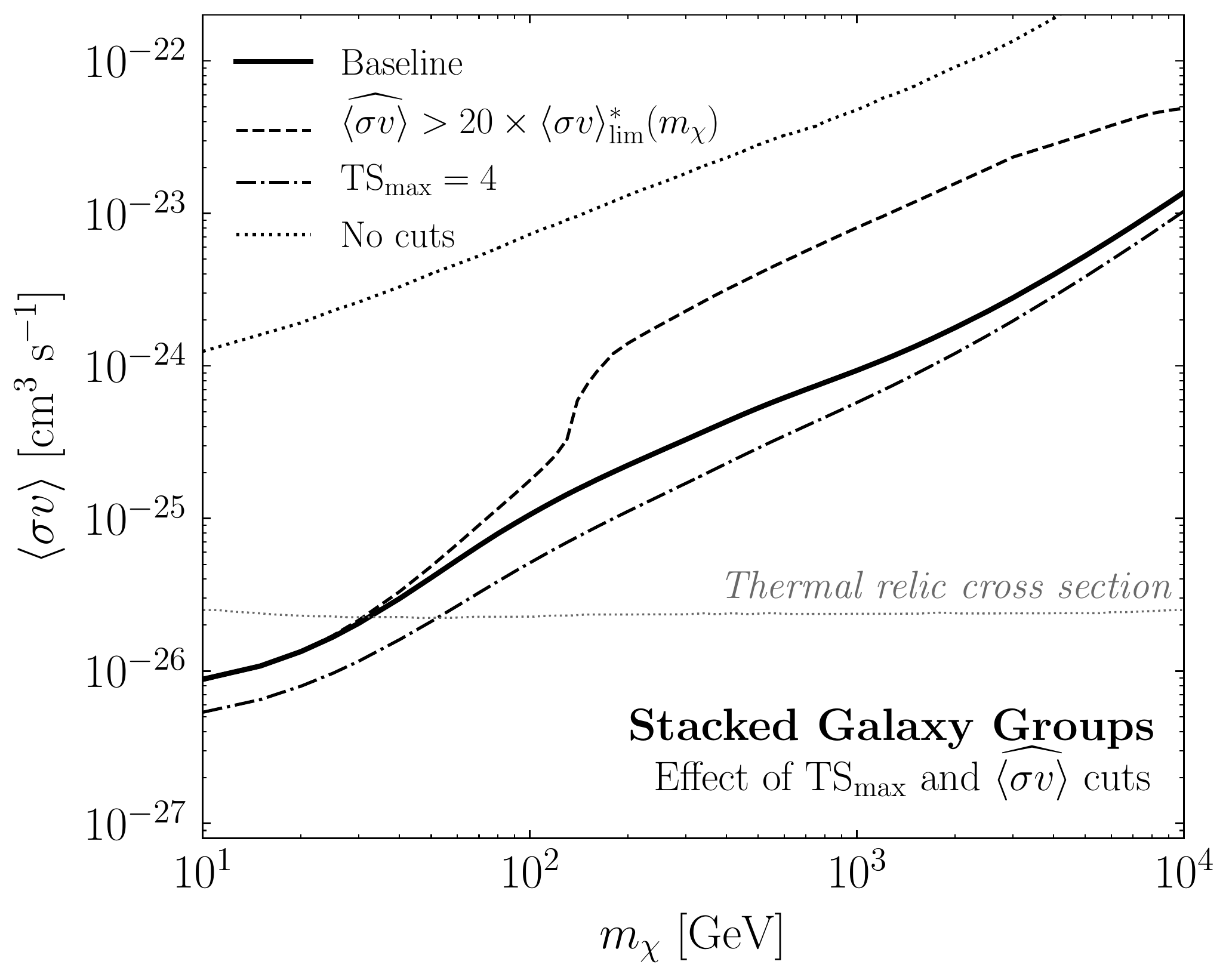} 
  \caption{The same as the baseline analysis shown in Fig.~\ref{fig:bounds1} of the main analysis, except varying several assumptions made in the analysis.  \textbf{(Left)} We show the effect of relaxing the overlapping halo criterion to $5^\circ$ (dashed), reducing the latitude cut to $|b|\geq 15^\circ$ (dot-dashed), excluding Virgo (dotted), and including Andromeda (gray).  The limit obtained when starting from an initial 10,000 halos is shown as the red line.  \textbf{(Right)} We show the effect of strengthening the cross section (dashed) or weakening the TS$_\text{max}$ (dot-dashed) selection criteria, as well as completely removing the TS$_\text{max}$ and cross section cuts (dotted). }
  \label{fig:cutsandhalos}
\end{figure*}

\noindent  {\bf Halo Selection Criteria.}  
Here, we demonstrate how variations on the halo selection conditions listed above affect the baseline results of Fig.~\ref{fig:bounds1}.  In the left panel of Fig.~\ref{fig:cutsandhalos}, the red line shows the limit that is obtained when starting with 10,000 halos instead of 1000, but requiring the same selection conditions.  Despite the modest improvement in the limit, we choose to use 1000 halos in the baseline study because systematically testing the robustness of the analysis procedure, as done in Ch.~\ref{ch:groups_sim}, becomes computationally prohibitive otherwise. In order to calibrate the analysis for higher halo numbers, it would be useful to use semi-analytic methods to project the sensitivity, such as those discussed in Ref.~\cite{Cowan:2010js,Edwards:2017mnf}, although we leave the details to future work.

Virgo is the object with the highest $J$-factor in the stacked sample. As made clear in the dedicated study of this object by the \emph{Fermi} Collaboration~\cite{Ackermann:2015fdi}, there are challenges associated with modeling the diffuse emission in Virgo's vicinity.  However, we emphasize that the baseline limit is not highly sensitive to any one halo, including the brightest in the sample.  For example, the dotted line in the left panel of Fig.~\ref{fig:cutsandhalos} shows the impact on the limit after removing Virgo from the stacking. Critically, we see that the limit is almost unchanged, highlighting that the stacked result is not solely driven by the object with the largest $J$-factor.

The effect of including Andromeda (M31) is shown as the gray solid line. We exclude Andromeda from the baseline analysis because of its large angular size, as discussed in detail above. Our analysis relies on the assumption that the DM halos are approximately point-like on the sky, which fails for Andromeda, and we therefore deem it to fall outside the scope of the systematic studies performed here.

The dashed line shows the effect of tightening the condition on overlapping halos from $2^\circ$ to $5^\circ$. Predictably, the limit is slightly weakened due to the smaller pool of available targets.  We also show the effect of decreasing the latitude cut to $b\geq 15^\circ$ (dot-dashed line). In this case, the number of halos included in the stacked analysis increases, but the limit is weaker---considerably so below $m_\chi \sim 10^3$~GeV.  The weakened limits are likely due to enhanced diffuse emission along the plane as well as contributions from unresolved point sources, both of which are difficult to accurately model. In cases with such mismodeling, the addition of a DM template can generically improve the quality of the fit, which leads to excesses at low energies, in particular.  The baseline latitude cut ameliorates  precisely these concerns.

The right  panel of Fig.~\ref{fig:cutsandhalos} illustrates the effects of changing, or removing completely, the cross section and TS$_\text{max}$ cuts on the halos.  Specifically, the dashed black line shows what happens when we require that a halo's excess be even more inconsistent with the limits set by other galaxy groups; specifically, requiring that $(\sigma v)_\text{best} > 20 \times (\sigma v)^*_\text{lim}$. The dot-dashed line shows the limit when we decrease the statistical significance requirement to $\text{TS}_\text{max} > 4$.  
Note that the two changes have opposite effects on the limits.  This is expected because more halos with excesses are included in the stacking procedure with the more stringent cross section requirement, which weakens the limit, whereas fewer are included if we reduce the TS$_\text{max}$ cut, strengthening the limit.  

\begin{figure*}[tb]
  \centering
  \includegraphics[width=.7\textwidth]{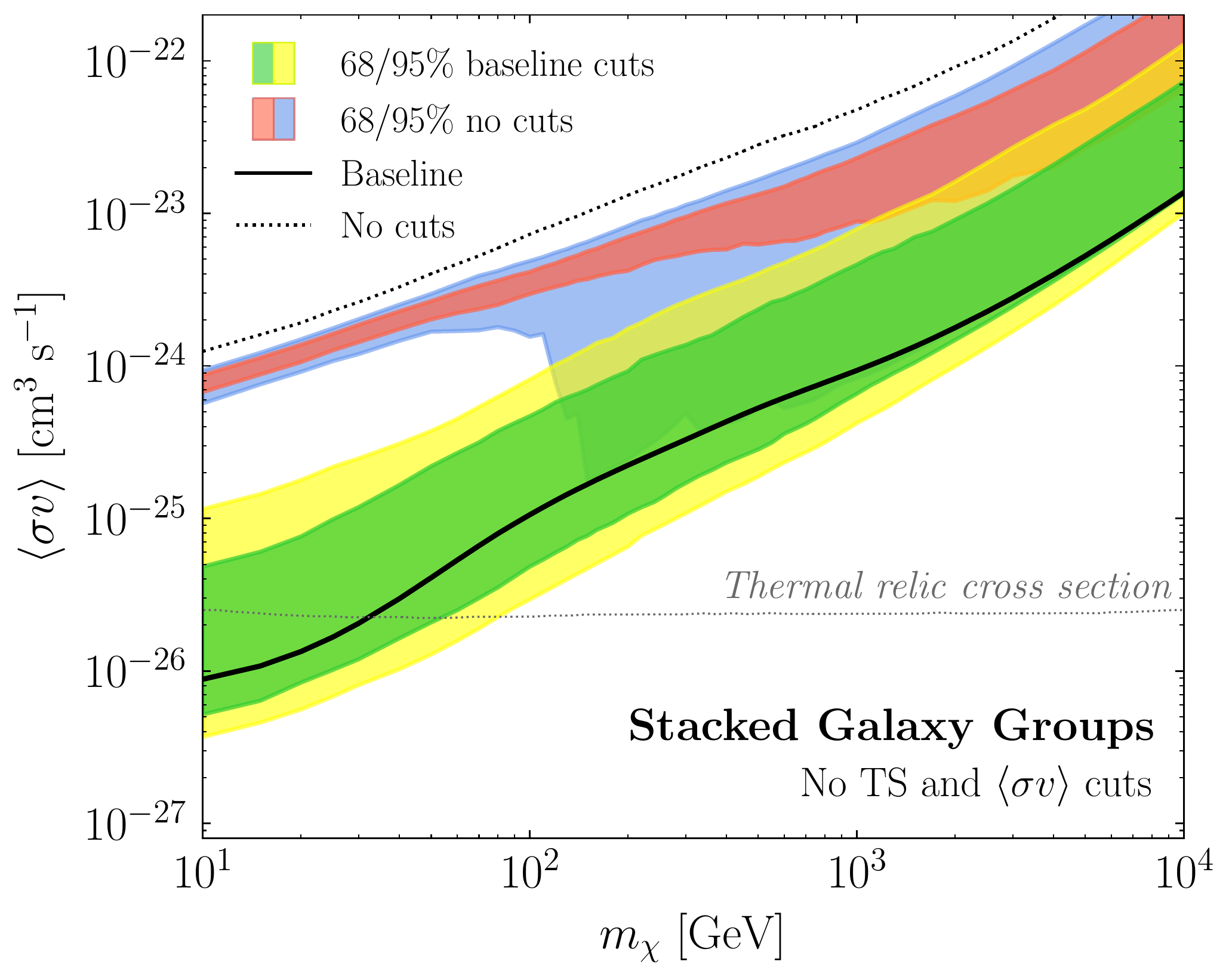}
  \caption{The results of the baseline analysis with the default cuts, as shown in Fig.~\ref{fig:bounds1}, compared to the corresponding result when no cuts are placed on the TS$_\text{max}$ or cross section of the halos in the catalog.  The significant offset between the limit obtained with no cuts (dotted line) and the corresponding expectation from random sky locations (red/blue band)  demonstrates that many of the objects that are removed by the TS$_\text{max}$ and cross section cuts are  legitimately associated with astrophysical emission. See text for details.}
  \label{fig:systematics_nots_cuts}
\end{figure*}

The dotted line in the right panel of Fig.~\ref{fig:cutsandhalos} shows  what happens when no requirement at all is placed on the TS$_\text{max}$ and cross section; in this case, the limit is dramatically weakened by several orders of magnitude.   We show the same result in Fig.~\ref{fig:systematics_nots_cuts} (dotted line), but with a comparison to the null hypothesis corresponding to no TS$_\text{max}$ and cross section cuts, which is shown as the 68\%~(95\%) red~(blue) bands.\footnote{We thank A.~Drlica-Wagner for suggesting this test.}    In the baseline case, the limit is consistent with the random sky locations---\emph{i.e.}, the solid black line falls within the green/yellow bands.  However, with no TS$_\text{max}$ and cross section cuts, this is no longer true---\emph{i.e.}, the dotted black line falls outside the red/blue bands.  Clear excesses are observed above the background expectation in this case, but they are inconsistent with a DM interpretation as they are strongly excluded by other halos in the stack.  When deciding on the TS$_\text{max}$ and cross section requirements that we used for the baseline analysis in Fig.~\ref{fig:bounds1}, our goal was to maximize the sensitivity reach while simultaneously ensuring that an actual DM signal would not be excluded.  We verified the selection criteria thoroughly by performing injected signal tests on the data (discussed above) as well as on mock data (discussed in Ch.~\ref{ch:groups_sim}).  Ideally, galaxy groups would be excluded from the stacking based on the specific properties of the astrophysical excesses that they exhibit, as opposed to the TS$_\text{max}$ and cross section requirements used here.  For example, one can imagine excluding groups that are known to host  AGN or galaxies with high amounts of star-formation activity.  We plan to study such possibilities in future work.        \vspace{0.1in}

\begin{figure*}[t]
  \centering
  \includegraphics[width=.45\textwidth]{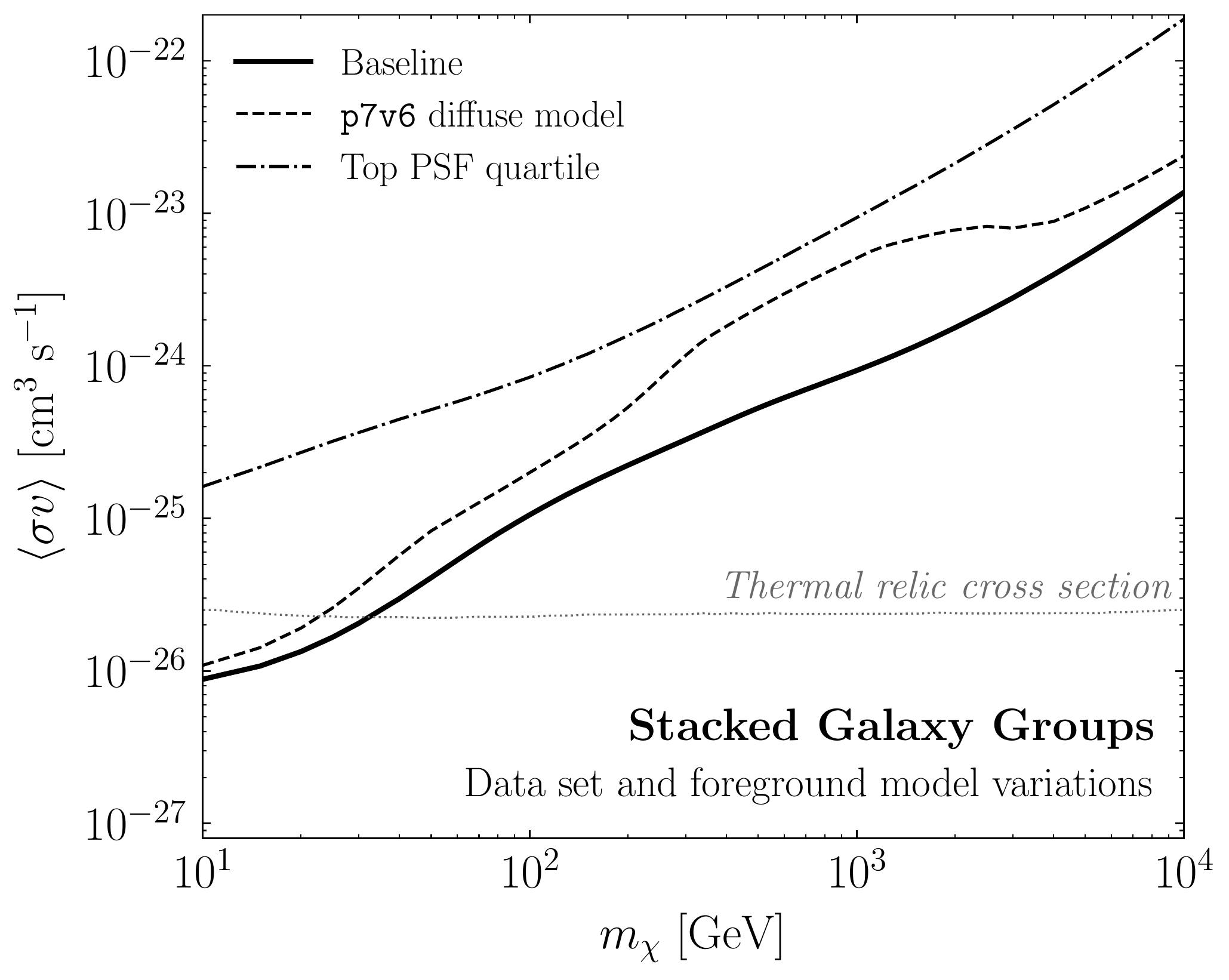}
  \includegraphics[width=.45\textwidth]{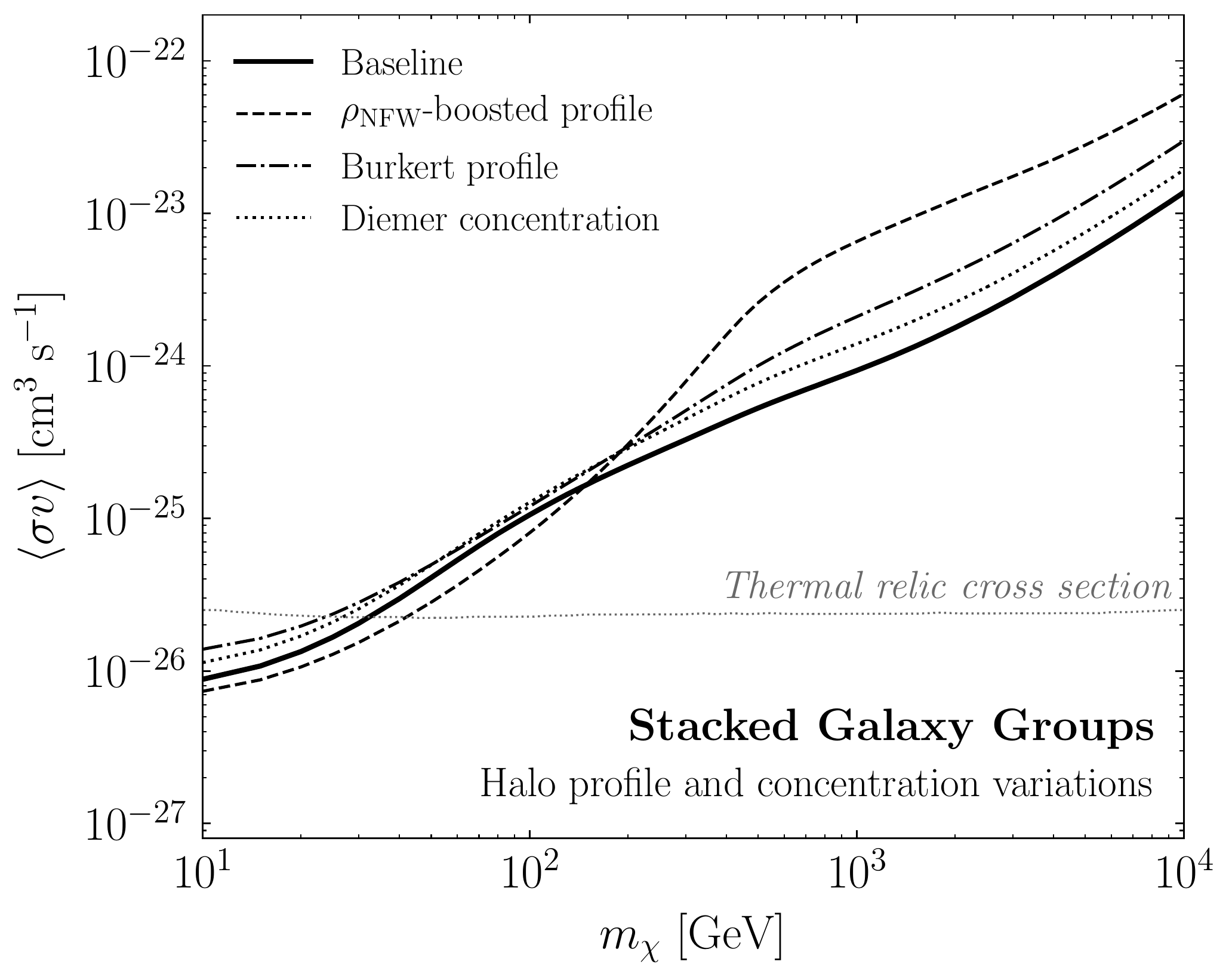}
  \caption{The same as the baseline analysis shown in Fig.~\ref{fig:bounds1} of the main analysis, except varying several assumptions made in the analysis.  \textbf{(Left)} We show the effect of using the top PSF quartile of the UltracleanVeto data set (dot-dashed) and the \texttt{p7v6} diffuse model (dashed). \textbf{(Right)} We show the effect of using the cored Burkert profile~\cite{Burkert:1995yz} (dot-dashed) and the Diemer~and~Kravtsov concentration model~\cite{Diemer:2014gba} (dotted).  The ``$\rho_\text{NFW}$-boosted profile'' (dashed) shows what happens when the annihilation flux from the subhalo boost is assumed to follow the NFW profile (as opposed to a squared-NFW profile). }
  \label{fig:systematics_data_profile}
\end{figure*}

\noindent  {\bf Data Set and Foreground Models.}  
In the results presented thus far, we have used all quartiles of the UltracleanVeto event class of the {\it Fermi} data.  Alternatively, we can restrict ourselves to the top quartile of events, as ranked by PSF.  Using this subset of data has the advantage of improved angular resolution, but the disadvantage of a $\sim$75\% reduction in statistics.  The  left panel of Fig.~\ref{fig:systematics_data_profile} shows the limit (dot-dashed line) obtained by repeating the analysis with the top quartile of UltracleanVeto data; the bounds are weaker than in the all-quartile case, as would be expected.  However, the amount by which the limit weakens is not completely consistent with the decrease in statistics.  Rather, it appears that when we lower the photon statistics, more halos that were previously excluded by the cross section and TS$_\text{max}$ criteria in the baseline analysis are allowed into the stacking and collectively weaken the limit.

Another choice that we made for the baseline analysis was to use the \texttt{p8r2} foreground model for gamma-ray emission from cosmic-ray processes in the Milky Way.   In this model, the bremsstrahlung and boosted pion emission are traced with gas column-density maps and the IC emission is modeled using \texttt{Galprop}~\cite{Strong:2007nh}.  After fitting the data with these three components, any `extended emission excesses' are identified and added back into the foreground model~\cite{Acero:2016qlg}.  To study the dependence of the results on the choice of foreground model, we repeat the analysis using the Pass~7 \emph{gal\_2yearp7v6\_v0.fits} (\texttt{p7v6}) model, which includes large-scale structures like Loop~1 and the \emph{Fermi} bubbles---in addition to the bremsstrahlung, pion, and IC emission---but does not account for any data-driven excesses as is done in \texttt{p8r2}.  The results of the stacked analysis using the \texttt{p7v6} model are shown in the left panel of Fig.~\ref{fig:systematics_data_profile} (dashed line).  The limit is somewhat weaker to that obtained using \texttt{p8r2}, though it is broadly similar to the latter.  This is to be expected for stacked analyses, where the dependence on mismodeling of the foreground emission is reduced because the fits are done on small, independent regions of the sky, so that offsets in the point-to-point normalizations of the diffuse model can have less impact. For more discussion of this point, see Ref.~\cite{Daylan:2014rsa,Linden:2016rcf,Narayanan:2016nzy,Cohen:2016uyg}.\vspace{0.1in}

\noindent  {\bf Halo Density Profile and Concentration.} Our baseline analysis makes two assumptions about the profiles of gamma-ray emission from the extragalactic halos.  The first assumption is that the DM profile of the smooth halo is described by an NFW profile:
\begin{equation}
\rho_{\rm NFW}(r) = \frac{\rho_s}{r/r_s\,(1+r/r_s)^2}\,,
\end{equation}
where $\rho_s$ is the normalization and $r_s$ the scale radius~\cite{Navarro:1996gj}.  The NFW profile successfully describes the shape of cluster-size DM halos in $N$-body simulations with and without baryons (see, {\it e.g.}, Ref.~\cite{Springel:2008cc,Schaller:2014uwa}).  However, some evidence exists pointing to cored density profiles on smaller scales ({\it e.g.}, dwarf galaxies), and the density profiles in these systems may be better described by the phenomenological Burkert profile~\cite{Burkert:1995yz}:
\begin{equation}
\rho_{\rm Burkert}(r) = \frac{\rho_B}{(1+r/r_B)(1+(r/r_B)^2)}\,,
\end{equation}
where $\rho_B$ and $r_B$ are the Burkert corollaries to the NFW $\rho_s$ and $r_s$, but have numerically different values. While it appears unlikely that the Burkert profile is a good description of the DM profiles of the cluster-scale halos considered here, using this profile provides a useful systematic variation because it predicts less annihilation flux than the NFW profile does.  The right panel of Fig.~\ref{fig:systematics_data_profile} shows the effect of using the Burkert profile to describe the halos in the T15 and T17 catalogs (dot-dashed line); the limit is slightly weaker, as expected.

The second assumption we made is that the shape of the gamma-ray emission from DM annihilation follows the projected integral of the DM-distribution squared.  This is likely incorrect because the contribution from the boost factor, which can be substantial, should have the spatial morphology of the distribution of DM subhalos.  Neglecting tidal effects, we expect the subhalos to follow the DM distribution (instead of the squared distribution).  Including tidal effects is complicated, as subhalos closer to the halo center are more likely to be tidally stripped, which both increases their concentration and decreases their number density.  We do not attempt to model the change in the spatial morphology of the subhalo distribution from tidal stripping and instead consider the limit where the annihilation flux from the subhalo boost follows the NFW distribution.  This gives a much wider angular profile for the annihilation flux for large clusters,  compared to the case where the boost is simply a multiplicative factor.  The dashed line in the right panel of Fig.~\ref{fig:systematics_data_profile} shows the effect on the limit of modeling the gamma-ray emission in this way (labeled ``$\rho_\text{NFW}$-boosted profile").  The extended spatial profile leads to a minimal change in the limit over most of the mass range, which is to be expected given that most of the galaxy groups can be well-approximated as point sources.

A halo's virial concentration is an indicator of its overall density and is defined as $c_\text{vir} \equiv r_\text{vir}/r_s$, where $r_\text{vir}$ is the virial radius and $r_s$ the NFW scale radius of the halo.  A variety of models exist in the literature that map from halo mass to concentration.  Our fiducial case is the Correa \emph{et al.} model from Ref.~\cite{Correa:2015dva}.  Here we show how the limit (dotted line) changes when we use the model of Diemer and Kravtsov~\cite{Diemer:2014gba}, updated with the Planck 2015 cosmology~\cite{Ade:2015xua}.  The change to the limit is minimal, which is perhaps a reflection of the fact that the change in the mean concentrations between the concentration-mass models is small compared to the statistical spread predicted in these models, which is incorporated into the $J$-factor uncertainties.  We have also verified that increasing the dispersion on the concentration for the Correa~\emph{et al.} model to 0.24~\cite{Bullock:1999he}, which is above the 0.14--0.19 range used in the baseline study, worsens the limit by a $\mathcal{O}(1)$ factor.\vspace{0.1in}

\begin{figure}[t]
  \centering
  \includegraphics[width=0.46\textwidth]{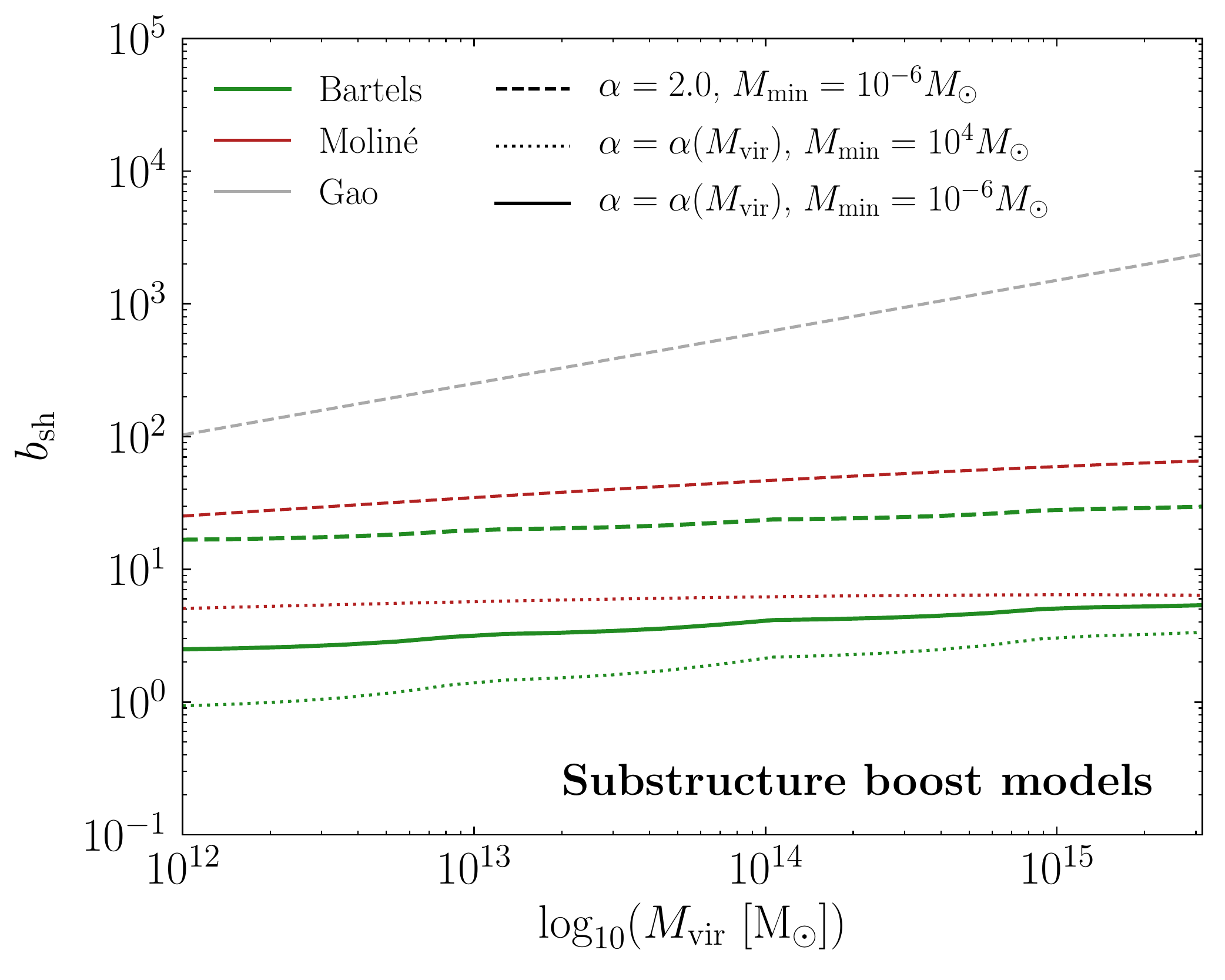} 
     \includegraphics[width=.45\textwidth]{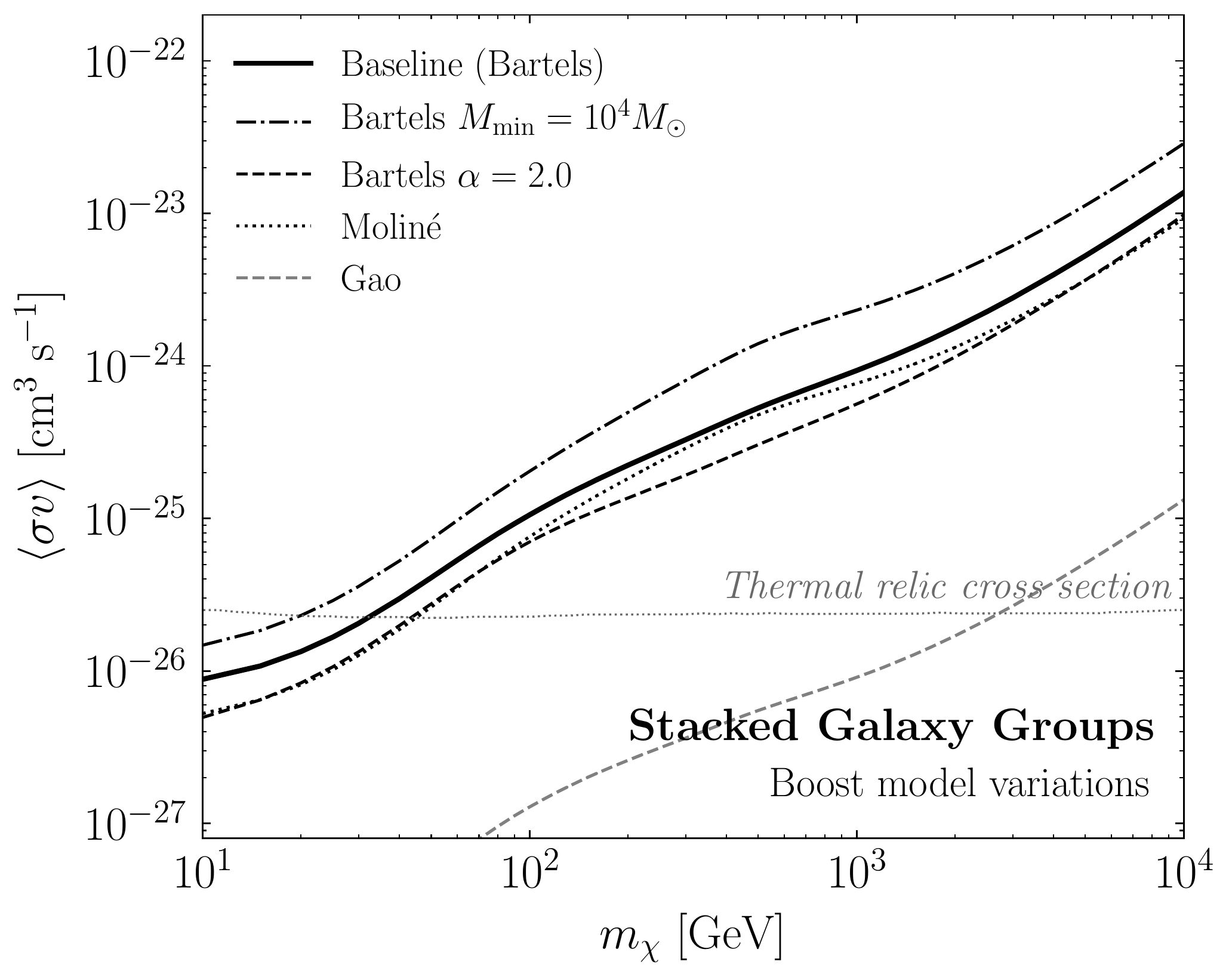}
  \caption{ \textbf{(Left)}) Examples of substructure boost models commonly used in the literature, reproduced from Ch.~\ref{ch:groups_sim}. Our fiducial model, based on Ref.~\cite{Bartels:2015uba} using $M_\text{min} = 10^{-6}$~M$_\odot$ and self-consistently computing $\alpha$, is shown as the thick green solid line. Variations on $M_\text{min}$ and $\alpha$ are shown with the dotted and dashed lines, respectively. Also plotted are the boost models of Molin\'e~\cite{Moline:2016pbm} (red) and Gao~\cite{Gao:2011rf} (grey).   \textbf{(Right)} The same as the baseline analysis shown in Fig.~\ref{fig:bounds1} of the main analysis, except varying the boost model.
  }
  \label{fig:systematics_boost}
\end{figure}

\noindent  {\bf Substructure Boost.}  
Hierarchical structure formation implies that larger structures can host smaller substructures, the presence of which can significantly enhance signatures of DM annihilation in host halos. Although several models exist in the literature to characterize this effect, the precise enhancement sensitively depends on the methods used as well as the astrophysical and particle physics properties that are assumed.  Phenomenological extrapolation of subhalo properties (\emph{e.g.}, the concentration-mass relation) over many orders of magnitude down to very small masses $\mathcal O(10^{-6}$)~M$_{\odot}$ lead to large enhancements of $\mathcal O(10^{2})$ and $\mathcal O(10^{3})$ for galaxy- and cluster-sized halos, respectively~\cite{Gao:2011rf}. Recent numerical simulations and analytic studies~\cite{Anderhalden:2013wd,Correa:2015dva,Ludlow:2013vxa} suggest that the concentration-mass relation flattens at smaller masses, yielding boosts that are much more modest, about an order-of-magnitude below phenomenological extrapolations~\cite{Nezri:2012tu,Sanchez-Conde:2013yxa}.  In addition, the concentration-mass relation for field halos cannot simply be applied to subhalos, because the latter undergo tidal stripping as they fall into and orbit their host.  Such effects tend to make the subhalos more concentrated---and therefore more luminous---than their field-halo counterparts, though the number-density of such subhalos is also reduced~\cite{Bartels:2015uba}.    

When taken together, the details of the halo formation process shape the subhalo mass function $dn/dM_\text{sh}\propto M_\text{sh}^{-\alpha}$, where $\alpha  \in \left[1.9, 2.0\right]$.  The mass function does not follow a power-law to arbitrarily low masses, however, because the underlying particle physics model for the DM can place a minimum cutoff on the subhalo mass, $M_\text{min}$.  For example, DM models with longer free-streaming lengths wash out smaller-scale structures, resulting in higher cutoffs.

The left panel of Fig.~\ref{fig:systematics_boost} shows a variety of boost models commonly used in DM studies. The fiducial boost model used here~\cite{Bartels:2015uba} is shown as the thick green solid line and variations on $M_\text{min}$ and $\alpha$ are also plotted. The right panel of Fig.~\ref{fig:systematics_boost} shows that the expected limit when $M_\text{min} = 10^4$~M$_\odot$ instead of $M_\text{min} = 10^{-6}$~M$_\odot$ (dot-dashed) is weaker across all masses.  While a minimum subhalo mass of  $10^{4}$~M$_\odot$ is likely inconsistent with bounds on the kinetic decoupling temperature of thermal DM, this example illustrates the importance played by $M_\text{min}$ in the sensitivity reach.  Additionally, Fig.~\ref{fig:systematics_boost} demonstrates the case where $\alpha=2.0$ (dashed line).  Increasing the inner slope of the subhalo mass function leads to a correspondingly stronger limit, however observations tend to favor a slope closer to $\alpha = 1.9$ (which is what the most massive halos correspond to in our fiducial case).

Ref.~\cite{Sanchez-Conde:2013yxa} derived a boost factor model that accounts for the flattening of the concentration-mass relation at low masses, but does not include the effect of tidal stripping.  They assume a minimum sub-halo mass of $10^{-6}$~M$_\odot$ and a halo-mass function $dN/dM \sim M^{-2}$.  This was updated by Ref.~\cite{Moline:2016pbm} to account for the effect of tidal disruption. This updated boost factor model, which takes $\alpha = 1.9$, gives the constraint shown in Fig.~\ref{fig:systematics_boost} labeled ``Molin\'e" (dotted).  This model is to be contrasted with the boost factor model of Ref.~\cite{Gao:2011rf}, labeled ``Gao" in Fig.~\ref{fig:systematics_boost} (grey-dashed), which uses a phenomenological power-law extrapolation of the concentration-mass relation to low sub-halo masses.  Because the annihilation rate increases with increasing concentration parameter, the model in Ref.~\cite{Gao:2011rf} predicts substantially larger boosts than other scenarios that take into account a more realistic flattening of the concentration-mass relation at low subhalo masses.\vspace{0.1in}

\begin{figure*}[t]
  \centering
  \includegraphics[width=.45\textwidth]{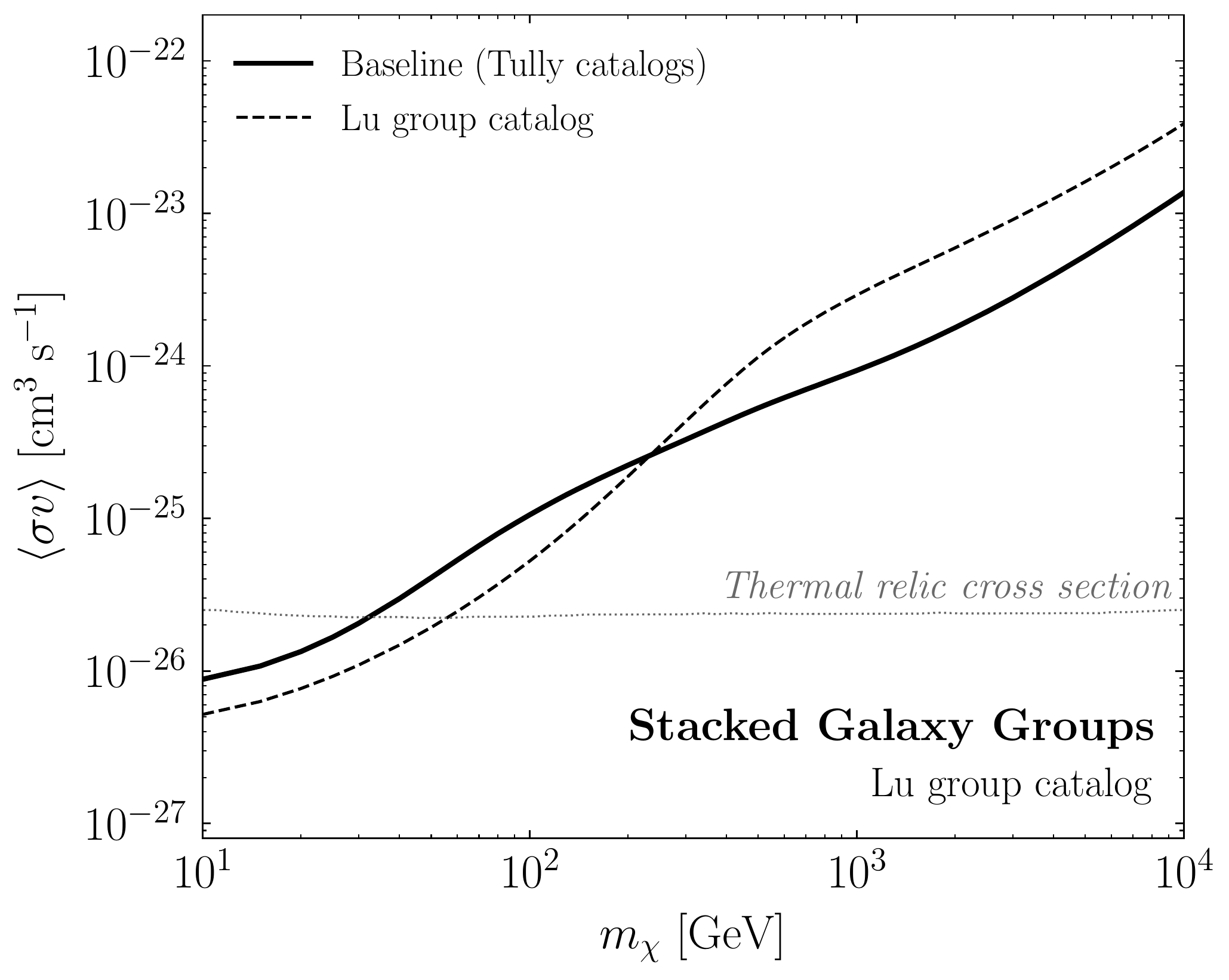}
   \includegraphics[width=.45\textwidth]{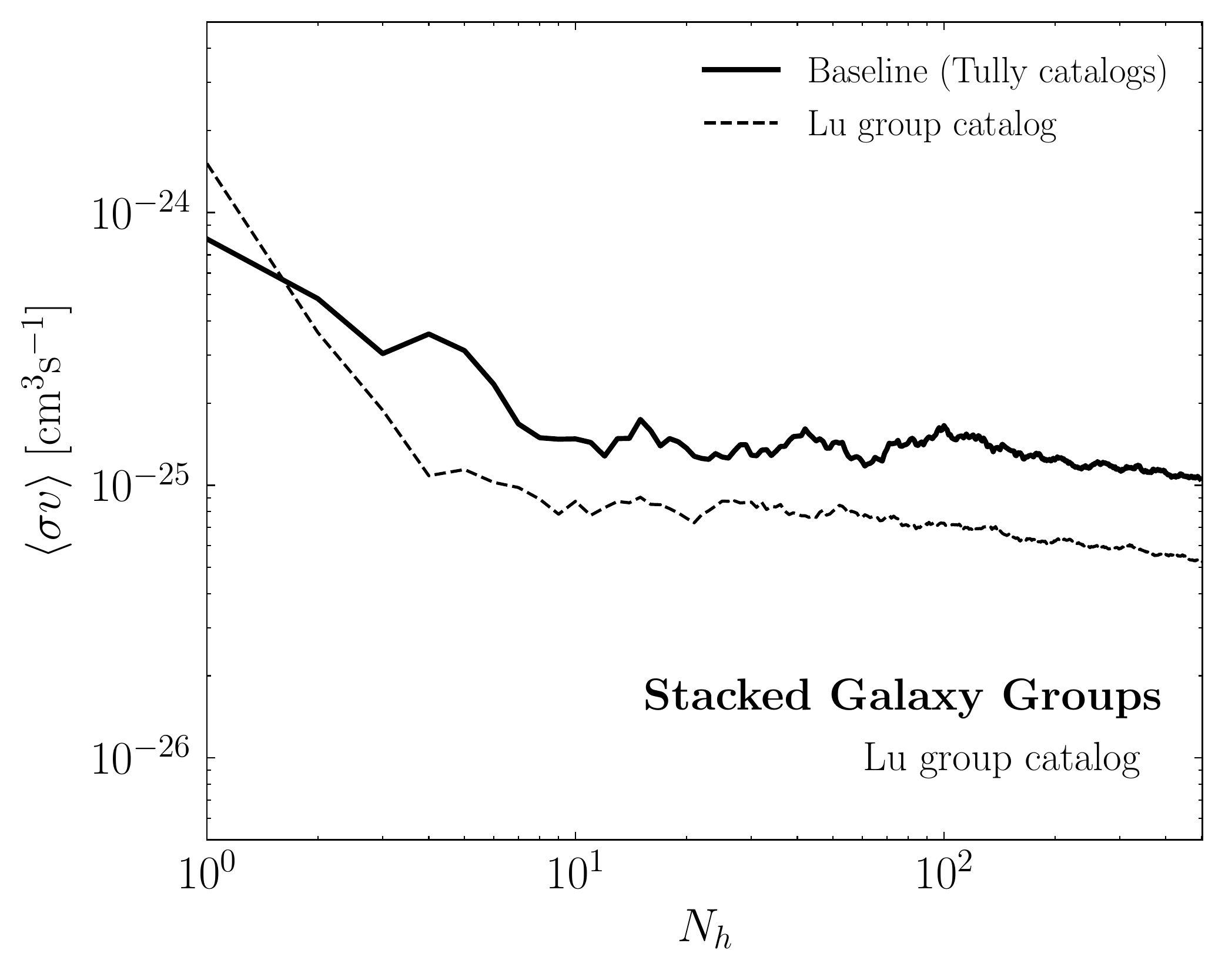} 
  \caption{The same as Fig.~\ref{fig:bounds1} of the main analysis, except using the Lu~\emph{et al.} galaxy group catalog~\cite{Lu:2016vmu} (dashed) instead of the T15 and T17 catalogs in the baseline analysis. }
  \label{fig:lucatalog}
\end{figure*}

\noindent  {\bf Galaxy Group Catalog.}  
We now explore the dependence of the results on the group catalog that is used to select the halos.  In this way, we can better understand how the DM bounds are affected by uncertainties on galaxy clustering algorithms and the inference of the virial mass of the halos.  The baseline limits are based on the T15 and T17 catalogs, but here we repeat the analysis using the Lu~\emph{et al.} catalog~\cite{Lu:2016vmu}, which solely relies on 2MRS observations.  The group-finding algorithm used by Ref.~\cite{Lu:2016vmu} is different to that of T15 and T17 in many ways, relying on a friends-of-friends algorithm as opposed to one based on matching group properties at different scales to $N$-body simulations. Lu~\emph{et al.} also use a different halo mass determination.  For these reasons, it provides a good counterpoint to T15 and T17 for estimating systematic uncertainties associated with the identification of galaxy groups. While T17 includes measured distances for nearby groups, the Lu catalog corrects for the effect of peculiar velocities following the prescription in Ref.~\cite{1996AJ....111..794K} and the effect of Virgo infall as in Ref.~\cite{2014ApJ...782....4K}. Figure~\ref{fig:lucatalog} is a repeat of Fig.~\ref{fig:bounds1} in the main analysis, except using the Lu~\emph{et al.} catalog.  Despite important differences between the group catalogs used, the Lu~\emph{et al.} results are very similar to the baseline case. \\

\noindent There are a variety of sources of systematic uncertainty beyond those described here that  deserve further study.  For example, a systematic bias in the $J$-factor determination due to offsets in either the mass inference or the concentration-mass relation can be a potential source of uncertainty. A better understanding of the galaxy-halo connection and the small-scale structure of halos is required to mitigate this. Furthermore, we assumed distance uncertainties to be subdominant in our analysis. While this is certainly a good assumption over the redshift range of interest---nearby groups have measured distances, while groups further away come with spectroscopic redshift measurements with small expected peculiar velocity contamination---uncertainties on these do exist. We have also assumed that our targets consist of virialized halos and have not accounted for possible out-of-equilibrium effects in modeling these~\cite{1993AJ....105.2035D}.

\sectionline

\singlespacing
\bibliographystyle{utphys}

\cleardoublepage
\ifdefined\phantomsection
  \phantomsection  
\else
\fi
\addcontentsline{toc}{chapter}{Bibliography}

\bibliography{ch-nptfit/NPTF,ch-igrb/fermi_igrb,ch-darksky/fermi_darksky,ch-clusters/fermi_clusters,ch-intro/introduction}

\sectionline

\end{document}